\documentclass[twocolumn]{aastex62}
\usepackage{natbib}
\usepackage{graphicx}
\usepackage{graphics}
\usepackage{amssymb}
\usepackage{amsmath}
\usepackage{mathtools}
\usepackage{multirow}
\usepackage{bm}
\usepackage{braket}
\usepackage[caption=false]{subfig}

\usepackage[utf8]{inputenc}
\usepackage[T1]{fontenc}

\def\dg{\hbox{$^\circ$}}
\def\arcmin{\hbox{$^\prime$}}

\def\utw{\smash{\rlap{\lower5pt\hbox{$\sim$}}}}
\def\udtw{\smash{\rlap{\lower6pt\hbox{$\approx$}}}}

\def\fs{\hbox{$.\!\!^{\rm s}$}}
\def\fdg{\hbox{$.\!\!^\circ$}}

\def\farcs{\hbox{$.\!\!^{\prime\prime}$}}

\def\ii{\textrm{i}}
\AtBeginDocument{\renewcommand{\d}{\textrm{d}}}

\shorttitle{PSDs of gamma-ray emission from blazars}
\shortauthors{Tarnopolski, \.Zywucka, Marchenko \& Pascual-Granado}

\DeclarePairedDelimiter\floor{\lfloor}{\rfloor}
\DeclarePairedDelimiter\ceil{\lceil}{\rceil}

\begin{document}

\title{A comprehensive power spectral density analysis of astronomical time series I:\\the \textit{Fermi}-LAT gamma-ray light curves of selected blazars}


\author[0000-0003-4666-0154]{Mariusz Tarnopolski}
\email{mariusz.tarnopolski@uj.edu.pl}
\affiliation{Astronomical Observatory, Jagiellonian University, Orla 171, 30--244, Krak\'ow, Poland}

\author[0000-0003-2644-6441]{Natalia \.Zywucka}
\email{n.zywucka@oa.uj.edu.pl}
\affiliation{Centre of Space Research, North-West University, Potchefstroom, South Africa}

\author[0000-0002-7175-1923]{Volodymyr Marchenko}
\email{volodymyr.marchenko@oa.uj.edu.pl}
\affiliation{Astronomical Observatory, Jagiellonian University, Orla 171, 30--244, Krak\'ow, Poland}

\author[0000-0003-0139-6951]{Javier Pascual-Granado}
\email{javier@iaa.es}
\affiliation{Instituto de Astrof\'isica de Andaluc\'ia -- CSIC, 18008 Granada, Spain}

\begin{abstract}

We present the results of the \textit{Fermi}-LAT 10-years-long light curves (LCs) modeling of selected blazars: six flat spectrum radio quasars (FSRQs) and five BL Lacertae (BL Lacs), examined in 7-, 10-, and 14-day binning. The LCs and power spectral densities (PSDs) were investigated with various methods: Fourier transform, Lomb-Scargle periodogram (LSP), wavelet scalogram, autoregressive moving average (ARMA) process, continuous-time ARMA (CARMA), Hurst exponent ($H$), and the $\mathcal{A}-\mathcal{T}$ plane. First, with extensive simulations we showed that parametric modeling returns unreliable parameters, with a high dispersion for different realizations of the same stochastic model. Hence any such analysis should be supported with Monte Carlo simulations. For our blazar sample, we find that the power law indices $\beta$ calculated from the Fourier and LSP modeling mostly fall in the range $1\lesssim\beta\lesssim 2$. Using the wavelet scalograms, we confirm a quasi-periodic oscillation (QPO) in PKS~2155$-$304 at a $3\sigma$ significance level, but do not detect any QPOs in other objects. The ARMA fits reached higher orders for 7-day binned LCs and lower orders for 10- and 14-days binned LCs for the majority of blazars, suggesting there might exist a characteristic timescale for the perturbations in the jet and/or accretion disk to die out. ARMA and CARMA modeling revealed breaks in their PSDs at timescales of few hundred days. The estimation of $H$ was performed with several methods. We find that most blazars exhibit $H > 0.5$, indicating long-term memory. Finally, the FSRQ and BL Lac subclasses are clearly separated in the $\mathcal{A}-\mathcal{T}$ plane.

\end{abstract}

\keywords{galaxies: active --- quasars: general --- BL Lacertae objects: general --- galaxies: jets --- gamma rays: galaxies --- methods: statistical --- BL Lacertae objects: individual (TXS~0506+056) --- quasars: individual (PKS~1830$-$211)}

\tableofcontents

\section{Introduction}
\label{sect::introduction}

Blazars are jetted radio-loud objects, which constitute a peculiar class of active galactic nuclei (AGNs). They are originally defined as extragalactic objects which possess a non-thermal continuum along the entire electromagnetic spectrum, exhibit high degree of radio-to-optical polarization, compact radio emission, and pointing their relativistic jets towards an observer \citep[see e.g.,][for a review]{Angel80,Urry95,Pado17}. Blazars show variability in different energy bands, often rapid with shortly lasting flares, from months down to minutes \citep{Wagn95,Rani10,Nils18}. They are commonly divided into two subgroups, i.e. flat spectrum radio quasars (FSRQs) and BL Lacertae (BL Lac) objects, based on characteristics visible in their optical spectra. FSRQs have prominent emission lines with the equivalent width $>$5~\r{A}, while spectra of BL Lacs are featureless (with a smooth continuum) or display weak lines only. This division can be also done based on different accretion regimes of AGNs, i.e. FSRQs have high accretion rates, while BL Lacs have low accretion rates \citep{Ghisellini11}. Moreover, taking into account the position of the synchrotron peak, $\nu_{\mathrm{peak}^s}$, of the spectral energy distribution (SED; in the $\nu - \nu F$ plane), BL Lacs are split into low-peaked (LBL), intermediate-peaked (IBL), and high-peaked (HBL) BL Lacs \citep{Abdo10a}. An additional group of extreme HBL, having $\nu_{\mathrm{peak}^s}\gtrsim 10^{17}\,{\rm Hz}$ is also considered \citep{Costamante01,Akiyama16}. 

Over the years, many multiwavelength as well as different single energy band studies have been conducted to investigate the physical processes causing variability visible in blazars' light curves (LCs), especially their power spectral densities (PSDs). In particular, the timescales at which the PSD changes its character, e.g. breaks, could give an indication of the emission regions, i.e. accretion disk in for instance FSRQs, miscoquasars, or X-ray binary systems, or jet in BL Lacs, and thus of the physical mechanism of observed variations caused in these regions. Several LC investigations were based on the high energy (HE; $E>100\,{\rm MeV}$) $\gamma$-ray \textit{Fermi}-Large Area Telescope (LAT) data. 11-months-long weekly-binned LCs of 104 blazars, including 58 FSRQs, 42 BL Lacs, and four objects with uncertain classification, were analyzed by \citet{Abdo10d}. Their PSDs were obtained with the structure function (SF) and fitted with a power law (PL), $1/f^\beta$, giving an index $1.1\lesssim\beta\lesssim 1.6$ for a majority of blazars. It was also found that the FSRQs, LBLs, and IBLs show larger variations than HBLs, wherein all studied blazars can be described as steady sources with a series of flares. \citet{sobolewska14} studied 4-years-long, adaptively-binned, densely-sampled \textit{Fermi}-LAT LCs of 13 well known and bright blazars, i.e. eight FSRQs and five BL Lacs. The LCs were modeled with the Ornstein–Uhlenbeck (OU) process, and a superposition of OU processes (sup-OU). It turned out that the majority (i.e., 10) of investigated objects are better described by the sup-OU model, with the PL index $\beta\sim$1. It is worth mentioning that the index does not depend on the redshift or blazar classification. Recently, \citet{ryan19} analyzed long-term (9.5 years) daily- and weekly-binned \textit{Fermi}-LAT LCs (taken from the Fermi Science Support Center\footnote{\url{http://fermi.gsfc.nasa.gov/ssc/data/access/lat/msl\_lc/}}) of 13 blazars previously studied by \citet{sobolewska14}. The LCs were analyzed with the continuous-time autoregressive moving average (CARMA) model \citep{kelly14}, indicating that the OU processes, i.e. CARMA(1,0), do not pick up the characteristics of variability in the data, while higher-order processes, i.e. CARMA(2,1), provide a much better description of the variability. The multiwavelength analysis of the famous source OJ~287 yielded CARMA$(1,0)$ as the best fit for the HE data, while optical and radio LCs (with much more measurements) exhibited higher-order CARMA models \citep{goyal18}.

Regarding the PSD study, quasi-periodic oscillations (QPOs) in AGN, where a QPO is defined as \textit{''concentration of variability power over a limited frequency range''} \citep{Vaug05b}, may help to establish the variability regions and physical processes responsible for variability as well as to place constraints on the black hole (BH) mass estimations as it is in the case of microquasars, neutron stars, and X-ray sources \citep[e.g.][]{Stel98,Muno99,Abra04,Toro05}. Most of the analyses aimed at searching for QPOs focus around different implementations of periodograms and scalograms. \citet{Espa08} analyzed X-ray XMM-Newton LCs of 10 bright AGNs. The observations were conducted in the energy range $0.75 - 10$~keV, while the analysis was performed with the continuous wavelet transform and the SF. A QPO was found for only one AGN, i.e. 3C~273, with a period of 55 minutes at the significance level $>3\sigma$, and a range of BH mass was estimated as $[7.3\times10^{6}$, $8.1\times10^{7}]\,M_{\odot}$. \citet{lachowicz09} looked for QPOs in the $0.3-10$~keV band observations of PKS 2155$-$304 by analyzing a 64~ks XMM-Newton LC with the multi-harmonic analysis of variance periodogram \citep{Schw96}, the SF, and the wavelet scalogram \citep{torrence98}. A QPO with a period of $\sim 4.6$~h was found in the data at $>3\sigma$ significance level, which allowed authors to estimate the BH mass to be within the range $[3.3\times10^{7}$, $2.1\times10^{8}]\,M_{\odot}$. \citet{Gupt09} collected a set of 20 V and R filter optical LCs of the blazar PKS~0716+714, lasting from 7.7~h up to 12.3~h. Each LC was studied separately with a wavelet plus randomization technique, giving a set of possible QPO timescales and corresponding BH mass estimates (see Table 1 in \citealt{Gupt09} for details). Subsequently, \citet{Rani10a} studied a 9.6-hour-long R-band optical LC of PKS~0716+714, delivered by the Physical Research Laboratory in India, using the SF analysis, Lomb-Scargle periodogram (LSP), as well as fitting a pure PL to the derived PSD. The authors found a QPO with a period of $\sim 3$~minutes with significance $>3\sigma$. The BH mass was estimated to be within $[1.5\times10^{6}$, $9.6\times10^{6}]\,M_{\odot}$. In the four examples of a QPO search in X-ray data presented above, the lower value of the BH mass estimation is based on a non-rotating BH model, while the higher value stems from the assumption of a maximally rotating BH. It was also assumed that a QPO is connected to fluctuations in the accretion disk and its time scale corresponds to an orbital time scale originating near the last stable orbit, i.e. the most inner parts of the accretion disk. On the other hand, the BH mass estimates of 3C~273 given by \citet{Espa08} do not agree with previous values obtained with the reverberation-mapping method or the correlation between host galaxy luminosity and BH mass. This suggests that the QPO found in X-ray data is rather not originating in the inner parts of the accretion disk as it can be explained in the case of microquasars, neutron stars, and X-ray sources. 

Regarding a search for QPOs in AGNs, especially blazars, in the HE $\gamma$-ray regime, \textit{Fermi}-LAT provides long-term, high quality, and densely sampled data of many bright and well known sources. For example, \citet{Acke15} analyzed the $\sim$6.9-year-long LC of blazar PG~1553+113 in the energy range $0.1-300$~GeV, with the LSP and the continuous wavelet transform. They found a possible QPO with a period of $798\pm 30$~days, where the $\gamma$-ray signal peaked at $>99\%$ confidence level. The authors stressed that the presence of a binary supermassive BH (SMBH) could initiate the observed long-term quasi-oscillations. \citet{prokhorov17} investigated 10 $\gamma$-ray sources, including three binary systems and seven blazars. The blazars' LCs, spanning over 7.8 years in the energy range $0.3-500$~GeV, were analyzed with the generalized LSP \citep{Zech09}. QPOs were found in three cases, i.e. PG~1553+113, PKS~2155$-$304, and BL~Lacertae, with periods of 798, 644, and 698 days, respectively; the QPOs reached the significance level $>5\sigma$. The remaining blazars from this sample are considered as candidates for blazars with QPO signals in their $\gamma$-ray emission as well as blazars having binary SMBH at their central regions. \citet{zhou18} analyzed 10-years-long data in the energy range $0.1-300$~GeV, using the weighted wavelet Z-transform (WWZ) and LSP. A QPO in PKS~2247$-$131 with a period of $\sim$~34 days with a significance of $\sim 4.6\sigma$ was found, supporting the suggestion that the flux quasi-oscillation is caused by the helical structure of the jet. \citet{tavani18} reanalyzed the same data of PG~1553+113 as \citet{Acke15}, extending the LC by two subsequent years. Using the continuous wavelet transform the authors confirmed the QPO and its period found by the previous study of this blazar and, again, interpreting the quasi-oscillation in the signal to be a result of the binary SMBH system.   

A bunch of possible interpretations of QPOs appearing in the blazar multiwavelength data was introduced and discussed in the recent literature. The short-term, monthly, periodic modulation may arise directly in the jet due to its helical structure, intrinsic rotation, or precession \citep[e.g.][]{Rieg04,Vlah04,Capr13} as well as by processes causing perturbations and instabilities in the accretion disk-jet system \citep[e.g.][]{Rome00,Piha13}. Another widely discussed possible origin of a QPO is the presence of a gravitationally bound binary SMBH at the center of a blazar, which might cause instabilities in the jet \citep[e.g.][]{Katz97,Valt08,Cava17,holgado18}. However, the origin of radio, optical, X-ray, and $\gamma$-ray QPOs is still uncertain. In particular, QPOs might as well arise from intrinsically aperiodic stochastic processes without a connection to any physical periodic behavior underlying the observed variability. QPOs naturally appear in many stochastic autoregressive (AR) models, starting with AR(2) as the simplest such model that allows a Lorentzian-like peak in its PSD. Therefore, an apparent quasiperiodicity need not be a real physical periodicity, but form due to particular autocorrelations present in the governing process.

We aim to conduct a comprehensive analysis of the temporal properties of blazar LCs, in particular constraining the shape and features of the PSD, by employing a wide range of methods, and performing extensive testing of their reliability. We start, in Sect.~\ref{sect::data}, by describing the preparation of data, in particular the processing of {\it Fermi}-LAT data for producing the LCs (Sect.~\ref{sect::2.1}), followed by justifying the nonlinear model that allows to transform flux values to amplitudes of the underlying stochastic process governing the variability (Sect.~\ref{sect::rms}), and outlining the algorithm for interpolating missing data (Sect.~\ref{sect::interpolation}). The overall purpose of this paper is threefold: first, we aim (in Sect.~\ref{methods}) to provide a possibly exhaustive, yet compact, mathematical description of the various time series analysis techniques employed further on. Sect.~\ref{methods} can be considered as a tutorial, explaining what the methods can and cannot do. Second, we validate them in Sect.~\ref{testing} by extensive benchmark testing. The design was to ascertain the reliability of outcomes returned by each method. We describe the levels to which they provide reliable results, and highlight, in particular, that owing to genuine randomness of stochastic processes, different realizations of the same process can lead to remarkably different results. Finally, in Sect.~\ref{results} we apply all these methods to $\gamma$-ray LCs of some of the brightest or otherwise famous blazars, bearing in mind the limitations uncovered in Sect.~\ref{testing}. Most of the chosen blazars have been intensively examined in the literature, what allows for a detailed comparison with our results. We also investigate some objects that recently gained more attention (TXS~0506+056, PKS~1830$-$211) in an attempt to provide insight into their driving mechanisms. In Sect.~\ref{summary_of_results} we discuss the results, comparing them with previous findings from the literature (Sect.~\ref{comparison}), putting them in context (Sect.~\ref{interpretation}), and listing several applications for the employed methodology (Sect.~\ref{applications}). This work can therefore be considered either from the point of view of a critical evaluation of some computational techniques (Sect.~\ref{methods} and \ref{testing}), or purely from the point of astrophysical research (Sect.~\ref{sect::data}, \ref{results}, \ref{discussion}). A brief summary of our findings is provided in Sect.~\ref{conclusions}. As a whole, it forms a comprehensive overview of variable astrophysical sources, blazars in particular.

\section{Data}
\label{sect::data}

\subsection{Processing of Fermi data}
\label{sect::2.1}

The \textit{Fermi}-LAT \citep{Atwo09} is a HE $\gamma$-ray telescope, sensitive to photons in the energy range from 20 MeV to 300 GeV and with a wide field of view of 2.4 sr. It allows to resolve individual sources, such as point-like AGNs, to detect transients, and to monitor variability.

We performed a standard binned maximum likelihood analysis\footnote{\url{https://fermi.gsfc.nasa.gov/ssc/data/analysis/scitools/binned\_likelihood\_tutorial.html}} using the latest 1.2.1 version of \textsc{Fermitools}, i.e. conda distribution of the Fermi ScienceTools\footnote{\url{https://github.com/fermi-lat/Fermitools-conda/wiki}}, and the \textsc{fermipy} \citep{Wood17} facilities. The binned likelihood analysis is recommended to analyze long time periods and large bins of bright known sources. It allows us to balance the accuracy of analyses with their execution time and computational power. All data analyzed here comes from the \textit{LAT 8-year Source Catalog} \citep[4FGL; ][]{Fermi19}, spanning the time range of 239557417---577328234 MET, i.e. 2008 August 04 to 2019 April 19, giving in total $\sim$11 years of data. In the binned analysis, we defined the spatial bin size to be 0\fdg1, and the number of energy bins per decade of 8. We considered the 128 event class, i.e. events having high probability of being photons, from the region of interest (ROI) of 10$^{\circ}$, centered on the selected blazars, and the event type equal to 3. We used the energy range between 100~MeV and 300~GeV. The high energy level was introduced to avoid the energy reconstruction oversaturation, while analyzing photons $<100$~MeV a point spread function gets large and the event reconstruction is not accurate enough \citep[see, e.g. ][for more details]{Atwo07,Prin18}. The zenith angle was set up to 90$^{\circ}$ to avoid the Earth's limb, which is a strong $\gamma$-ray background source. Additionally, we made sure that the data quality is good enough and the time intervals are proper by choosing \texttt{DATA\_QUAL$>$0} and \texttt{LAT\_CONFIG==1} options. Finally, we used the instrument response function \texttt{P8R3\_SOURCE\_V2} and the latest Pass~8 background models, i.e. the Galactic diffuse emission (\texttt{gll\_iem\_v07.fits}) and the extragalactic isotropic diffuse emission (\texttt{iso\_P8R3\_SOURCE\_V2\_v1.txt}) models, including also all known point-like foreground/background sources in the ROI. 

\begin{figure*}
\centering
\includegraphics[width=\textwidth]{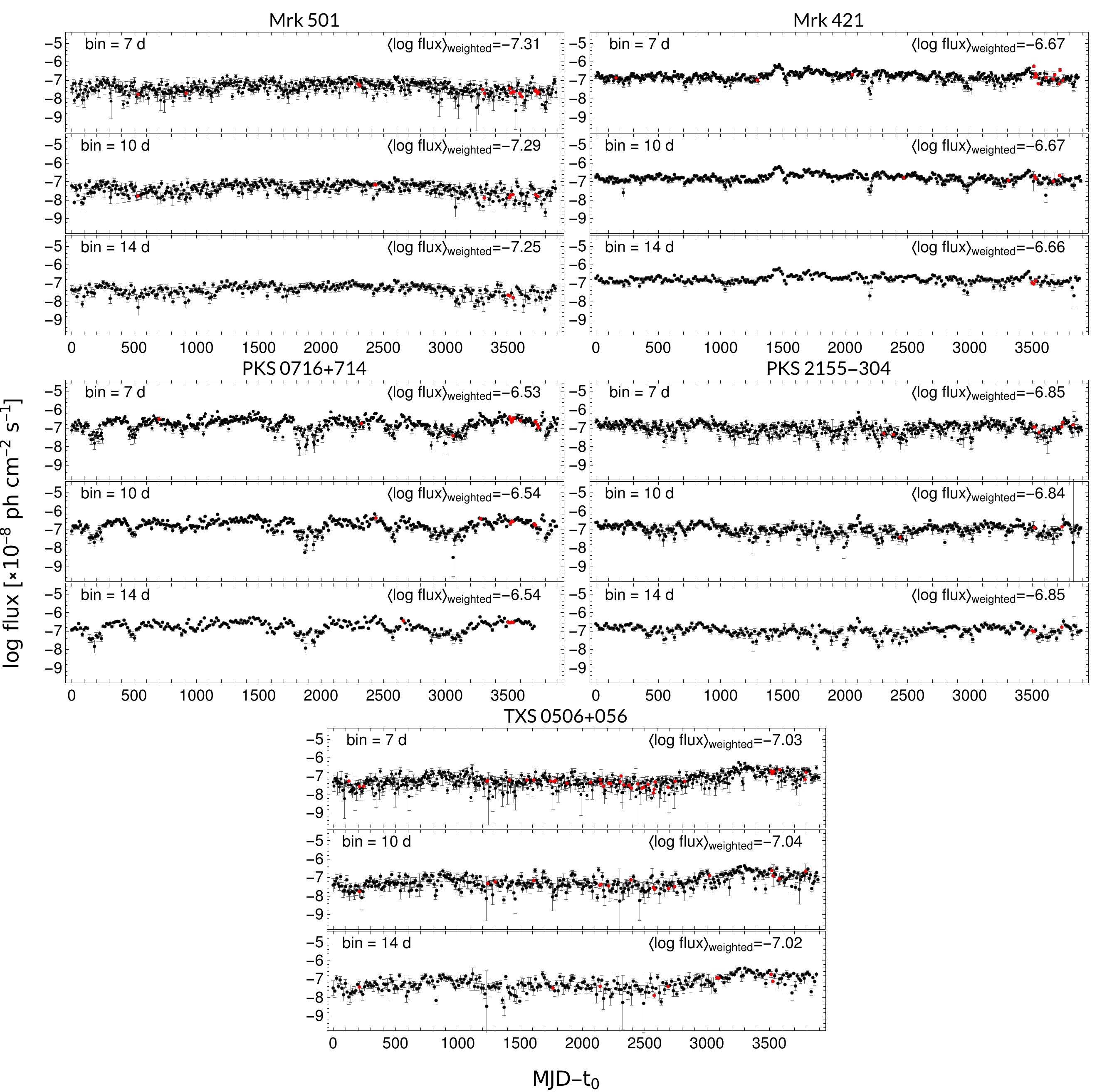}
\caption{Logarithmic LCs of BL Lacs. The red points are the interpolations done with MIARMA. The uncertainties of the weighted means are $\leqslant 0.01$. }
\label{fig_LCs_BLLAC}
\end{figure*}

\begin{figure*}
\centering
\includegraphics[width=\textwidth]{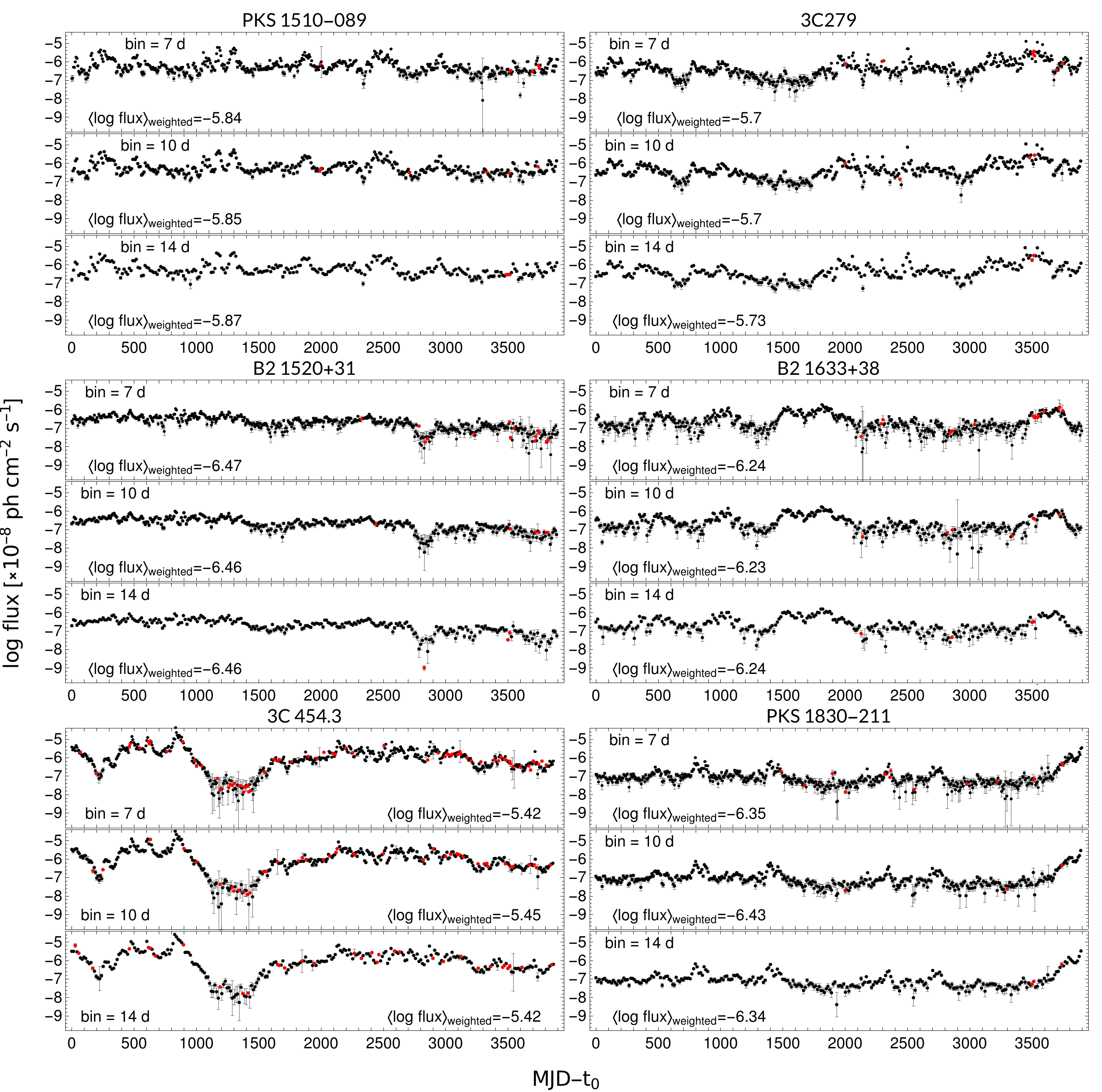}
\caption{Same as Fig.~\ref{fig_LCs_BLLAC}, but for FSRQs. }
\label{fig_LCs_FSRQ}
\end{figure*}

We generated a set of three LCs for each blazar (see Sect.~\ref{sect::2.4} for a description of the objects), using three time bins, i.e. 7, 10, and 14~days, and including observations with the test statistic $TS>25$ (significance of $\gtrsim 5\sigma$). They are displayed in Fig.~\ref{fig_LCs_BLLAC} and \ref{fig_LCs_FSRQ}.


\subsection{Nonlinear model}
\label{sect::rms}

\begin{figure*}
\centering
\includegraphics[width=0.49\textwidth]{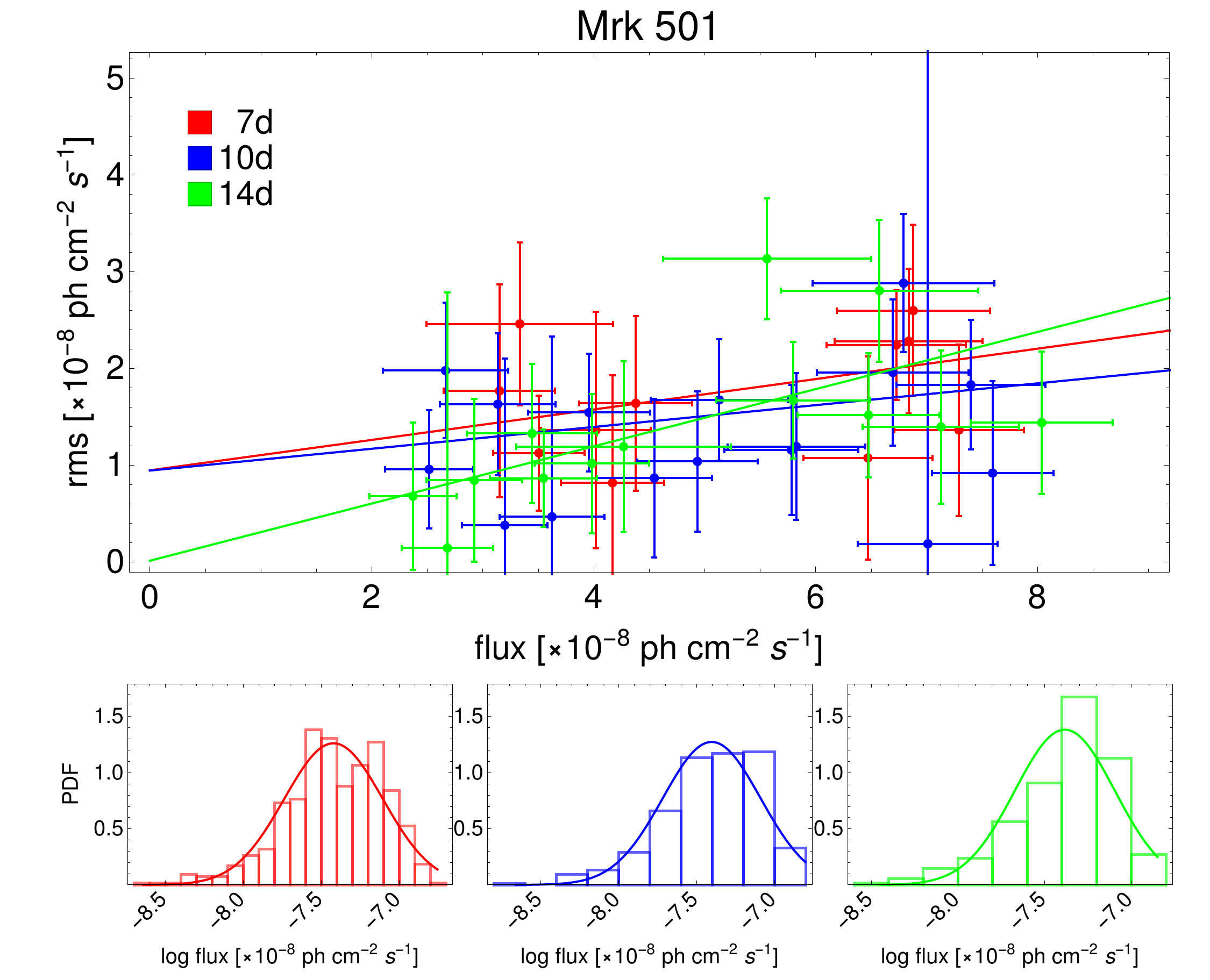}
\includegraphics[width=0.49\textwidth]{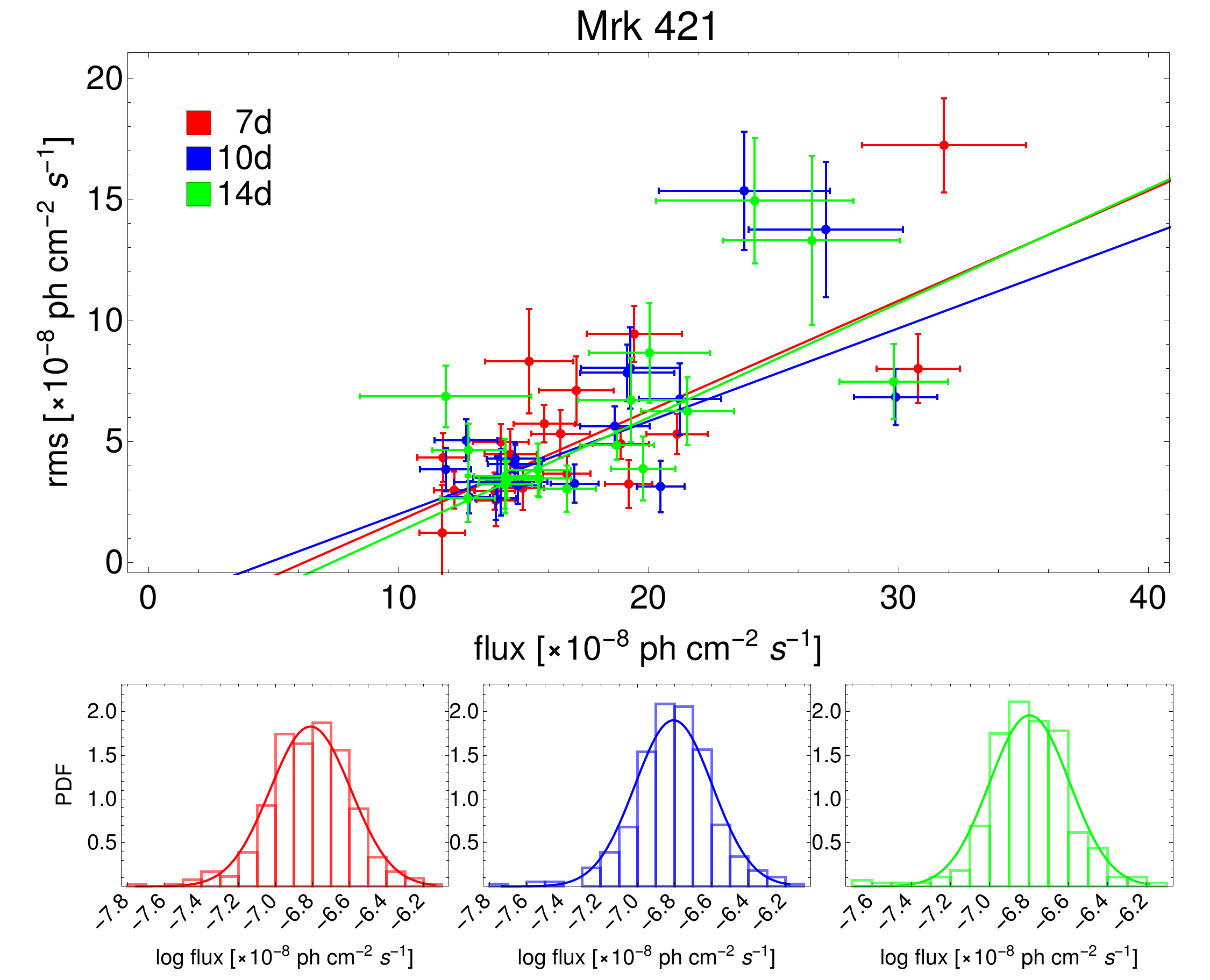}\\
\includegraphics[width=0.49\textwidth]{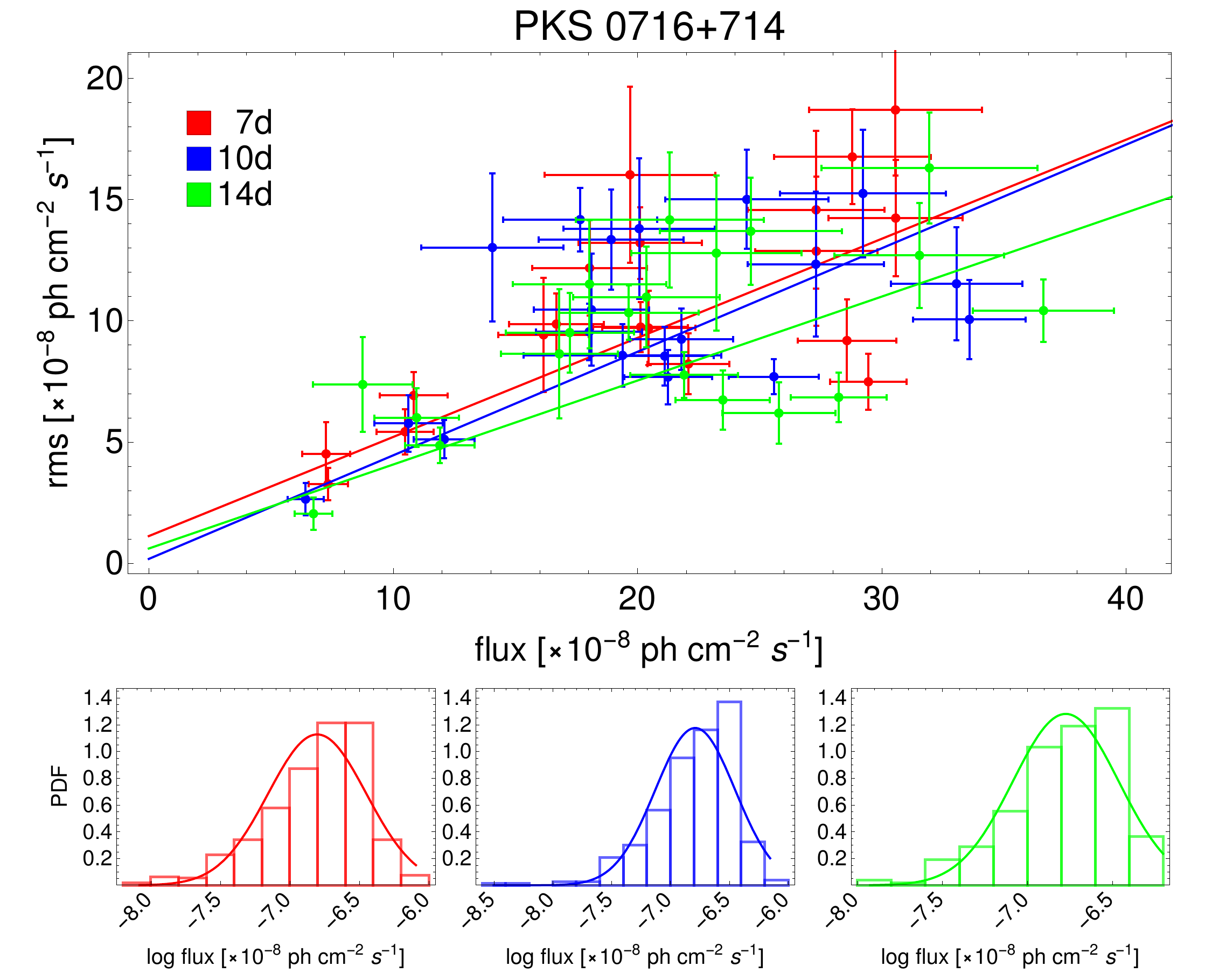}
\includegraphics[width=0.49\textwidth]{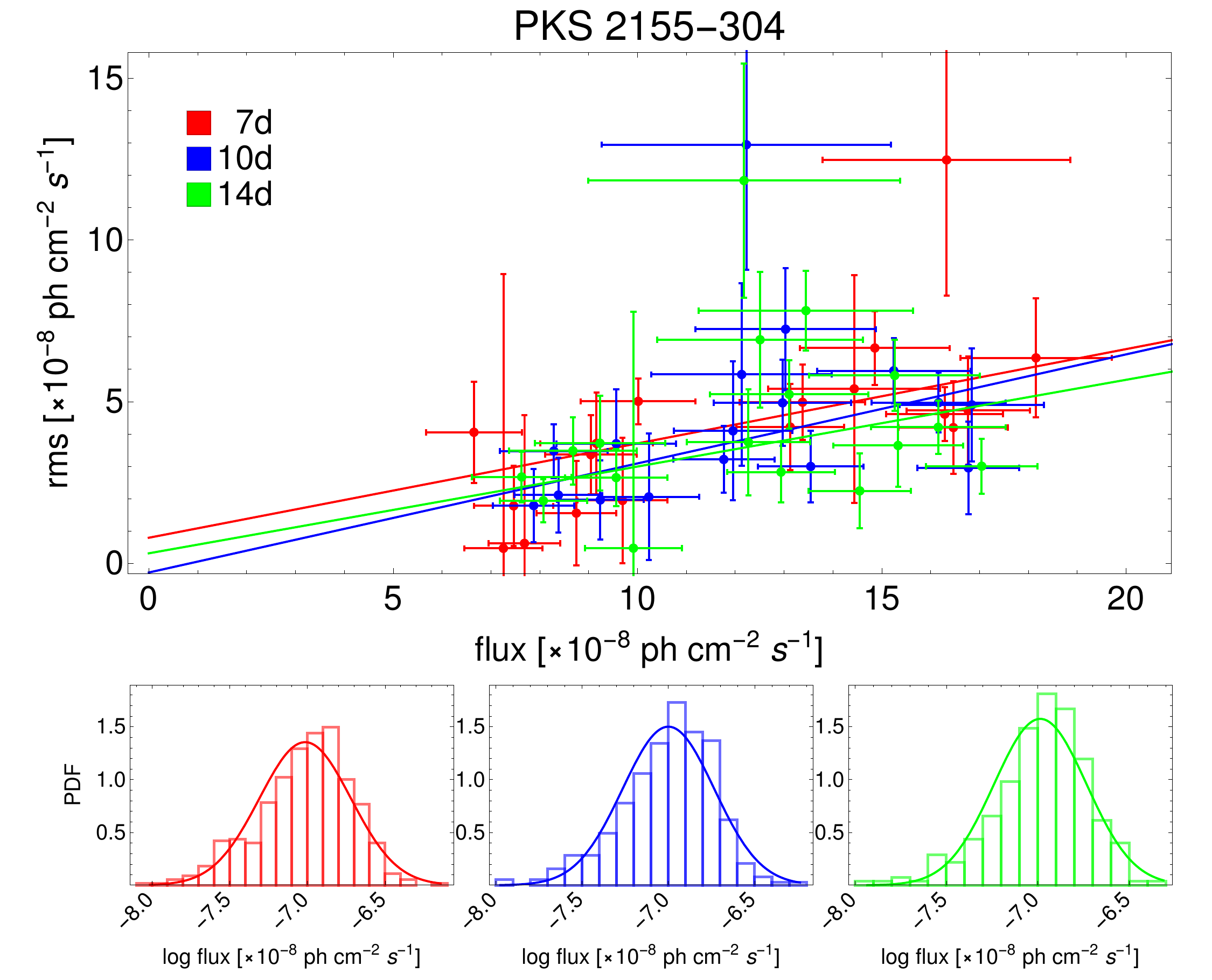}\\
\includegraphics[width=0.49\textwidth]{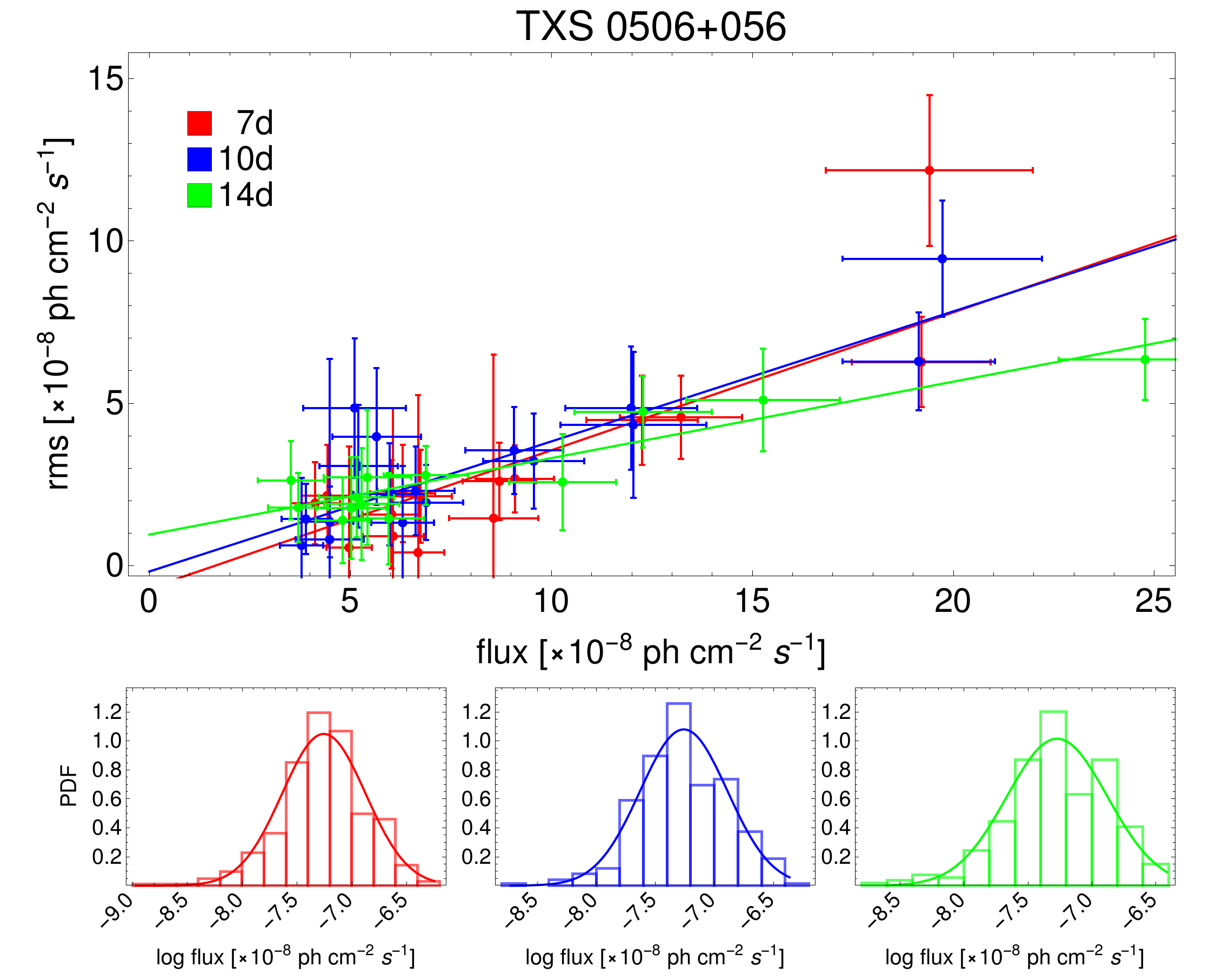}
\caption{Rms-flux relations and logarithmic distribution of fluxes for BL Lacs. }
\label{fig_RMS_BLLAC}
\end{figure*}

\begin{figure*}
\centering
\includegraphics[width=0.49\textwidth]{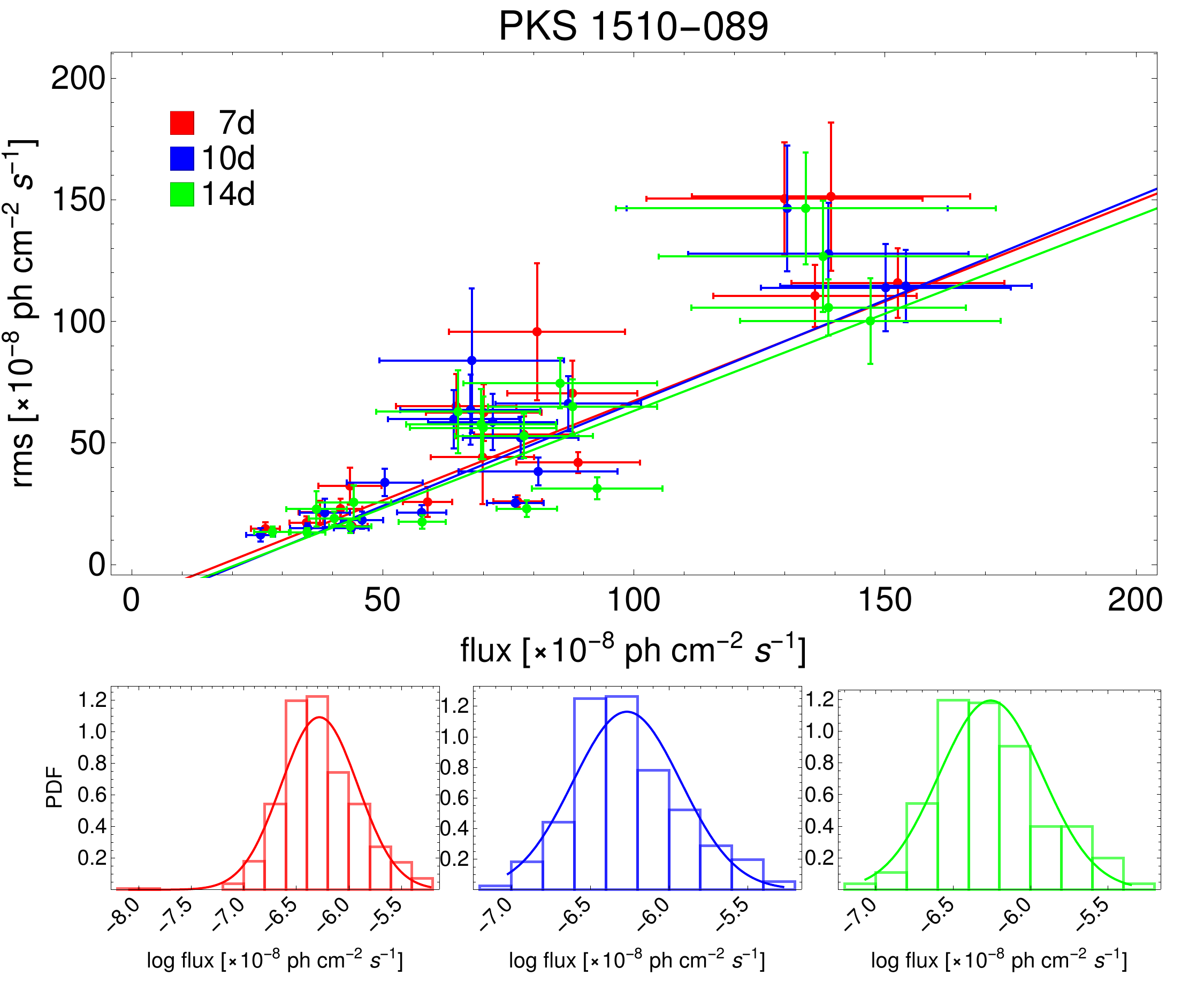}
\includegraphics[width=0.49\textwidth]{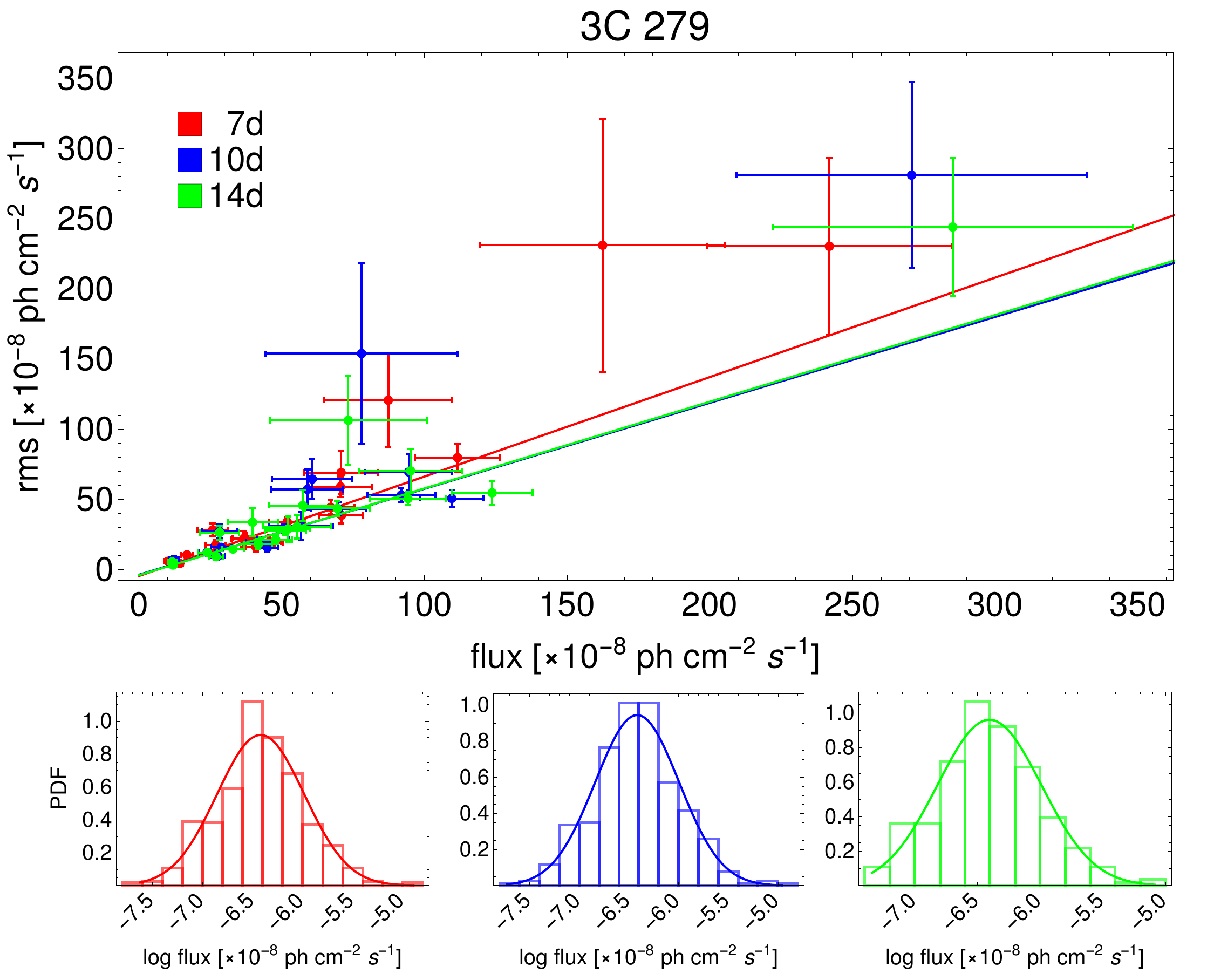}\\
\includegraphics[width=0.49\textwidth]{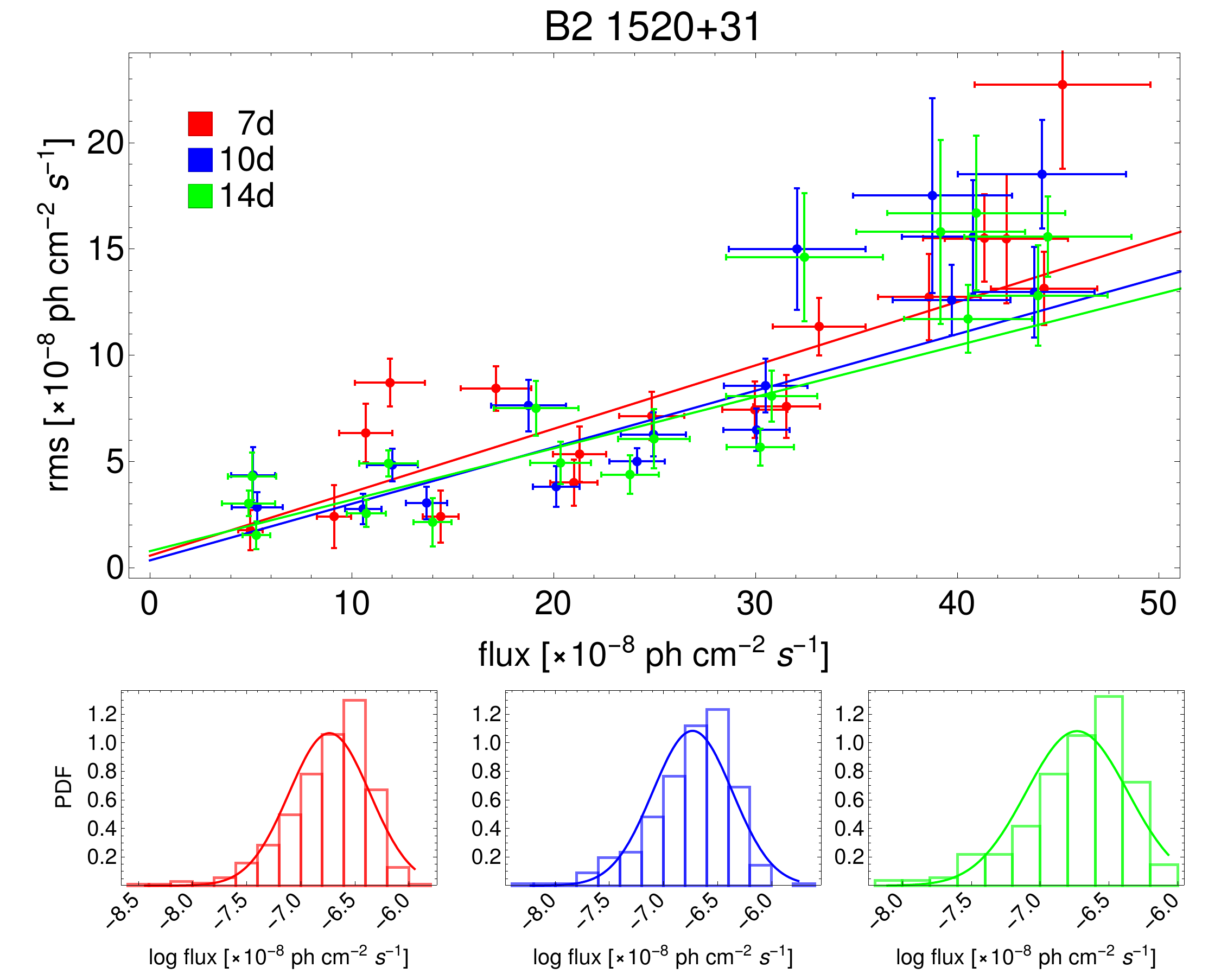}
\includegraphics[width=0.49\textwidth]{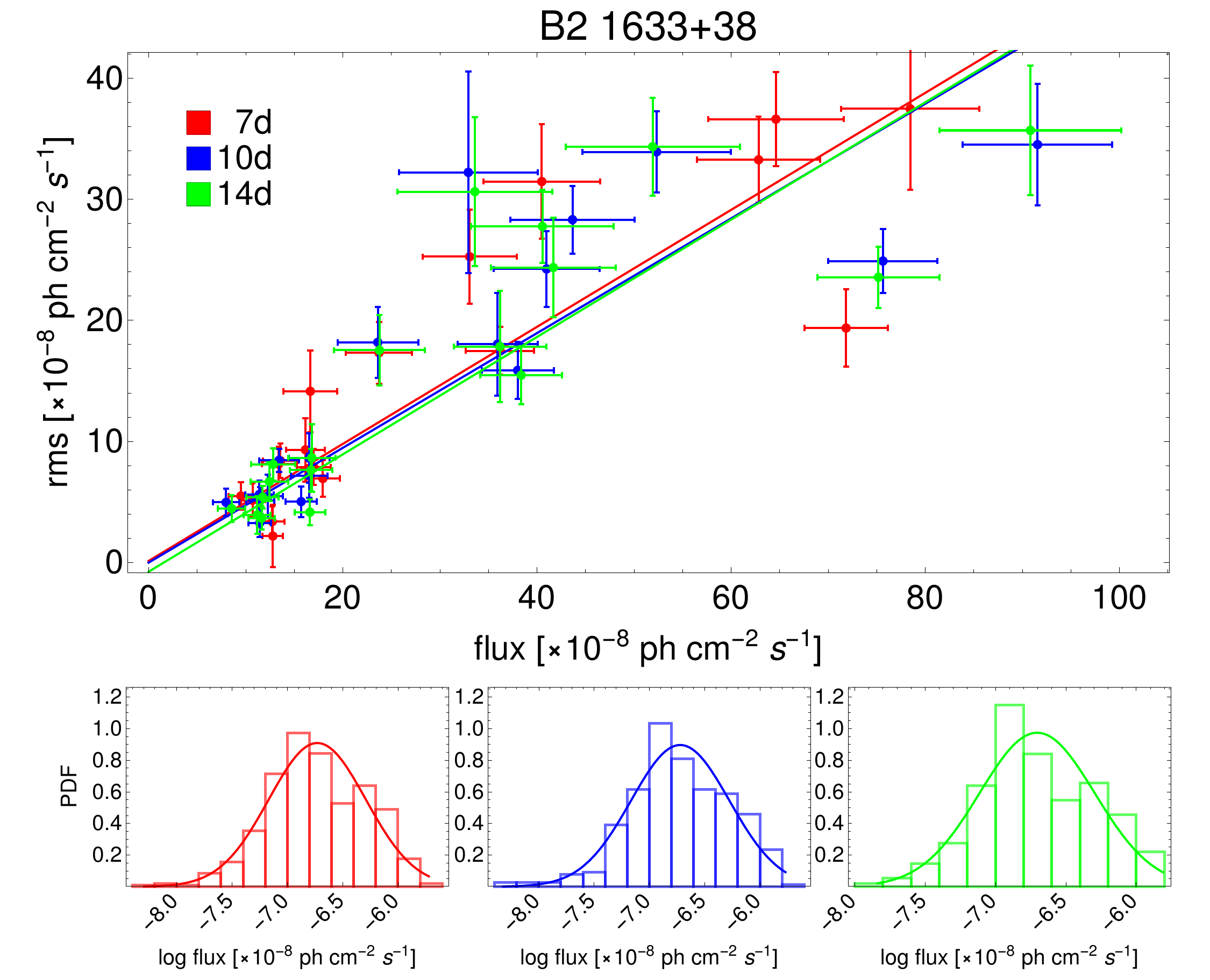}\\
\includegraphics[width=0.49\textwidth]{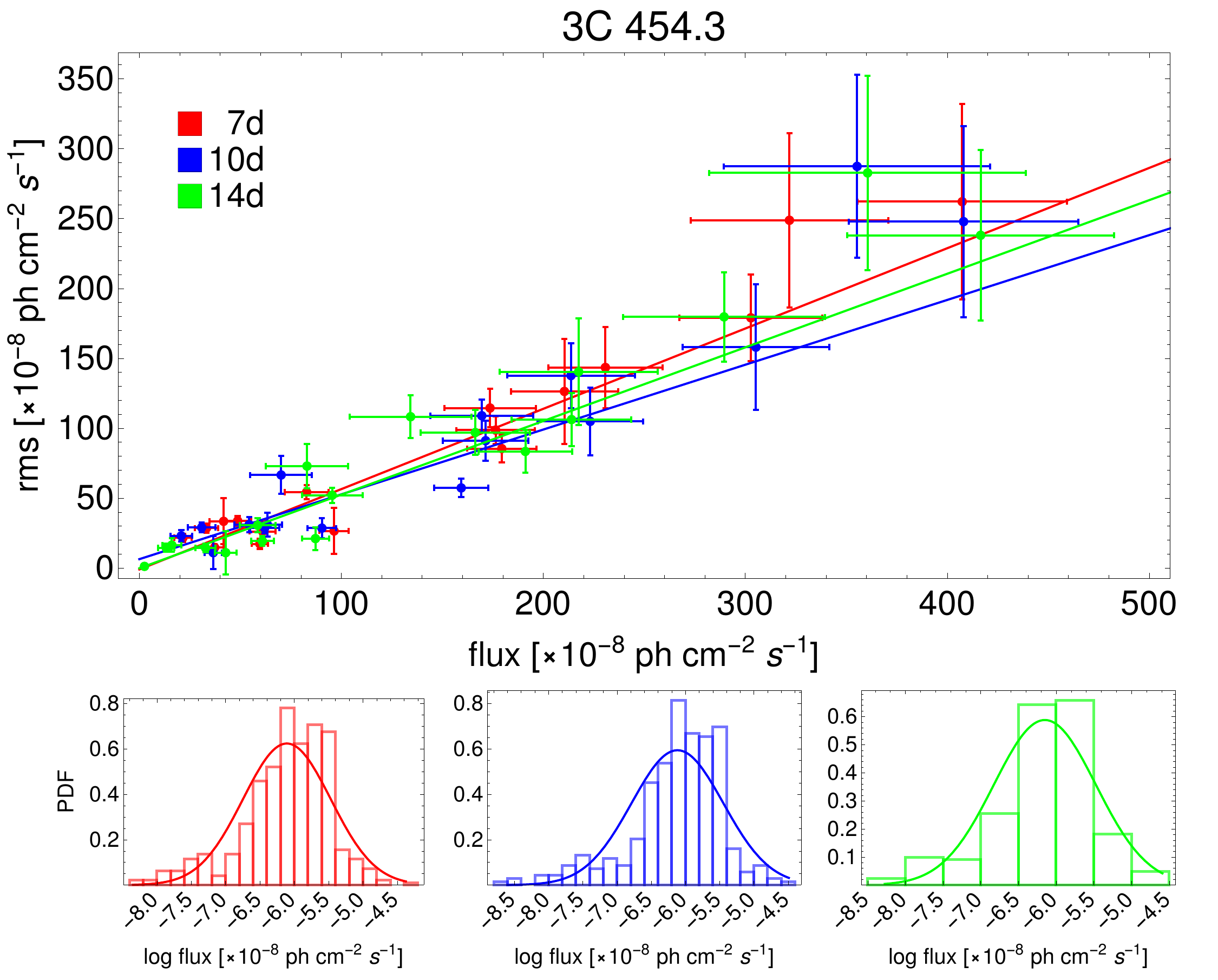}
\includegraphics[width=0.49\textwidth]{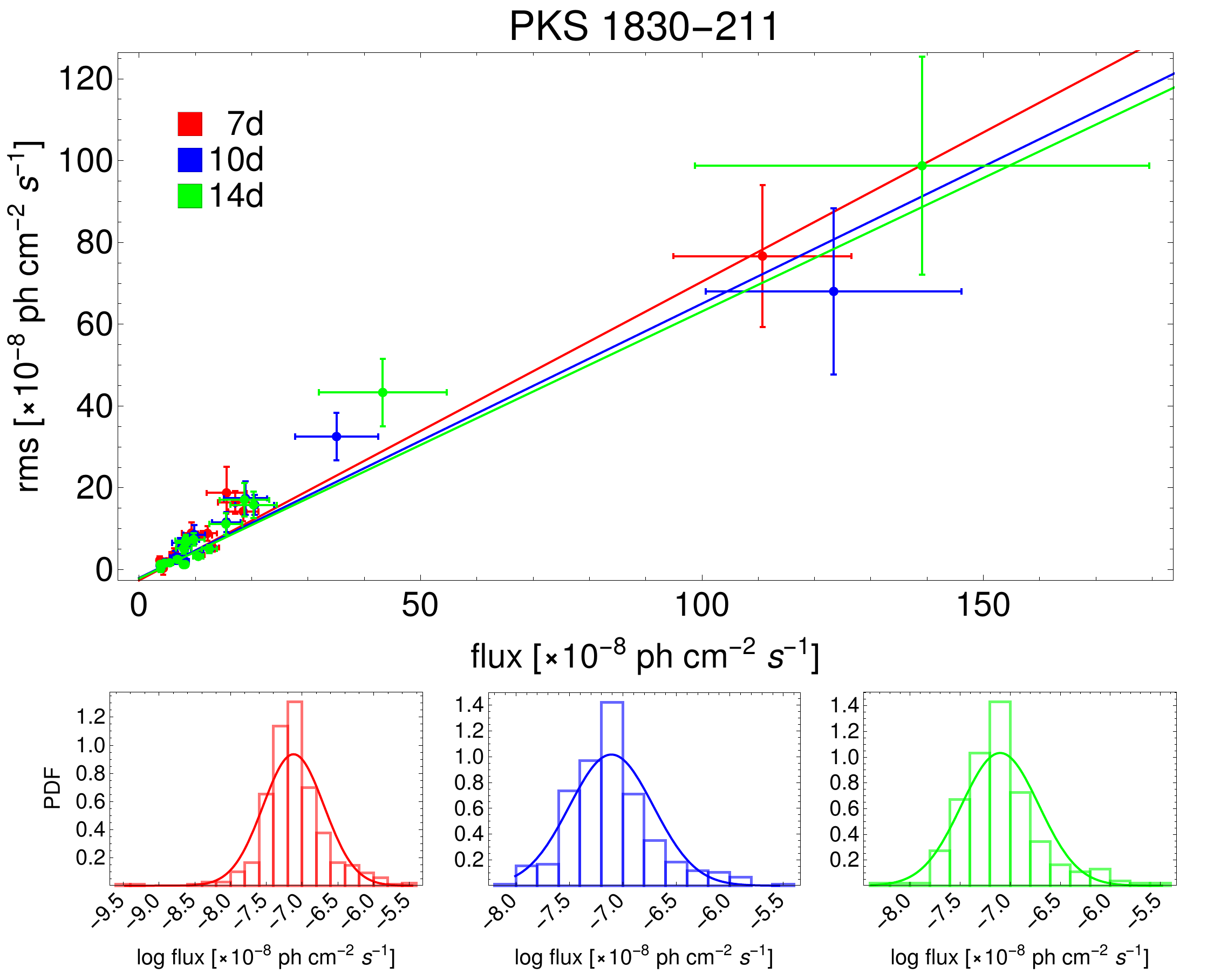}
\caption{Same as Fig.~\ref{fig_RMS_BLLAC}, but for FSRQs. }
\label{fig_RMS_FSRQ}
\end{figure*}

Blazars often exhibit flares during which their brightness increases by several orders of magnitude. Such occurrences make the observed LCs highly nonstationary, making their analysis difficult. The most straightforward method of converting a nonstationary time series to a stationary one is to differentiate the series, i.e. to investigate the consecutive differences of the original data. Such an approach is natural for linear processes (Sect.~\ref{sect::farima}). However, taking the logarithms of the fluxes is another way, and an astrophysically motivated one. Indeed, one can write
\begin{equation}
f(t) = \exp\left[ l(t) \right],
\label{eq1}
\end{equation}
where $f(t)$ is the LC, and $l(t)$ is some stochastic process underlying the observed variability. Eq.~\eqref{eq1} carries meaning when \citep{uttley05}:
\begin{enumerate}
\item the distribution of the observed fluxes is lognormal, and
\item the rms-flux relation is linear.
\end{enumerate}
The root-mean-square (rms) is the amplitude of the variability, i.e. it is the standard deviation of the time series' values. To examine how does the rms depend on the flux, one proceeds as follows:
\begin{enumerate}
\item Divide the LC into $k$ non-overlapping segments, each consisting of $N_j$ points, $j=1,\ldots,k$.
\item Compute the mean, $\bar{f}_j$, of fluxes $\{ f_i \}_j$ in the $j$th segment.
\item In each segment, compute the rms. The variance in the LC is a sum of the true variability and the uncertainties of the flux measurements. Therefore, one should consider the excess variance \citep{vaughan03}:
\begin{equation}
\sigma_{xs,j}^2=S_j^2-\overline{\sigma_{\rm err}^2}_{,j},
\label{eq2}
\end{equation}
where
\begin{equation}
S_j^2 = \frac{1}{N_j-1}\sum\limits_{i=1}^{N_j} \left( f_i - \bar{f}_j \right)^2
\label{eq3}
\end{equation}
is the variance in the $j$th segment, and
\begin{equation}
\overline{\sigma_{\rm err}^2}_{,j} = \frac{1}{N_j} \sum\limits_{i=1}^{N_j} \Delta f_i^2
\label{eq4}
\end{equation}
is the mean square error. Thence, the rms is $\sigma_{xs,j}$.
\item The standard error of $\bar{f}_j$ is the standard error of the mean, i.e. $\Delta\bar{f}_j = \sigma_j/\sqrt{N_j}$, where $\sigma_j$ is the standard deviation of the fluxes.
\item The error of the rms is computed herein via the law of error propagation:
\begin{equation}
\Delta\sigma_{xs,j} = \frac{1}{2\sigma_{xs,j}} \sqrt{\Delta S_j^2 + \Delta\overline{\sigma_{\rm err}^2}_{,j}},
\label{eq5}
\end{equation}
for which one needs to know the errors of $S_j^2$ and $\overline{\sigma_{\rm err}^2}_{,j}$.
\item The error of $S_j^2$ is given as \citep{rao73}
\begin{equation}
\Delta S_j^2 = \sqrt{\frac{1}{N_j}\left( \mu_{4,j} - \frac{N_j-3}{N_j-1} S_j^4 \right)},
\label{eq6}
\end{equation}
where $\mu_{4,j}$ is the fourth central moment. 
\item Because $\overline{\sigma_{\rm err}^2}_j$ is the mean of squares, therefore its error, $\Delta\overline{\sigma_{\rm err}^2}_{,j}$, will be the standard error of the mean (calculated as in point 4.)
\item One finally obtains $k$ pairs $\left( \bar{f}_j, \sigma_{xs,j} \right)$, each equipped with uncertainties $\left( \Delta\bar{f}_j, \Delta\sigma_{xs,j} \right)$.
\end{enumerate}
In Fig.~\ref{fig_RMS_BLLAC} and \ref{fig_RMS_FSRQ} the rms-flux relations are shown for all objects considered herein. Each object is analyzed in three binnings, indicated by the colors. The LCs were divided into $k\sim 20$ segments. Fitting a straight line when each point possesses uncertainties in both $x$ and $y$ directions is performed via weighted orthogonal regression\footnote{\url{https://mathematica.stackexchange.com/a/13122/22013}} \citep{york66}. In all instances, the slopes were positive when taking into account their errors---even for the dimmest among the considered sources, i.e. Mrk~501, for which the linearity of the rms-flux relation is ambiguous. Displayed are also the probability density functions (PDFs) of the logarithmic fluxes, with best-fitting Gaussians overlayed. We emphasize that the distribution fitting was performed by maximizing the loglikelihood over the data, not by a regression to the histograms, which are displayed for illustration only. The results for Mrk~421, PKS~1510$-$089, and B2~1520+31 are consistent with \citet{Kushwaha17}. Overall, we conclude that utilizing Eq.~(\ref{eq1}) is justified. Therefore, hereinafter we investigate the underlying process $l(t)$ by analyzing the (decimal) logarithms of the fluxes. Note, however, that steep PSDs ($\beta\gtrsim 1$, i.e. nonstationary processes) might lead to flux distributions being neither normal nor lognormal \citep{Morris19}.

\subsection{Interpolation}
\label{sect::interpolation}

The fraction of missing data varies from one LC to another, ranging in our sample from 0.4\% up to 13\%. Interpolation of gaps provides homogenization of the sample. Furthermore, gaps might potentially hamper any variability study due to the resulting irregularly sampled time series. For instance, the spectral window function associated to observation times introduces spurious peaks in the PSD \citep{deeming75}. Also, gaps make the calculation of autocorrelation or cross-correlation functions challenging, since these functions are only properly defined for evenly sampled time series.

\citet{granado18} showed that the frequency content extracted from PSD estimation might be biased when simple interpolation, like polynomial, is used for filling the gaps in time series. We therefore use the method of interpolation by autoregressive
and moving average \citep[MIARMA, ][]{granado15} which is aimed at preserving the original frequency content irrespective of the spectral properties of a given signal. Since it is based on ARMA models, it can fit not only analytic, i.e. deterministic, components but also stochastic signals, so it is suitable for application to blazar LCs.

The MIARMA algorithm is based on a forward-backward weighted prediction based on ARMA models fitted to the segments bracketing the gaps. The optimal order is estimated and fixed at the first step of the algorithm sequence using Akaike coefficients. ARMA models are fitted for each gap independently, that is, a local prediction is obtained. This allows weakly nonstationary signals to be interpolated, too. 

To generate the estimates of errors for the missing data interpolated with MIARMA, we draw the missing errors from the empirical distribution of the uncertainties at hand, and complement the interpolated values with such obtained errors. We therefore sustain the distribution of errors, and employ one such realization for subsequent computations.

\subsection{Akaike and Bayesian Information Criteria}
\label{sect::aic}

Different fits to the same data set are compared using the small sample Akaike\footnote{It is advised \citep{hurvich89,burnham} to employ the small sample correction when $N/p<40$, which is the case when fitting the PSDs herein.} and Bayesian Information Criteria ($AIC_c$ and $BIC$), given by
\begin{equation}
AIC_c=2p-2\mathcal{L}+\frac{2(p+1)(p+2)}{N-p-2}
\label{eq_aic_c}
\end{equation}
and
\begin{equation}
BIC=p\ln N-2\mathcal{L},
\label{eq_bic}
\end{equation}
where $\mathcal{L}$ is the loglikelihood, with $N$ data points and $p$ parameters \citep{akaike,schwarz,hurvich89,burnham}. A preferred model is one that minimizes $AIC_c$ or $BIC$. $AIC_c$ is liberal, and has a tendency to overfit, i.e. it might point at an overly complicated model in order to follow the data better. $BIC$ is much more stringent, and tends to underfit, i.e. it prefers models with a smaller number of parameters. Therefore, when the two $IC$ point at different models, the truth lies somewhere in between.

What is essential in assesing the relative goodness of a fit in the $AIC_c$ method is the difference, $\Delta_i=AIC_{c,i}-AIC_{c,\rm min}$, between the $AIC_c$ of the $i$th model and the one with the minimal $AIC_c$. If $\Delta_i<2$, then there is substantial support for the $i$th model (or the evidence against it is worth only a bare mention), and the proposition that it is a proper description is highly probable. If $2<\Delta_i<4$, then there is strong support for the $i$th model. When $4<\Delta_i<7$, there is considerably less support, and models with $\Delta_i>10$ have essentially no support \citep{burnham}.

In case of $BIC$, $\Delta_i=BIC_i-BIC_{\rm min}$, and the support for the $i$th model (or evidence against it) also depends on the differences: if $\Delta_i<2$, then there is substantial support for the $i$th model. When $2<\Delta_i<6$, then there is positive evidence against the $i$th model. If $6<\Delta_i<10$, the evidence is strong, and models with $\Delta_i>10$ yield a very strong evidence against the $i$th model \citep[essentially no support;][]{kass}.

\subsection{Notes on individual objects}
\label{sect::2.4}

We provide here basic information on the examined blazars. Their properties are summarized in Table~\ref{BlazarSummary}.

\subsubsection{Markarian 501}

Markarian 501 (Mrk~501; $\alpha=16^{\mathrm{h}}53^{\mathrm{m}}$52\fs22, $\delta=39$\dg45\arcmin36\farcs61 in the J2000 epoch) is one of the most intensively monitored blazars from radio up to very high energy (VHE; $E>100\,{\rm GeV}$) $\gamma$-rays. The source was identified in X-rays based on observations conducted by the High Energy Astronomy Observatory~1 \citep{Schwartz78} and is one of the first HBLs detected in VHE $\gamma$-rays in the observing program of nearby BL Lacs by the Whipple Observatory \citep[$z<0.1$, ][]{Quinn96}. \citet{Ulri75} measured the spectroscopic redshift of $z=0.034$, and showed polarimetric variability of Mrk~501. Mass of the central BH was estimated as $8.5\times10^{8}\,M_{\odot}$ based on the stellar velocity dispersion method \citep{Falo02}. Recently, \citet{Bhat19} found a QPO in $\sim$10-years-long \textit{Fermi}-LAT data, with a period of $\sim$330~days. Moreover, this HBL is considered as a candidate HE neutrino emitter \citep{Righi19}. 

\subsubsection{Markarian 421}

Markarian 421 (Mrk~421; $\alpha=11^{\mathrm{h}}04^{\mathrm{m}}$27\fs31, $\delta=38$\dg12\arcmin31\farcs80 in the J2000 epoch), located at $z=0.031$ \citep{Ulri75} and with the BH mass of $3.2\times10^{8}\,M_{\odot}$ \citep{Falo02}, is one of the nearest and intensively monitored HBLs in various energy bands. Its optical LC is one of the longest, starting in 1899 \citep{Miller75}, showing rapid flux and polarization variability \citep[e.g.][]{Fraija17}, and bright flares in different epochs. Extreme flux variability was also detected in X-ray data \citep[see, e.g. ][]{Kataoka16,Kapanadze18}. Mrk~421 is the first BL Lac detected in HE $\gamma$-rays by the Energetic Gamma Ray Experiment Telescope \citep[EGRET;][]{Lin92} and VHE $\gamma$-rays by the Whipple Observatory \citep{Punch92}. Moreover, it is highly variable \citep{Blazejowski05,Kushwaha17} and is actively flaring across the entire electromagnetic spectrum, showing also correlations between particular energy bands, i.e. X-rays and $\gamma$-rays \citep{Fossati08,Aleksic15,Ahnen16}; near-infrared (IR), optical, ultraviolet (UV), and $\gamma$-rays \citep{Carnerero17}; radio and $\gamma$-rays \citep{Hovatta15}. Similarly to Mrk~501, Mrk~421 is considered a promising HE neutrino emitter candidate \citep{Turley16,Organokov19}. No QPO was reported by \citet{gupta19} in 7-years-long, and by \citet{Sandrinelli17} in 8-years-long \textit{Fermi}-LAT LC. Recently, \citet{bhat20} found a QPO in $\sim$10-years-long \textit{Fermi}-LAT data, with a period of $\sim$285~days.

\subsubsection{PKS 0716+714}
PKS~0716+714 ($\alpha=07^{\mathrm{h}}21^{\mathrm{m}}$53\fs45, $\delta=$71\dg20\arcmin36\farcs36, J2000 epoch) is an IBL blazar \citep{Giom99} particularly known for its intraday and long-term variability at different frequencies \citep[e.g.][]{Villata08,Rani13,Liao14} as well as short flaring events \citep{Chandra15}. Since no emission/absorption lines were observed in IR, optical, and UV spectra, \citet{Nilsson08} estimated its redshift to be $z=0.31$ based on an average luminosity of the host galaxy, and \citet{Danforth13} confirmed this value by analyzing high-resolution far-UV data. \citet{Sandrinelli17} reported no QPOs in HE $\gamma$-rays and the optical band. On the contrary, \citet{prokhorov17} analyzed 9-years-long \textit{Fermi}-LAT data, using the generalized LSP to find a QPO with a period of $\sim$340 days, with $>$5$\sigma$ confidence level against white noise. Also, \citet{li18} analyzed $0.1-200$~GeV data of PKS~0716+714 with the Jurkevich method \citep{Jurk71}, LSP, and the red-noise spectra \citep[REDFIT38 software; ][]{Schu02}. A tentative QPO with a period of 344 $\pm$ 16 days was found at a significance level of $>1.96\sigma$. Recently, \citet{bhat20} found a QPO in $\sim$10-years-long \textit{Fermi}-LAT data, with a period of $\sim$346~days at a significance level of 3.9$\sigma$. \citet{Kaur18} estimated a BH mass of $5.6\times10^{8}\,M_{\odot}$, while \citet{Liu19} gives separate values for the Schwarzchild case, $2.7\times10^{8}\,M_{\odot}$, and the Kerr BH, $8.1\times10^{8}\,M_{\odot}$. \citet{covino19} examined $\gamma$-ray LCs of, among others, PKS~0716+714, finding a PL index $\beta = 1.11\pm0.12$ of its PSD. 

\subsubsection{PKS 2155$-$304}
PKS~2155$-$304 ($\alpha=21^{\mathrm{h}}58^{\mathrm{m}}$52\fs07, $\delta=-$30\dg13\arcmin32\farcs12, J2000 epoch), at $z=0.116$ \citep{Falomo93} and a BH mass of $1.3\times10^{7}\,M_{\odot}$ \citep{Lian03}, was detected in VHE $\gamma$-rays by the University of Durham Mark 6 $\gamma$-ray telescope \citep{Chadwick99}. Since then, comprehensive monitoring in VHE energies revealed short and bright flares  as well as long- and short-term variability \citep[e.g.][]{Aharonian05,Foschini07,Ghisellini08}. Moreover, \citet{Zhang14} discovered a QPO with a period of 317 days in 35-years-long optical data, confirmed by \citet{Sandrinelli14} who found a period of 315 days, analyzing different data sets. \citet{Sandrinelli14} also studied a 6-years-long \textit{Fermi}-LAT LC between 0.1 and 300 GeV, noticing $\sim$2 times the optical period, i.e. a periodicity peak at 650--660 days. \citet{Sandrinelli16} confirmed the previous result, obtaining 642 days with $3\sigma$ significance, and revealed two more peaks, at 61 and 52 days, with 2.58$\sigma$ significance. Hints of a possible QPO with a period of $\sim 670-700$~days were seen in a 5.5-years-long \textit{Fermi}-LAT LC as well \citep{HESS17}. Subsequently, \citet{zhang17a} reported a periodicity of $\sim$635~days with a 4.9$\sigma$ significance, based on an 8-years-long \textit{Fermi}-LAT LC, while \citet{bhat20} found a period of $\sim$610~days with a 4.5$\sigma$ significance, using $\sim$10-years-long \textit{Fermi}-LAT data. Most recently, \citet{penil20} analyzed a 9-years-long \textit{Fermi}-LAT LC, and identified, with several methods, a QPO with a period of $\sim$620~days, on a significance level 2--4$\sigma$. PKS~2155$-$304 is also the first AGN with a QPO detection in optical polarization, showing two peaks, at 13 and 30 minutes \citep{Pekeur16}. \citet{covino19} found a PL index $\beta = 1.11\pm0.16$ of its PSD.

\subsubsection{TXS 0506+056}

TXS~0506+056 ($\alpha=05^{\mathrm{h}}09^{\mathrm{m}}$25\fs96, $\delta=$ 05\dg41\arcmin35\farcs33, J2000 epoch), located at $z=0.3365$ \citep{Paiano18}, was first detected in radio at 5 GHz by the Arecibo telescope \citep{Lawr83}, and then in HE $\gamma$-rays by EGRET \citep{Lamb97} and \textit{Fermi}-LAT \citep{Abdo10c}. TXS~0506+056 is a bright BL Lac, occupying the transition region between IBLs and HBLs, considered lately as a counterpart of the IceCube neutrino event \citep[e.g.][]{Icec18,Anso18,Pado18,Halz19}. Its BH mass was recently estimated to be $3.0\times10^{8}\,M_{\odot}$ \citep{Pado19}.

\subsubsection{PKS 1510$-$089}

PKS~1510$-$089 ($\alpha=15^{\mathrm{h}}12^{\mathrm{m}}$50\fs53, $\delta=-$09\dg05\arcmin59\farcs83, J2000 epoch) is a bright and highly variable FSRQ located at $z=0.361$ \citep{Burbidge66,Thompson90}, with a central BH mass of $1.6\times10^{8}\,M_{\odot}$ \citep{Lian03}. The blazar was detected in HE $\gamma$-rays by EGRET \citep{Hartman99}, and then in VHE regime by the High Energy Stereoscopic System \citep{HESS13}. Several multiwavelength observational campaigns were conducted to study its variability and to search for correlations between different energy bands. \citet{Abdo10b} reported a complex variability in $\gamma$-rays and optical/UV bands, showing a strong correlation of $\gamma$-rays and optical fluxes. \citet{Foschini13} notified the shortest flux variability ever detected in HE, with the time scale of $\sim$20 minutes. \citet{Aleksic14} concluded that HE and VHE emission originates in the same region, presumably situated behind the broad-line region, finding also similar variability patterns in HE and radio (at 137~MHz) emission. More recently, \citet{Sandrinelli16} were looking for QPOs in multiwavelength data of PKS~1510$-$089, uncovering a HE $\gamma$-ray periodicity with a period of 115 days at the 2.58$\sigma$ significance level, and two optical peaks at 345 and 575 days. On the other hand, neither \citet{castignani17} nor \citet{gupta19} found any QPOs in 8-years-long and 7-years-long \textit{Fermi}-LAT LCs, respectively.

\subsubsection{3C 279}

3C~279 ($\alpha=12^{\mathrm{h}}56^{\mathrm{m}}$11\fs17, $\delta=-$05\dg47\arcmin21\farcs53, J2000 epoch), located at $z=0.536$ \citep{Burbidge65} and with a BH mass of $2.7\times10^{8}\,M_{\odot}$ \citep{Woo02}, is one of the well monitored and studied FSRQs at multiwavelength \citep{Collmar10,Hayashida15,Alvarez18}. It is one of the first blazars discovered in $\gamma$-rays by EGRET \citep{Hartman92}, and subsequently in VHE by the Major Atmospheric Gamma-ray Imaging Cherenkov (MAGIC) telescope \citep{MAGIC08}. Several outbursts were monitored in broad-band spectrum, i.e. $\gamma$-ray flares in 2013 \citep{Asano15}, 2014 \citep{Paliya15}, and 2015 \citep{Ackermann16}. \citet{Sandrinelli16} found QPOs with peaks at 24 and 39 days in 6-years-long \textit{Fermi}-LAT data, as well as several peaks in near-IR and optical LCs. Interestingly, \citet{Qian19} reported a possible binary SMBH system as an engine of 3C 279, examining 31 superluminal components in the jets. 

\subsubsection{B2 1520+31}

B2~1520+31 ($\alpha=15^{\mathrm{h}}22^{\mathrm{m}}$09\fs99, $\delta=$ 31\dg44\arcmin14\farcs38, J2000 epoch) is a high-redshift \citep[$z=1.489$;][]{Shaw12} FSRQ with a BH mass of $8.3\times10^{8}\,M_{\odot}$ \citep{Sbar12}, detected in HE $\gamma$-rays by \textit{Fermi}-LAT in the three months sky survey \citep{Abdo09}. \citet{Kushwaha17} noticed a lognormal distribution in 7-years-long \textit{Fermi}-LAT LCs, while its PSD is typical for an accretion-powered compact source with a PL profile showing hints of a break. A QPO in a $\sim$4-years-long and 3-days-binned \textit{Fermi}-LAT LC with a period of $\sim$71 days was reported by \citet{gupta19}.

\subsubsection{B2 1633+38}

B2~1633+38 ($\alpha=16^{\mathrm{h}}35^{\mathrm{m}}$15\fs49, $\delta=$ 38\dg08\arcmin04\farcs5, J2000 epoch) is an FSRQ type blazar, located at $z=1.814$ \citep{Stri74} and with a BH mass of $2.3\times10^{9}\,M_{\odot}$ \citep{Sbar12}, first detected in HE range by the EGRET instrument \citep[Compton observatory;][]{Matt93}. This blazar is highly variable at radio and optical frequencies (classified as an optically violent variable blazar) as well as X-ray and HE $\gamma$-ray energies \citep[e.g., ][]{Bart95,Rait12}. \citet{Alga18} analyzed, among others, its $\gamma$-ray LC and fitted its PSD with a PL with $\beta=1.7$. Recently, \citet{Oter20} reported possible QPOs in optical R and V bands, with periods between 657 and 705 days, at a significance of $2-5\sigma$, polarized optical data at 654--701 days ($2-4\sigma$) and, finally, in a 10-years-long \textit{Fermi}-LAT data at 581--646 days ($2-4\sigma$).  

\subsubsection{3C 454.3}

3C~454.3 ($\alpha=22^{\mathrm{h}}53^{\mathrm{m}}$57\fs75, $\delta=$ 16\dg08\arcmin53\farcs56, J2000 epoch) is the brightest in $\gamma$-rays \citep{Vercellone11} and relatively distant FSRQ with $z=0.859$ \citep{Lynds67}, a BH mass of $8.3\times10^{8}\,M_{\odot}$ \citep{Woo02}, and known for its  high variability in multiwavelength as well as prominent flaring activity \citep[e.g.][]{Kushwaha17,Gupta17}. The 2010 outburst, which made the blazar the brightest in HE in the sky \citep{Abdo11}, was particularly well studied. \citet{Sasada14} analyzed mid-IR and optical data, spanning over 4 years, finding large-amplitude outburst lasting months, and rapid flares appearing and disappearing within a few days. \citet{Diltz16} modeled the broad-band SED generated with data gathered during the 2010 flare, concluding that the one zone lepto-hadronic model provides merely poor fits. Interestingly, \citet{Weaver19} discovered a QPO with a characteristic period of 36 minutes, analyzing R-band optical data of the 2016 flare. 

\subsubsection{PKS 1830--211}

PKS~1830$-$211 ($\alpha=18^{\mathrm{h}}33^{\mathrm{m}}$39\fs92, $\delta=$ -21\dg03\arcmin39\farcs9, J2000 epoch) is a well-known distant gravitationally lensed quasar \citep{Rao88,Subr90}, with $z=2.507$ \citep{Lidm99} and a BH mass of $5.0\times10^{8}\,M_{\odot}$ \citep{Nair05}; the lens is a spiral galaxy located at $z=0.89$ \citep{Wikl96}. \citet{Jaun91} reported a surprisingly bright radio Einstein ring, which is several times brighter than other phenomena of this type. PKS~1830$-$211 is the brightest high-redshift $\gamma$-ray blazar detected by \textit{Fermi}-LAT with a large flaring activity in HE $\gamma$-rays \citep{Abdo15}. Moreover, this blazar seems to be variable in both sub-millimiter flux density and linear polarimetry \citep{Mart19}. Since PKS~1830$-$211 is a distant $\gamma$-ray emitter, it can be used to study extragalactic background light and to examine cosmological models.

\begin{deluxetable*}{ccccccc}
\tabletypesize{\footnotesize}
\tablecolumns{7}
\tablewidth{0pt}
\tablecaption{Summary of the literature parameters of individual blazars. \label{BlazarSummary}}
\tablehead{
\colhead{Object} & \colhead{blazar type} & \colhead{$z$} & \colhead{$M_{\mathrm{BH}} [M_{\odot}]$} & \colhead{$\beta$} & \colhead{QPO [days]} & \colhead{QPO significance} \\
\small{(1)} & \small{(2)} & \small{(3)} & \small{(4)} & \small{(5)} & \small{(6)} & \small{(7)} }
\startdata
Mrk 501 & HBL & 0.034 & $8.5\times10^{8}$ & $0.99\pm0.01^{a}$ & $\sim$330$^{g}$ & 2.58$\sigma$ \\
Mrk 421 & HBL & 0.031 & $3.2\times10^{8}$ & $1.20\pm0.11^{b}$ & $\sim$285$^{h}$ & 3.6$\sigma$ \\
PKS 0716+714 & IBL & 0.310 & $5.6\times10^{8}$ & $1.11\pm0.12^{c}$ & $\sim$346$^{h}$ & 3.9$\sigma$ \\
PKS 2155$-$304 & HBL & 0.116 & $1.3\times10^{7}$ & $1.11\pm0.16^{c}$ & $\sim$610$^{h}$ & 4.5$\sigma$\\
TXS 0506+056 & IBL/HBL & 0.337 & $3.0\times10^{8}$ & --- & --- & --- \\ 
\hline
PKS 1510$-$089 & FSRQ & 0.361 & $1.6\times10^{8}$ & $1.10\pm0.30^{d}$ & 115$^{i}$ & 2.58$\sigma$ \\
3C 279 & FSRQ & 0.536 & $2.7\times10^{8}$ & $1.08\pm0.25^{d}$ & 39 and 24$^{i}$ & 3$\sigma$\\
B2 1520+31 & FSRQ & 1.489 & $8.3\times10^{8}$ & $1.15\pm0.09^{b}$ & $\sim$71$^{j}$ & 3$\sigma$ \\
B2 1633+38 & FSRQ & 1.814 & $2.3\times10^{9}$ & $1.70\pm0.20^{e}$ & 581--646$^{k}$ & 2--4$\sigma$ \\
3C 454.3 & FSRQ & 0.859 & $8.3\times10^{8}$ & $1.50\pm0.16^{d}$ & --- & --- \\
PKS 1830$-$211 & FSRQ & 2.507 & $5.0\times10^{8}$ & $1.25\pm0.12^{f}$ & --- & ---\\
\enddata
\tablecomments{Columns: (1) source name; (2) detailed classification of the object; (3) redshift; (4) BH mass; (5) literature value of the power-law $\beta$ for $\gamma$-ray PSD (\textit{Fermi}-LAT data); (6) QPOs in \textit{Fermi}-LAT data (if `---', then no QPO search was done for this object); (7) significance of QPO, where 1.96$\sigma$ is 95\%, 2.58$\sigma$ is 99\%, and 3$\sigma$ is 99.73\%. \\PSD indices $\beta$ taken from: $^{a}$\citet{Bhat19} based on analysis of 10-years-long data, $^{b}$\citet{Kushwaha17} --- 7-years-long data, $^{c}$\citet{covino19} --- 10-years-long data, $^{d}$\citet{Naka13} --- 4-years-long data, $^{e}$\citet{Alga18} --- 3-years-long data, and $^{f}$\citet{Abdo15} --- 3-years-long data. \\QPO periods taken from: $^{g}$\citet{Bhat19}, $^{h}$\citet{bhat20}, $^{i}$\citet{Sandrinelli16}, $^{j}$\citet{gupta19}, $^{k}$\citet{Oter20}.}
\end{deluxetable*}

\section{Methods}
\label{methods}

We denote the values of a time series by $x$. For discrete data, we use the notation $x_k$, $k\in\{1,\ldots,N\}$, measured at times $t_k$, and $\Delta x_k$ denote the measurement uncertainties. For a continuous process we use $x(t)$ or $x_t$ for simplicity, $t\in\mathbb{R}$. Obviously, $x_k(t_k)$ is a sampled set of observations of the underlying $x(t)$. 

\subsection{Fourier spectra}
\label{methods::fourier}

The discrete Fourier transform (DFT) of the set $\{x_k\}_{k=1}^N$ is
\begin{equation}
{\rm DFT}(f_s) = \sum\limits_{k=1}^N x_k \exp\left[ -2\pi \ii f_s t_k \right]
\label{eq7}
\end{equation}
for a discrete set of frequencies $f_s = \frac{s-1}{N\delta t}$, \mbox{$s\in\{1,\ldots,N\}$}, where $\delta t = t_{k+1} - t_k$ is a constant time interval between consecutive observations, its inverse being the sampling rate, ${\rm SR} = 1/\delta t$. The first point, ${\rm DFT}(f_{s=1})={\rm DFT}(0)=\sum_{k=1}^N x_k$, is the zero-frequency value (so-called DC value). Subtracting from the time series its mean, $\bar{x}$, removes this spurious component from the DFT and the subsequent PSD. The Nyquist frequency $f_{\rm Nyq} = \frac{1}{2\delta t}$ is half the SR, i.e. it corresponds to twice the sampling interval, and denotes the maximal frequency (minimal period) that can be meaningfully inferred from the observed time series. 

The Fourier PSD is defined herein to be
\begin{equation}
P(f_s) = \frac{2\delta t}{N}|DFT(f_s)|^2.
\label{eq8}
\end{equation}
For the range $f_s\in[f_{\rm Nyq}, 2\cdot f_{\rm Nyq}]$ it is a symmetric reflection of the PSD in $f_s\in[0,f_{\rm Nyq}]$, hence only half of the $N$ frequencies are physically meaningful.

The Poisson noise level, coming from the statistical noise due to uncertainties in the measurements, $\Delta x_k$, is given by
\begin{equation}
P_{\rm Poisson} = \frac{2\delta t}{N} \sum\limits_{k=1}^N \Delta x_k^2.
\label{eq9}
\end{equation}

For fitting in the log-log space a power law (PL),
\begin{equation}
P(f) = \frac{P_{\rm norm}}{f^\beta},
\label{eq10}
\end{equation}
or a PL plus Poisson noise (PLC) power spectrum,
\begin{equation}
P(f) = \frac{P_{\rm norm}}{f^\beta}+C,
\label{eq11}
\end{equation}
where $P_{\rm norm}$ is a normalizing constant, and $C$ is an estimate of $P_{\rm Poisson}$, one needs to take into account two things. First, the evenly spaced frequencies $f_s$ are no longer uniformly spaced when logarithmized, i.e. their density is greatly increased at higher $f_s$ values (where the Poisson noise can be expected to be significant). A straightforward least squares fitting would then rely mostly on points clustered in one region of the $\log f_s$ values. To circumvent this, binning is applied. The values of $\log f_s$ are binned into equal-width bins, with at least two points in a bin, and the representative frequencies are computed as the geometric mean in each bin. The PSD value in a bin is taken as the arithmetic mean of the logarithms of the PSD \citep{papadakis93,isobe15}. Second, the logarithm of a PSD is biased in the sense that the expected value of the logarithmic PSD is shifted upwards by $\gamma/\ln(10) \approx 0.250682$ from the computed PSD, where $\gamma$ is the Euler-Mascheroni constant \citep{vaughan05}. Hence the bias-corrected logarithmic PSD is obtained by subtracting $0.250682$ from the raw logarithmic PSD.

\subsection{Lomb-Scargle periodogram}
\label{methods::ls}

The Lomb-Scargle periodogram \citep[LSP; ][]{lomb,scargle,press89,vanderplas18} for arbitrarily spaced data (i.e., unevenly sampled time series) is computed as
\begin{equation}
\begin{split}
P_{LS}(\omega) = \frac{1}{2\sigma^2} &\left[ \frac{\left(\sum\limits_{k=1}^N (x_k-\bar{x}) \cos[\omega(t_k-\tau)]\right)^2}{\sum\limits_{k=1}^N \cos^2[\omega(t_k-\tau)]} \right. \\
+& \left. \frac{\left(\sum\limits_{k=1}^N (x_k-\bar{x}) \sin[\omega(t_k-\tau)]\right)^2}{\sum\limits_{k=1}^N \sin^2[\omega(t_k-\tau)]}  \right]
\end{split}
\label{eq12}
\end{equation}
where $\omega = 2\pi f$ is the angular frequency, $\tau\equiv\tau(\omega)$ is
\begin{equation}
\tau(\omega)=\frac{1}{2\omega} \arctan \left[ \frac{\sum\limits_{k=1}^N\sin(2\omega t_k)}{\sum\limits_{k=1}^N\cos(2\omega t_k)} \right],
\label{eq13}
\end{equation}
and $\bar{x}$ and $\sigma^2$ are the sample mean and variance.

The lower limit for the sampled frequencies is $f_{\rm min} = 1/(t_{\rm max}-t_{\rm min})$, corresponding to the length of the time series. The upper limit, $f_{\rm max}$,  would be the Nyquist frequency, the same as in the Fourier spectrum (Sect.~\ref{methods::fourier}), if the data were uniformly sampled. For unevenly spaced data, the common choices for a pseudo-Nyquist frequency are somewhat arbitrary \citep{vanderplas18}. The result of \citet{eyer99} gives a proper and meaningful way to assess the high-end frequency limit in case of non-uniform sampling: $f_{\rm Nyq} = \frac{1}{2p}$, where $p$ is the smallest value that satisfies $t_k=t_1+n_k p$, $n_k\in\mathbb{N}$. Such $p$ is a kind of the greatest common divisor of the set $\{t_k-t_1\}_{k=2}^N$. This, however, may not exist if $t_k$ are irrational; strictly speaking, $p$ is then equal to zero, hence $f_{\rm Nyq}=\infty$. In reality, the Nyquist frequency is limitted by the temporal precision of the measurements, giving a large, but not infinite, value.

The frequency grid should not be too sparse because it would then miss spectral structures between the grid points. On the other hand, a too fine grid will require a longer computation time with no value added. A general approach is to choose a grid so that every peak in the periodogram is sampled $n_0=5-10$ times \citep{vanderplas18}. This leads to a total number of sampling frequencies being
\begin{equation}
N_P = n_0\frac{f_{\rm max}}{f_{\rm min}}.
\label{eq15}
\end{equation}
We employ $n_0 = 10$ hereinafter.

Finally, to adjust the Poisson noise level from Eq.~(\ref{eq9}) to the normalization that is used in Eq.~(\ref{eq12}), a conversion is employed:
\begin{equation}
P_{\rm Poisson,LS} = \frac{1}{2\sigma^2} \frac{1}{2\delta t} P_{\rm Poisson}.
\label{eq9_LS}
\end{equation}

\subsection{Wavelet scalogram}

\subsubsection{Wavelets}
\label{sect::wavelets}

A wavelet $\psi(t)$ \citep{farge92,flandrin92,wojtaszczyk97,torrence98,addison02,mallat09,martinez09,kirby13} is a short, temporaly and spectraly localized oscillation with finite energy (i.e., normalized in the $L^2$ norm, $\braket{\psi(t),\psi(t)}=1$) and zero mean\footnote{The zero-mean criterion is equivalent to the admissibility criterion: $$C_g = \int\limits_0^{\infty}\frac{\left| \hat{\psi}(f) \right|^2}{f}{\rm d}f < \infty,$$ where $\hat{\psi}(f)$ is the Fourier transform of the mother wavelet. It follows that $\hat{\psi}(0)=0$, otherwise the above integral would blow up. $C_g$ is called the admissibility constant. For the Morlet wavelet, the admissibility criterion is satisfied for $\omega > 5$.}, $\braket{\psi(t)}=0$. A mother wavelet generates the dictionary, or child wavelets, forming the basis:
\begin{equation}
\psi_{s,l}(t) = \frac{1}{\sqrt{s}}\psi\left( \frac{t-l}{s} \right).
\label{eq16}
\end{equation}
The term $l\in\mathbb{R}$ refers to translation, hence location of the wavelet, and $s\in\mathbb{R}_+$ is the scale, corresponding to dilation. The factor $1/\sqrt{s}$ ensures energy normalization across different scales, $\braket{\psi_{s,l}(t),\psi_{s,l}(t)}=1$. A continuous wavelet transform (CWT) allows to decompose the signal $x(t)$ into a combination of $\psi_{s,l}(t)$ using the wavelet coefficients:
\begin{equation}
W(s,l) = \braket{x(t),\psi_{s,l}(t)} = \int\limits_t x(t) \psi_{s,l}^*(t) {\rm d}t.
\label{eq17}
\end{equation}
The signal can therefore be reconstructed as
\begin{equation}
x(t) = \sum\limits_{s,l} W(s,l) \psi_{s,l}(t).
\label{eq18}
\end{equation}

A discrete wavelet transform (DWT) is a CWT computed for a discrete set of translations and scales:
\begin{equation}
d_{j,k} = \braket{x(t),\psi_{j,k}} = \int\limits_t x(t) \psi_{j,k}(t) {\rm d}t
\label{eq19}
\end{equation}
with child wavelets
\begin{equation}
\psi_{j,k}(t) = \frac{1}{2^{j/2}}\psi \left( \frac{t}{2^j}-k \right),
\label{eq20}
\end{equation}
where $j\in\mathbb{Z}$ represents the octave (time-scale) and $k\in\mathbb{Z}$ the position of the wavelet. The octaves are divided into an integer number of voices. By comparison with Eq.~(\ref{eq16}), $2^j$ corresponds to the scale $s$, and $k\cdot 2^j$ to the location $l$. For a discrete realization of a time series, $\{x_t\}$, the CWT and DWT can be computed by changing integration to summation in Eq.~(\ref{eq17}) and (\ref{eq19}). 

\begin{figure}
\centering
\includegraphics[width=0.8\columnwidth]{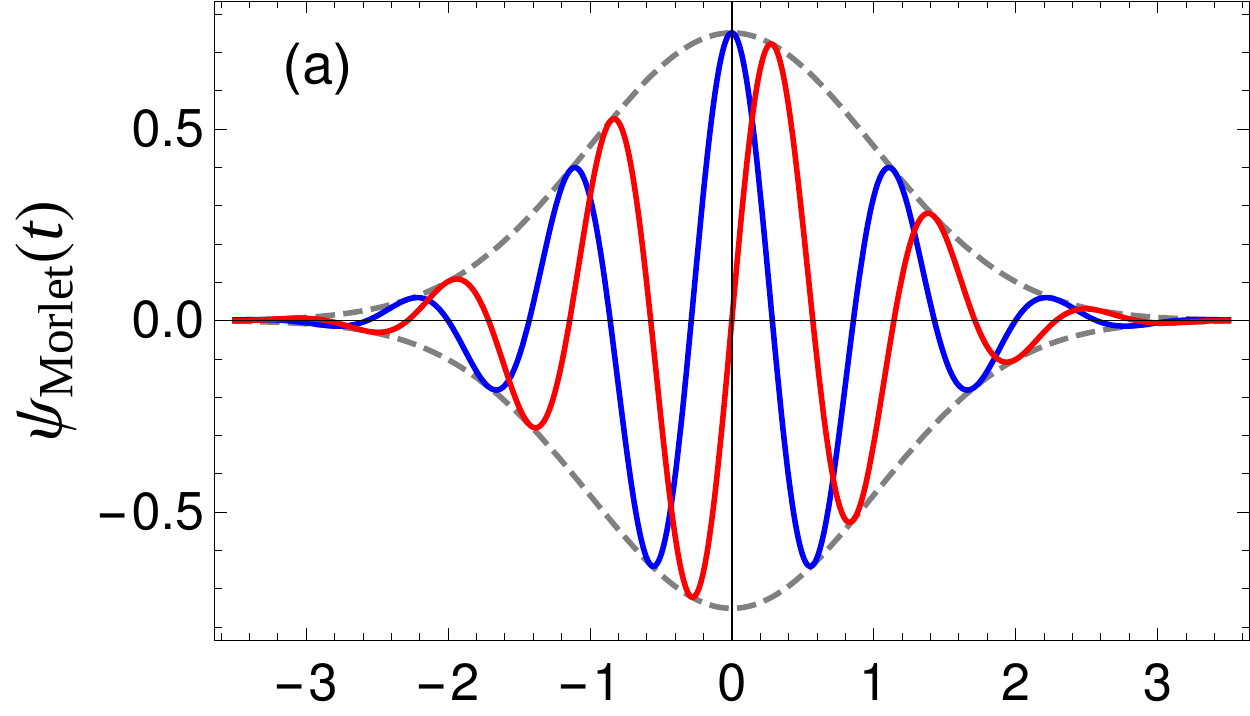} \\
\includegraphics[width=0.8\columnwidth]{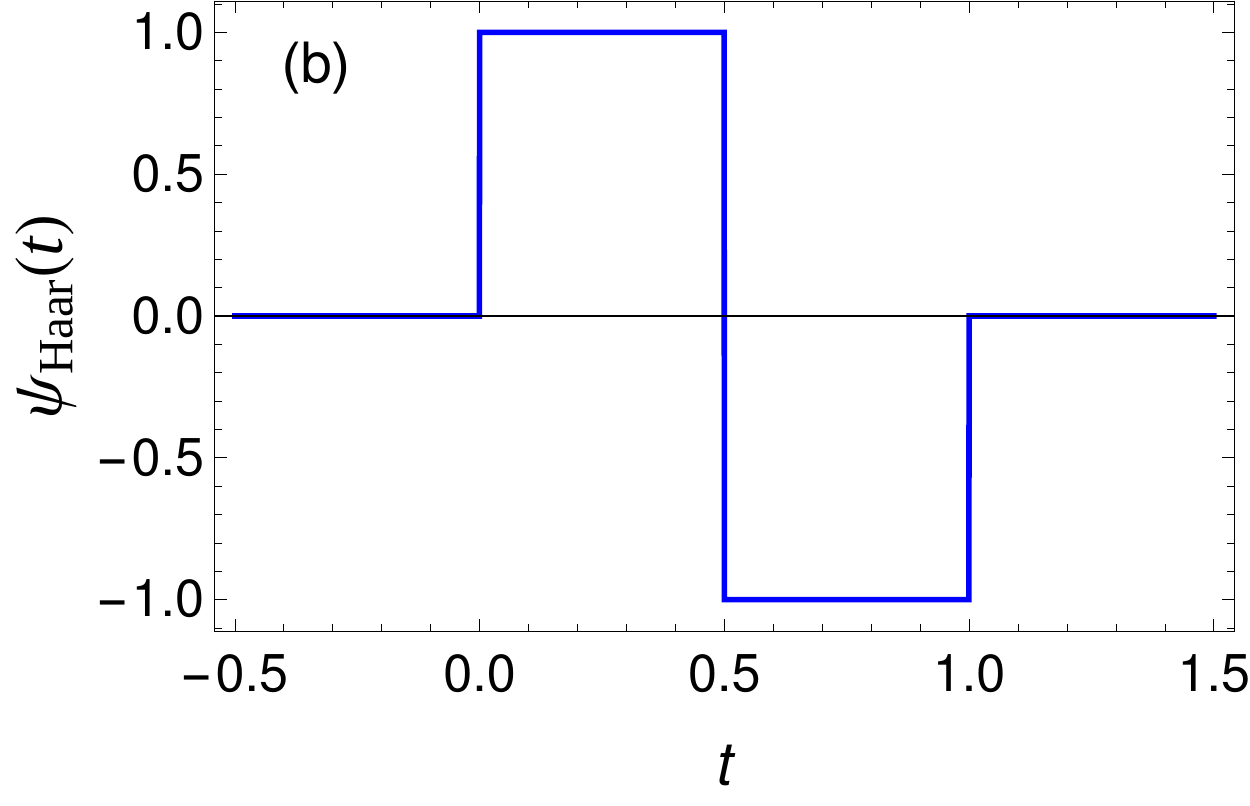}
\caption{(a) The real (blue) and imaginary (red) parts of the Morlet wavelet from Eq.~(\ref{eq21}), and its Gaussian envelope (gray dashed). (b) The Haar wavelet from Eq.~(\ref{eq22}).}
\label{fig_morlet_haar}
\end{figure}

Among the many wavelet families, for CWT we use throughout this work the complex-valued Morlet wavelet (Fig.~\ref{fig_morlet_haar}),
\begin{equation}
\psi_{\rm Morlet}(t) = \frac{1}{\pi^{1/4}} \left[\exp{\left( \ii\omega t \right)} - \exp{\left( -\frac{\omega^2}{2} \right)}\right] \exp{\left(-\frac{t^2}{2}\right)},
\label{eq21}
\end{equation}
with $\omega=5.5$, so that the admissibility criterion is satisfied. For the DWT, we employ the Haar wavelet (Fig.~\ref{fig_morlet_haar}),
\begin{equation}
\psi_{\rm Haar}(t) = \left\{ \begin{array}{cc} 
                1  & \quad 0 \leqslant t < \frac{1}{2} \\
                -1 & \quad \frac{1}{2} \leqslant t < 1 \\
                0  & \quad {\rm otherwise} \\
                \end{array} \right.
\label{eq22}
\end{equation}

\subsubsection{Scalogram}

The scalogram---wavelet periodogram---is a two-dimensional, time-frequency $(l,s)$ representation of the energy-density map,
\begin{equation}
P_{\rm wav}(s,l) = |W(s,l)|^2,
\label{eq23}
\end{equation}
i.e. it shows the temporal localization of a (possibly transient) frequency present in the signal $x(t)$. This is significantly different from the Fourier spectra or LSPs, where only global components are investigated, without their time evolution. Such a global, wavelet periodogram can be obtained as an average over the observed period for each scale $s$: 
\begin{equation}
P_{\rm wav}(s) = \frac{1}{N}\sum\limits_{l=1}^{N} |W(s,l)|^2.
\label{eq24}
\end{equation}

\subsubsection{Cone of Influence}

Because of the finiteness of $\{x_k\}$ and due to the underlying assumption of a cyclic data set, near the end points ($l=1$ and $l=N$) the scalogram is prone to errors. This erroneous zone is called the cone of influence (COI), and its border is obtained as the distance from the highest peak of the autocorrelation of wavelet's power to the point where it decays to ${\rm e}^{-1}$ of the peak's height. For the Morlet wavelet, the border of the COI extends $s\sqrt{2}\approx 1.41s$ from the edges of the scalogram \citep{torrence98}, for each scale $s$. A more conservative choice, employed hereinafter, is to extend the COI to $3s$.

\subsubsection{Irregularly sampled time series}

The above considerations lead to a straightforward implementation in case of regularly sampled time series. For data with irregular sampling, such as the LCs herein, we employ the Welch overlapping segment averaging (WOSA) method, implemented in the package \textsc{wavepal}\footnote{\url{https://github.com/guillaumelenoir/WAVEPAL}} \citep{lenoir18a,lenoir18b}. WOSA consists of segmenting the time series into overlapping segments, tapering them, taking the periodogram on each segment, and taking the average of all the periodograms. The COI is precisely defined as those $\{l,s\}$ that fulfill
\begin{equation}
\left| l-t_m \right| \leqslant 3\omega s,
\label{eq25}
\end{equation}
where $m={\rm min}$ and $m={\rm max}$. The maximum scale that can be probed is therefore
\begin{equation}
s_{\rm max} = \frac{t_{\rm max} - t_{\rm min}}{6\omega}.
\label{eq26}
\end{equation}
The gaps in the data obviously prevent any inference within them, and therefore they constitute the Shannon-Nyquist exclusion zone (SNEZ), excluded due to local aliasing issues. The SNEZ is computed based on the local Shannon-Nyquist theorem. See \citep{lenoir18a,lenoir18b} for mathematical details of the whole procedure.

To test the significance of the detected features in the scalograms, they are tested against a CARMA stochastic model (see Sect.~\ref{carma_def}). This is a more general family of noise than the easily tested white noise, or commonly considered colored noise \citep{uttley02}. We employ the significance testing to search for QPOs at the level of at least $3\sigma$ (99.73\% confidence level). 

\subsection{ARMA processes}
\label{sect::arma}

The autoregressive moving average process of order $(p,q)$, denoted ARMA$(p,q)$, is a (preferably stationary) stochastic process obeying the difference equation \citep{scargle81,box94,brockwell,brockwell2,moreno19}
\begin{equation}
x_t = c + \sum\limits_{k=1}^p \varphi_k x_{t-k} + \sum\limits_{k=1}^q \theta_k \varepsilon_{t-k} + \varepsilon_t,
\label{eq27}
\end{equation}
where $\varepsilon_t$ denotes white noise, $\varepsilon_t\sim \mathcal{N}(0,\sigma^2)$, $\varphi_k$ are the autoregression (AR), and $\theta_k$ are the moving average (MA) coefficients. The current value, $x_t$, of an ARMA$(p,q)$ process depends on a linear combination of $p$ previous ones, $\{x_{t-k}\}_{k=1}^p$, a linear combination of $q$ previous noise terms, $\{\varepsilon_{t-k}\}_{k=1}^q$, and the current noise term $\varepsilon_t$. When $q=0$, an ARMA$(p,q)$ becomes an autoregressive AR$(p)$ process of order $p$, with no dependence on previous noise terms $\varepsilon_{t-k}$, except for the current one, $\varepsilon_t$. Likewise, when $p=0$, ARMA$(p,q)$ is a moving average MA$(q)$ process with no dependence on the previous $x_{t-k}$ values at all.

A convenient formulation of an ARMA$(p,q)$ is through the backshift operator $B$, which acts like
\begin{equation}
By_t = y_{t-1},
\label{eq28}
\end{equation}
so that $B^k y_t = y_{t-k}$, giving a concise form of Eq.~(\ref{eq27}) as
\begin{equation}
\varphi(B) x_t = c + \theta(B) \varepsilon_t,
\label{eq29}
\end{equation}
with polynomial operators
\begin{equation}
\begin{aligned}
\varphi(B) = 1-\sum\limits_{k=1}^p \varphi_k B^k, \\
\theta(B) = 1+\sum\limits_{k=1}^q \theta_k B^k.
\end{aligned}
\label{eq30}
\end{equation}
With these definitions, the constraint on the $\varphi_k$ coefficients for an ARMA$(p,q)$ process to be stationary is that $\varphi(z)\neq 0\, \forall z\in\mathbb{C} \land |z|\leqslant 1$ (i.e., the roots of the autoregression polynomial lie outside the unit circle).

A second-order stationary process can be written as a linear process (after subtracting the deterministic component)
\begin{equation}
x_t = \sum\limits_{k=-\infty}^{\infty} \psi_k \varepsilon_{t-k} \equiv \psi(B) \varepsilon_t,
\label{eq31}
\end{equation}
where
\begin{equation}
\psi(B) = \sum\limits_{k=-\infty}^{\infty} \psi_k B^k
\label{eq32}
\end{equation}
and $\sum\limits_{k=-\infty}^{\infty} |\psi_k|<\infty$. The identity $\psi(z)=\theta(z)/\varphi(z)$ links the coefficients $\varphi_k,\theta_k,\psi_k$. If $\psi_k=0$ for $j<0$, Eq.~(\ref{eq31}) is an MA$(\infty)$ process.

The constant $c$ is related to the mean $\mu$ of the ARMA process values $\{x_t\}$ via
\begin{equation}
\mu = \frac{c}{1-\sum\limits_{k=1}^p \varphi_k}.
\label{eq33}
\end{equation}
The autocovariances $\gamma_j$ can be obtained by solving the system of recurrent equations: 
\begin{equation}
\begin{aligned}
\gamma_0 = \sum\limits_{k=1}^p \varphi_k\gamma_k + \sigma^2 \left( 1-\sum\limits_{k=1}^q \theta_k\psi_k \right), \\
\gamma_j = \sum\limits_{k=1}^p \varphi_j\gamma_{k-j} - \sigma^2 \sum\limits_{k=0}^q \theta_{j-k}\psi_k,
\end{aligned}
\label{eq34}
\end{equation}
for $j\in\{1,\ldots,p\}$ and with $\theta_0=-1$. The variance of $\{x_t\}$ is therefore $\gamma_0$.

The PSD of an ARMA$(p,q)$ process is
\begin{equation}
P_{\rm ARMA}(\lambda) = \frac{\sigma^2}{2\pi}\frac{\left| \theta\left( {\rm e}^{-\ii\lambda} \right) \right|^2}{\left| \varphi\left( {\rm e}^{-\ii\lambda} \right) \right|^2},\quad -\pi\leqslant\lambda\leqslant\pi,
\label{eq35}
\end{equation}
with $\lambda$ (continuous) corresponding in range to the Fourier frequencies \citep{brockwell,brockwell2}, $-\pi<f_j\equiv 2\pi j/N\leqslant\pi$, $j\in\mathbb{Z}$, and the PSDs for the AR$(p)$ and MA$(q)$ process are obtained by setting $\theta(z)=1$ or $\varphi(z)=1$ (i.e., $q=0$ or $p=0$), respectively.

A further generalization of an ARMA$(p,q)$ process is described in Sect.~\ref{sect::farima}. Fittings are routinely performed by the method of moments or maximum likelihood estimation \citep{beran,box94,brockwell,brockwell2,beran2}. 

\subsection{CARMA modeling}
\label{carma_def}

CARMA \citep{brockwell2,moreno19}, is a stochastic process that obeys the stochastic differential equation:
\begin{equation}
\begin{split}
\frac{\d^p x(t)}{\d t^p} + \alpha_{p-1}\frac{\d^{p-1} x(t)}{\d t^{p-1}} + \ldots + \alpha_0 x(t) = \\
\beta_q \frac{\d^q \varepsilon(t)}{\d t^q} + \beta_{q-1} \frac{\d^{q-1} \varepsilon(t)}{\d t^{q-1}} + \ldots + \varepsilon(t),
\end{split}
\label{eq36}
\end{equation}
with $\alpha_p = 1$ and $\beta_0 = 1$. Stationarity is ensured if $p>q$. The PSD of a CARMA$(p,q)$ is
\begin{equation}
P_{\rm CARMA}(\lambda) = \sigma^2\frac{\left| \sum\limits_{j=0}^q \beta_j (2\pi\ii\lambda)^j \right|^2}{\left| \sum\limits_{k=0}^p \alpha_k (2\pi\ii\lambda)^k \right|^2},\quad -\pi\leqslant\lambda\leqslant\pi.
\label{eq37}
\end{equation}
When CARMA$(1,0)$ is considered, i.e. the Ornstein-Uhlenbeck process \citep{koen05,kelly09,kelly11,sobolewska14}, the PSD simplifies to
\begin{equation}
P_{\rm OU}(\lambda) = \frac{\sigma^2}{\alpha_0^2+(2\pi\lambda)^2},\quad -\pi\leqslant\lambda\leqslant\pi,
\label{eq38}
\end{equation}
i.e. it is a Lorentzian centered at zero and with a break frequency at $\alpha_0/(2\pi)$. The general PSD from Eq.~(\ref{eq37}) can be expressed as a weighted sum of such Lorentzian functions, hence allows to flexibly model variability in different time scales, detect QPOs, and to determine variability-based classification of astrophysical sources. The PSD can allow for several breaks, corresponding to a number of zero-centered Lorentzians, whose widths are the locations of the breaks. Non-zero-centered Lorentzians are used to model QPOs. CARMA accounts for irregular sampling and error measurements, and fits an LC in the time domain using the publicly available\footnote{\url{https://github.com/bckelly80/carma_pack}} Markov Chain Monte Carlo (MCMC) sampler developed by \citet{kelly14}. In short, Eq.~(\ref{eq36}) is translated into a state space representation:
\begin{eqnarray}
\left\{ \begin{array}{l}
x(t) = \bm{b} \bm{y}(t) + \delta(t)\\
\d \bm{y}(t) = A\bm{y}(t)\d t + \bm{e}\d W(t) \\
\end{array} \right.
\label{eq39}
\end{eqnarray}
where $W(t)$ is a Wiener process [i.e., a continuous Brownian motion, whose stochastic derivative forms the white noise $\varepsilon(t)$], $\delta(t)$ represents measurement errors, $\bm{b} = \left( \beta_0,\ldots,\beta_{p-1} \right)$, $\beta_j=0$ for $j>q$, $\bm{e} = \left( 0,\ldots,0,\sigma \right)^\top$ is $p$-dimensional, and the matrix $A$ has elements $A_{p,j} = -\alpha_{j-1}$, $A_{i,i+1}=1$, and zero otherwise. Instead of a costly maximization of the corresponding likelihood function, a Kalman filter algorithm is applied. This requires converting the solution of Eq.~(\ref{eq39}),
\begin{equation}
\bm{y}(t) = \bm{y}_0\exp\left( At \right) + \int\limits_0^t \exp\left[ A(t-u) \right] \bm{e} \d W(u),
\label{eq40}
\end{equation}
into that of the discrete time process evaluated at the sampled time stamps by integrating between consecutive measurements. The maximum likelihood estimation is performed with Bayesian inference via an adaptive Metropolis algorithm as the Markov Chain Monte Carlo sampler (see \citealt{kelly14} for details on the implementaion).

\subsection{Hurst exponent}
\label{sect::hurst}

The Hurst exponent $H$ \citep{hurst,mandel68,flandrin92,beran,katsev,tarnopolski16d,knight17} measures the statistical self similarity of a time series $x(t)$. It is said that $x(t)$ is self-similar (or self-affine) if it satisfies
\begin{equation}
x(t)\stackrel{\textbf{\textrm{.}}}{=}\lambda^{-H}x(\lambda t),
\label{eq41}
\end{equation}
where $\lambda>0$ and $\stackrel{\textbf{\textrm{.}}}{=}$ denotes equality in distribution. Self similarity is connected with long range dependence (memory) of a process via the autocorrelation function at lag $k$
\begin{equation}
\rho(k)=\frac{1}{2}\left[ (k+1)^{2H}-2k^{2H}+(k-1)^{2H} \right].
\label{eq42}
\end{equation}
$\rho(k)$ decays to zero as $k\rightarrow\infty$ so slowly that its accumulated sum does not converge (i.e., $\rho (k) \propto|k|^{-\delta}$, $0<\delta<1$) when $H>0.5$, and $x(t)$ is then called a persistent process.

The meaning of $H$ can be understood as follows: for a persistent stochastic process, if some measured quantity attains relatively high values, the system prefers to keep them high. The process is, however, probabilistic \citep{grech04}, and hence at some point the observed quantity will eventually drop to oscillate around some relatively low value. But the process still has long-term memory (being a global feature), therefore it prefers to stay at those low values until the transition occurs randomly again. In case of $H<0.5$, the process is anti-persistent, and it possesses short-term memory, meaning that the observed values frequently switch from relatively high to relatively low (more precisely, the autocorrelations $\rho(k)$ decay fast enough so that their sums converge to a finite value), and there is no preference among the increments. This is a so-called mean-reverting process.

\begin{figure*}
\includegraphics[width=\textwidth]{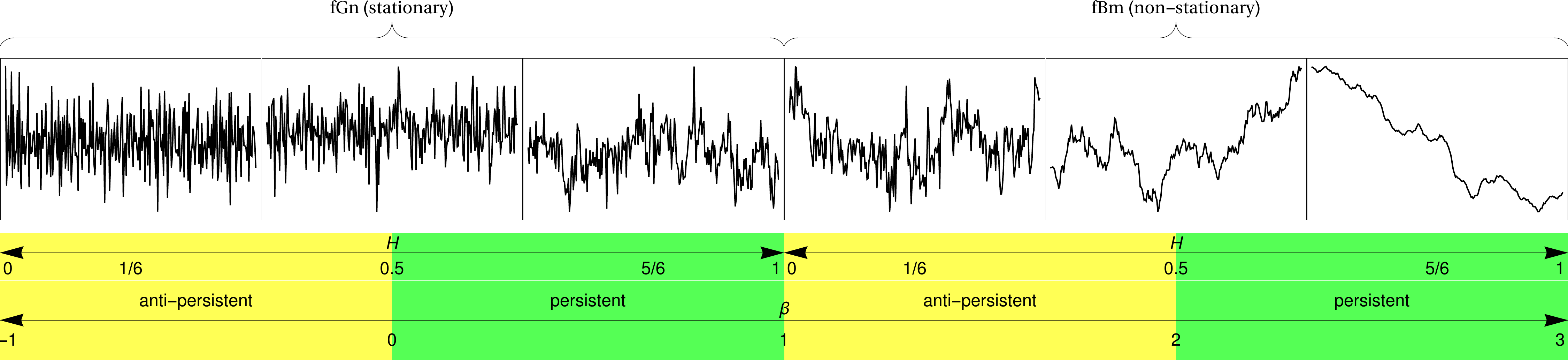}
\caption{The discontinuity of the Hurst exponent on the border between fGn and fBm. The high-$H$ fGn and low-$H$ fBm, both with PL index $\beta\approx 1$, can be easily misidentified. (Figure based on \citealt{gilfriche18}.)}
\label{fig_Hurst_disc}
\end{figure*}

The properties of $H$ can be summarized as:
\begin{enumerate}
\item $0<H<1$,
\item $H=0.5$ for an uncorrelated process (white noise or Brownian motion),
\item $H>0.5$ for a persistent (long-term memory, correlated) process,
\item $H<0.5$ for an anti-persistent (short-term memory, anti-correlated) process.
\end{enumerate}
The archetypal processes with long range dependence are the fractional Gaussian noise (fGn, a stationary process) and fractional Brownian motion (fBm, a nonstationary process). Their PSD is of a PL type, with $\beta_{\rm fBm} = 2H+1$ and $\beta_{\rm fGn} = 2H-1$ \citep{serinaldi10}. The increments of fBm constitute an fGn. There is a discontinuity of $H$ at the border between the two, i.e. for $\beta\approx 1$, where an fGn with $H\lesssim 1$ is visually similar to an fBm with $H\gtrsim 0$, as illustrated in Fig.~\ref{fig_Hurst_disc}, and the two can be easily misidentified. Sect.~\ref{testing::hurst} is devoted to testing the reliability of various algorithms (described in the following sections) for extraction of $H$ and assessing regions of the $(0,1)$ interval where misidentification is likely to occur.

The employed algorithms for computing $H$ rely on the power law or exponential scaling of $\sigma$ vs. $t$, where $\sigma$ is a measure of the dispersion in the analyzed time series, hence the slope of the appropriate linear regression in the log-log or semi-log spaces yields the seeked value of $H$.

\subsubsection{Extraction from a PSD}

A self-affine process exhibiting long-range dependence is characterized by a PL PSD \citep{malamud99,gao03,serinaldi10,tarnopolski16d}:
\begin{equation}
P(f) \propto \frac{1}{f^\beta}.
\label{eq43}
\end{equation}
The exponent $\beta$ is linearly related to the value of $H$ via
\begin{itemize}
\item $H=\frac{\beta+1}{2}$ for an fGn-like process, i.e. with $\beta\in (-1,1)$,
\item $H=\frac{\beta-1}{2}$ for an fBm-like process, i.e. with $\beta\in (1,3)$.
\end{itemize}
This property makes it easy to simulate processes with a given $H$, but also to obtain an estimate of $H$ for time series exhibiting a PLC spectrum from Eq.~(\ref{eq11}), where the Poisson noise level $C$ results from fluctuations present in the data, and the PL part of the PSD is considered to carry meaningful information about the examined system and the underlying stochastic process governing it. It should be pointed out that some processes, like ARMA from Sect.~\ref{sect::arma}, might have PL regions in their PSDs, but associating an $H$ with their exponent is unjustified and can lead to spurious detections. Therefore, a justification of long-range dependence in an examined system should be provided before any attempt to extract $H$ is undertaken. A fundamental first step is to establish whether the signal is stationary (fGn-like) or nonstationary (fBm-like).

\subsubsection{Rescaled range algorithm---R/S}

The $R/S$ algorithm is described herein after \citep{mandel69,suyal}\footnote{We utilize the implementation from \url{http://demonstrations.wolfram.com/HurstExponentOfStockPrice/}.}. Let $\{x_k\}_{k=1}^N$ be an evenly spaced time series. For a temporal window $w$ such that $w_0\leqslant w\leqslant N$, where $w_0$ is the smallest window size, consider the subset of the original dataset $\left\{X_j(t_0,w)\right\}=\{x_j\}_{j=t_0}^{t_0+w-1}$, where $1\leqslant t_0\leqslant N-w+1$. For simplicity, one can rename the index $j\rightarrow j'$ for each subset so that $j'$ goes from 1 to $w$. Then the mean of each of these subsets is
\begin{equation}
\overline{X}(t_0,w)=\frac{1}{w}\sum\limits_{j'=1}^w X_{j'}(t_0,w).
\label{eq44}
\end{equation}
Similarly one calculates the standard deviation $S$ corresponding to the above means:
\begin{equation}
S(t_0,w)=\Big[\frac{1}{w-1}\sum\limits_{j'=1}^w \big(X_{j'}(t_0,w)-\overline{X}(t_0,w)\big)^2\Big]^{\frac{1}{2}}.
\label{eq45}
\end{equation}
We hereafter adopt $w_0=4$. The datasets are then rescaled by their mean
\begin{equation}
Y_{j'}(t_0,w)=X_{j'}(t_0,w)-\overline{X}(t_0,w)
\label{eq46}
\end{equation}
and new variables $y_k$ are defined by calculating the accumulative sum
\begin{equation}
y_k(t_0,w)=\sum\limits_{j'=1}^k Y_{j'}(t_0,w).
\label{eq47}
\end{equation}
The range $R$ is defined as the difference between the maximal and minimal value in each set $\{y_k\}_{k=1}^w$:
\begin{equation}
R(t_0,w)=\max\{y_k(t_0,w)\}-\min\{y_k(t_0,w)\}.
\label{eq48}
\end{equation}
Finally, the rescaled range $R/S$ is defined as
\begin{equation}
(R/S)(t_0,w)=\frac{R(t_0,w)}{S(t_0,w)}.
\label{eq49}
\end{equation}
Taking $t_0$ to run from 1 to $N-w+1$, one calculates the $R/S$ for each temporal window $w$ as the average of those values:
\begin{equation}
(R/S)(w)=\frac{1}{N-w+1}\sum\limits_{t_0=1}^{N-w+1}(R/S)(t_0,w).
\label{eq50}
\end{equation}
In practice, $w$ is uniformly spaced on a logarithmic grid. The slope of the linear regression of $\log (R/S)(w)$ versus $\log w$ gives an estimate for $H$.

\subsubsection{Detrended Fluctuation Analysis---DFA}

In the DFA algorithm \citep{peng94,peng95,hu01,grech05}, one starts with calculating the accumulative sum
\begin{equation}
X_t = \sum\limits_{k=1}^t \big( x_k-\bar{x} \big)
\label{eq51}
\end{equation}
which is next partitioned into non-overlapping segments of length $s$ each. In each segment, the corresponding part of the time series $X_t$ is replaced with its linear fit\footnote{This is the first-order DFA. Fitting a polynomial of order $q$ constitutes the DFA of order $q$.}, resulting in a piecewise-linear approximation of the whole $X_t$, denoted by $X_{\rm lin}(t)$. The fluctuation as a function of the segment length $s$ is defined as
\begin{equation}
F(s) = \Big[\frac{1}{N}\sum\limits_{t=1}^N \big(X_t-X_{\rm lin}(t)\big)^2\Big]^{\frac{1}{2}}.
\label{eq52}
\end{equation}
The slope $\alpha$ of the linear regression of $\log F(s)$ versus $\log s$ is an estimate for $H$: $H=\alpha$ if $\alpha\in(0,1)$, i.e. for fGn-like signals, and $H=\alpha-1$ if $\alpha\in(1,2)$, i.e. for fBm-like signals. 

\subsubsection{Wavelets}

\paragraph{AWC}

We utilize the averaged wavelet coefficient (AWC) method of \citet{simonsen98}. It relies on the scaling in Eq.~(\ref{eq41}) and employs the CWT as from Eq.~(\ref{eq17}). Applying the CWT to Eq.~(\ref{eq41}), one arrives at
\begin{equation}
W(\lambda s, \lambda l) = \lambda^{H+1/2} W(s,l).
\label{eq53}
\end{equation}
The AWC is defined as the standard arithmetic mean over the translations $l$ at a given scale $s$:
\begin{equation}
W(s) = \braket{\left| W(s,l) \right|}_l.
\label{eq54}
\end{equation}
By a linear regression of $W(s)$ vs. $s$ in a log-log plot, $H$ can be obtained from the slope $\alpha_{\rm fBm}$ or $\alpha_{\rm fGn}$ via $H=\alpha_{\rm fBm}-1/2$ for an fBm-like process, and $H=\alpha_{\rm fGn}+1/2$ for an fGn-like one.

\paragraph{DWT}

$H$ can be obtained with the DWT \citep{veitch99,maclachlan13,tarnopolski15c,tarnopolski16d,knight17} using, e.g., the Haar wavelet as the basis (see Sect.~\ref{sect::wavelets}). The basis is obtained from a mother wavelet according to Eq.~(\ref{eq16}). The relation between the variance of the wavelet transform coefficients $d_{j,k}$ from Eq.~(\ref{eq19}) and the scale $j$ can be obtained as
\begin{equation}
\log_2 {\rm var}(d_{j,k})=\alpha\cdot j+{\rm const.}
\label{eq55}
\end{equation}
The slope $\alpha$ is obtained by fitting a line to the linear part of the $\log_2 {\rm var}(d_{j,k})$ vs. $j$. The relation between $\alpha$ and $H$ is $H=\frac{\alpha-1}{2}$ when $\alpha\in (1,3)$ and $H=\frac{\alpha+1}{2}$ when $\alpha\in (-1,1).$\footnote{Two most common instances, related to fractional Brownian and Gaussian noises; see \citealt{maclachlan13,tarnopolski16d,knight17} and references therein for additional details.}

\subsubsection{FARIMA process}
\label{sect::farima}

The fractional autoregressive integrated moving average process, FARIMA$(p,d,q)$ \citep{granger80,hosking81,beran,brockwell2,beran2}, is an ARIMA$(p,d,q)$ process \citep{scargle81,box94,brockwell,brockwell2} with $d$ allowed to be fractional. An ARIMA$(p,d,q)$ process, in turn, is a generalization of an ARMA$(p,q)$ process (see Sect.~\ref{sect::arma}), in which the observed values are replaced with their consecutive differences, and this differencing operation is repeated $d$ times.

Using the backshift operator $B$, an ARIMA$(p,d,q)$ process, $d\in\mathbb{N}$, obeys the difference equation
\begin{equation}
\varphi(B) (1-B)^d x_t = c + \theta(B) \varepsilon_t
\label{eq56}
\end{equation}
with $\varphi(z)$ and $\theta(z)$ as defined in Sect.~\ref{sect::arma}. This is a nonstationary process for all $d\neq 0$, and reduces to the ARMA$(p,q)$ process when $d=0$. If $x_t$ is an ARIMA$(p,d,q)$ process, then $y_t:=(1-B)^d x_t$ is an ARMA$(p,q)$ one; hence differencing $d$ times removes the nonstationarity.\footnote{Note that $(1-B)x_t = x_t - x_{t-1}$.}

FARIMA$(p,d,q)$ is described by the same Eq.~(\ref{eq56}) as ARIMA$(p,d,q)$, but with the constraint on $d$ relaxed so that $d$ can take fractional values. To define fractional differencing, one uses the identities
\begin{equation}
\begin{aligned}
(1-B)^d = \sum\limits_{k=0}^d {d\choose k} (-B)^k, \\
{d\choose k} = \frac{d!}{k!(d-k)!} = \frac{\Gamma(d+1)}{\Gamma(k+1)\Gamma(d-k+1)}.
\end{aligned}
\label{eq57}
\end{equation}
Extending $d$ to be any real number, formally the definition of the fractional differencing operator becomes
\begin{equation}
(1-B)^d = \sum\limits_{k=0}^{\infty} \frac{\Gamma(d+1)}{\Gamma(k+1)\Gamma(d-k+1)} (-B)^k.
\label{eq58}
\end{equation}
For $-0.5<d<0.5$, FARIMA$(p,d,q)$ is stationary; instances with $d>0.5$ can be reduced to $|d|<0.5$ by differencing. $d$ is related to $H$ via $H=d+0.5$, hence $d\in(0,0.5)$ are signatures of long range dependence, i.e. a persistent process \citep{abry2000}. Likewise, $d\in(-0.5,0)$ signifies an anti-persistent process, with short-term memory. The PSD of a FARIMA$(p,d,q)$ process is
\begin{equation}
P_{\rm FARIMA}(\lambda) = \frac{\sigma^2}{2\pi}\frac{\left| \theta\left( {\rm e}^{-\ii\lambda} \right) \right|^2}{\left| \varphi\left( {\rm e}^{-\ii\lambda} \right) \right|^2} \left| 1-{\rm e}^{-\ii\lambda} \right|^{-2d},
\label{eq59}
\end{equation}
where $-\pi\leqslant\lambda\leqslant\pi$ correspond to Fourier frequencies.

\subsection{The $\mathcal{A-T}$ plane}
\label{sect::ATplane}

\begin{figure}
\centering
\includegraphics[width=\columnwidth]{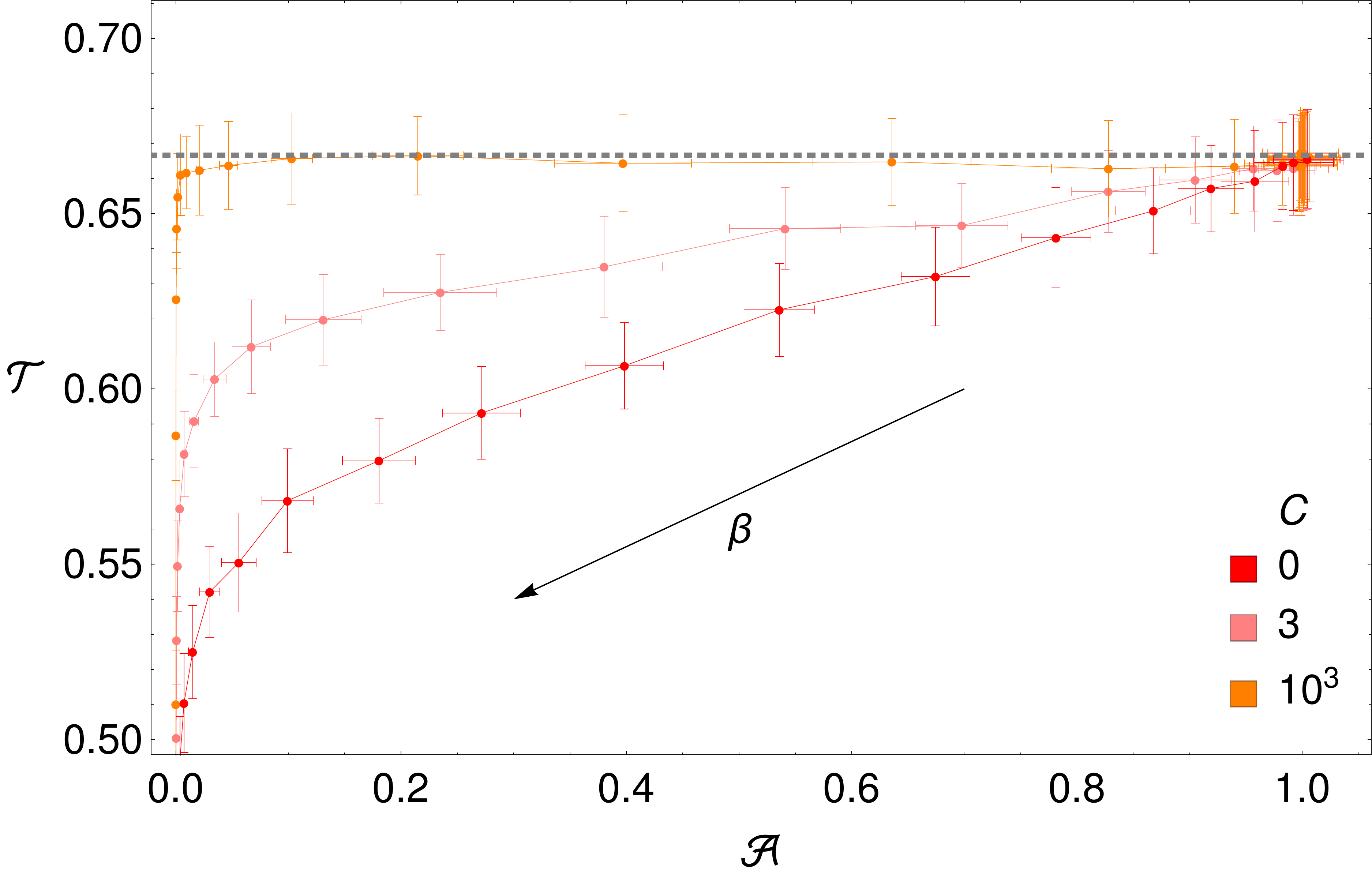}
\caption{Locations in the $\mathcal{A-T}$ plane of the PL plus Poisson noise spectra of the form $P(f)\propto 1/f^\beta+C$, with $\beta\in\{0,0.1,\ldots,3\}$. For each PSD, 100 realizations of the time series were generated, and the displayed points are the mean locations of them. The error bars depict the standard deviation of $\mathcal{A}$ and $\mathcal{T}$ over these 100 realizations. The case $\beta=0$ is a pure white noise, with $(\mathcal{A},\mathcal{T})=(1,2/3)$. The generic PL case ($C=0$) is the lowest curve (red); with an increasing level of the Poisson noise, $C$, the curves are raised and shortened, as the white noise component starts to dominate over the PL part.}
\label{fig_Zunino}
\end{figure}

The $\mathcal{A-T}$ plane was initially introduced to provide a fast and simple estimate of the Hurst exponent \citep{tarnopolski16d}. It is however also able to differentiate between different types of colored noise, $P(f)\propto 1/f^\beta$, characterized by different PL indices $\beta$ \citep{zunino17}. In Figure~\ref{fig_Zunino} the locations in the $\mathcal{A-T}$ plane of PLC spectra of the form $P(f)\propto 1/f^\beta+C$ are shown.

Three consecutive data points, $x_{k-1},x_k,x_{k+1}$, can be arranged in six ways; in four of them, they will create a peak or a valley, i.e. a turning point \citep{kendall,brockwell,tarnopolski16d,tarnopolski19b}. The probability of finding a turning point in such a subset is therefore $2/3$, and the expected value for a random data set is $\mu_T=\frac{2}{3}(N-2)$ --- the first and last points cannot form turning points. Let $T$ denote the number of turning points in a time series, and $\mathcal{T}=T/N$ be their fraction in a time series. $\mathcal{T}$ is assymptotically equal to $2/3$ for a purely random time series (white noise). A process with $\mathcal{T}>2/3$ (i.e., with raggedness exceeding that of a white noise) will be more noisy than white noise. Similarly, a process with $\mathcal{T}<2/3$ will be ragged less than white noise.

The Abbe value \citep{neumann2,neumann,williams,kendall1971,mowlavi,tarnopolski16d,tarnopolski19b} is defined as
\begin{equation}
\mathcal{A}=\frac{\frac{1}{N-1}\sum\limits_{i=1}^{N-1}(x_{i+1}-x_i)^2}{\frac{2}{N}\sum\limits_{i=1}^N (x_i-\bar{x})^2}.
\label{}
\end{equation}
It quantifies the smoothness of a time series by comparing the sum of the squared differences between two successive measurements with the standard deviation of the time series. It decreases to zero for time series displaying a high degree of smoothness, while the normalization factor ensures that $\mathcal{A}$ tends to unity for a purely noisy time series (more precisely, for a white noise process).

\section{Benchmark testing}
\label{testing}

\subsection{Fourier spectra and Lomb-Scargle periodograms}
\label{sect::FourLS}

\begin{figure*}
\centering
\includegraphics[width=0.45\textwidth]{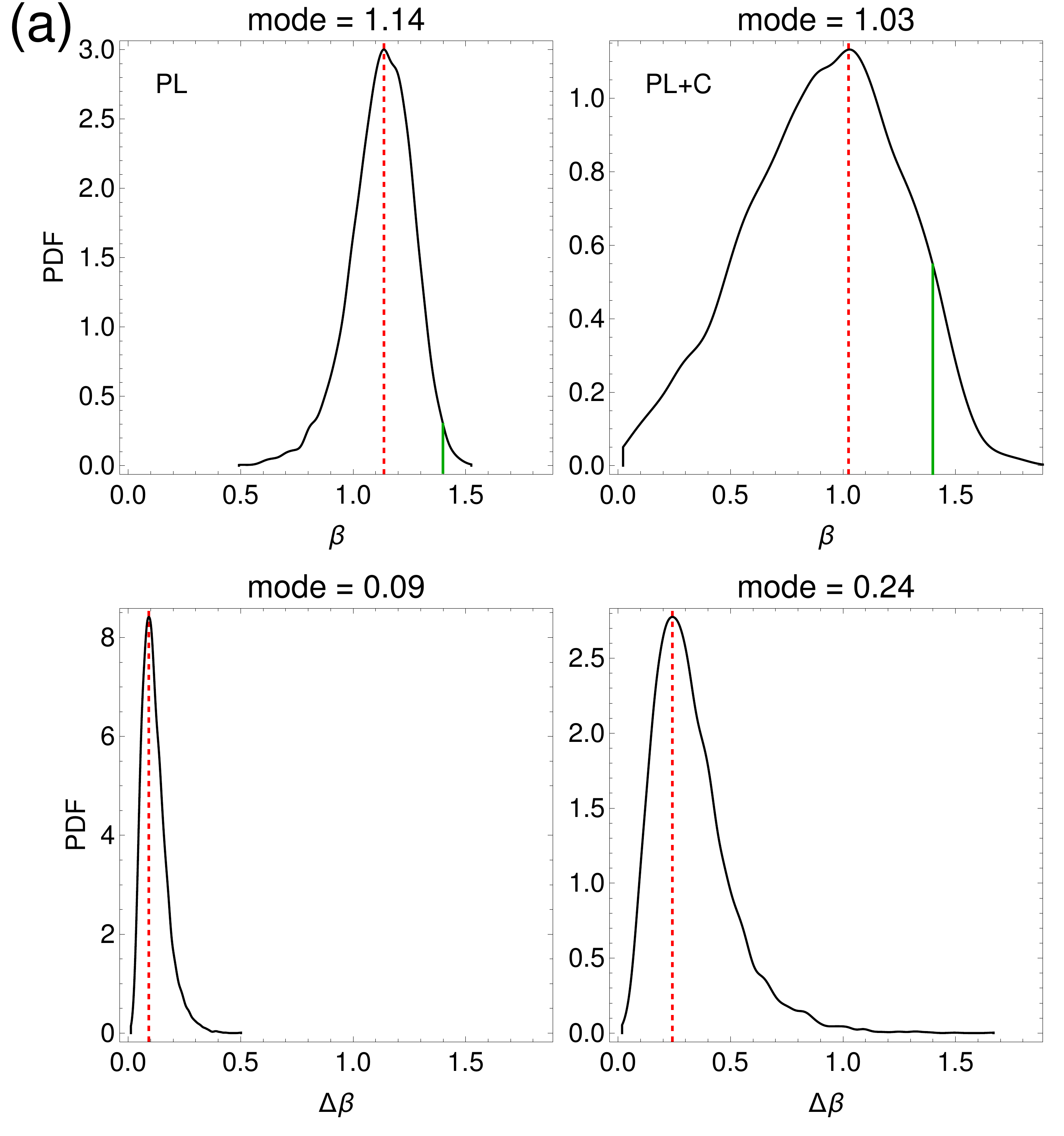}
\includegraphics[width=0.45\textwidth]{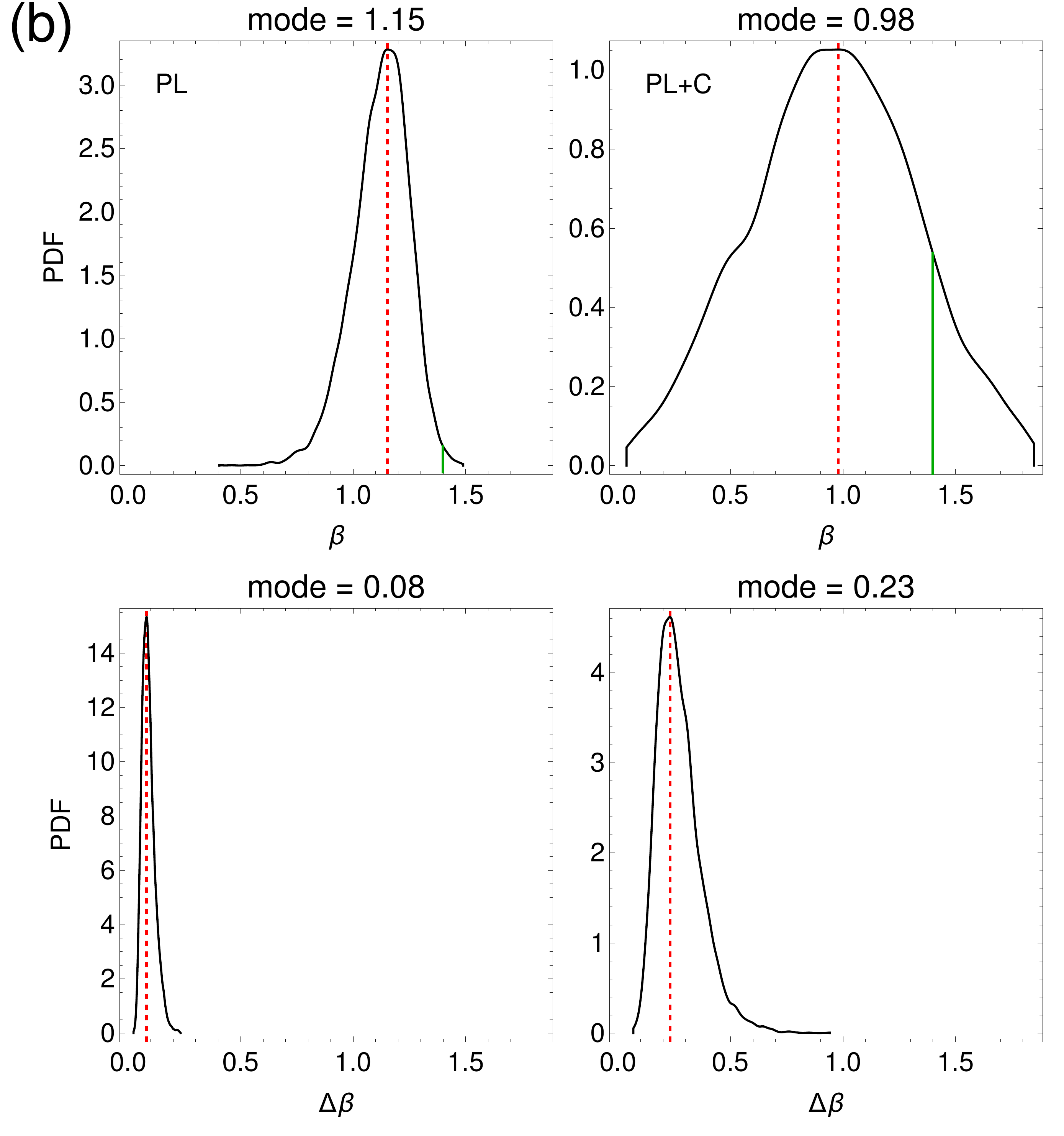}
\caption{The distributions of the PL index $\beta$ ({\it upper row}) and its error ({\it bottom row}) from fitting to (a) Fourier spectra, and (b) LSPs. Both panels display the pure PL with $N=512$ ({\it left column}), and PLC with $N=512$ ({\it right column}). The red dashed lines mark the modes of the distributions, and the solid green line in the upper row denotes the input value of $\beta = 1.4$.}
\label{fig_MC_Fourier_LSP}
\end{figure*}

We generated 6000 time series\footnote{Increasing the number by a 1000 until we reached convergence of the distributions of interest.} of length $N=512$ (i.e., comparable to our actual LCs) with a time step $\delta t = 7\,{\rm d}$ with the PLC spectrum from Eq.~(\ref{eq11}), with $\beta=1.4$ (representative value based on the blazar results; see Sect.~\ref{results::fourier} and \ref{results::LSP}) and $P_{\rm norm}/C=0.005/3.07$ (giving a clear transition from the PL part of the PSD to the white noise region that covers a significant range of frequencies), computed their Fourier PSDs, binned them according to Sect.~\ref{methods::fourier}, and fitted Eq.~(\ref{eq11}). Additionally, we considered the pure PL with $C=0$ and the same $\beta = 1.4$. The distributions of the obtained indices $\beta$ are displayed in Fig.~\ref{fig_MC_Fourier_LSP}(a). In case of a pure PL, the results are close to $\beta = 1.4$, although systematically smaller. On the other hand, only in 23.5\% of simulations the fitted $\beta$ is consistent with 1.4 within the standard error of the fit. For the PLC ($C\neq 0$) the retrieved indices span a significantly wider range $0.1-2$, and the typical error is five times larger than when $C=0$. This high standard error is the reason why 42.9\% indices are equal to 1.4 within error, nearly twice as many as in the pure PL case [see bottom row of Fig.~\ref{fig_MC_Fourier_LSP}(a)]. We also performed the same simulation with $N=2048$ (in order to simulate a longer/better sampled, yet still observationally plausible gamma-ray LC) and observed that the mode is much closer to the true value $\beta = 1.4$, and the distribution is only slightly narrower. A reasonable expectation would be that the longer the time series, the more accurate the estimation of the spectral index would be, however the convergence appears to be slow.

The analysis was repeated with the LSPs. The results, displayed in Fig.~\ref{fig_MC_Fourier_LSP}(b), are very similar to the DFT ones. For PLC, the distributions of $\beta$ are more symmetric than in case of the DFT approach, and due to a larger number of frequencies being tested [five times more than Fourier frequencies, according to Eq~(\ref{eq15})], the distributions of standard errors are narrower.

Overall, we conclude that assessing the shape of the spectrum via binning and fitting---be it either the DFT, or LSP---is unrealiable, especially in the PLC case, as the output is barely related to the input true index $\beta$. The situation is somewhat better when the Poisson noise component is absent, and the spectrum can be considered a pure PL. The scatter then gives a range of possible true values of $\beta$ yielding the observed PSD, however it it relatively narrow, and the resulting value might be used as a rough estimate. A systematic effect is that the fitted $\beta$ is likely underestimated.

\subsection{Wavelet scalogram}

\begin{figure*}
\centering
\includegraphics[width=0.46\textwidth]{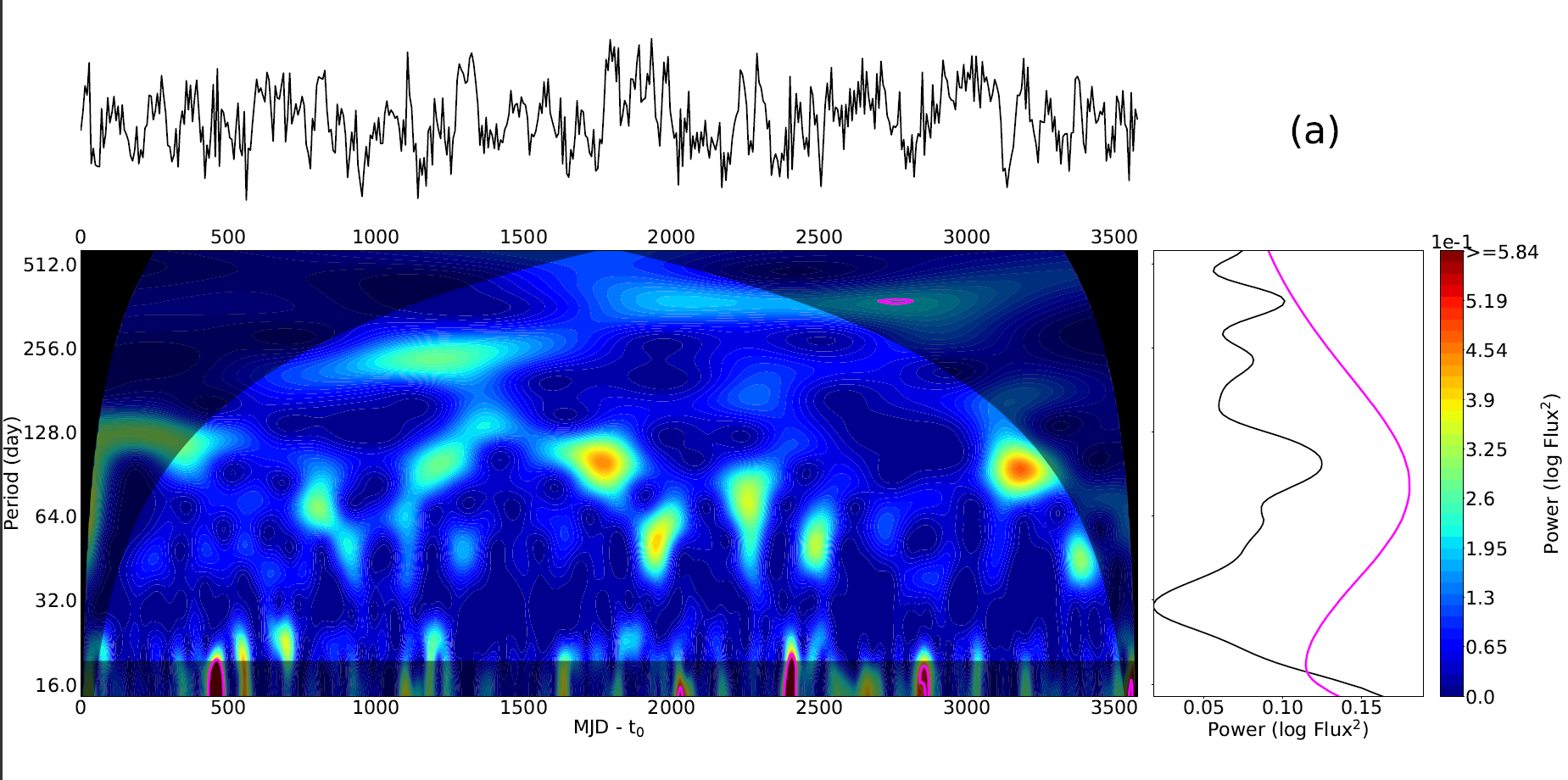}
\includegraphics[width=0.46\textwidth]{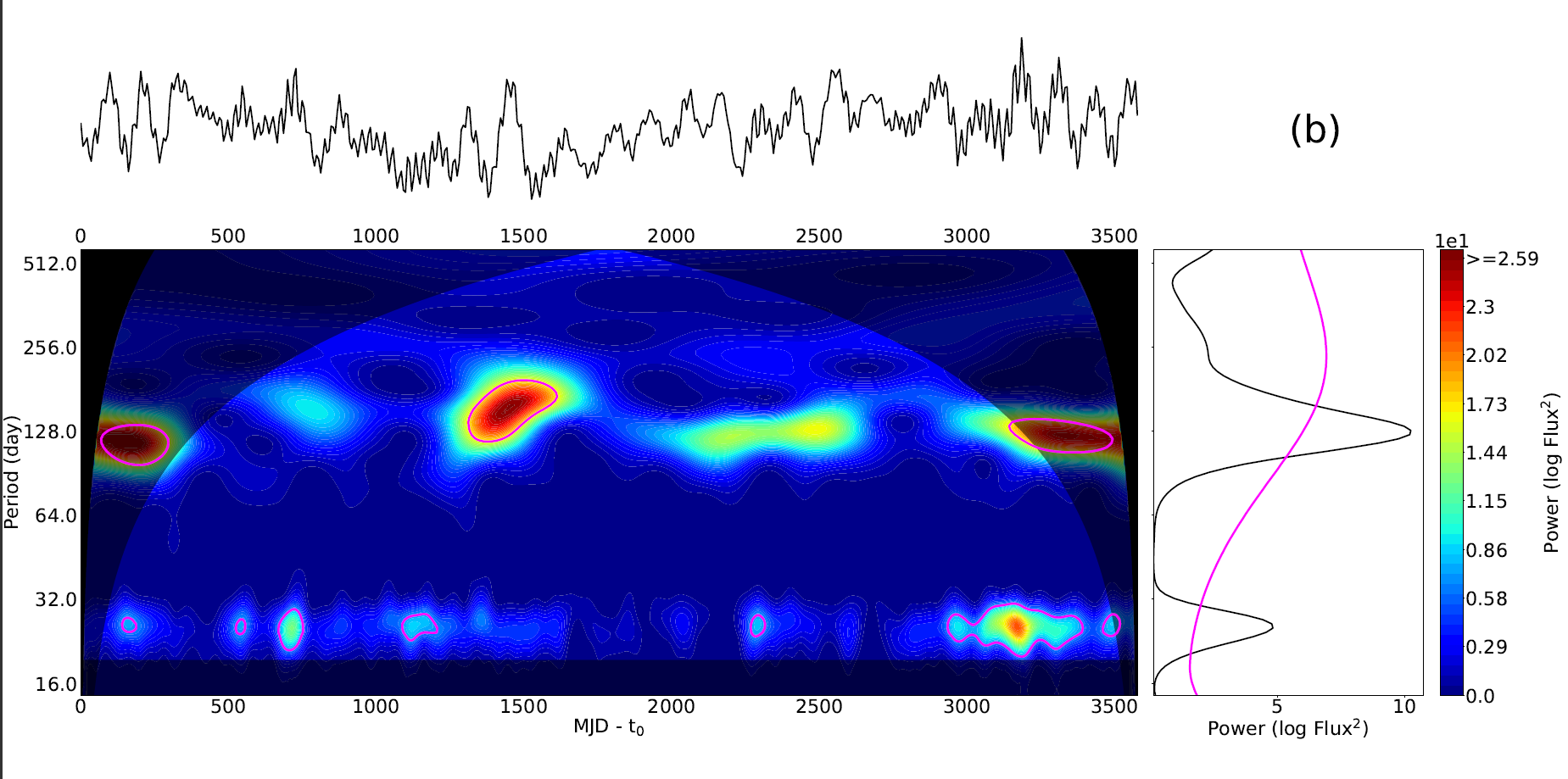}\\
\includegraphics[width=0.46\textwidth]{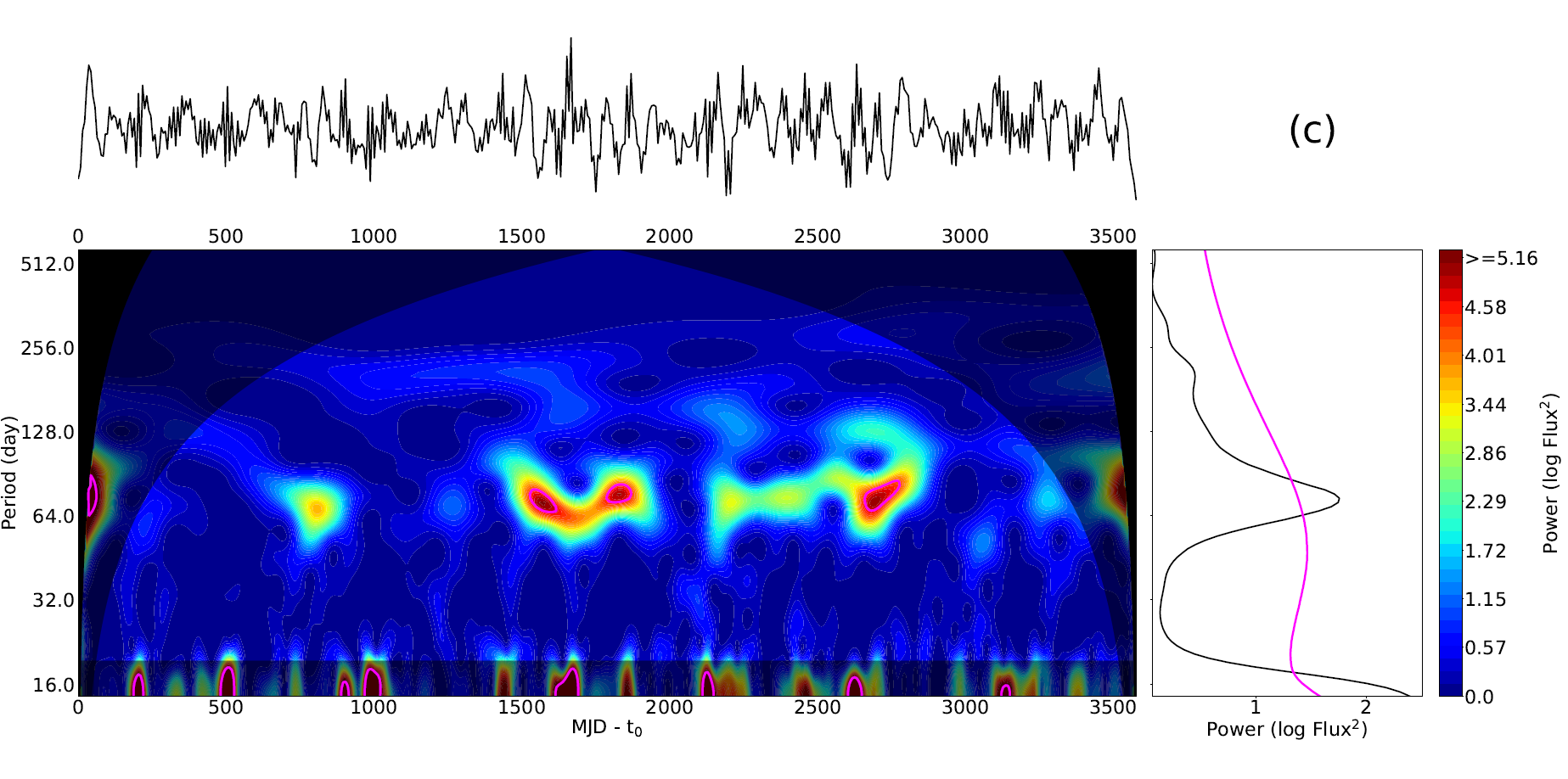}
\includegraphics[width=0.46\textwidth]{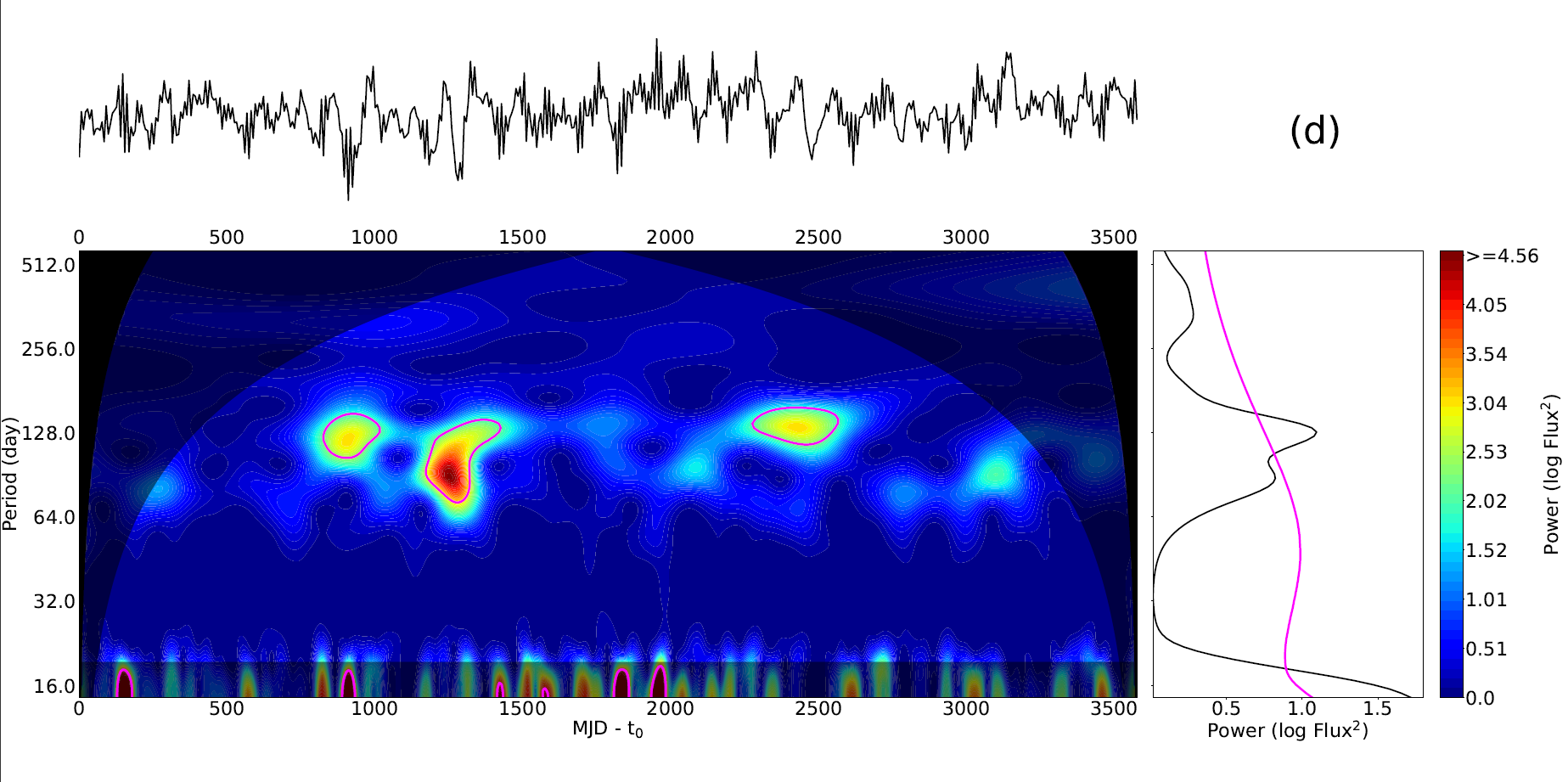}\\
\includegraphics[width=0.46\textwidth]{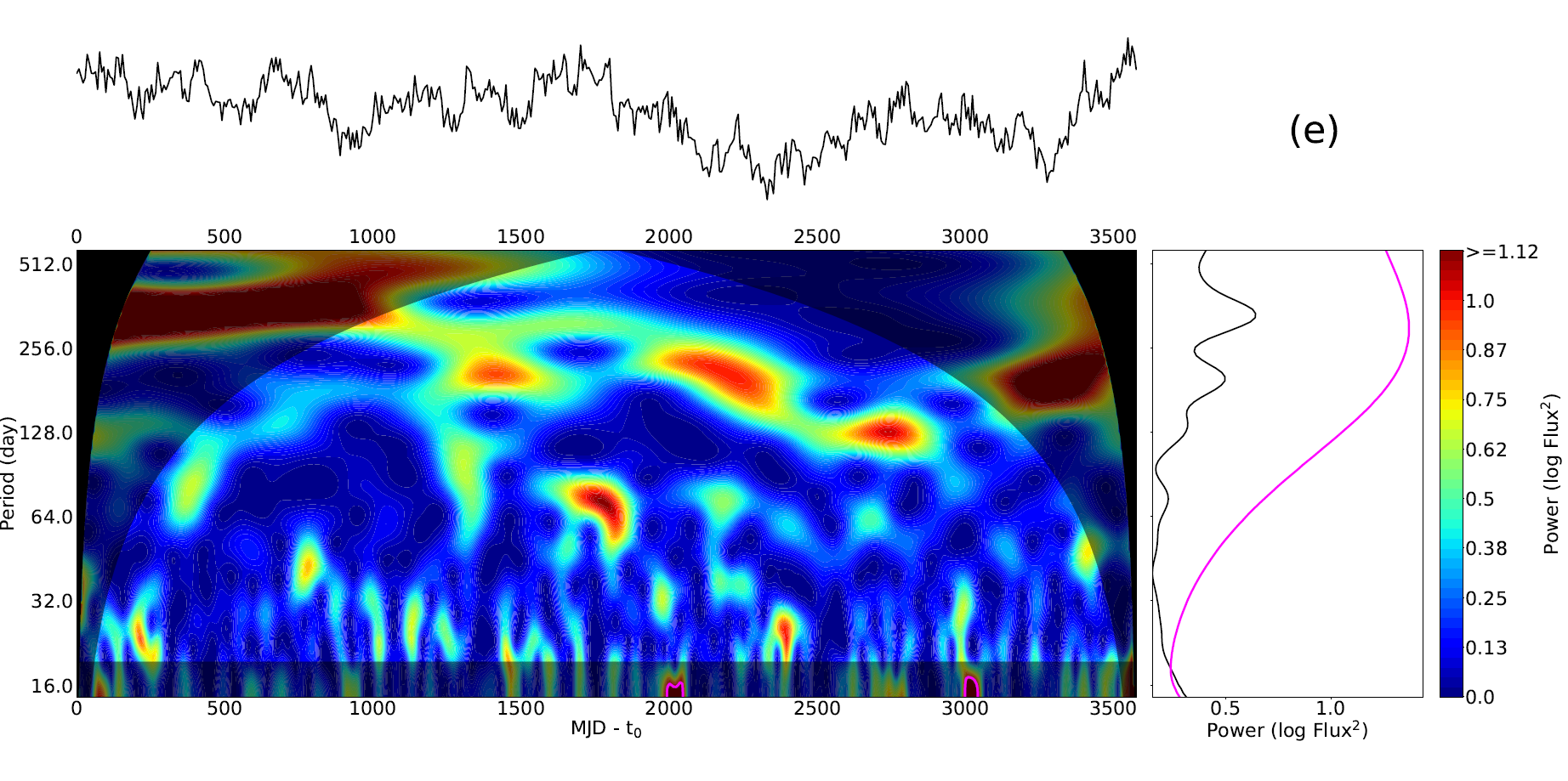}
\includegraphics[width=0.46\textwidth]{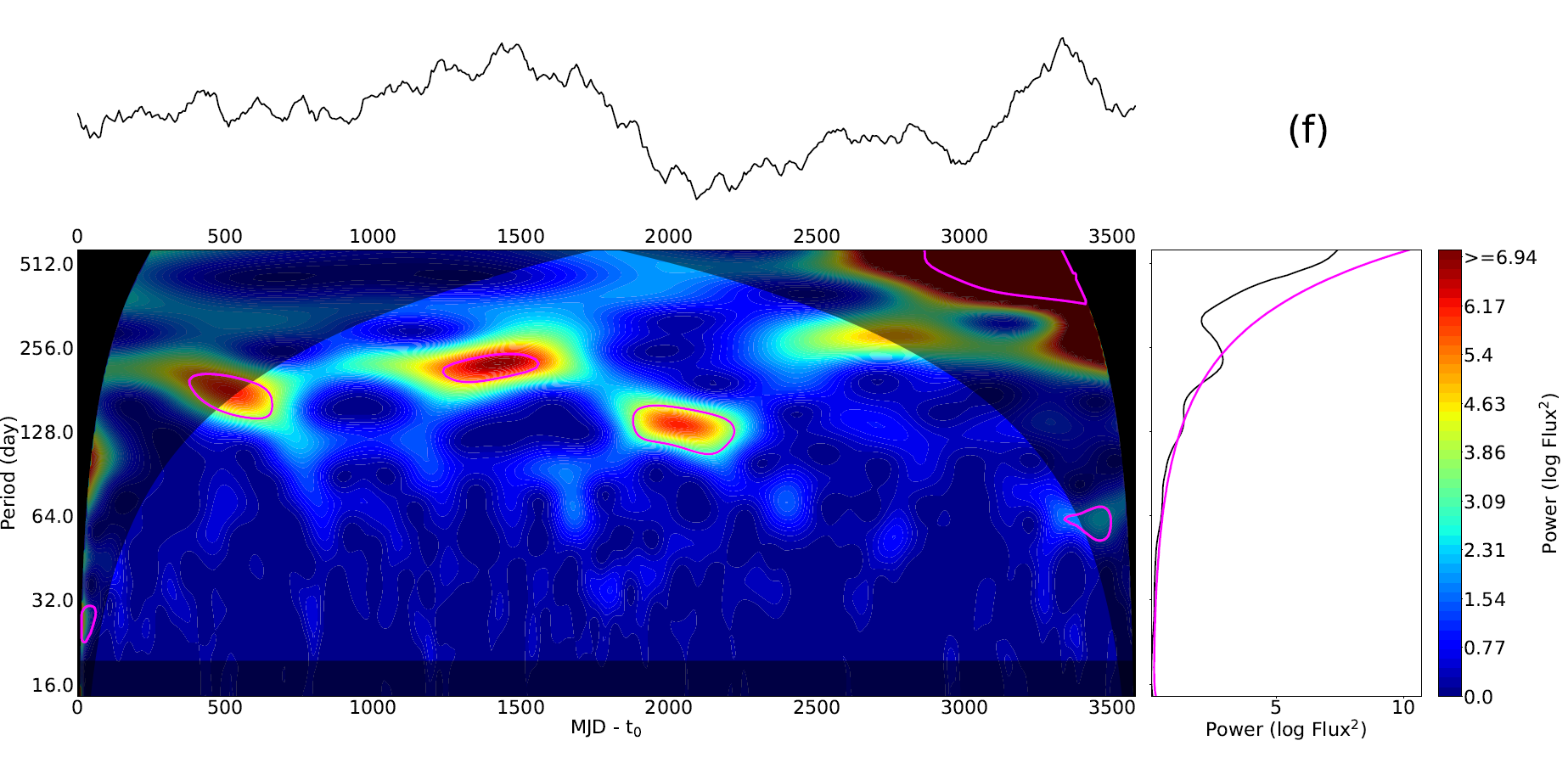}\\
\includegraphics[width=0.46\textwidth]{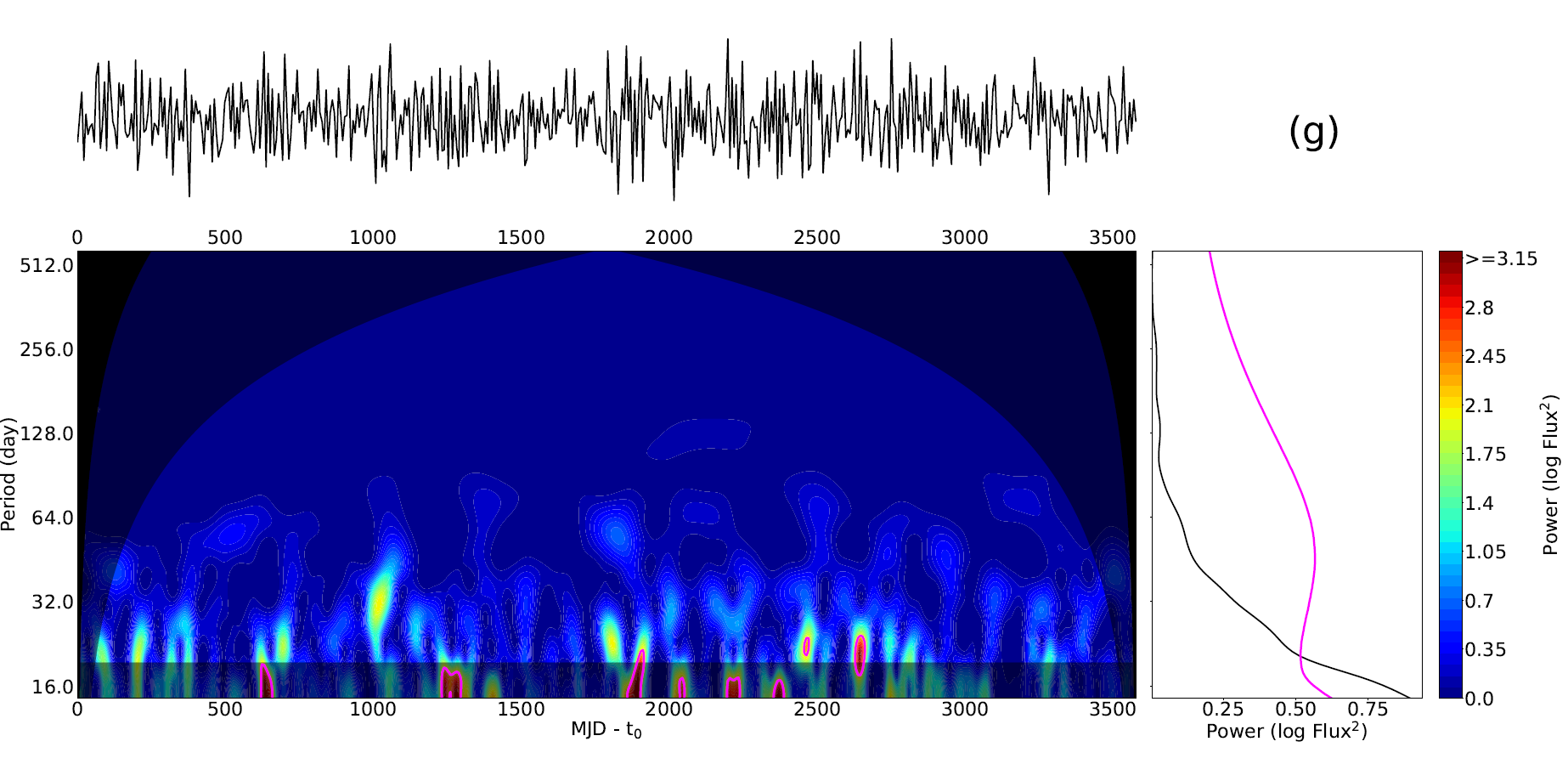}
\includegraphics[width=0.46\textwidth]{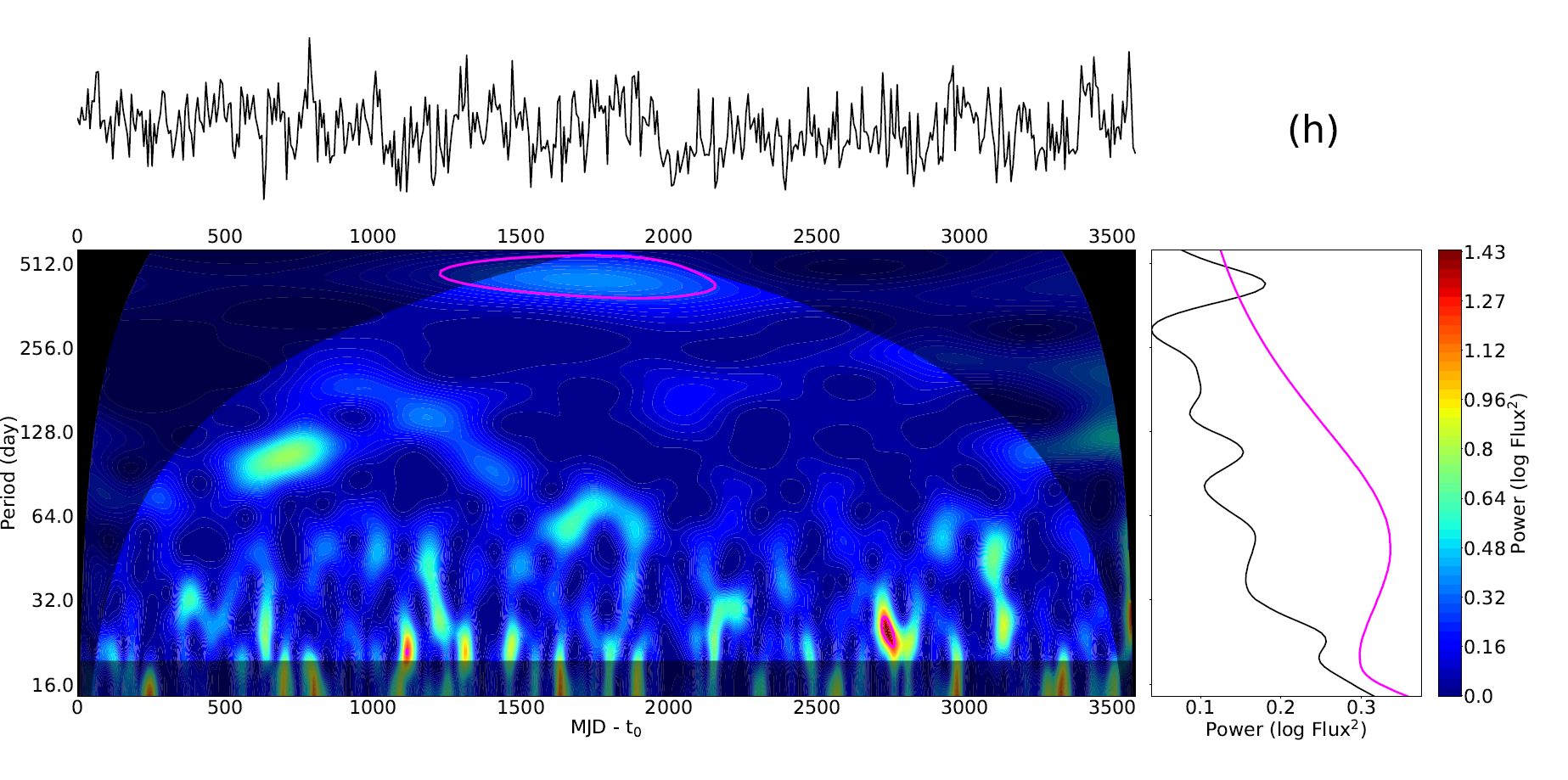}\\
\includegraphics[width=0.46\textwidth]{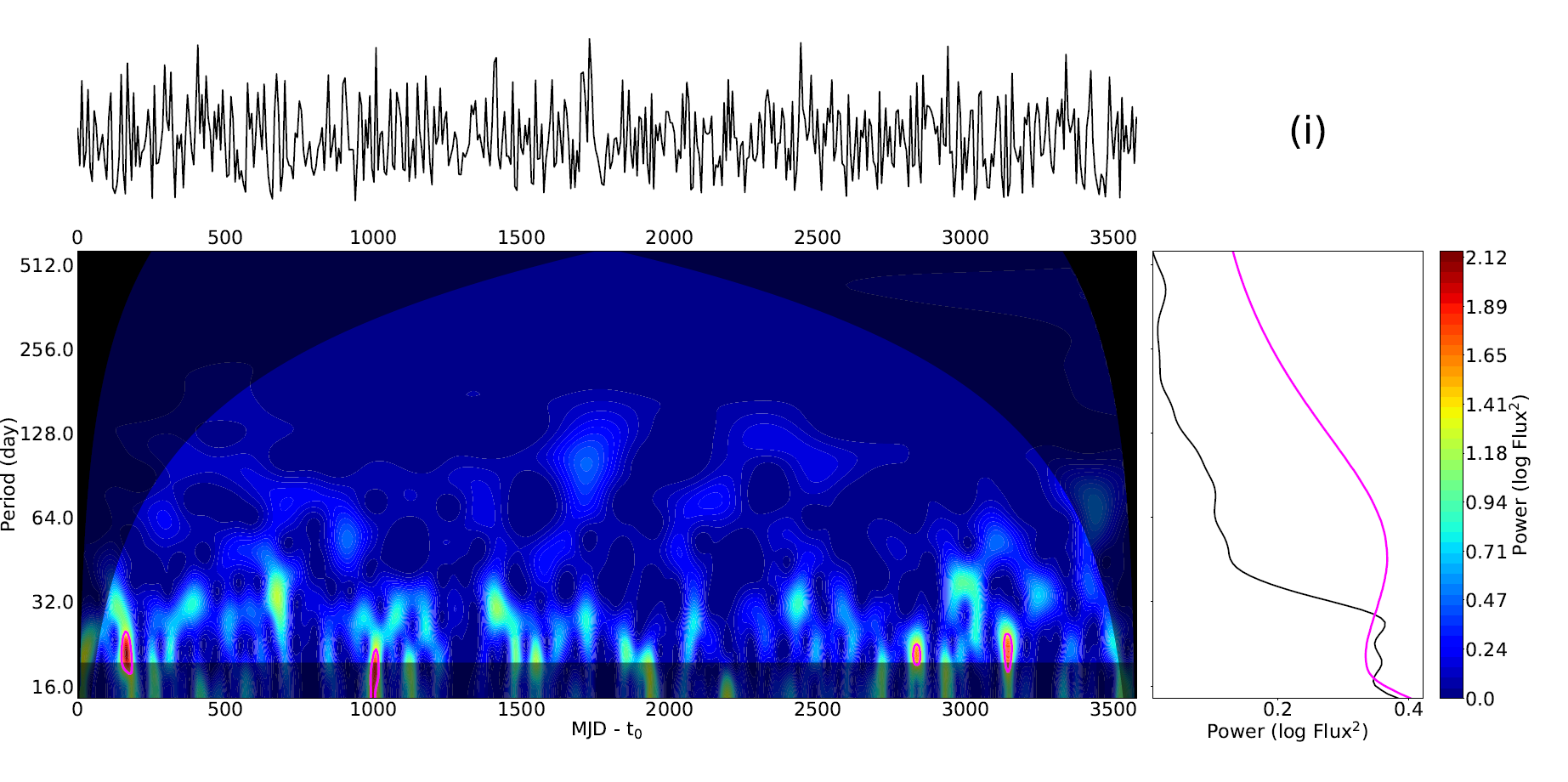}
\includegraphics[width=0.46\textwidth]{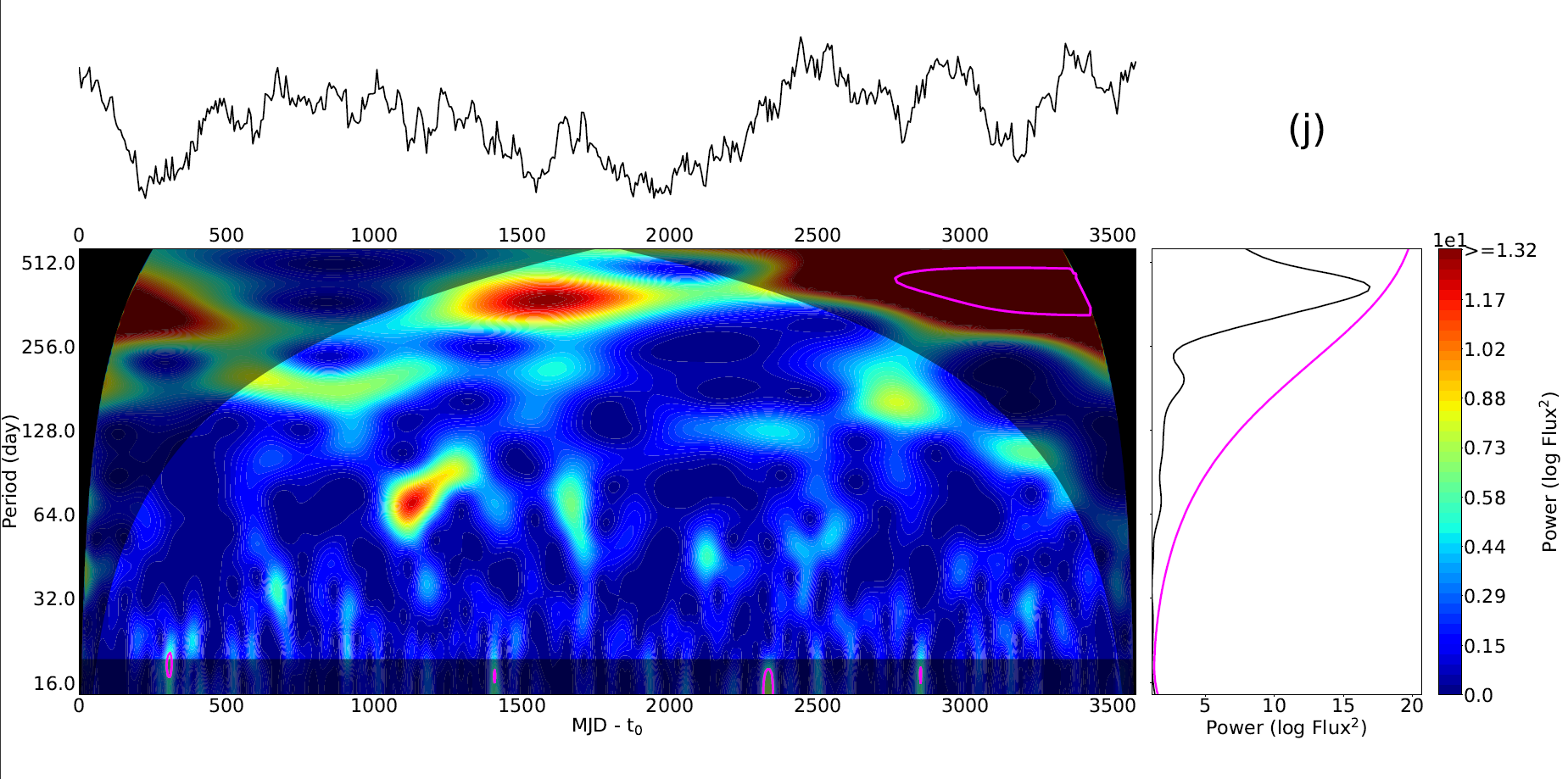}
\caption{Exemplary wavelet scalograms and global periodograms for different stochastic models: (a) ARMA, (b) CARMA, (c) FARIMA with $H=0.25$, (d) FARIMA with $H=0.75$, (e) fBm with $H=0.25$, (f) fBm with $H=0.75$, (g) fGn with $H=0.25$, (h) fGn with $H=0.75$, (i) PLC corresponding to $H=0.25$, (j) PLC corresponding to $H=0.75$. }
\label{fig_out1}
\end{figure*}

To investigate the typical features of different types of processes in the time-frequency domain, the wavelet scalograms for various time series (of length $N=500$) were generated: ARMA\footnote{\label{note1}ARMA and FARIMA were realized with parameters generated in Sect.~\ref{sect::arma_bench}.} and CARMA (both expected to have $H=0.5$), FARIMA\footnotemark[\getrefnumber{note1}] with $d=\pm 0.25$, i.e. $H=0.25$ and $H=0.75$, fBm and fGn with $H=0.25$ and $H=0.75$, PLC with $\beta=1$, i.e. borderline between high-$H$ fGn and low-$H$ fBm (see also Fig.~\ref{fig_Hurst_disc}), and $\beta=2$, i.e. corresponding to $H=0.5$. The values $H=0.25$ and $0.75$ are representative for short- and long-term memory processes, respectively. For each of the 10 process types, 100 random realizations were generated. In Fig.~\ref{fig_out1} we present some typical examples and their scalograms.

\begin{figure}[ht!]
\centering
\includegraphics[width=\columnwidth]{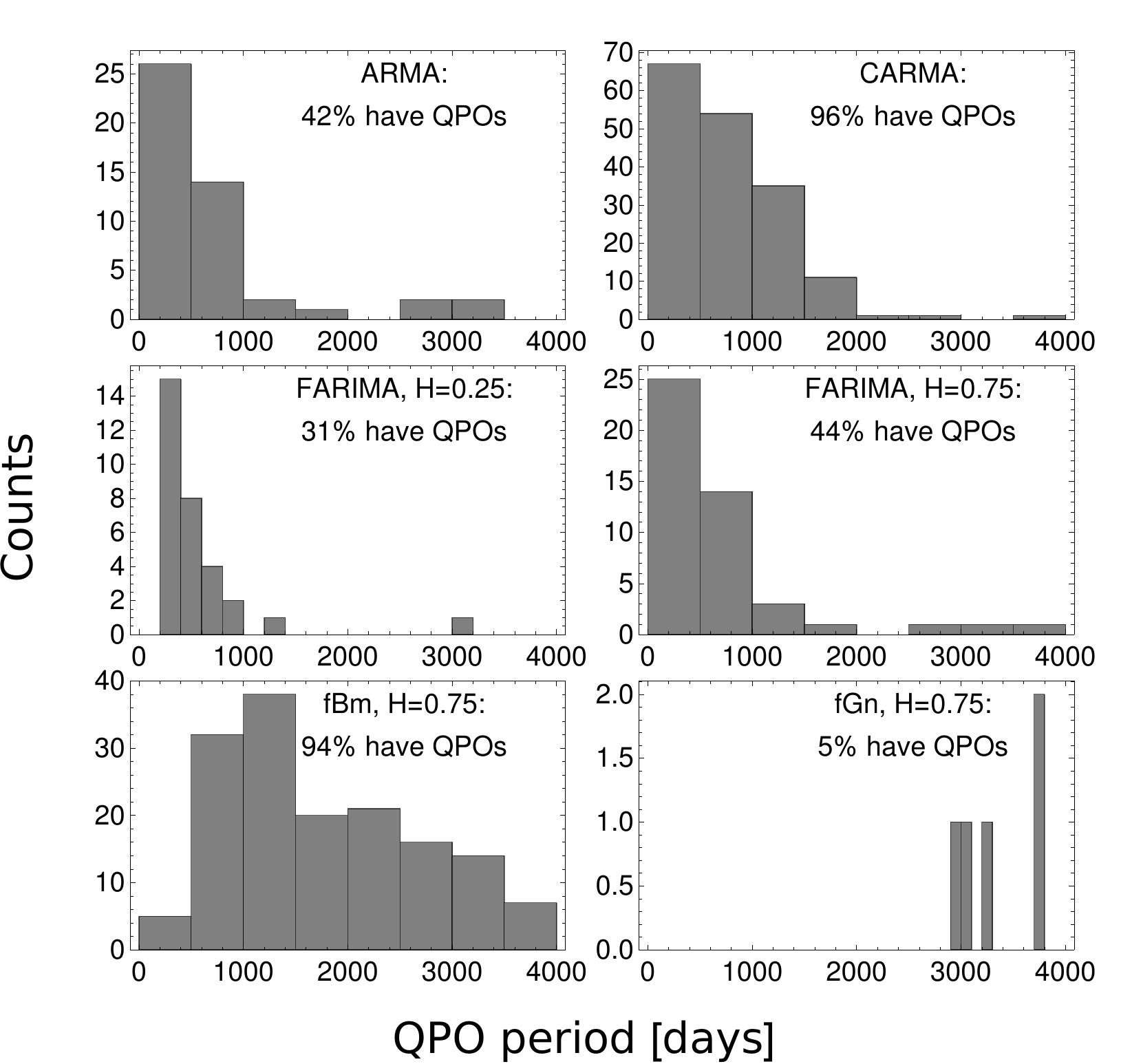}
\caption{Distributions of QPO periods of various simulated stochastic processes found with the wavelet scalograms. For fBm and fGn with $H=0.25$, and PLC with $\beta=1$ and $\beta=2$ the fraction is 0\%, and hence are not displayed here.}
\label{fig_200}
\end{figure}

We observe that fBm with $H=0.75$ is dominated by low frequency components, i.e. short-term variations are negligible compared to the long-term behavior. In fact, in a majority of the realizations the whole global PSD lies above the $3\sigma$ confidence level. Therefore, CARMA models that the scalograms are tested against are poor models for persistent, highly nonstationary fBm. In case of $H=0.25$, the frequencies are more spread out, and a QPO rarely emerges. Usually no structure at all is observed for fGn regardless of $H$. Likewise for PLC processes, for both $\beta=1$ and $\beta=2$. ARMA and FARIMA often ($\sim 30-40\%$ of instances) exhibit QPO-like features. CARMA$(5,3)$ models were set to explicitly contain a QPO, which is nearly always detected by the scalogram, but with a very wide spread of the corresponding periods. This is due to the fact that the testing is made against a CARMA$(1,0)$ background, as justified by the results of CARMA fittings to the blazar LCs of interest (see Sect.~\ref{results::carma} for details). We often record tentative QPOs at very high frequencies in the global wavelet PSD, resulting from individual spikes in the time series that contribute significantly to the total power at a given scale. Therefore, in Fig.~\ref{fig_200} we show the histograms of only QPO periods exceeding 200~days, i.e. focusing on often reported in the literature $\sim$year-long periods. Overall, QPOs arise in ARMA, FARIMA and CARMA models, in whose PSDs they can be immanent, yet stochastic, features.

\subsection{ARMA and FARIMA processes}
\label{sect::arma_bench}

To investigate how accurate is the fitting of ARMA and FARIMA processes, we perform MC simulations in the following manner:
\begin{enumerate}

\item We choose the order of an ARMA$(p,q)$ process to be $(5,3)$ and keep it fixed throughout. It is high enough to allow for nontrival features in the PSDs, but low enough to be computationally efficient.

\item We generate randomly a set of $(\varphi,\theta,\sigma)$ parameters, and verify that it leads to a stationary ARMA process. $\varphi$ and $\theta$ are drawn from a $\mathcal{U}(-1,1)$, and $\sigma$ from a $\mathcal{U}(0.1,1)$ distribution. 

\item We simulate 150 time series of length $N=512$ with the above set of parameters.

\item For every time series, we fit a total of 100 ARMA$(p,q)$ processes, with $p,q\in\{0,\ldots,9\}$. We record their $AIC_c$ and $BIC$ values, based on which we choose the order of the best fitting process. As we have generated 150 time series from the same parameters, we have 150 such best fitting orders, from which we construct a density map in the $(p,q)$ space.

\item We choose the order that is the commonest as the most probable, and record it for further use.

\item We perform points 4. and 5., fitting a FARIMA$(p,d,q)$ process. Additionally, we record the 150 values of $d$ for each set of $(\varphi,\theta,\sigma)$ parameters, and construct an empirical probability density of $d$. 

\item We repeat the above procedure 250 times\footnote{We verified that for 500 iterations the results are practically the same.}, i.e. for 250 different sets of $(\varphi,\theta,\sigma)$ parameters.

\item We obtain therefore sets of density maps for ARMA and FARIMA fittings, and for both fitted models we have 250 such maps based on $AIC_c$, and 250 maps based on $BIC$. In other words, we obtain 250 best fitting ARMA orders based on $AIC_c$ for each set of parameters, 250 best fitting orders based on $BIC$, and similarly two sets of best fitting orders (also 250 values each) for the fitted FARIMA processes.

\item From these four sets of 250 orders each, we construct final density maps.

\end{enumerate}

We next employ exactly the same procedure, but generating the time series (to which we again fit both ARMA and FARIMA processes) from a FARIMA$(5,d,3)$ process. The set of input parameters $(\varphi,\theta,\sigma)$ is hence complemented with $d$, which we choose to be $d=0.3$, i.e. corresponding to $H=0.8$.

We perform such testing (for mock time series generated from both ARMA and FARIMA processes) in two instances:
\begin{enumerate}
\item for {\it clean} data---generated exactly as described above;
\item for {\it noisy} data---values are obtained as drawn from $\mathcal{N}\left(x_k,({\rm max}\{|x_k|\}/5)^2\right)$ distributions, where $x_k$ are the {\it clean} data points, and the fluctuations' standard deviation is set to 20\% of the most extremal value.
\end{enumerate}

\begin{figure*}
\centering
\includegraphics[width=0.49\textwidth]{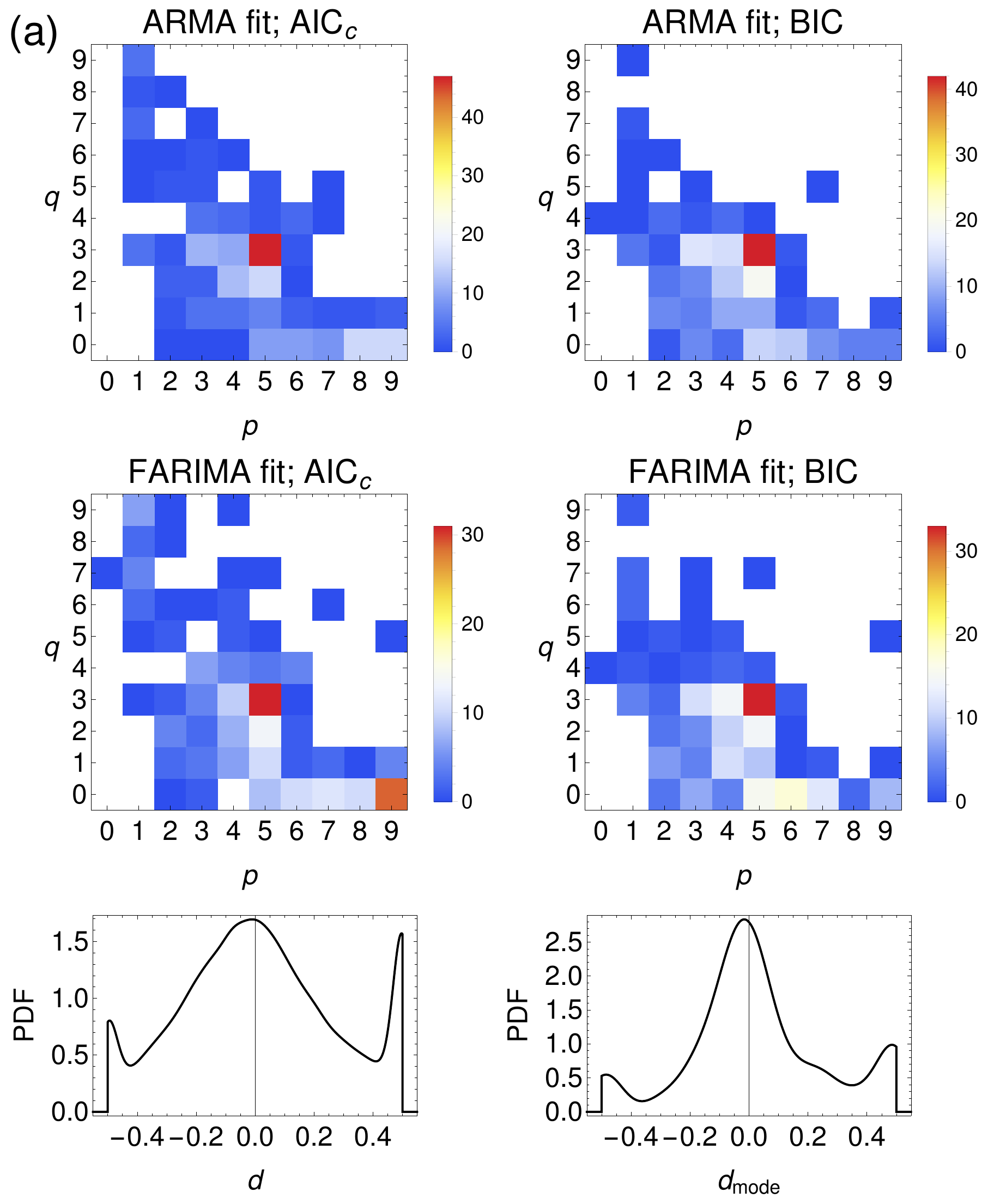}
\includegraphics[width=0.49\textwidth]{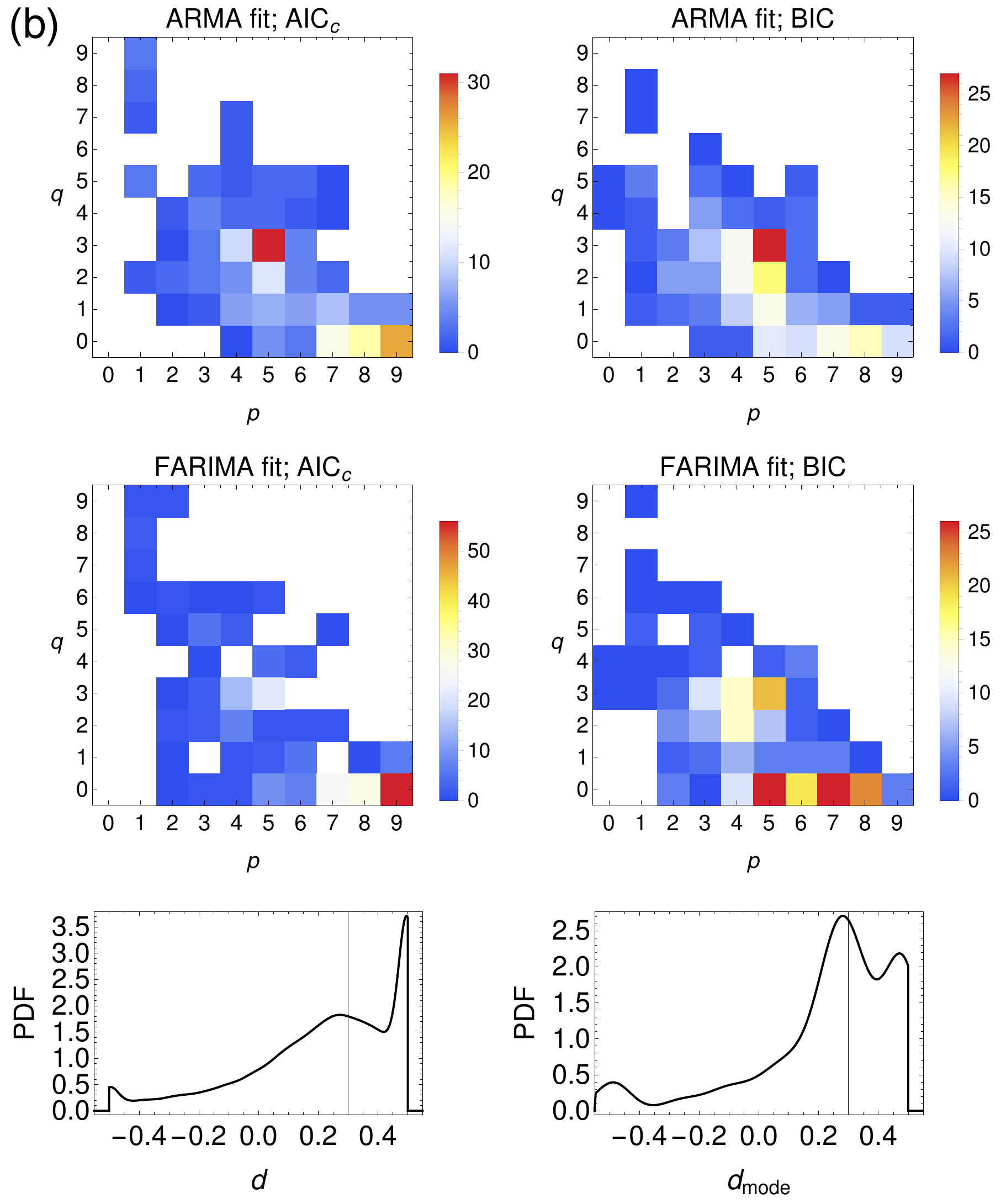}\\
\includegraphics[width=0.49\textwidth]{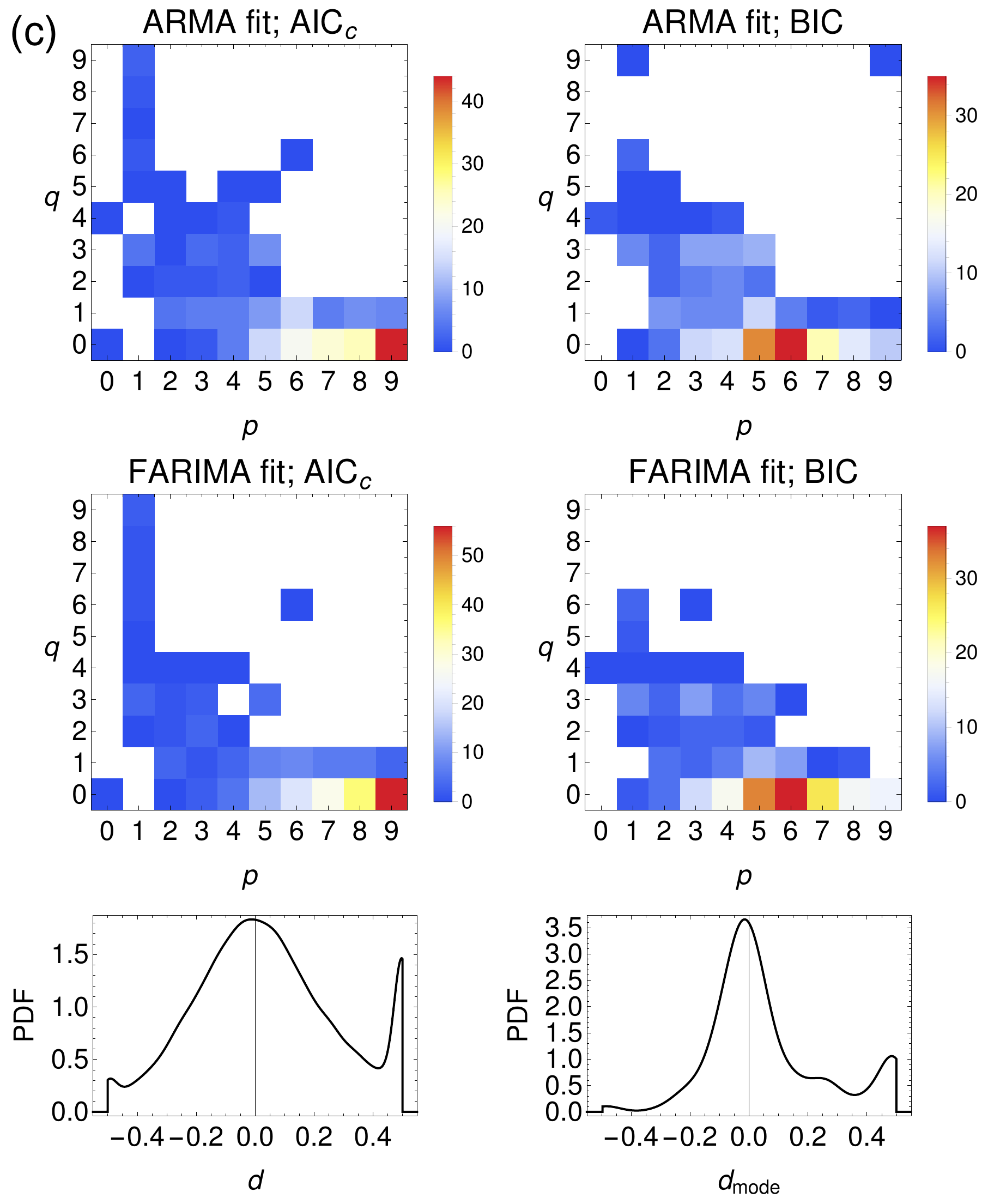}
\includegraphics[width=0.49\textwidth]{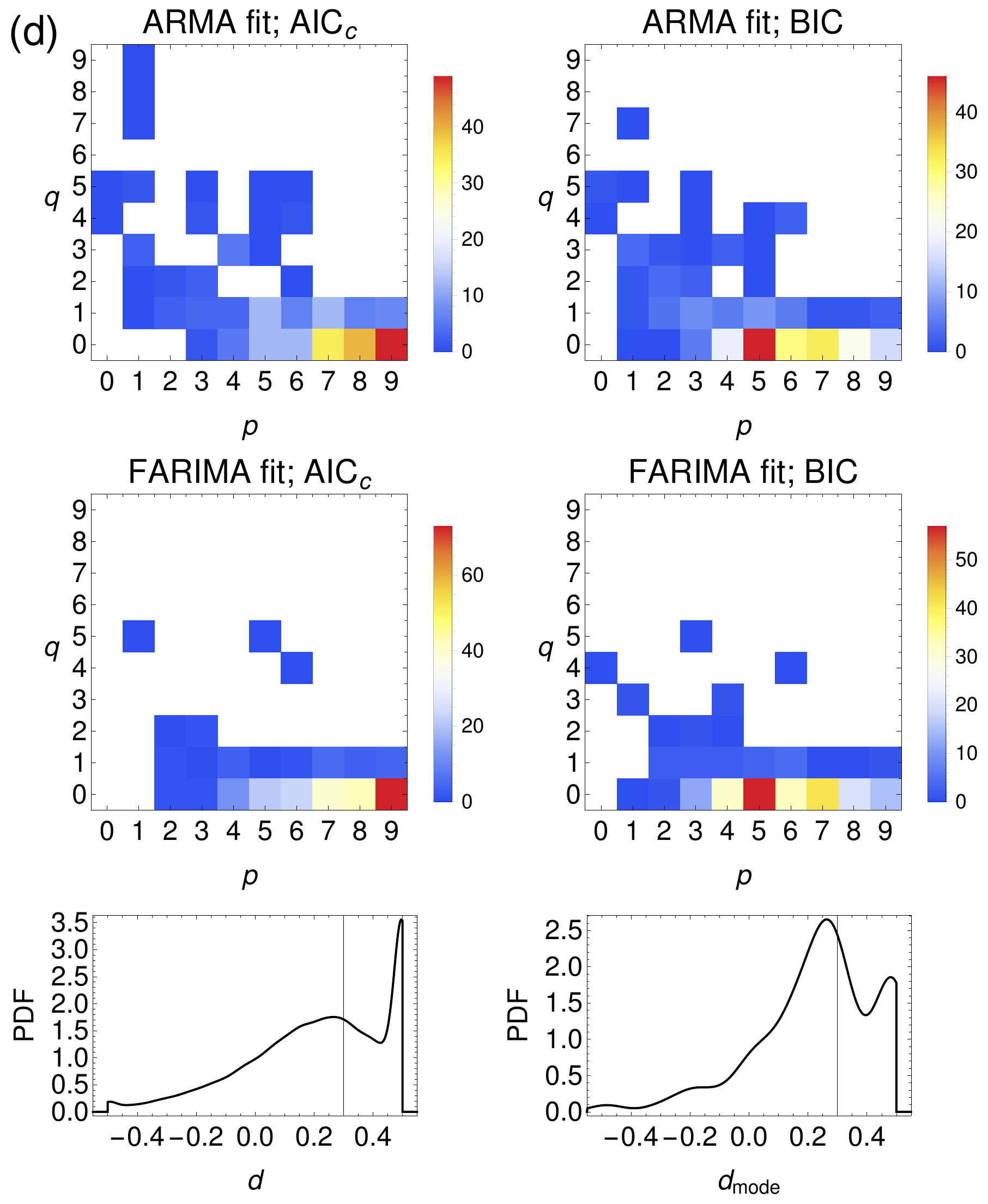}
\caption{Benchmark results for {\it clean} time series generated from (a) an ARMA$(5,3)$, and (b) FARIMA$(5,0.3,3)$ process: density maps of the ARMA$(p,q)$ and FARIMA$(p,d,q)$ best-fit orders according to $AIC_c$ and $BIC$, and distributions of the differencing parameter $d$ and the mode $d_{\rm mode}$ obtained for each set of parameters $(\varphi,\theta,\sigma)$. Vertical lines denote the underlying true value. Panels (c) and (d) are the same as (a) and (b), respectively, but for {\it noisy} data.}
\label{clean_noise}
\end{figure*}

The results for {\it clean} data are displayed in Fig.~\ref{clean_noise}(a) and (b): distributions of the best orders $(p,q)$ and the differencing parameter $d$, generated from an ARMA [Fig.~\ref{clean_noise}(a)] and FARIMA [Fig.~\ref{clean_noise}(b)] processes. Diplayed are the distribution of all $250\times 150$ recorded individual values of $d$, and the distribution of the 250 modes $d_{\rm mode}$, obtained from the distributions of $d$ for each set of parameters $(\varphi,\theta,\sigma)$. In most cases, the true order $(5,3)$ is correctly identified, although in some instances an overabundance of high order AR$(p)$ processes is obtained. This is particularly the case when a FARIMA model was fitted to data generated from a FARIMA process [see Fig.~\ref{clean_noise}(b)]. The distributions of the differencing parameter $d$ are peaked around the correct values (i.e., $d=0$ and $d=0.3$ for ARMA and FARIMA processes, respectively). A prominent peak, however, is observed at $d=0.5$ (and to a smaller extent at $d=-0.5$). This is a sign of nonstationarity present in the data. We stress that all generated time series come from stationary processes, altough some particular realizations might be spuriously identified as nonstationary. Based on the positions of local minima of the distributions, an outcome falling in the range $d\in(-0.4,0.4)$ should be a safe estimate. Values outside this range could be verified by differencing the time series, i.e. examining the set of differences $\{ x_k - x_{k-1} \}$ instead of the original data $\{ x_k \}$. This turns a nonstationary process with a given $d$ to one with $d-1$. Second differences can be employed if the issue persists (although it is rare, and if occurs a careful verification of the quality of the data is strongly suggested).

A less appealing picture is painted when we consider {\it noisy} data [Fig.~\ref{clean_noise}(c) and (d)]. While the distributions of $d$ are of similar shapes as in case of {\it clean} data sets, the order ($p,q)$ is systematically misidentified: a high order AR$(p)$ process is pointed at instead of the correct $(5,3)$. We see that for both {\it clean} and {\it noisy} data, the MA component of the underlying ARMA or FARIMA process is spuriously interpreted as nonstationarity, or adds to the order of the resulting AR process.

The above simulations were repeated for $N=100$ and $N=2000$. A reasonable expectation is that the scatter in the retrieved order would be larger for shorter time series, which is indeed the case---the correct order (within $\Delta p,\Delta q = \pm 1$) was retrieved $\sim 10$ times more often for $N=2000$ than for $N=100$. Moreover, while the retrieval of AR and MA coefficients is satisfactory, with a concentration around the $y=x$ line, where $x$ are the input parameters, and $y$ are the fitted ones, its root-mean square error, as expected, also considerably decreases when $N$ increases (from $~\sim 0.37$ for $N=100$, to $\sim 0.30$ and $\sim 0.27$ for $N=512$ and $N=2000$, respectively.

\subsection{CARMA modeling}

\citet{ryan19} performed very detailed simulations to assess the reliability of fitting CARMA$(1,0)$ and $(2,1)$ models, focusing on retrieving the break time scales and local PL indices $\beta$. They found that CARMA$(2,1)$ is flexible enough to correctly recover $\beta$ even from a pure PL, and identifies the breaks with sufficient accuracy. They also gave a way of assessing the reality of a break: it is likely a physical feature if it is consistently detected by both lower- and higher-order CARMA fits.

\begin{figure}
\centering
\includegraphics[width=0.8\columnwidth]{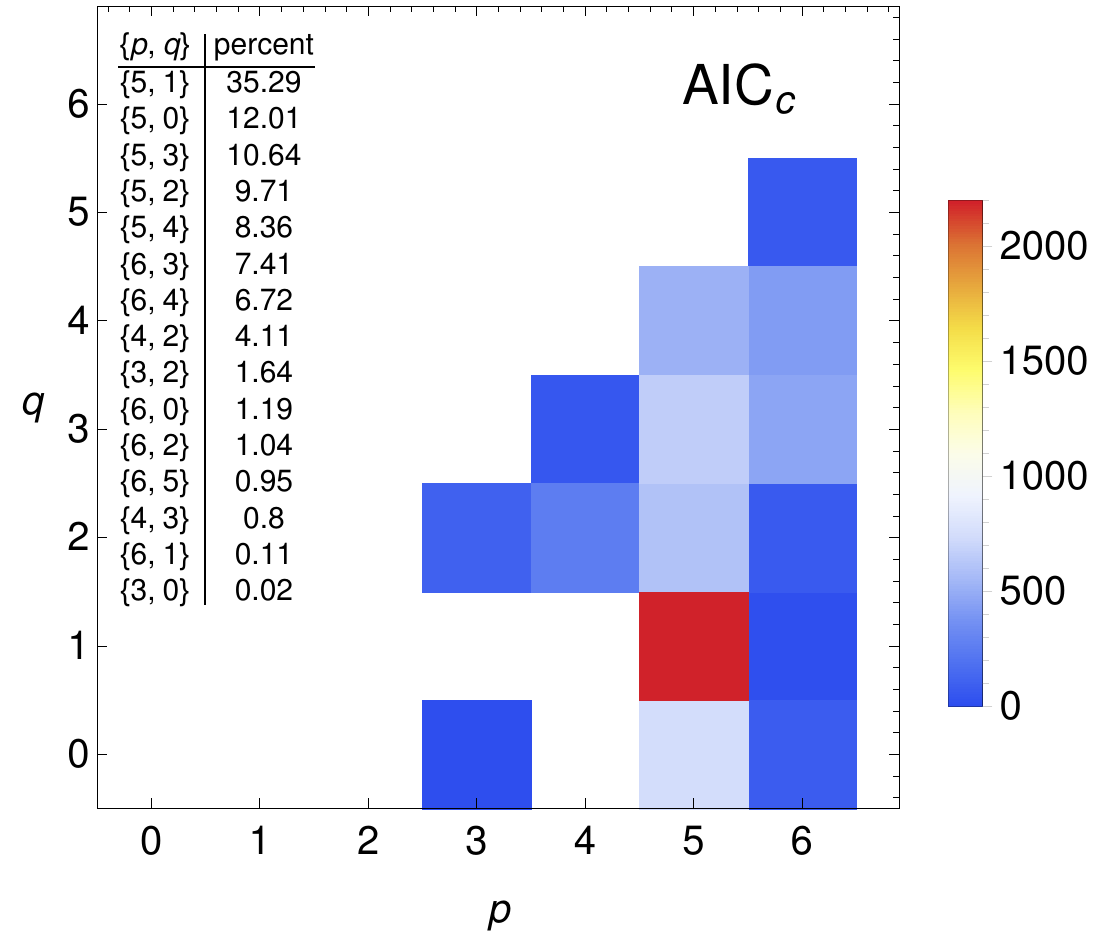}\\
\includegraphics[width=0.8\columnwidth]{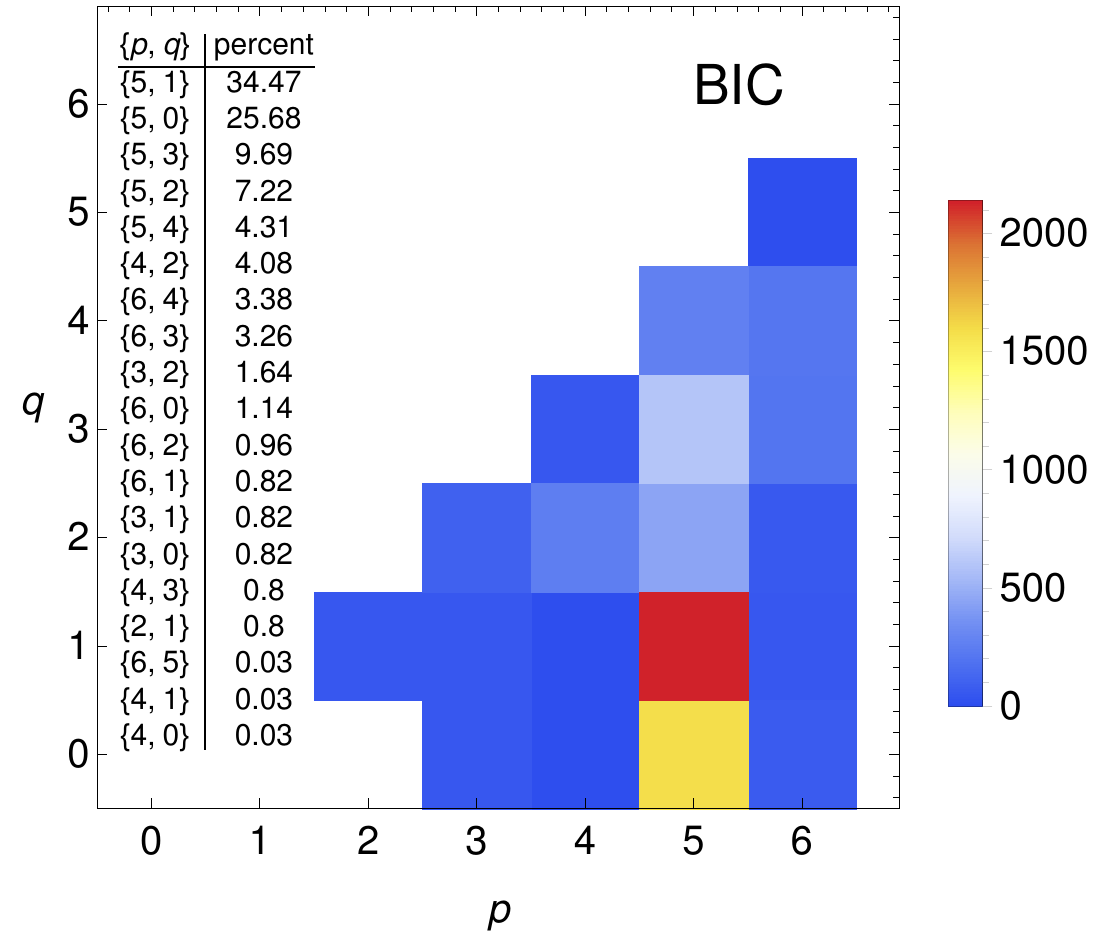}
\caption{Density maps of CARMA$(p,q)$ best-fit orders, fitted to 6222 realizations of CARMA$(5,3)$ process, according to $AIC_c$ and $BIC$. }
\label{fig_carma_pq}
\end{figure}

Herein, we investigate the accuracy of fitting a CARMA model to data in the sense of recovering the correct order $(p,q)$. To do so, we perform MC simulations according to the following procedure:
\begin{enumerate}
\item We choose a relatively high order, $(p,q)=(5,3)$, to challenge the algorithm, and for correspondence with Sect.~\ref{sect::arma_bench}.
\item We choose the values of $\sigma$, the MA parameters $\beta_i$, and the roots of the AR polynomial---in this way we ensure the simulated process to be stationary.
\item We choose the total length of the time series to be $N_{\rm tot}=550$ points; next, we randomly discard 50 of them, creating an unevenly sampled time series with $N=500$ observations (i.e., we end with $\sim 9\%$ of the data missing).
\item We then simulate 100 realizations of the CARMA$(5,3)$ stochastic process with the above described characteristics. The final time series are subject to random flucuations drawn from a truncated Gaussian distribution---these are to mimic observational errors.
\item For each realization, we perform fitting of CARMA models with $1\leqslant p \leqslant 6$ and corresponding $0\leqslant q < p$. For each of these $(p,q)$ pairs we compute the $AIC_c$ and $BIC$, based on which we choose the best process order---the one with the lowest $AIC_c$ or $BIC$. Note that all 100 realizations come from the same underlying process, so we expect to obtain $(5,3)$ or nearby orders in a majority of instances.
\item We build a density map of the obtained best $(p,q)$ pairs in the $p-q$ plane.
\end{enumerate}
Using the above procedure, with realistically spaced irregular data resembling that of the {\it Fermi}-LAT LCs of blazars in our sample, we constrain the validity and reliability of the CARMA fitting. The results are displayed in Fig.~\ref{fig_carma_pq}. About 35\% of realizations are recognized as CARMA$(5,1)$ by both IC, hence the order of the MA part is underestimated. In case of $BIC$, the second most common order is $(5,0)$ in 25\%. The input order $(5,3)$ is returned in 10\% of instances. In total, models with the correct $p=5$ are returned in 76\% and 81\% cases by employing $AIC_c$ and $BIC$, respectively. The lowest order $(1,0)$ is never returned, likewise $(2,1)$ with $AIC_c$, and is among the rarest pointed at by $BIC$. Hence, even when the fitting missed the value $p=5$, higher orders, relatively close to the correct one, were preferred.  Overall, we conclude that the procedure is indeed sufficiently reliable and robust in assessing the order of the fit, and we utilize $BIC$ hereinafter.

\subsection{Hurst exponent}
\label{testing::hurst}

\begin{figure*}
\centering
\includegraphics[width=0.95\textwidth]{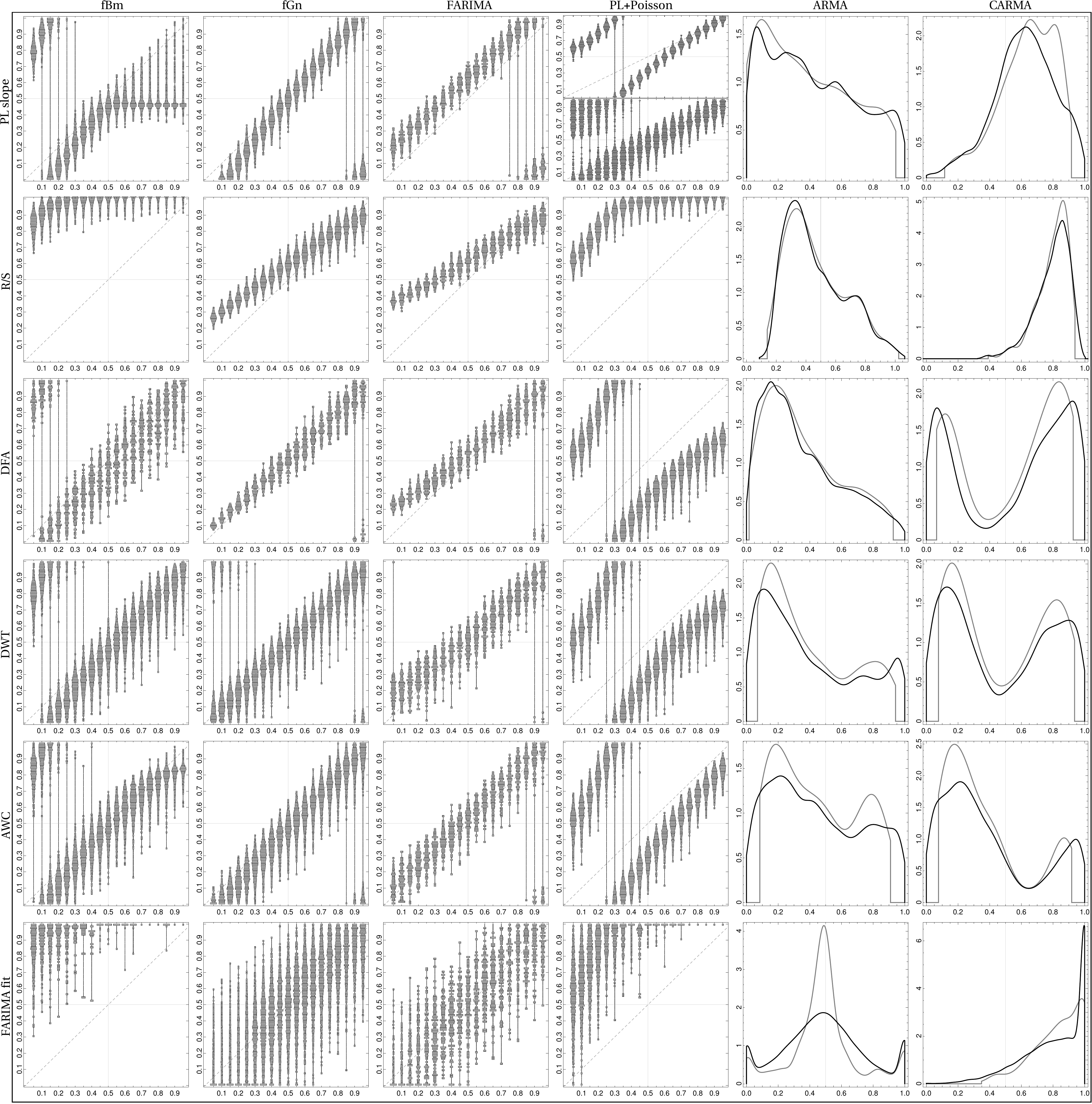}
\caption{Results, in form of violin plots (with $[0,1]\times [0,1]$ range) and PDFs (with a $[0,1]$ support), of the benchmark testing of various algorithms ({\it rows}) for estimating $H$, for different types of stochastic processes ({\it columns}). For the four columns from the left (i.e., labeled fBm, fGn, FARIMA and PLC) the horizontal axis denotes the input $H_{\rm in}$ values, and the vertical axis marks the values $H_{\rm out}$ returned by the respective algorithms (labeling the rows). The diagonal dashed line denotes the equality line. The PLC model, in the PL slope approach, was evaluated in two ways: by fitting a pure PL (the upper panel in the double plot) and a PLC (lower panel). For the two columns on the right (labeled ARMA and CARMA), the plots show PDFs of the obtained values of $H$; for these two processes, $H_{\rm in}=0.5$ (marked with the vertical thin line) is expected. The black line in each plot denotes the distribution for all $250\times 100$ time series, while the gray line is for the 250 medians for each parameter set.}
\label{fig_H}
\end{figure*}

We test the $H$ estimation algorithms from Sect.~\ref{sect::hurst} on the following time series, all of length $N=512$:
\begin{enumerate}
\item fBm and fGn, with $H\in\{0.05,0.1,\ldots,0.95\}$---1000 realizations for each $H$;
\item FARIMA$(5,d,3)$ for a few randomly chosen sets of parameters generated in Sect.~\ref{sect::arma_bench}, with $d\in\{-0.45,-0.4,\ldots,0.45\}$---100 realizations for each $d$; 
\item generated from a PLC spectrum, with $\beta\in\{1.1,1.2,\ldots,2.9\}$---1000 realizations for each $\beta$;
\item ARMA$(5,3)$ for all 250 sets of parameters from Sect.~\ref{sect::arma_bench}---100 realizations for each set;
\item CARMA$(5,3)$ for 250 sets of parameters (different than for ARMA and FARIMA)---100 realizations for each set.
\end{enumerate}
The results are gathered in Fig.~\ref{fig_H}. For fBm, fGn, FARIMA, and PLC---for which an input $H$ value can be explicitly set---we show the resulting distributions as distribution charts (violin plots), i.e. for each input $H_{\rm in}\in\{0.05,0.1,\ldots,0.95\}$ (depicted on the horizontal axis) a distribution of the resulting output $H_{\rm out}$ is shown, that spans the vertical direction on the plots. For a nearly ideal algorithm, the distributions should be centered at $H_{\rm out}=H_{\rm in}$ and with a small dispersion. For ARMA and CARMA, which in turn are expected to give $H=0.5$, the PDFs are plotted of all obtained $H$ values, and for the medians of each parameter set. For the PLC time series we compared estimates from two approaches: one is a straightforward fit of a pure PL, and second is fitting the proper underlying PLC model. In the first approach one should expect the outcomes to be underestimated, because the flat region of the PSD (the high frequency Poisson noise component) will flatten the fitted PSD model and yield lower indices than the input $\beta$ values. For the second approach we expect the results to be roughly appropriate, however from Sect.~\ref{sect::FourLS} we already know that the expected scatter of the calculated value of $\beta$ is substantial, hence the distributions should be wide. This is indeed the case, as shown in the double panel in the first row of Fig.~\ref{fig_H}. This clearly illustrates the importance of validating the presence of a PL component in the PSD of any examined time series---we stress that the algorithms for extracting $H$ are not black boxes: the long-range dependence manifests itself through a PL PSD.   

No single algorithm turns out to be best; some work well for a particular stochastic process, but fail miserably for others, e.g. fitting a PL rarely gives the correct estimate for fBm, but works quite well for fGn. The R/S method significantly overestimates $H$ for all examined processes except for ARMA ones, and fails completely for nonstationary fBm \citep{north1994,gilmore02}. Likewise, it overestimates for CARMA processes. DFA, DWT and AWC give consistent and roughly correct results for processes with a nearly PL PSD (i.e. fBm, fGn, and FARIMA), but miss the correct $H=0.5$ in case of ARMA and CARMA. Fitting a FARIMA model and inferring the value of $H$ from the obtained $d$ works well for fGn, FARIMA and ARMA, although with a big dispersion around the correct value. However, for nonstationary processes it clusters at $d=0.5$. This issue might be resolved by differencing, though, as discussed in Sect.~\ref{sect::arma_bench}. In many cases one can observe that very high and very low $H$ values lead to improper characterization of the time series, resulting in drastic misclassification, e.g. the DFA tends to return a very high $H$ for processes with very low input $H$. Finally, the PSD of CARMA process [Eq.~(\ref{eq37})] cannot be considered to be a PL even approximately (for the employed sets of parameters and frequency ranges of the PSDs), hence the long-term dependence is difficult to assess.

We therefore recommend to employ simultaneously a few algorithms when computing $H$ and examine the consistency of the obtained estimates. The shape of the investigated time series' PSD should be also evaluated to justify the presence of long-term dependence before any attempts to extract $H$ are undertaken.

\subsection{$\mathcal{A-T}$ plane}

In Sect.~\ref{sect::ATplane}, the $\mathcal{A}-\mathcal{T}$ plane was described, and the locations of different processes with PSDs of PLC form were displayed in Fig.~\ref{fig_Zunino}. Herein we investigate what regions of the $\mathcal{A}-\mathcal{T}$ plane is occupied by some other stochastic processes considered in this work.

\begin{figure*}
\centering
\includegraphics[width=\textwidth]{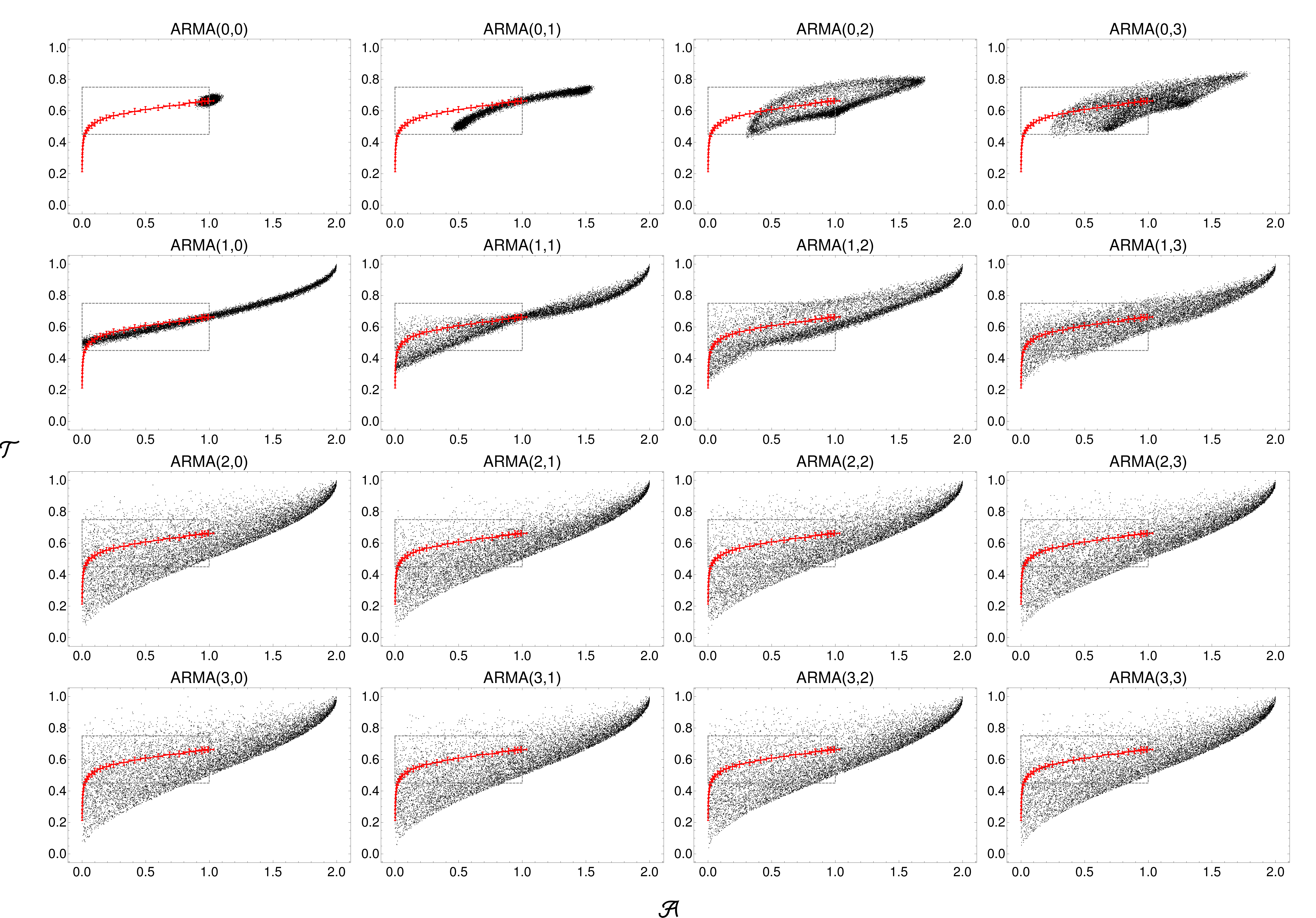}
\caption{Stationary ARMA$(p,q)$ processes, with $0\leqslant p\leqslant 3,\,0\leqslant q\leqslant 3$, in the $\mathcal{A-T}$ plane. The gray dashed box encloses the region from Fig.~\ref{fig_Zunino}; the red line is the pure PL with $0\leqslant\beta\leqslant 3$ (i.e., red line from Fig.~\ref{fig_Zunino}). }
\label{fig_ARMA_AT}
\end{figure*}

We examine the family of ARMA$(p,q)$ processes, with $0\leqslant p\leqslant 3,\,0\leqslant q\leqslant 3$. We generate $10^4$ sets of AR and MA coefficients, drawn uniformly from the regions of the parameter space of the AR coefficients $\varphi_k$ that ensure stationarity, with MA coefficients drawn from $\mathcal{U}\left(-1,1\right)$ and compute the locations $\left(\mathcal{A},\mathcal{T}\right)$. Results are displayed in Fig.~\ref{fig_ARMA_AT}. We also checked whether the regions occupied by each process change when the coefficients are drawn from a $\mathcal{U}\left(-5,5\right)$ distribution, and find that the boundaries are the same.
The ARMA$(0,0)$ process is just white noise, so it clusters around $\left(\mathcal{A},\mathcal{T}\right) = \left( 1, 2/3 \right)$. The PSD of ARMA$(1,0)$ is a close approximation to a PL, hence its $\left(\mathcal{A},\mathcal{T}\right)$ values closely follow the true PL line depicted in the plots, deviating only for $\beta \gtrsim 2$. We emphasize that the pure PL line follows the fBm line \citep{tarnopolski16d,zunino17,tarnopolski19b}. Other ARMA models spread out into larger parts of the $\left(\mathcal{A},\mathcal{T}\right)$ plane, with less internal structure for higher orders. See \citep{tarnopolski19b} for an analytic treatment of the regions of availability for ARMA processes, as well as fBm and fGn.

\section{Results}
\label{results}

\subsection{Fourier spectra}
\label{results::fourier}

\begin{figure*}
\centering
\includegraphics[width=\textwidth]{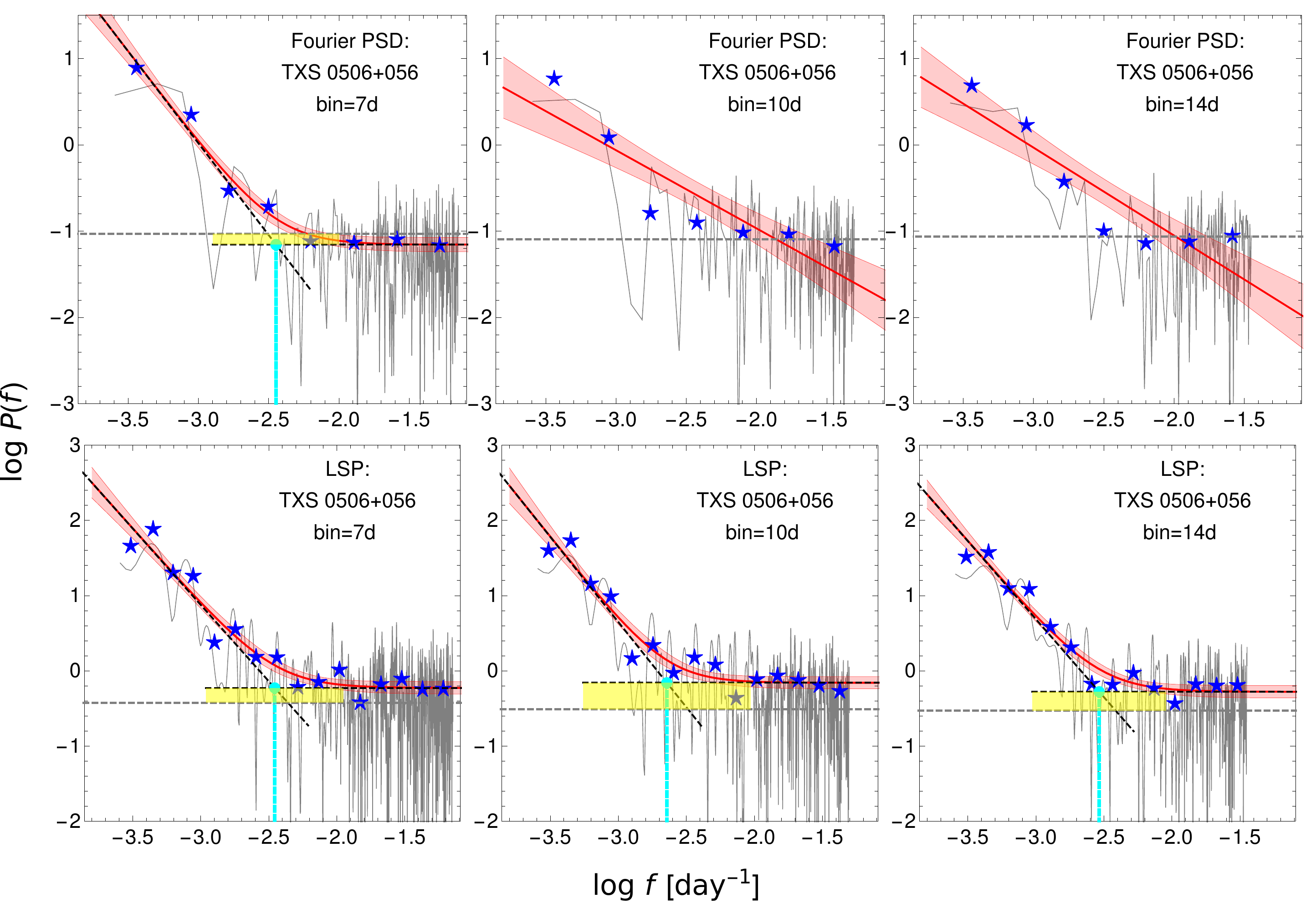}
\caption{PSD fits with PL and PLC: Fourier spectra ({\it top row}) and LSPs ({\it bottom row}) of TXS~0506+056. The gray line is the raw LSP, and the blue stars are the binned periodogram to which fitting was performed. The red solid line is the best fit, with the lighter red region around it marking the 68\% confidence interval. The black dashed lines are the PL component and $C$ level [the parameter from Eq.~(\ref{eq11}) returned by the fitting] whose intersection is marked with the cyan point. The vertical cyan line denotes the frequency $f_0$ above which Poisson noise dominates over the PL component. It is calculated as $P_{\rm norm}/f_0^\beta = C \Rightarrow f_0 = (P_{\rm norm}/C)^{1/\beta}$, and the error propagation law is used to estimate the uncertainty. The width of the yellow rectangle denotes the standard error of $\log f_0$. The horizontal gray dashed lines are the Poisson noise levels inferred from data as per Eq.~(\ref{eq9}) and (\ref{eq9_LS}). (The complete figure set (11 images) is available in the online journal.)}
\label{fig_Fourier_results}
\end{figure*}

The Fourier spectra were generated and binned, as described in Sect.~\ref{methods::fourier}, for each of the 7, 10, and 14~d binnings. Next, Eq.~(\ref{eq10}) and (\ref{eq11}) were fitted in each case\footnote{We checked also the smoothly broken PL, where a transition between PLs with low- and high-frequency indices $\beta_1$ and $\beta_2$ occurs at a break frequency $f_{\rm break}$ (see \citealt{zywucka20}), but it was not competing in any instance.}. The better model was chosen based on the $AIC_c$ values; if they obeyed $\Delta_i<2$, we chose the pure PL model as the adequate one since it is simpler. The resulting $\beta$ indices and their standard errors are gathered in Table~\ref{table1}; plots of the best fits are displayed in Fig.~\ref{fig_Fourier_results}. The PSDs of FSRQs are steeper than those of BL Lacs. Most of the objects gave consistent $\beta$ estimates among the three bins, except for Mrk 501\footnote{This source is dim, hence its variations are dominated by the observational (Poisson) noise.}, which yielded a wide range: 1.96, 2.53, and 1.73. Nevertheless, the PSDs are systematically flatter than in a red noise case, $1/f^2$, with a mean of 1.19 and standard deviation of 0.43. One BL Lac, PKS 2155$-$304, exhibits $\beta\lesssim 1$, i.e. is significantly flatter than pink noise, $1/f$. On the other hand, two FSRQs, B2 1520+31 and B2 1633+38, are steeper than pink noise, yielding $\beta\in(1.2,1.7)$. The remaining objects, both BL Lacs and FSRQs, are very close to a pink noise description.
\begin{deluxetable}{cccc}
\tabletypesize{\footnotesize}
\tablecolumns{4}
\tablewidth{0pt}
\tablecaption{PL indices $\beta$ from Fourier PSDs. \label{table1}}
\tablehead{
\colhead{Source} & \colhead{7 d} & \colhead{10 d} & \colhead{14 d} \\
\tiny{(1)} & \tiny{(2)} & \multicolumn{1}{c}{\tiny{(3)}} & \tiny{(4)} }
\startdata
\multicolumn{4}{c}{BL Lacs}\\\hline
Mrk 501 & $1.96\pm 0.45^B$ & $2.53\pm 1.00^B$ & $1.73\pm 0.36^B$ \\
Mrk 421 & $0.93\pm 0.12^A$ & $0.94\pm 0.13^A$ & $0.84\pm 0.17^A$ \\
PKS 0716+714 & $0.97\pm 0.21^A$ & $0.94\pm 0.25^A$ & $0.94 \pm 0.33^A$\\
PKS 2155$-$304 & $0.67\pm 0.12^A$ & $0.73\pm 0.10^A$ & $0.78\pm 0.12^A$ \\
TXS 0506+056 & $2.13\pm 0.24^B$ & $0.91\pm 0.21^A$ & $1.02\pm 0.22^A$ \\\hline
\multicolumn{4}{c}{FSRQs}\\\hline
PKS 1510$-$089 & $1.03\pm 0.08^A$ & $1.09\pm 0.09^A$ & $1.06\pm 0.11^A$ \\
3C 279 & $1.12\pm 0.07^A$ & $1.14\pm 0.07^A$ & $1.17\pm 0.08^A$ \\
B2 1520+31 & $1.69\pm 0.06^B$ & $1.38\pm 0.09^A$ & $1.34\pm 0.10^A$ \\
B2 1633+38 & $1.22\pm 0.15^A$ & $1.25\pm 0.18^A$ & $1.03\pm 0.16^A$ \\
3C 454.3 & $1.68\pm 0.13^A$ & $1.73\pm 0.13^A$ & $1.79\pm 0.14^A$ \\
PKS 1830$-$211 & $1.05\pm 0.15^A$ & $1.03\pm 0.16^A$ & $1.14\pm 0.20^A$ \\
\enddata
\tablecomments{$^A$Pure PL.\\$^B$PLC.\\ Columns: (1) source name; (2) $\beta$ for 7 d binning; (3) $\beta$ for 10 d binning; (4) $\beta$ for 14 d binning. }
\end{deluxetable}

As demonstrated in Sect.~\ref{sect::FourLS}, extracting the parameters of Eq.~(\ref{eq11}) from a number of realizations of the same process leads to a wide range of results, obscuring the true underlying process. Hence, the obtained values of $\beta$ are not entirely reliable. On the other hand, except for Mrk 501, employing different binnings gave highly consistent results, and in most cases it turned out that the Poisson noise did not influence the PSDs much, i.e. a pure PL was chosen as the better description. To test the robustness, from each LC we generated 1000 MC realizations by drawing the observations randomly from $\mathcal{N}\left(\mu,\sigma^2\right)$, where $\mu$ is the observed value, and $\sigma$ is its standard error. That is, we produced 1000 stochastic realizations of the LCs allowed by the observable uncertainties. The results of such procedure are gathered in Table~\ref{table2}, where we display the mode and standard deviation of the resulting distributions of best-fit indices $\beta$, and the mode of their standard errors $\Delta\beta$, which we interpret as a typical error.

\begin{deluxetable*}{cccccccccc}
\tabletypesize{\footnotesize}
\tablecolumns{10}
\tablewidth{0pt}
\tablecaption{Distributions' characteristics of PL indices $\beta$ and their typical errors, $\Delta\beta$, from Fourier PSDs---MC realization of LCs by varying their values within the error bars. \label{table2}}
\tablehead{
Source & \multicolumn{3}{c}{7 d} & \multicolumn{3}{c}{10 d} & \multicolumn{3}{c}{14 d} \\
       & mode & std & typ. err. & mode & std & typ. err. & mode & std & typ. err. \\
\tiny{(1)} & \tiny{(2)} & \tiny{(3)} & \tiny{(4)} & \tiny{(5)} & \tiny{(6)} & \tiny{(7)} & \tiny{(8)} & \tiny{(9)} & \tiny{(10)} }
\startdata
\multicolumn{10}{c}{BL Lacs}\\ \hline
Mrk 501        & 1.61 & 0.17 & 0.25 & 2.07 & 0.27 & 0.52 & 1.74 & 0.33 & 0.32 \\
Mrk 421        & 0.80 & 0.03 & 0.15 & 0.84 & 0.04 & 0.23 & 0.74 & 0.06 & 0.24 \\
PKS 0716+714   & 0.89 & 0.02 & 0.20 & 0.88 & 0.04 & 0.24 & 0.87 & 0.04 & 0.31 \\
PKS 2155$-$304 & 0.57 & 0.07 & 0.11 & 0.52 & 0.11 & 0.09 & 0.64 & 0.08 & 0.11 \\ 
TXS 0506+056   & 1.94 & 0.42 & 0.20 & 0.75 & 0.67 & 0.18 & 0.82 & 0.39 & 0.21 \\\hline
\multicolumn{10}{c}{FSRQs}\\ \hline
PKS 1510$-$089 & 0.97 & 0.06 & 0.08 & 1.03 & 0.02 & 0.08 & 1.02 & 0.02 & 0.11 \\
3C 279         & 1.05 & 0.02 & 0.07 & 1.12 & 0.04 & 0.15 & 1.13 & 0.03 & 0.08 \\
B2 1520+31     & 1.72 & 0.13 & 0.11 & 1.26 & 0.19 & 0.11 & 1.28 & 0.15 & 0.11 \\
B2 1633+38     & 1.05 & 0.06 & 0.14 & 1.08 & 0.08 & 0.17 & 1.18 & 0.04 & 0.18 \\
3C 454.3       & 1.50 & 0.04 & 0.15 & 1.54 & 0.08 & 0.16 & 1.59 & 0.06 & 0.16 \\
PKS 1830$-$211 & 0.90 & 0.07 & 0.14 & 0.94 & 0.06 & 0.16 & 1.03 & 0.07 & 0.19 \\
\enddata
\tablecomments{Columns: (1) source name; (2) mode of the $\beta$ distribution for 7 d binning; (3) standard deviation of the $\beta$ distribution for 7 d binning; (4) typical error, i.e. mode of the $\Delta\beta$ distribution for 7 d binning; (5) mode of the $\beta$ distribution $\beta$ for 10 d binning; (6) standard deviation of the $\beta$ distribution for 10 d binning; (7) typical error, i.e. mode of the $\Delta\beta$ distribution for 10 d binning; (8) mode of the $\beta$ distribution $\beta$ for 14 d binning; (9) standard deviation of the $\beta$ distribution for 14 d binning; (10) typical error, i.e. mode of the $\Delta\beta$ distribution for 14 d binning. }
\end{deluxetable*}

The values of $\beta$ from both Tables~\ref{table1} and \ref{table2} are in agreement, hence the uncertainties of the flux do not play an important role in estimating $\beta$. However, it is crucial to emphasize that this only means that the estimates of the PL parameters for {\it a given LC} can be trustworthy when it comes to impact of the errors on the outcome. Still, this particular time series might be a peculiar realization of the process governing the observed variability, and need not represent the {\it true} parameters ultimately. In fact, as implied by the benchmark testing from Sect.~ \ref{sect::FourLS}, the mismatch between the (unknown) true value and the extracted one can be severe, especially when Poisson noise constitutes a significant component in the PSD.

\subsection{Lomb-Scargle periodograms}
\label{results::LSP}

The same analysis as in Sect.~\ref{results::fourier} was performed with the LSP (see Sect.~\ref{methods::ls}). The fitted $\beta$ indices are gathered in Table~\ref{table3}; plots of the best fits are displayed in Fig.~\ref{fig_Fourier_results}. The values of $\beta$ are slightly lower, and their errors are smaller than obtained by fitting Fourier spectra. The latter is caused by the ability of the LSP to generate PSDs with larger number of points, hence constraining the fits better. In Table~\ref{table4} the results of MC realizations drawn by varying the LC values within the error bars are displayed. We arrive at a similar conclusion as in Sect.~\ref{results::fourier}, i.e. the uncertainties of the flux do not affect the estimated PSD shape significantly. We repeat, though, that the PL parameters depend strongly on the particular realization of a time series, as was demonstrated in Sect.~\ref{sect::FourLS}.

\begin{deluxetable}{cccc}
\tabletypesize{\footnotesize}
\tablecolumns{4}
\tablewidth{0pt}
\tablecaption{PL indices $\beta$ from LSPs. \label{table3}}
\tablehead{
\colhead{Source} & \colhead{7 d} & \colhead{10 d} & \colhead{14 d} \\
\tiny{(1)} & \tiny{(2)} & \multicolumn{1}{c}{\tiny{(3)}} & \tiny{(4)} }
\startdata
\multicolumn{4}{c}{BL Lacs}\\\hline
Mrk 501 & $1.83\pm 0.20^B$ & $2.39\pm 0.22^B$ & $2.08\pm 0.22^B$ \\
Mrk 421 & $0.96\pm 0.07^A$ & $0.95\pm 0.08^A$ & $0.88\pm 0.08^A$ \\
PKS 0716+714 & $0.97\pm 0.09^A$ & $0.97\pm 0.11^A$ & $1.03\pm 0.12^A$ \\
PKS 2155$-$304 & $0.68\pm 0.10^A$ & $0.74\pm 0.10^A$ & $0.78\pm 0.11^A$ \\
TXS 0506+056 & $2.03\pm 0.26^B$ & $2.26\pm 0.34^B$ & $2.07\pm 0.25^B$ \\\hline
\multicolumn{4}{c}{FSRQs}\\\hline
PKS 1510$-$089 & $0.96\pm 0.07^A$ & $0.99\pm 0.09^A$ & $0.98\pm 0.10^A$ \\
3C 279 & $1.17\pm 0.09^A$ & $1.11\pm 0.11^A$ & $1.16\pm 0.12^A$ \\
B2 1520+31 & $1.37\pm 0.14^B$ & $1.20\pm 0.08^A$ & $1.22\pm 0.09^A$ \\
B2 1633+38 & $1.21\pm 0.08^A$ & $1.62\pm 0.18^B$ & $1.29\pm 0.10^A$ \\
3C 454.4 & $1.95\pm 0.19^B$ & $2.12\pm 0.21^B$ & $2.05\pm 0.24^B$ \\
PKS 1830$-$211 & $1.18\pm 0.10^A$ & $1.19\pm 0.11^A$ & $1.23\pm 0.12^A$ \\
\enddata
\tablecomments{$^A$Pure PL.\\$^B$PLC.\\ Columns: (1) source name; (2) $\beta$ for 7 d binning; (3) $\beta$ for 10 d binning; (4) $\beta$ for 14 d binning. }
\end{deluxetable}

\begin{deluxetable*}{cccccccccc}
\tabletypesize{\footnotesize}
\tablecolumns{10}
\tablewidth{0pt}
\tablecaption{Same as Table~\ref{table2}, but from LSPs. \label{table4}}
\tablehead{
Source & \multicolumn{3}{c}{7 d} & \multicolumn{3}{c}{10 d} & \multicolumn{3}{c}{14 d} \\
       & mode & std & typ. err. & mode & std & typ. err. & mode & std & typ. err. \\
\tiny{(1)} & \tiny{(2)} & \tiny{(3)} & \tiny{(4)} & \tiny{(5)} & \tiny{(6)} & \tiny{(7)} & \tiny{(8)} & \tiny{(9)} & \tiny{(10)} }
\startdata
\multicolumn{10}{c}{BL Lacs}\\ \hline
Mrk 501        & 1.86 & 0.27 & 0.23 & 2.42 & 0.33 & 0.26 & 2.12 & 0.37 & 0.30 \\
Mrk 421        & 0.85 & 0.08 & 0.07 & 0.83 & 0.11 & 0.09 & 0.77 & 0.30 & 0.09 \\
PKS 0716+714   & 0.90 & 0.02 & 0.09 & 0.91 & 0.04 & 0.11 & 0.97 & 0.03 & 0.12 \\
PKS 2155$-$304 & 0.56 & 0.05 & 0.09 & 0.56 & 0.26 & 0.10 & 0.62 & 0.08 & 0.12 \\ 
TXS 0506+056   & 2.09 & 0.20 & 0.32 & 2.12 & 0.30 & 0.34 & 1.96 & 0.24 & 0.29 \\\hline
\multicolumn{10}{c}{FSRQs}\\ \hline
PKS 1510$-$089 & 0.91 & 0.06 & 0.07 & 0.95 & 0.03 & 0.08 & 0.94 & 0.03 & 0.10 \\
3C 279         & 1.05 & 0.02 & 0.09 & 1.12 & 0.04 & 0.11 & 1.13 & 0.03 & 0.12 \\
B2 1520+31     & 1.36 & 0.16 & 0.19 & 1.07 & 0.12 & 0.08 & 1.09 & 0.12 & 0.09 \\
B2 1633+38     & 1.49 & 0.15 & 0.16 & 1.58 & 0.12 & 0.18 & 1.61 & 0.21 & 0.19 \\
3C 454.3       & 1.94 & 0.09 & 0.25 & 2.08 & 0.10 & 0.24 & 2.08 & 0.13 & 0.24 \\
PKS 1830$-$211 & 1.03 & 0.05 & 0.10 & 1.09 & 0.04 & 0.11 & 1.13 & 0.04 & 0.12 \\
\enddata
\tablecomments{Columns: (1) source name; (2) mode of the $\beta$ distribution for 7 d binning; (3) standard deviation of the $\beta$ distribution for 7 d binning; (4) typical error, i.e. mode of the $\Delta\beta$ distribution for 7 d binning; (5) mode of the $\beta$ distribution $\beta$ for 10 d binning; (6) standard deviation of the $\beta$ distribution for 10 d binning; (7) typical error, i.e. mode of the $\Delta\beta$ distribution for 10 d binning; (8) mode of the $\beta$ distribution $\beta$ for 14 d binning; (9) standard deviation of the $\beta$ distribution for 14 d binning; (10) typical error, i.e. mode of the $\Delta\beta$ distribution for 14 d binning. }
\end{deluxetable*}

We also note that the PSDs of Mrk 501 behave in an unorganized manner---i.e., the spread of their $\beta$ is wide, and the values do not overlap within errors. In particular, for the 10 d binning the PSD is steepest. This is most likely due to it being a dim object, where the variability is obscured by observational fluctuations and statistical noise. Despite visually similar LCs among the three binnings, we clearly observe that the fitted form of the PSD is strongly dependent on the particular stochastic realization, in agreement with our findings from Sect.~\ref{sect::FourLS}.

\subsection{Wavelet scalograms}
\label{results::wavepal}

\begin{figure*}
\includegraphics[width=\textwidth]{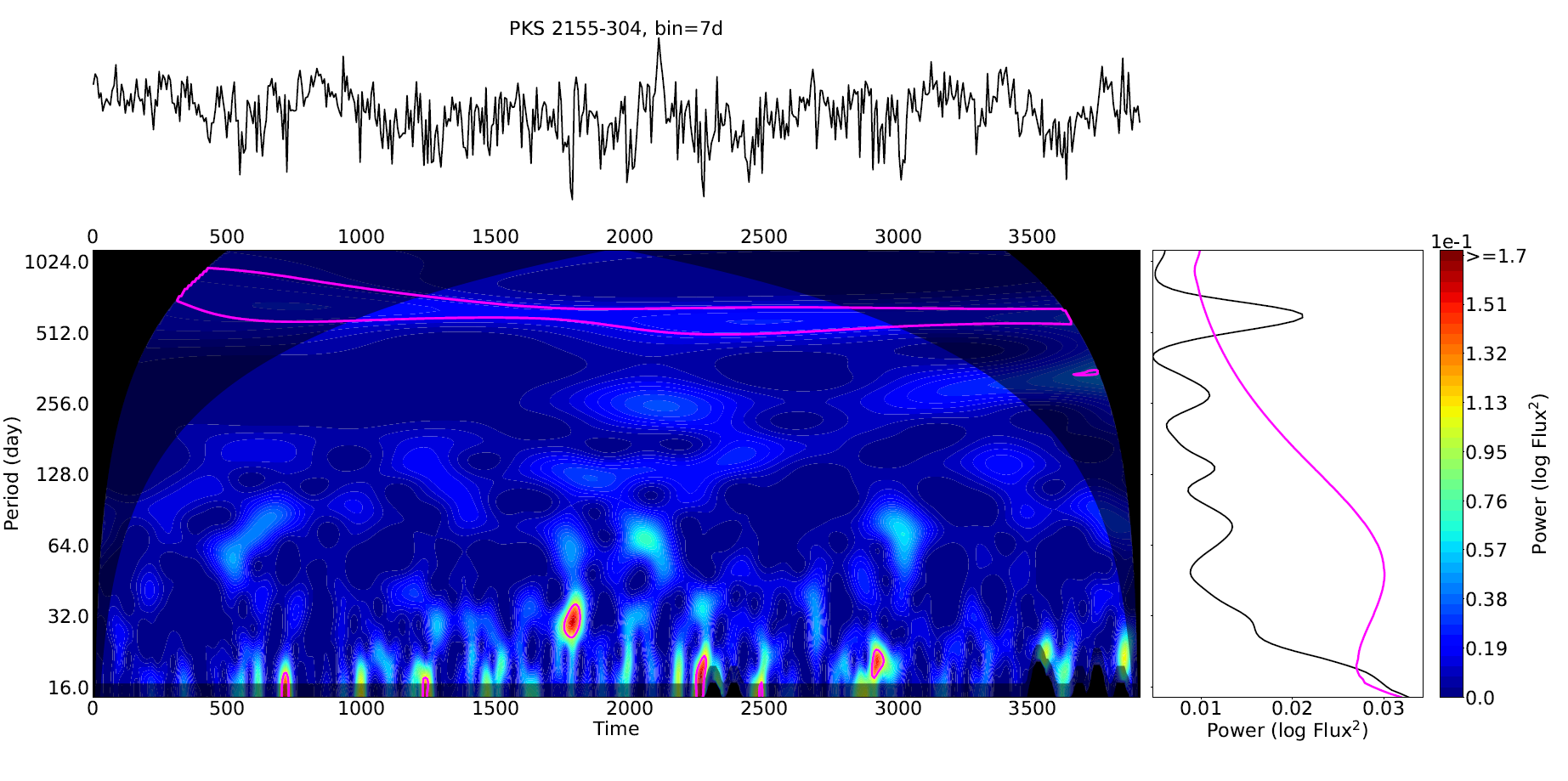}
\includegraphics[width=\textwidth]{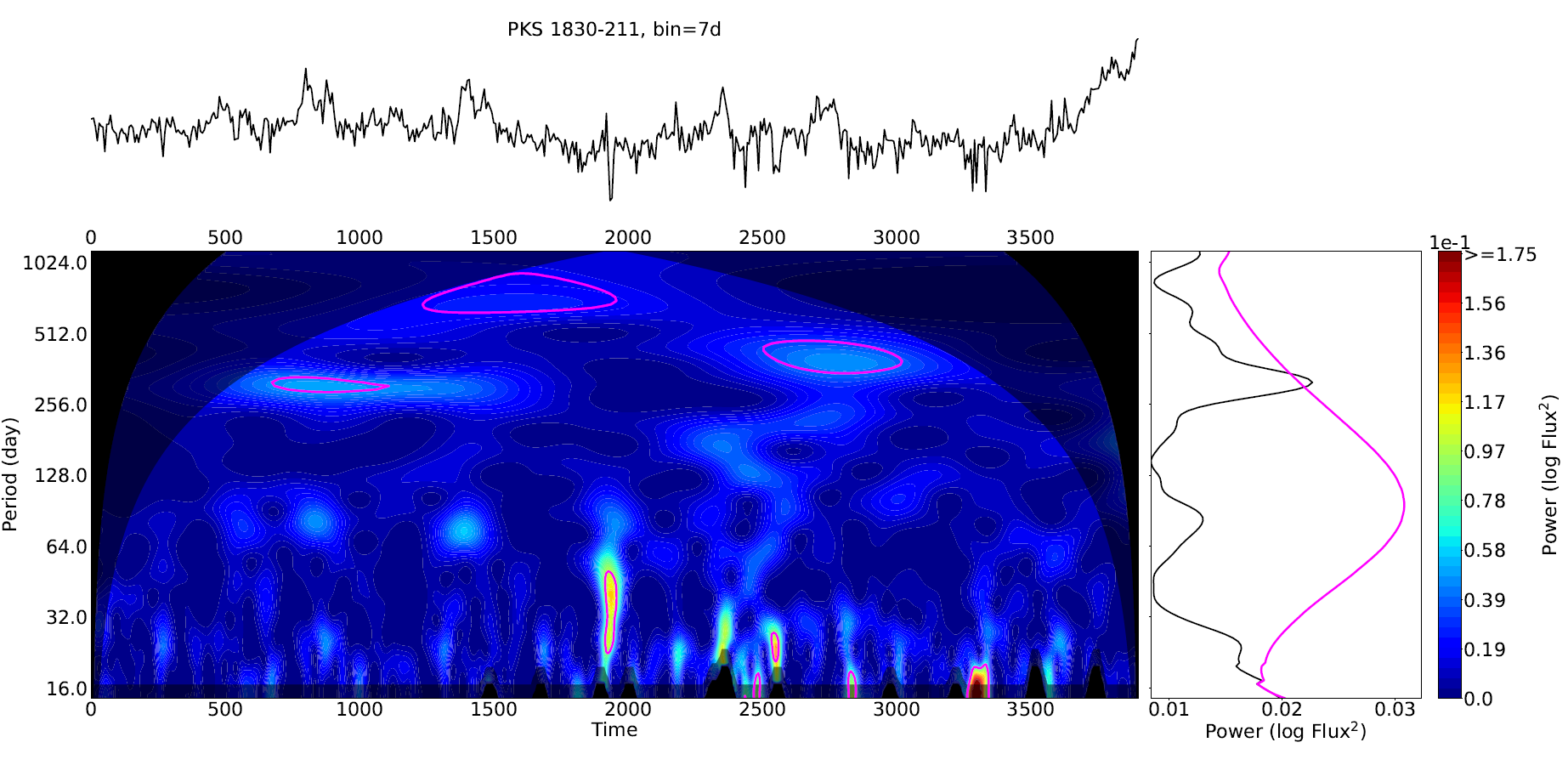}
\caption{Wavelet scalograms of the 7~d binned logarithmic LCs of PKS~2155$-$304 ({\it upper panel}) and PKS~1830$-$211 ({\it bottom panel}). The magenta lines and contours denote $3\sigma$ global and local confidence levels, respectively. (The complete figure set (33 images) is available in the online journal.)}
\label{fig_QPO}
\end{figure*}

We detect a well-known QPO in PKS~2155$-$304, at a period of $612\pm 42$~days\footnote{For comparison with previous works \citep{Sandrinelli14,Sandrinelli16,zhang17a} a Gaussian form was fitted to the corresponding peak in the LSP of the 7~d binned LC, and the error is the half width at half maximum. Applying the same procedure to the peak in the global wavelet periodogram yielded $605\pm 120$~days. (Fig.~\ref{fig_QPO}). This period is smaller than reported by previous works (650--660~days by \citealt{Sandrinelli14}; 642~days by \citealt{Sandrinelli16}; 635~days by \citealt{zhang17a}), but consistent with more recent ones (610~days by \citealt{bhat20}). \citealt{penil20} reported 620~days in a 28-day binned LC. Overall, a linear decrease over the years can be observed. The QPO is unambiguous in the 7~d binned LC examined herein, partially disappears in the 10~d binned LC, but is not present at all in the 14~d binned LC.} Additionally, when considering the global wavelet periodograms alone, PKS~1830$-$211 in the 7~d binning shows a marginal peak just above the $3\sigma$ confidence level. This peak, however, can be attributed to two short-lasting contours of $>3\sigma$ significance in the scalogram, at the borders of the COI. These contours correspond to apparently different periods. One can associate these contours with short, two-peak flux variations in the LC. They are however so short-lasting that cannot be called QPOs---if so, they would consist of only two cycles each, which is likely just a chance occurence owing to stochastic fluctuations, and definitely does not constitute a significant QPO detection. There is also a $3\sigma$ contour at an even higher period, $\gtrsim 500$~days, but its duration is comparable to its period. Overall, there is no convincing evidence to claim a QPO in PKS~1830$-$211, especially when compared to the picture painted by PKS~2155$-$304.

There is a peculiar feature in B2~1633+38, with a $3\sigma$ contour starting at $P\sim 500$~days, and increasing over the course of the LC to $P>1000$~days. The two peaks in the global periodograms, at 634 and 1066 days, are spurious in the sense that they arise due to this one structure that evolves strongly in time. This might be called a QPO, with an emphasis of the {\it quasi} term, as no single leading period can be associated with it.

\subsection{ARMA and FARIMA modeling}
\label{results::arma}

\begin{figure*}
\centering
\includegraphics[width=\textwidth]{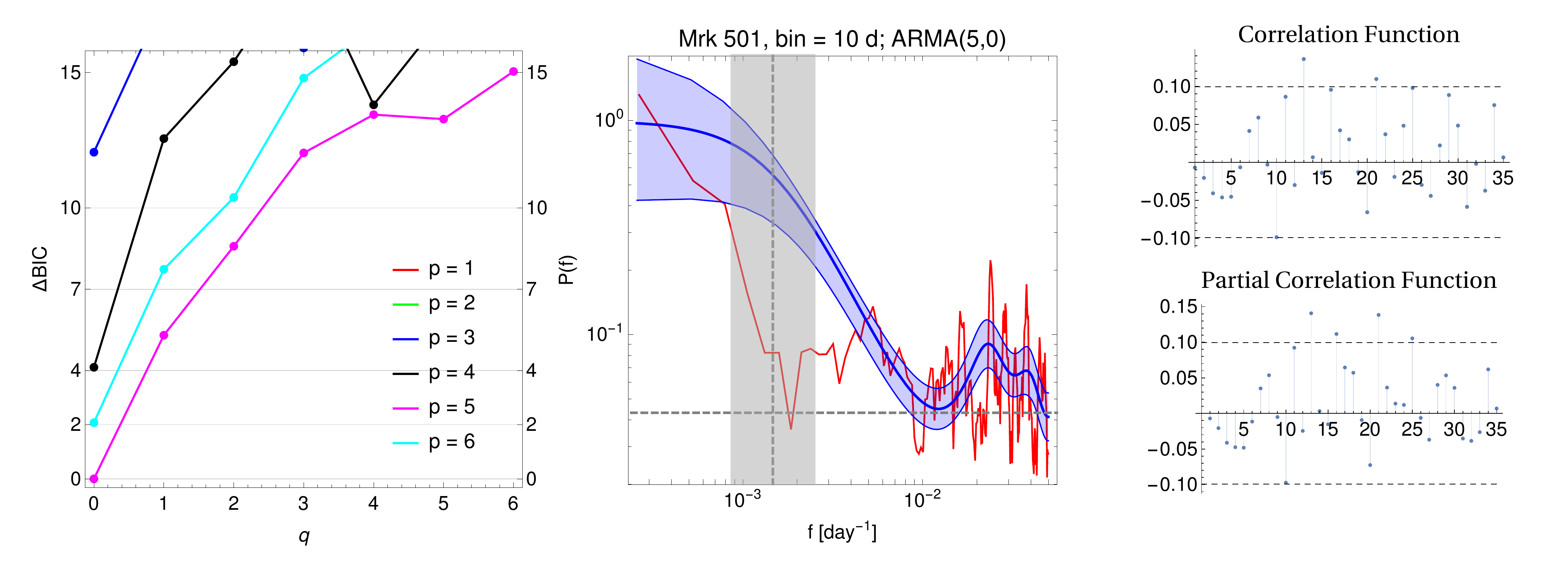}\\
\includegraphics[width=\textwidth]{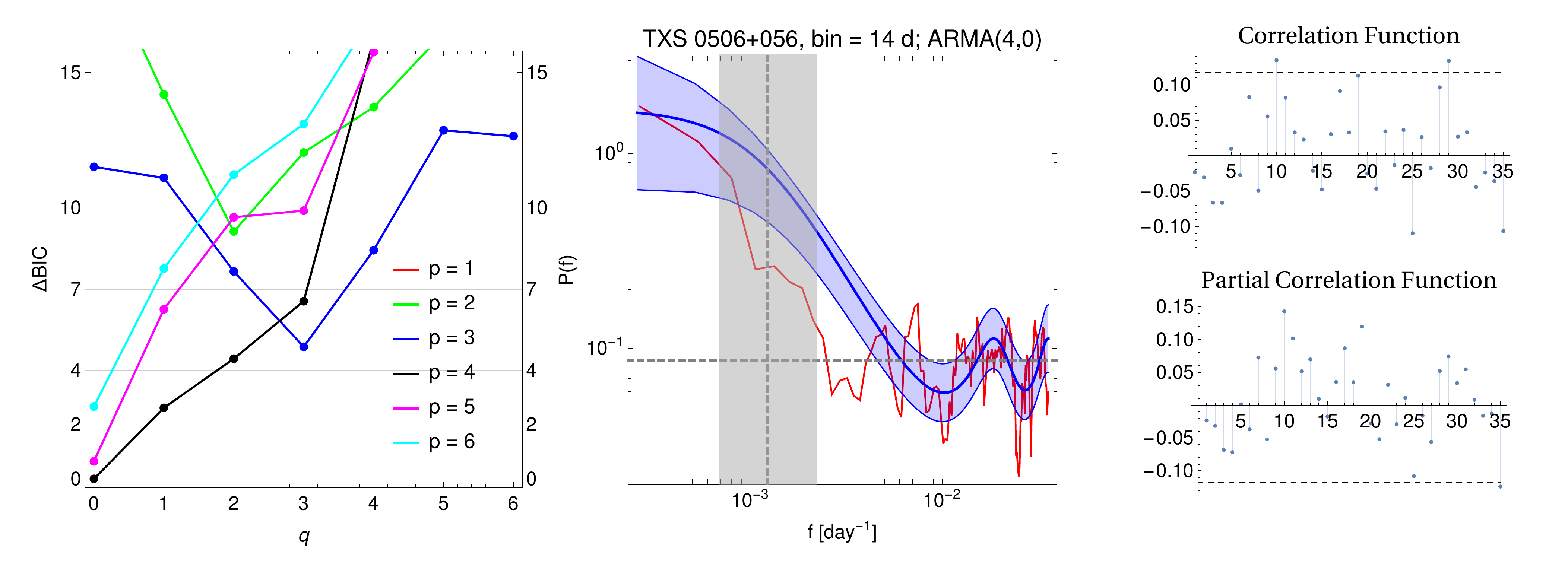}\\
\includegraphics[width=\textwidth]{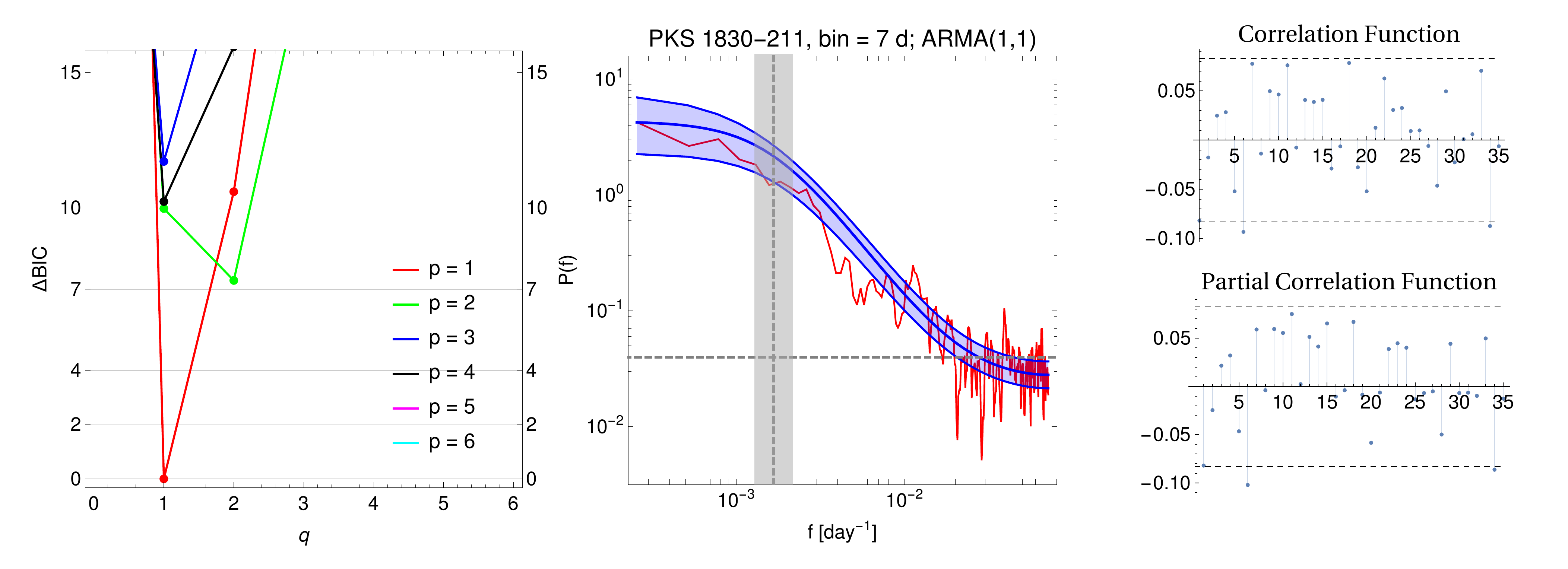}
\caption{Results of ARMA fitting. The blue line in the middle column is the PSD of the best-fit ARMA model, surrounded by a light blue 68\% confidence band. The horizontal dashed line is the Poisson noise level inferred from data. The vertical dashed line marks the break frequency, corresponding to $T_{\rm break,L}$ from Table~\ref{table5}, and the width of the gray rectangle symbolizes its standard error. The red line is the Fourier PSD, smoothed with a 5-point moving average, to show relation between the ARMA and Fourier features. The left column displays the $\Delta BIC$ scores of the ARMA$(p,q)$ fits. The right column contains the correlation and partial correlation functions of the fits' residuals; the dashed lines enclose the 95\% confidence region. (The complete figure set (33 images) is available in the online journal.)}
\label{fig_ARMA_examples}
\end{figure*}

To estimate the long time scales breaks of the ARMA fits, we first observe that the obtained PSDs' shapes resemble the Lorentzian from Eq.~(\ref{eq38}), i.e. they are flat at low frequencies, and then transition to an approximately PL decay (see the middle panels in Fig.~\ref{fig_ARMA_examples}). The general PSD of an ARMA process, Eq.~(\ref{eq35}), is not given by a combination of Lorentzian peaks\footnote{For some values of the parameters the PSD can be completely non-Lorentzian, e.g. monotonically increasing.}, though, so we proceed numerically in a way driven by an analogy with the simple Lorentzian shape. It is straightforward to check that $P_{\rm OU}(f)$ has its break at the same frequency as $fP_{\rm OU}(f)$ has its local maximum. We therefore locate the first local maximum of $fP_{\rm ARMA}(f)$ and identify it with the break frequency. To compute the uncertainties, the parametric bootstrap is employed \citep[][and references therein]{tarnopolski15a}: we generate 500 PSDs by varying randomly the coefficients of the best-fit ARMA model within their errors, and take the standard deviation of the distribution of breaks as the standard error.

The ARMA$(p,q)$ models were fitted using the method of moments first. For choosing the best fit, we employed $BIC$, because, as implied by the benchmark testing from Sect.~\ref{sect::arma_bench}, it is more likely to return if not $q$, then at least $p$ close to the real value when noise is present in the observations. In particular, the best model is chosen to be the one with smallest $p$ within the band $\Delta_i<2$. The outcomes of this method are gathered in Table~\ref{table5}. However, as demonstrated in Appendix~\ref{appendixA}, the maximum likelihood estimation is prone to underestimate the order. In most instances the results do not match: $(1,1)$ was the most commonly returned order, with occasional $(2,1)$, $(2,0)$, ($1,0)$, or $(1,2)$. On one hand, the method of moments is more likely to point at a plausible model as indicated by the testing from Sect.~\ref{sect::arma_bench}; on the other, the higher-order models, e.g. ARMA$(5,0)$ in case of Mrk~501 or TXS~0505+056 (Fig.~\ref{fig_ARMA_examples}), but also B2 1510+31, B2 1633+38, PKS 2155$-$304, are modeling the variations at the Poisson noise level, hence are likely overfitting, i.e. are modeling the fluctuations related to uncertainties in the data, and hence their order is superficially elevated to take this into account, and does not entirely reflect the features of the underlying process that governs the variability.

Note that $AIC_c$, as noted in Sect.~\ref{sect::aic}, is more liberal than $BIC$, hence points at unnecessarily complex models. In case of the examined blazars, it returned $(p,q)$ as high as $(8,8)$ or $(9,7)$. All in all, the combination of method of moments and $BIC$ gives the most reliable description of the LCs at hand, and hence we proceed to discuss them.

\begin{deluxetable*}{ccccccc}
\tabletypesize{\footnotesize}
\tablecolumns{7}
\tablewidth{0pt}
\tablecaption{Best-fit (method of moments; chosen based on the $BIC$) ARMA$(p,q)$ orders and break time scales. \label{table5}}
\tablehead{
 & \multicolumn{2}{c}{7 d} & \multicolumn{2}{c}{10 d} & \multicolumn{2}{c}{14 d} \\
\colhead{Source} & $(p,q)$ & $T_{\rm break,L}$ & $(p,q)$ & $T_{\rm break,L}$ & $(p,q)$ & $T_{\rm break,L}$ \\
 & & [day] & & [day] & & [day] \\
\tiny{(1)} & \tiny{(2)} & \tiny{(3)} & \tiny{(4)} & \tiny{(5)} & \tiny{(6)} & \tiny{(7)} }
\startdata
\multicolumn{7}{c}{BL Lacs}\\\hline
Mrk 501 & $(5,0)$ & $288\pm 113$ & $(5,0)$ & $593\pm 291$ & $(5,0)$ & $681\pm 371$ \\
Mrk 421 & $(1,1)$ & $169\pm 42$ & $(2,0)$ & $238\pm 78$ & $(2,0)$ & $207\pm 87$ \\
PKS 0716+714 & $(4,0)$ & $372\pm 203$ & $(2,0)$ & $268\pm 105$ & $(1,1)$ & $488\pm 178$ \\
PKS 2155$-$304 & $(3,0)$ & $169\pm 38$ & $(3,0)$ & $212\pm 60$ & $(2,0)$ & $200\pm 62$ \\
TXS 0506+056 & $(6,0)$ & $772\pm 491$ & $(5,0)$ & $509\pm 294$ & $(4,0)$ & $813\pm 477$ \\\hline
\multicolumn{7}{c}{FSRQs}\\\hline
PKS 1510$-$089 & $(1,0)$ & $145\pm 20$ & $(1,0)$ & $216\pm 35$ & $(1,0)$ & $251\pm 45$ \\
3C 279 & $(1,2)$ & $374\pm 97$ & $(1,1)$ & $352\pm 84$ & $(1,0)$ & $461\pm 96$ \\
B2 1520+31 & $(5,0)$ & $1199\pm 1501$ & $(4,0)$ & $1438\pm 1930$ & $(1,1)$ & $921\pm 351$ \\
B2 1633+38 & $(4,0)$ & $570\pm 405$ & $(3,0)$ & $581\pm 350$ & $(2,0)$ & $327\pm 106$ \\
3C 454.3 & $(1,1)$ & $1806\pm 724$ & $(1,0)$ & $1568\pm 1918$ & $(1,0)$ & $1350\pm 463$ \\
PKS 1830$-$211 & $(1,1)$ & $600\pm 167$ & $(1,1)$ & $660\pm 195$ & $(1,1)$ & $654\pm 187$ \\
\enddata
\tablecomments{Columns: (1) source name; (2,4,6) orders for 7, 10 and 14 d binnings; (3,5,7) corresponding long break time scales. }
\end{deluxetable*}

In the defining Eq.~(\ref{eq27}) of the ARMA process, the observation $x_t$ is directly influenced by $p$ preceding ones. For example, let $p=4$ in an LC with a 7 d binning. This means that the observations from the previous 28 days (in the observer frame) affect the current value. For an LC with a 14 d binning one can then expect $p=2$ in order to maintain the same time scale $\tau$. We see exactly this pattern in case of B2~1633+38 (Table~\ref{table5}).

The PSDs are presented in Fig.~\ref{fig_ARMA_examples}. Mrk~501 (in the 10~d binning) yields ARMA$(5,0)$. At higher frequencies (corresponding to timescales $\lesssim 90\,{\rm d}$) we observe local maxima above the Poisson noise level---these might be either QPOs, or just fluctuations---especially that wavelet scalograms show no signs of QPOs at all (Sect.~\ref{results::wavepal}). In case of TXS~0506+056 (for all binnings) the high-frequency variations are exactly at the Poisson noise level. PKS~1830$-$211 consistently exhibits a prominent break in its PSD at a time scale of several hundred days (see Table~\ref{table5}). On the other hand, Fourier spectra and LSPs do not hint at any break (compare with Sect.~\ref{results::fourier} and \ref{results::LSP}).

Overall, we observed breaks at time scales of a few hundred days in all but the following objects: B2~1520+31 (for which we obtain very big uncertainties), and 3C~454.3 (for which the breaks correspond to very long time scales, as the PSD of this source is highly consistent with a pure PL). The most pronounced and well-constrained breaks occur in Mrk~421, PKS~2155$-$304, PKS~1510$089$, and 3C~279. The break time scales are generally consistent with break timescales obtained by CARMA modeling (see Sect.~\ref{results::carma}). This equivalence is to be expected, as CARMA is just a continuous-time version of ARMA, and the data herein are nearly uniformly sampled, with not many missing points, hence the two models ought to give similar results. 

\subsection{CARMA modeling}
\label{results::carma}

\begin{figure}
\centering
\includegraphics[width=\columnwidth]{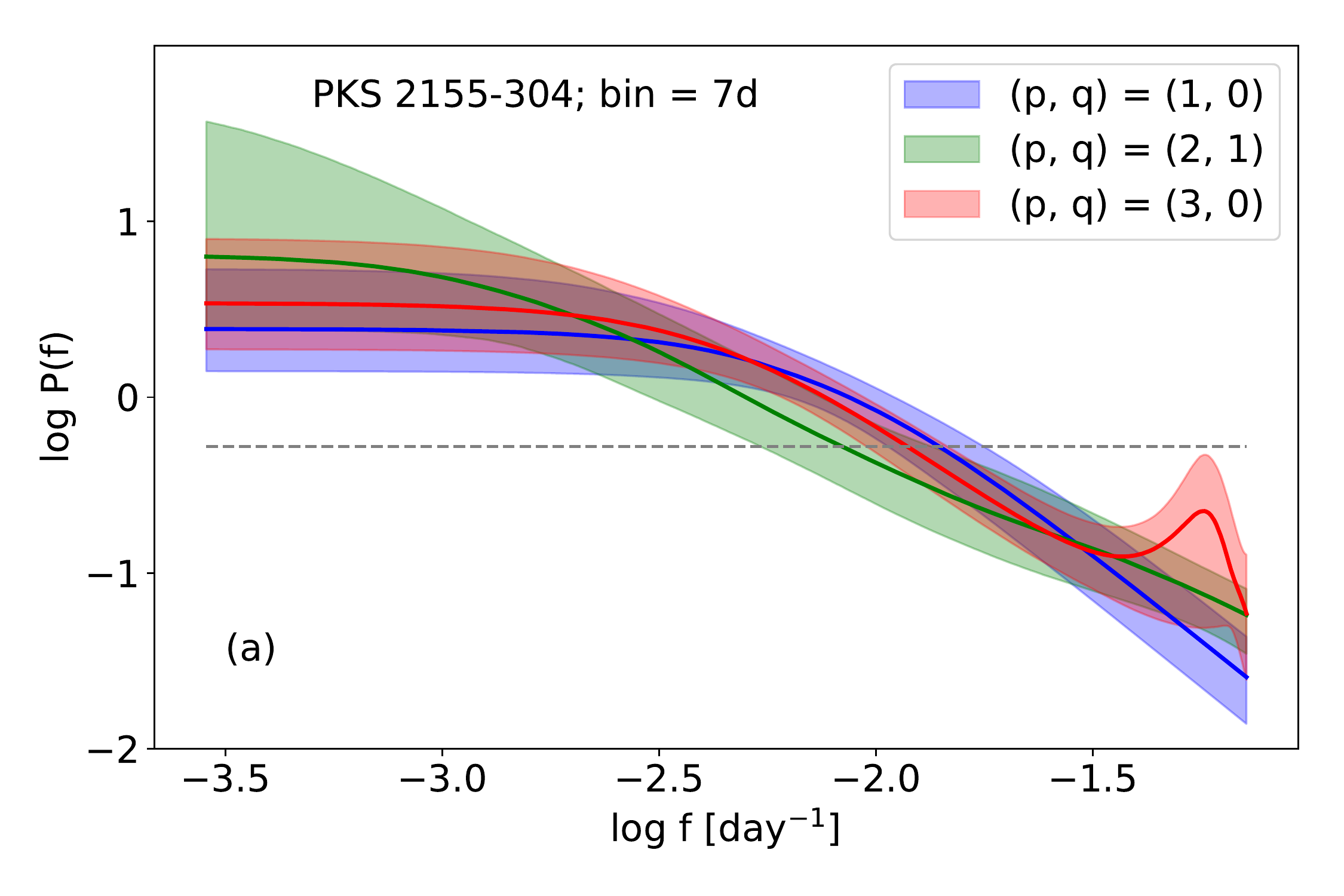}
\includegraphics[width=\columnwidth]{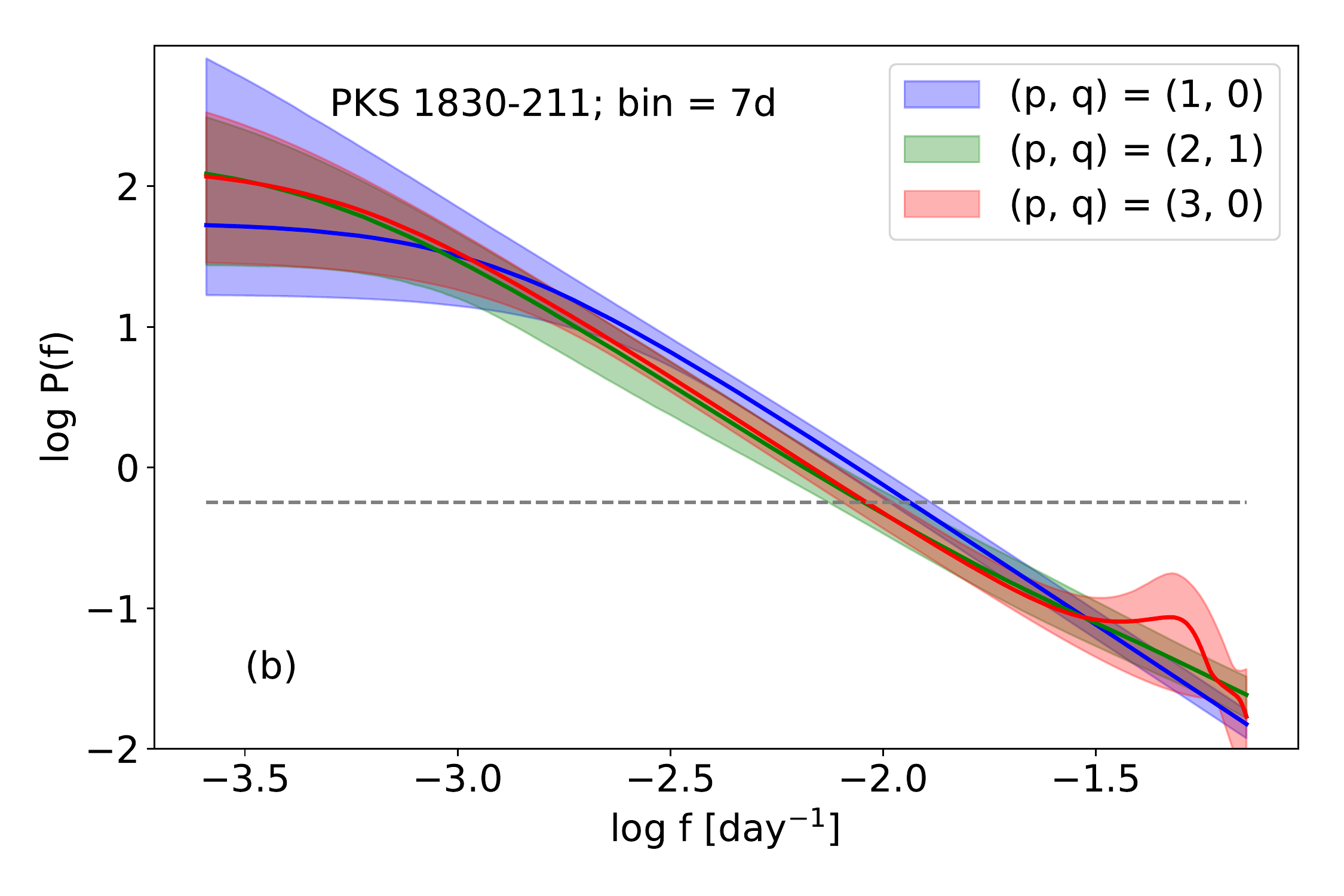}
\caption{Comparison of the high-order CARMA model obtained as the best fit according to $BIC$, and the simple ones with orders $(1,0)$ and $(2,1)$, for (a) the 7~d binned LC of PKS~2155$-$304, and (b) PKS~1830$-$211 in the 7~d binning. The horizontal dashed lines mark the Poisson noise level inferred from data.
The two other cases (B2~1520+31 with 7~d and 14~d binning---see Table~\ref{table7}) have similarly overlapping PSDs (but with no signs of the high-frequency peak visible here below the Poisson noise level), hence are not displayed.}
\label{fig_CARMA_compare}
\end{figure}

\citet{ryan19} recently investigated daily and weekly LCs of eight blazars examined herein as well, except for Mrk~501, TXS 0506+056, and PKS 1830$-$211, by fitting CARMA$(p,p-1)$ models, $1\leqslant p\leqslant 5$, and found that in all cases $p=1$ or $p=2$ give the best fits. We performed a broader search for $1\leqslant p\leqslant 7$ and $0\leqslant q\leqslant 6$. We employ $BIC$, and choose the simplest model (i.e., with lowest $p$) within the band $\Delta_i<2$ as the best fit. Our results, summarized in Table~\ref{table7}, are generally in agreement with \citet{ryan19}. In the case of 7~d binning of PKS 2155$-$304, when restricted only to CARMA$(p,p-1)$, we also obtain $p=2$ as the best, and $p=1$ as the simplest model within $\Delta_i<10$. Likewise for PKS~1830$-$211 in a 7~d binning. In these two fittings, however, the orders exceeding $(2,1)$ appear to be spuriously elevated to take into account subtle variations {\it below} the Poisson noise level, hence imply overfitting. Indeed, when the higher-order and simple models are compared (Fig.~\ref{fig_CARMA_compare}), they all exhibit PSDs broadly consistent with each other within the confidence intervals.

\begin{figure*}
\centering
\includegraphics[width=\textwidth]{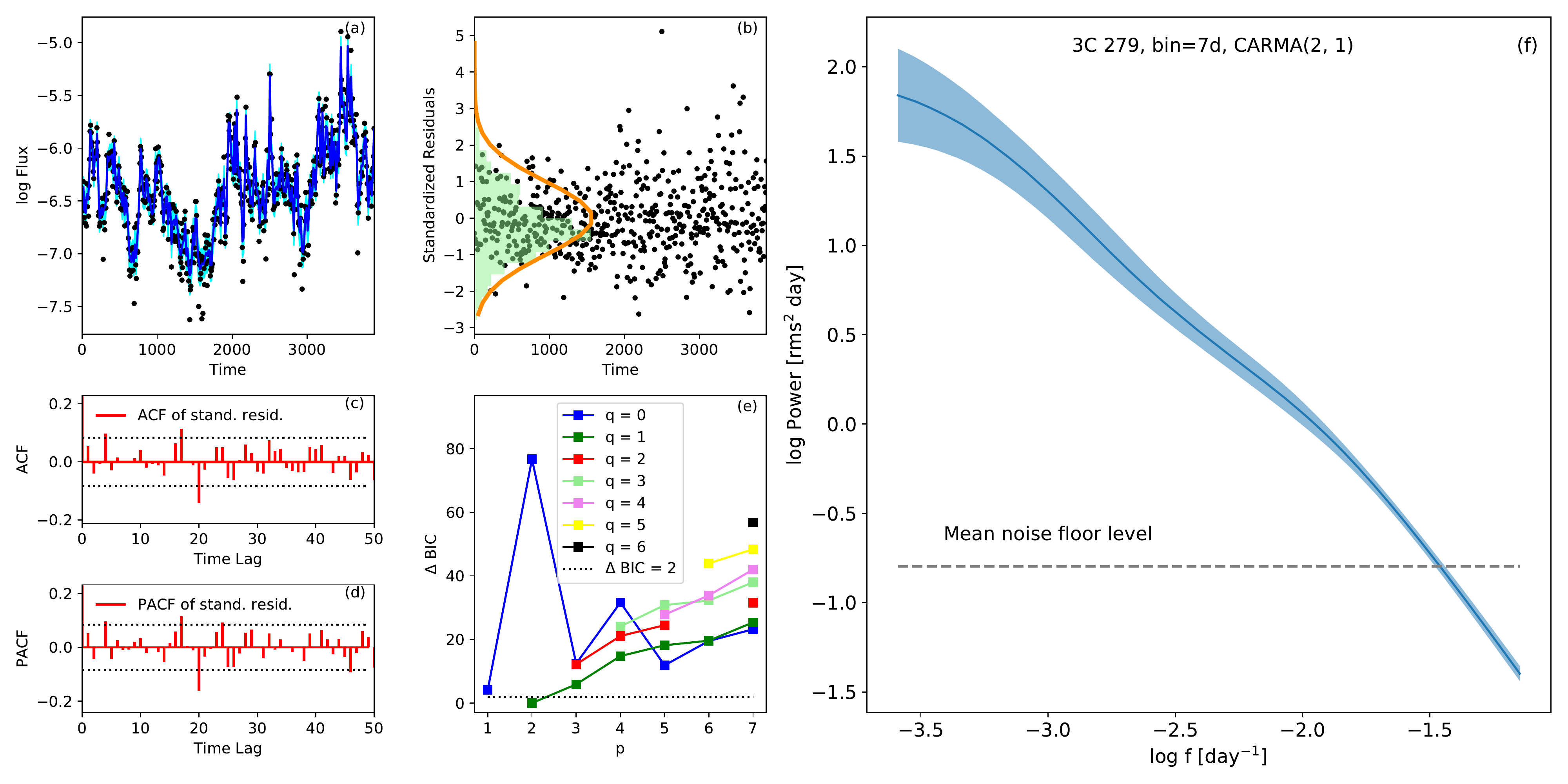}
\caption{Results of CARMA fitting. (a) The best fit to the LC. (b) Standardized residuals of the fit. (c)--(d) Autocorrelation and partial autocorrelation functions of fit's residuals. (e) The $\Delta BIC$ scores of the ARMA$(p,q)$ fits. (f) The blue line is the PSD of the best-fit CARMA model, surrounded by a light blue 68\% confidence band. The horizontal dashed line is the Poisson noise level inferred from data. (The complete figure set (33 images) is available in the online journal.)}
\label{fig_CARMA_examples}
\end{figure*}

In particular, almost all fits are in fact consistent with the simplest, CARMA$(1,0)$ model, as in the case of CARMA$(2,1)$ fits the second break in the PSD is mostly located below the Poisson noise level. The only exception is 3C~279, with breaks at $\sim 2000$ and $\sim 100$ days (see Fig.~\ref{fig_CARMA_examples}). Curiously, the 8.5-years-long {\it Fermi}-LAT LC of the famous source OJ~287 exhibits a break at $\sim$150~days in a CARMA$(1,0)$ best fit \citep{goyal18}.

\begin{deluxetable*}{cccccccccc}
\tabletypesize{\footnotesize}
\tablecolumns{10}
\tablewidth{0pt}
\tablecaption{Best-fit CARMA$(p,q)$ orders and break time scales. \label{table7}}
\tablehead{
 & & \colhead{7 d} & & & \colhead{10 d} & & & \colhead{14 d} & \\
\colhead{Source} & $(p,q)$ & $T_{\rm break,L}$ & $T_{\rm break,S}$ & $(p,q)$ & $T_{\rm break,L}$ & $T_{\rm break,S}$ & $(p,q)$ & $T_{\rm break,L}$ & $T_{\rm break,S}$\\
 & & [day] & [day] & & [day] & [day] & & [day] & [day]\\
\tiny{(1)} & \tiny{(2)} & \tiny{(3)} & \tiny{(4)} & \tiny{(5)} & \tiny{(6)} & \tiny{(7)} & \tiny{(8)} & \tiny{(9)} & \tiny{(10)} }
\startdata
\multicolumn{10}{c}{BL Lacs}\\\hline
Mrk 501 & $(2,1)$ & $1907\pm 1355$ & $22\pm 14^b$ & $(2,1)$ & $2727\pm 1137$ & $17\pm 6^b$ & $(2,1)$ & $2140\pm 1995$ & $35\pm 22^b$ \\
Mrk 421 & $(1,0)$ & $389\pm 107$ & --- & $(1,0)$ & $502\pm 158$ & --- & $(1,0)$ & $497\pm 168$ & --- \\
PKS 0716+714 & $(1,0)$ & $262\pm 64$ & --- & $(1,0)$ & $322\pm 74$ & --- & $(1,0)$ & $360\pm 92$ & --- \\
PKS 2155$-$304 & $(3,0)^a$ & $532\pm 333$ & $17\pm 11^b$ & $(1,0)$ & $240\pm 65$ & --- & $(1,0)$ & $345\pm 111$ & --- \\
TXS 0506+056 & $(2,1)$ & $2978\pm 975$ & $13\pm 5^b$ & $(2,1)$ & $2739\pm 1214$ & $21\pm 9^b$ & $(2,1)$ & $2885\pm 1063$ & $24\pm 9^b$ \\\hline
\multicolumn{10}{c}{FSRQs}\\\hline
PKS 1510$-$089 & $(1,0)$ & $209\pm 34$ & --- & $(1,0)$ & $260\pm 48$ & --- & $(1,0)$ & $308\pm 65$ & --- \\
3C 279 & $(2,1)$ & $2015\pm 1556$ & $99\pm 61$ & $(1,0)$ & $387\pm 87$ & --- & $(1,0)$ & $581\pm 187$ & --- \\
B2 1520+31 & $(3,1)^a$ & $3087\pm 961$ & $102\pm 33^b$ & $(1,0)$ & $1447\pm 1131$ & --- & $(5,0)^c$ & $2304\pm 1761$ & --- \\
B2 1633+38 & $(1,0)$ & $802\pm 563$ & --- & $(1,0)$ & $1148\pm 756$ & --- & $(1,0)$ & $1284\pm 985$ & --- \\
3C 454.3 & $(1,0)$ & $1967\pm 1610$ & --- & $(1,0)$ & $2671\pm 2414$ & --- & $(1,0)$ & $3073\pm 2639$ & --- \\
PKS 1830$-$211 & $(3,0)^a$ & $2110\pm 1126$ & $20\pm 15^b$ & $(1,0)$ & $2483\pm 2434$ & --- & $(1,0)$ & $4217\pm 4061$ & --- \\
\enddata
\tablecomments{Columns: (1) source name; (2,5,8) orders for 7, 10 and 14 d binnings; (3,6,9) long time scale breaks; (4,7,10) short time scale breaks.\\
$^a$This was returned as the best order, but the given $T_{\rm break,L}$ and $T_{\rm break,S}$ refer to a CARMA$(2,1)$ fit.\\
$^b$Below the Poisson noise level.\\
$^c$For CARMA$(2,1)$, $T_{\rm break,S}=120\pm 355\,{\rm days}$, effectively implying no break. 
}
\end{deluxetable*}

The fits to Mrk~501, TXS~0506+056, PKS~1830$-$211 are featureless, with $T_{\rm break}$ at very long time scales, making the PSDs consistent with a PL description. In particular, there are no signs of a QPO in case of PKS~1830$-$211, contrary to what the global wavelet periodogram (7 d binning, Sect.~\ref{results::wavepal}) might suggest. In conclusion, we did not detect any peculiar features in the PSDs obtained with fitting CARMA models to our data. The break time scales that are above the Poisson noise level are either at a few hundred, or several thousands of days.

\subsection{Hurst exponents}

\begin{figure*}
\centering
\includegraphics[width=1.3\textwidth,angle=90]{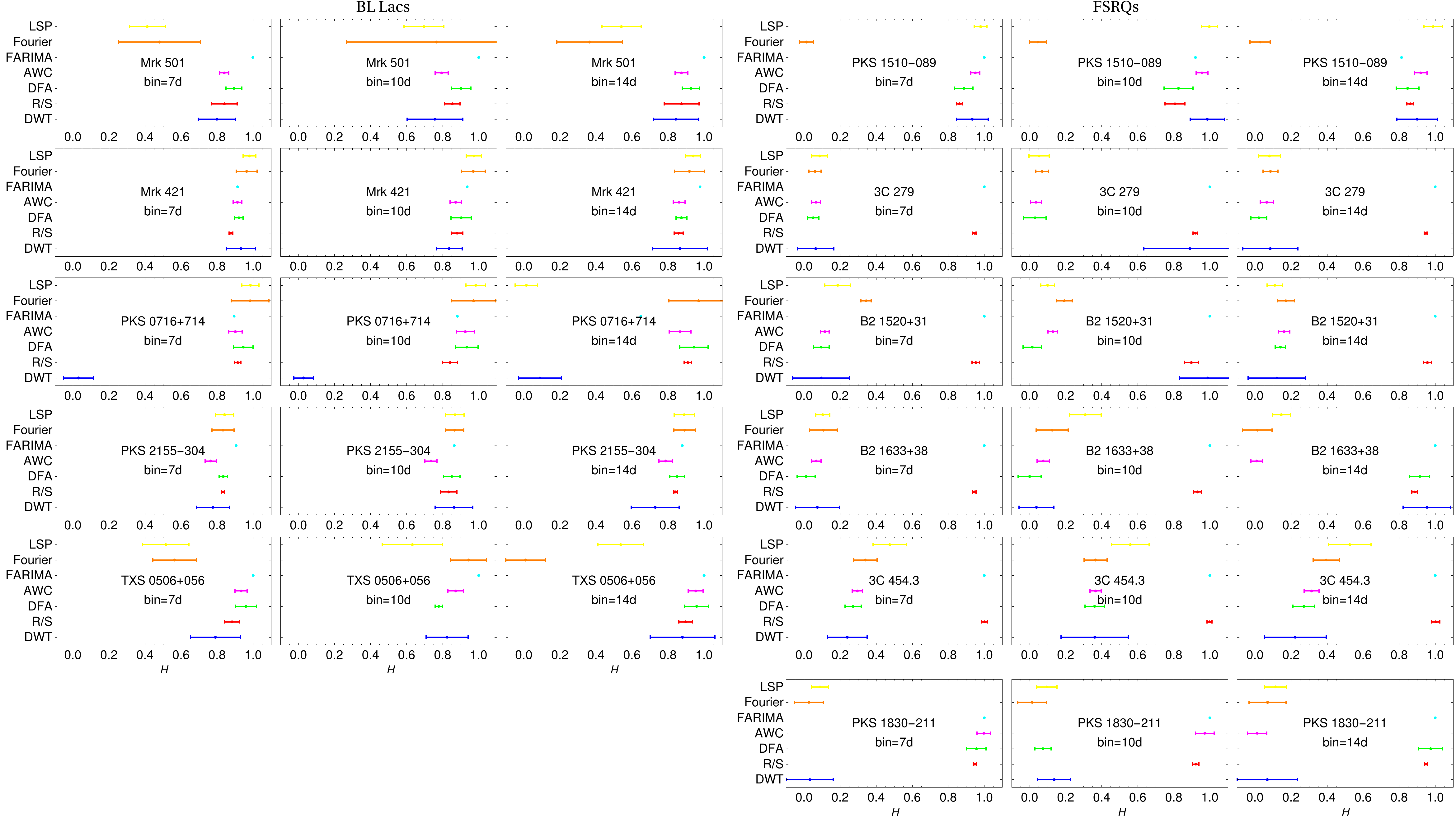}
\caption{Estimates of $H$ for BL Lac type objects and FSRQs. }
\label{fig_Hurst_results}
\end{figure*}

We used all methods from Sect.~\ref{sect::hurst} to estimate the Hurst exponents. The results are summarized graphically in Fig.~\ref{fig_Hurst_results}. In some instances, fitting a FARIMA model returned $d=0.5$ (leading to $H=1$), which implies nonstationarities present in the respective LCs. As illustrated in Fig.~\ref{fig_H}, FARIMA fitting is biased towards higher $H$ when applied to PLC (and even more biased in case of fBm); it has a great dispersion when used at fGn and FARIMA models; and while usually recovers correctly $H\approx 0.5$ for ARMA, it can drift towards $H=0$ or $1$; finally, for CARMA models it consistently returns $H$ close to $1$. Therefore, not much meaning can be given to the $H$ estimates from FARIMA fitting \citep{torre07}.

On the other hand, the Fourier spectra and LSPs yielded for many objects $\beta\approx 1$, which due to the discoutinuity of $H$ at the border between stationary and nonstationary processes (see Fig.~\ref{fig_Hurst_disc}) makes it uncertain whether $H\gtrsim 0$ or $H\lesssim 1$. Only Mrk 501, PKS 2155$-$304, B2 1520+31, and B2 1633+38 are far enough from the border to allow to confidently infer $H$ from $\beta$ (Tables~\ref{table1} and \ref{table3}). However, as Mrk 501 gives inconsistent $\beta$ for different binnings, the corresponding $H$ span the range $0.4-0.8$. For PKS 1510$-$089, Fourier spectra lead to $H\approx 0$, while LSPs yield $H\approx 1$.

The AWC often drifts to $H>1$, which is a nonsensical value. Overall, it systematically returns $H>0.5$. Its performence for PLC (see Fig.~\ref{fig_H}) is poor, and it greatly misidentifies $H\approx 0$ as $H\approx 1$ when nonstationarities are present, e.g. in the fBm case, so technically also for $\beta > 1$. The performance of the R/S algorithm is also poor and hence unreliable. What is left is only DFA and DWT.

\begin{figure*}
\centering
\includegraphics[width=0.49\textwidth]{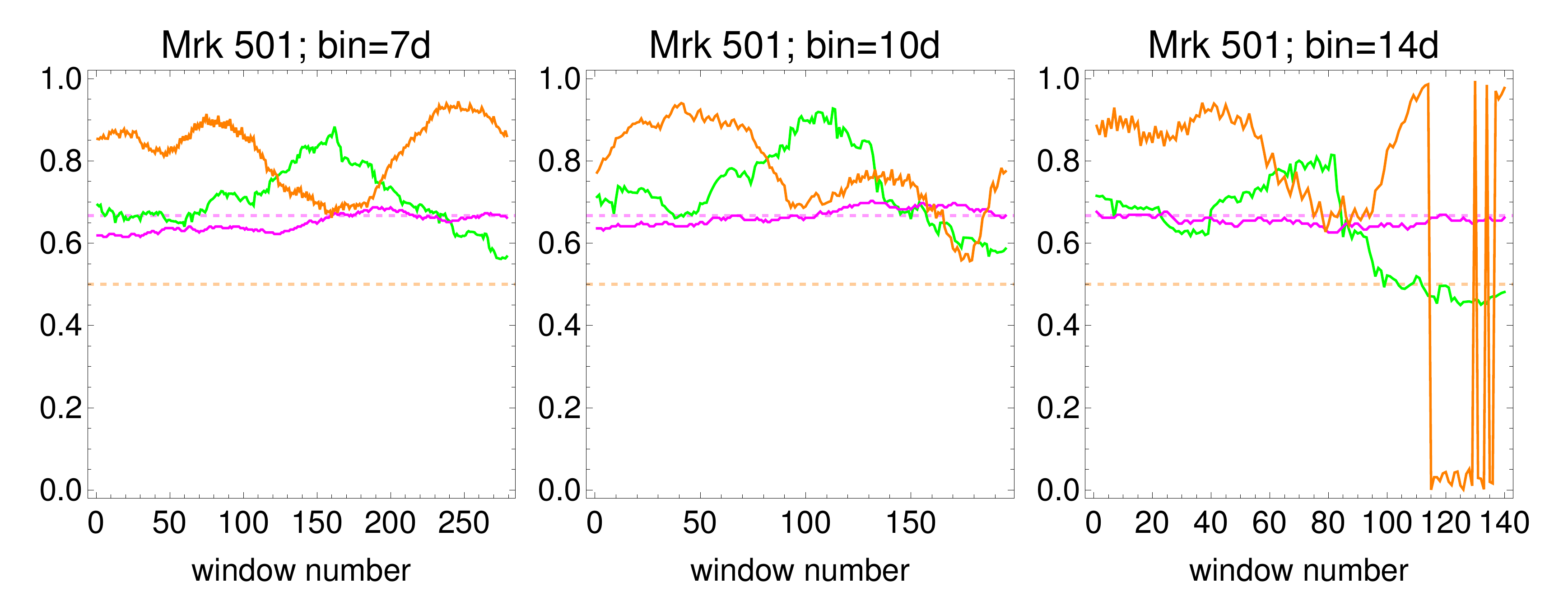}
\includegraphics[width=0.49\textwidth]{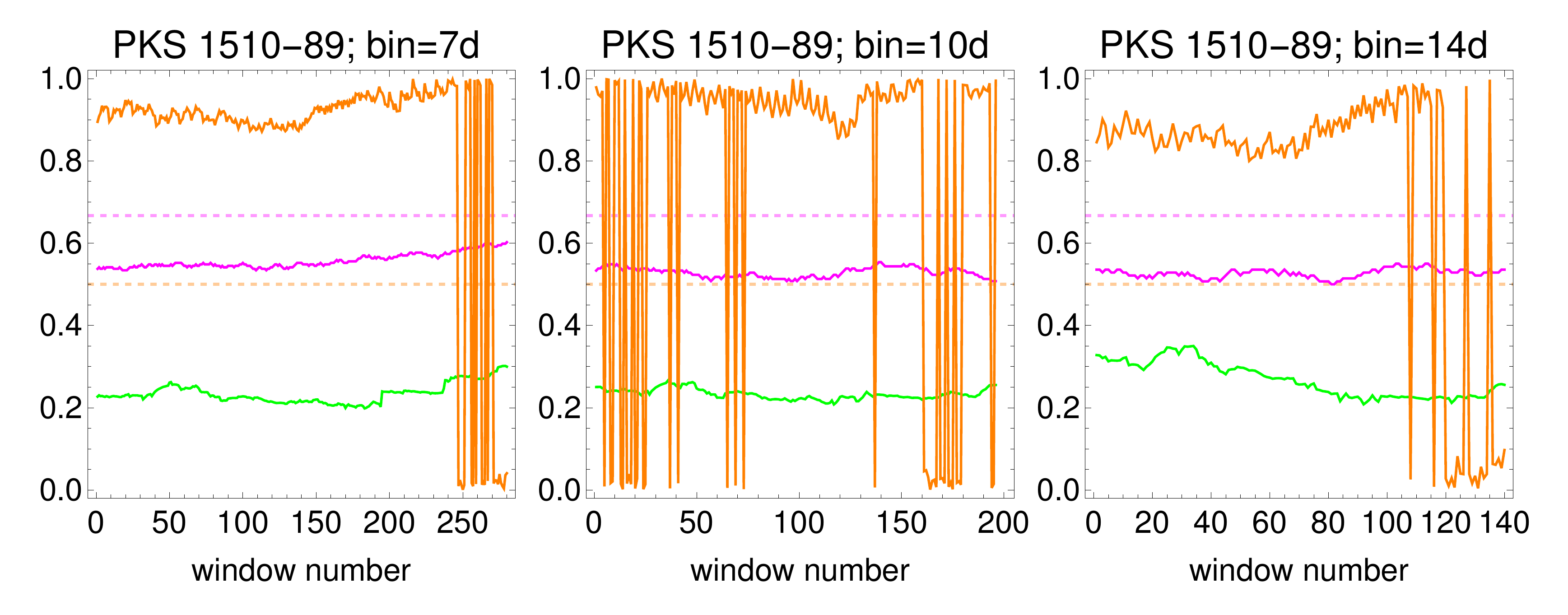}\\
\includegraphics[width=0.49\textwidth]{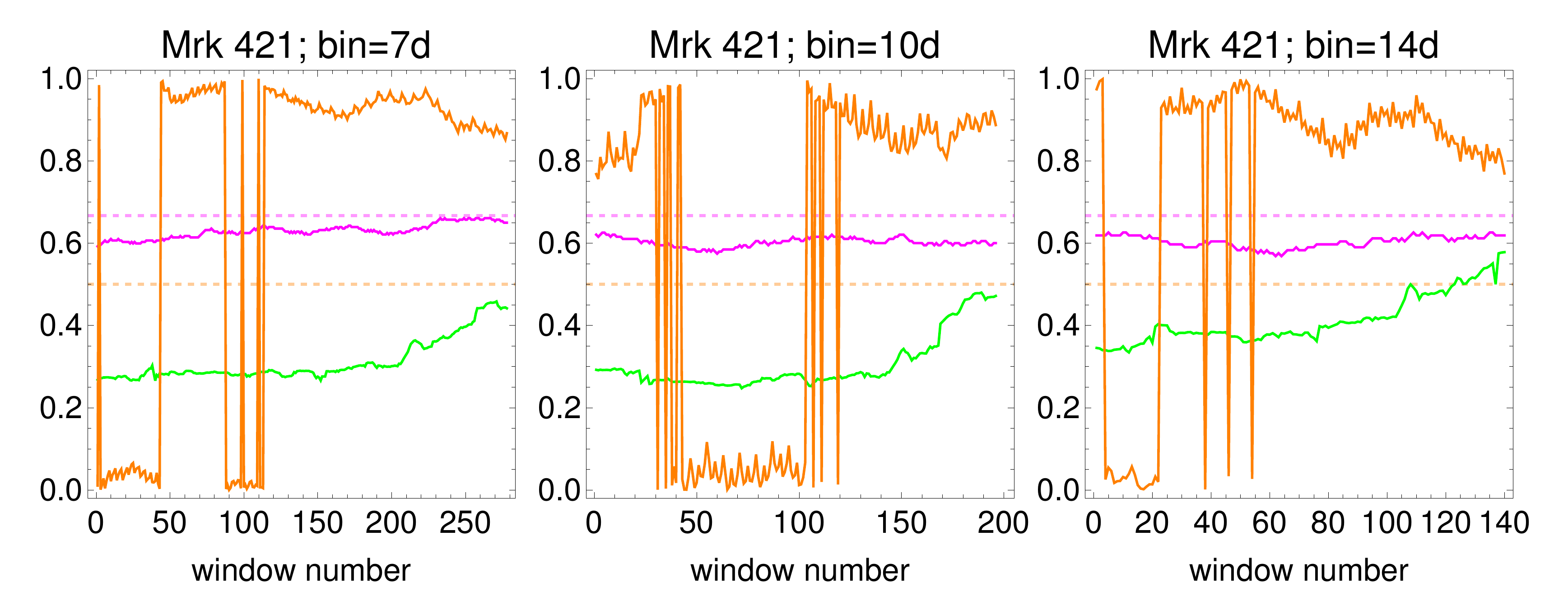}
\includegraphics[width=0.49\textwidth]{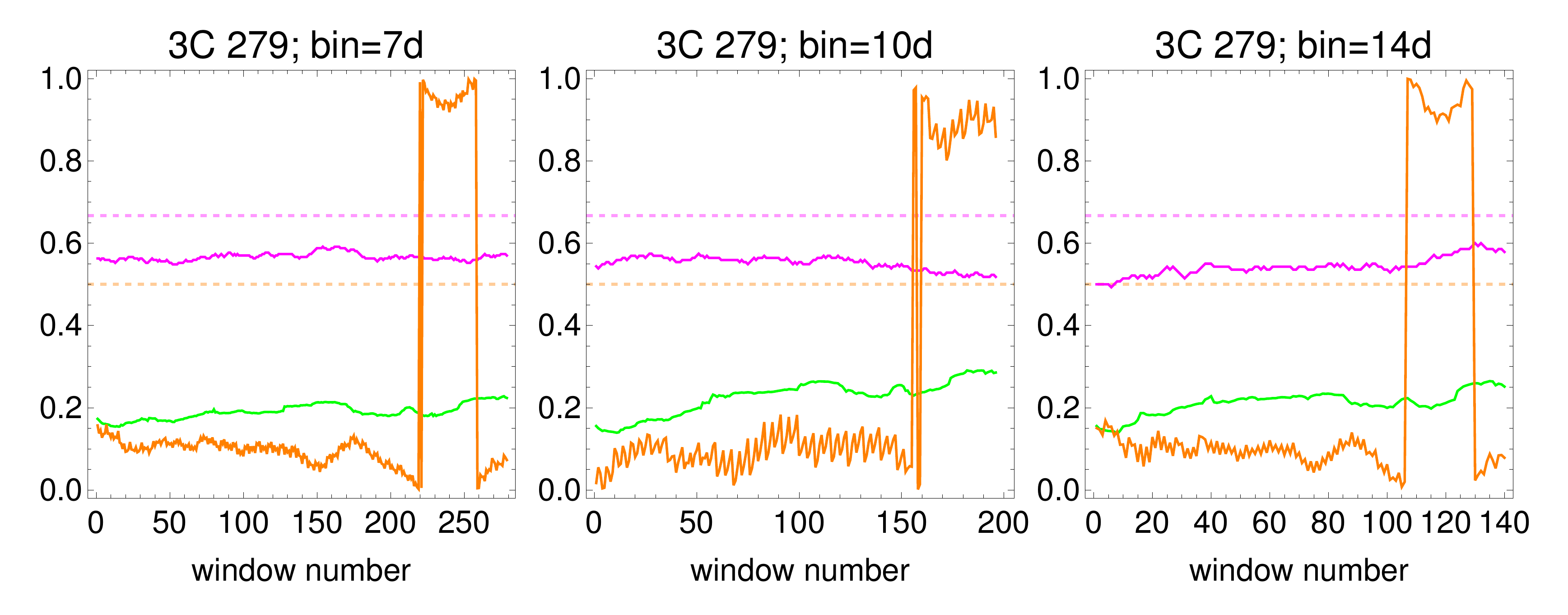}\\
\includegraphics[width=0.49\textwidth]{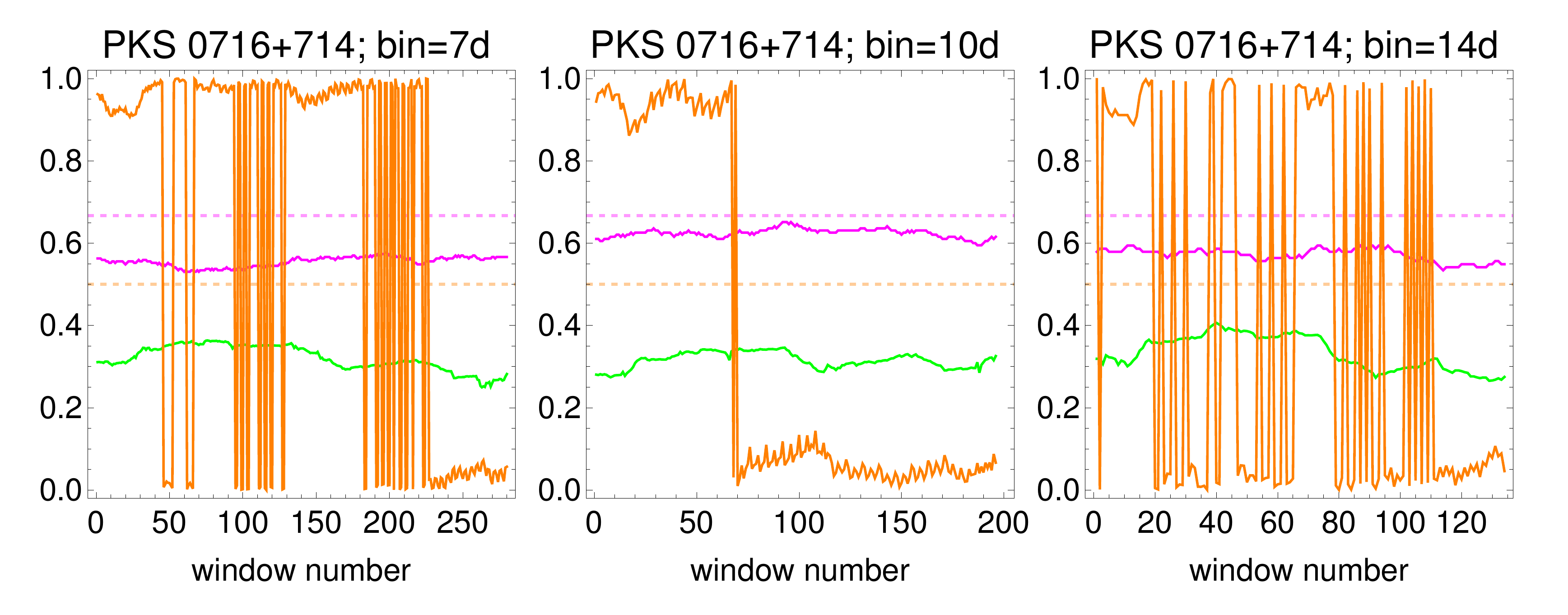}
\includegraphics[width=0.49\textwidth]{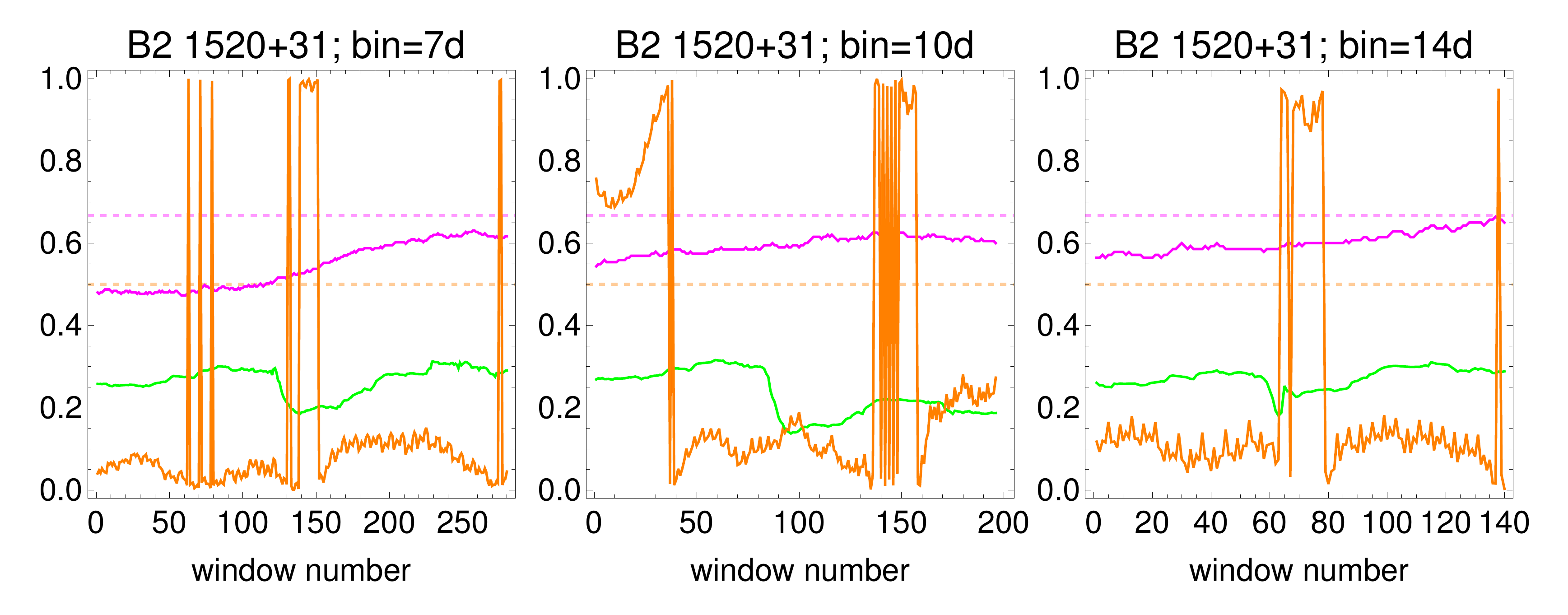}\\
\includegraphics[width=0.49\textwidth]{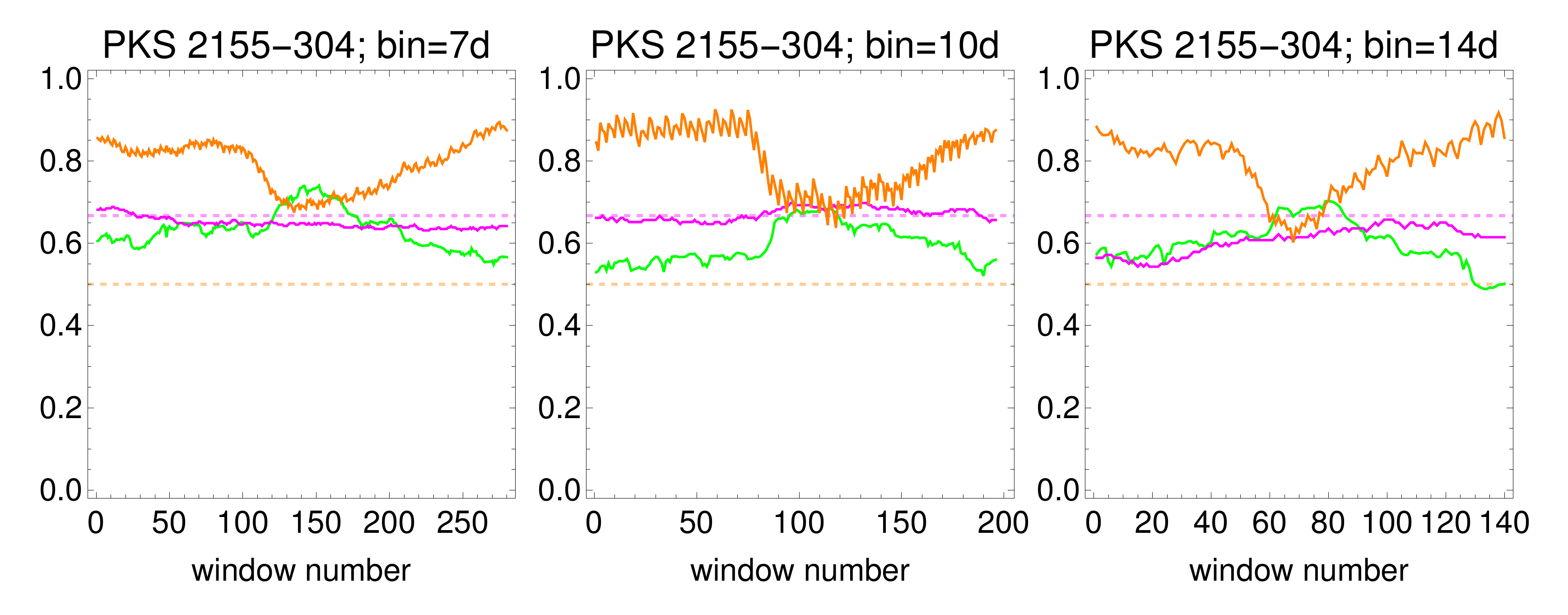}
\includegraphics[width=0.49\textwidth]{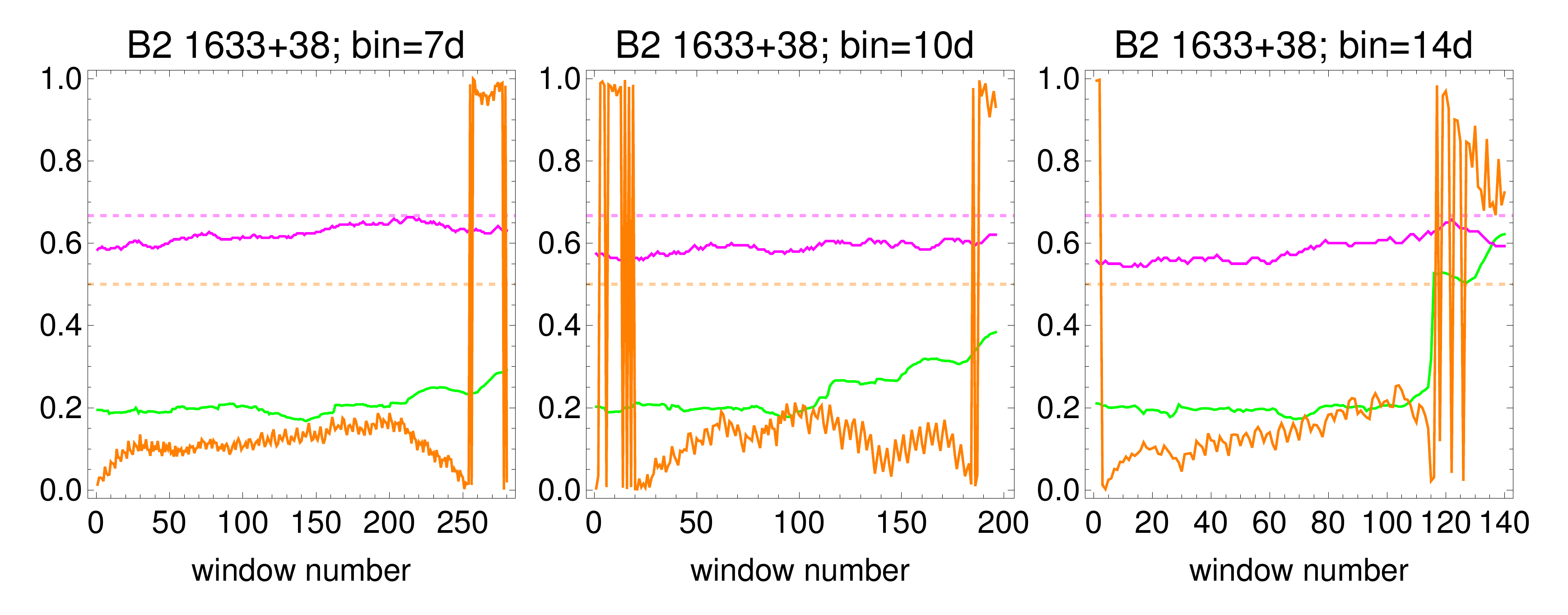}\\
\includegraphics[width=0.49\textwidth]{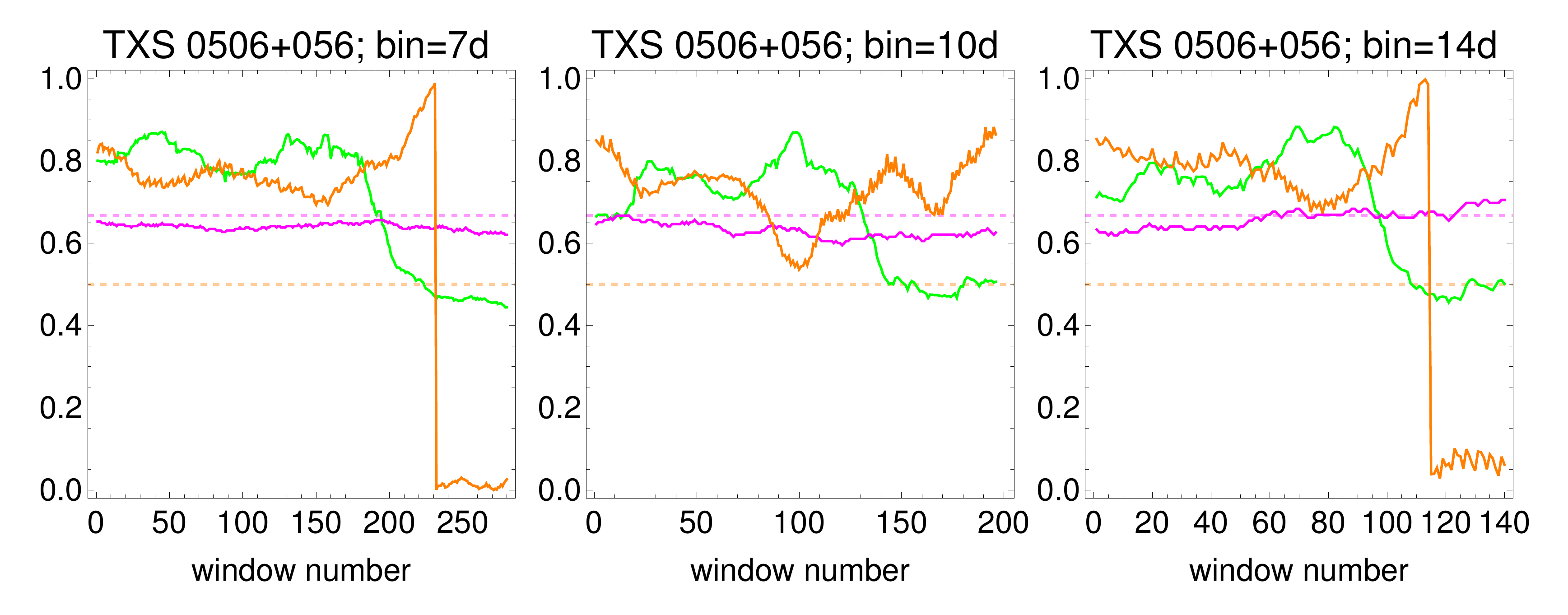}
\includegraphics[width=0.49\textwidth]{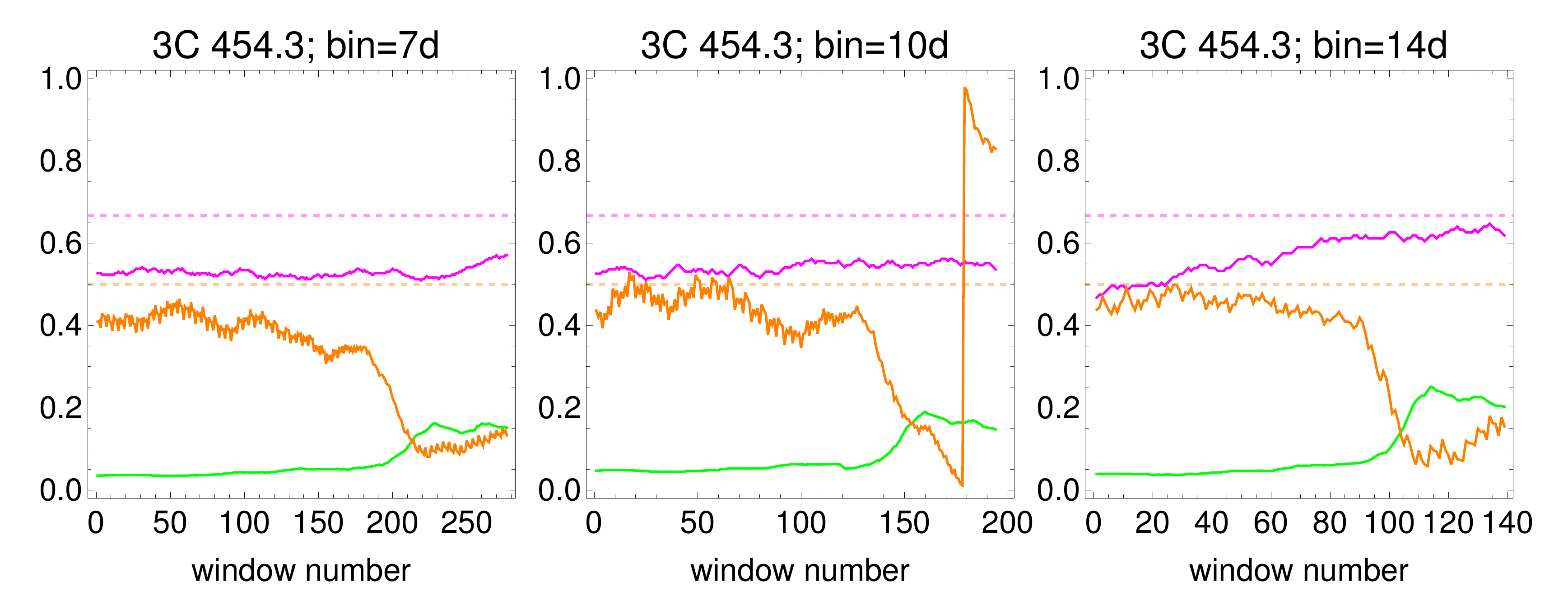}\\
\includegraphics[width=0.49\textwidth]{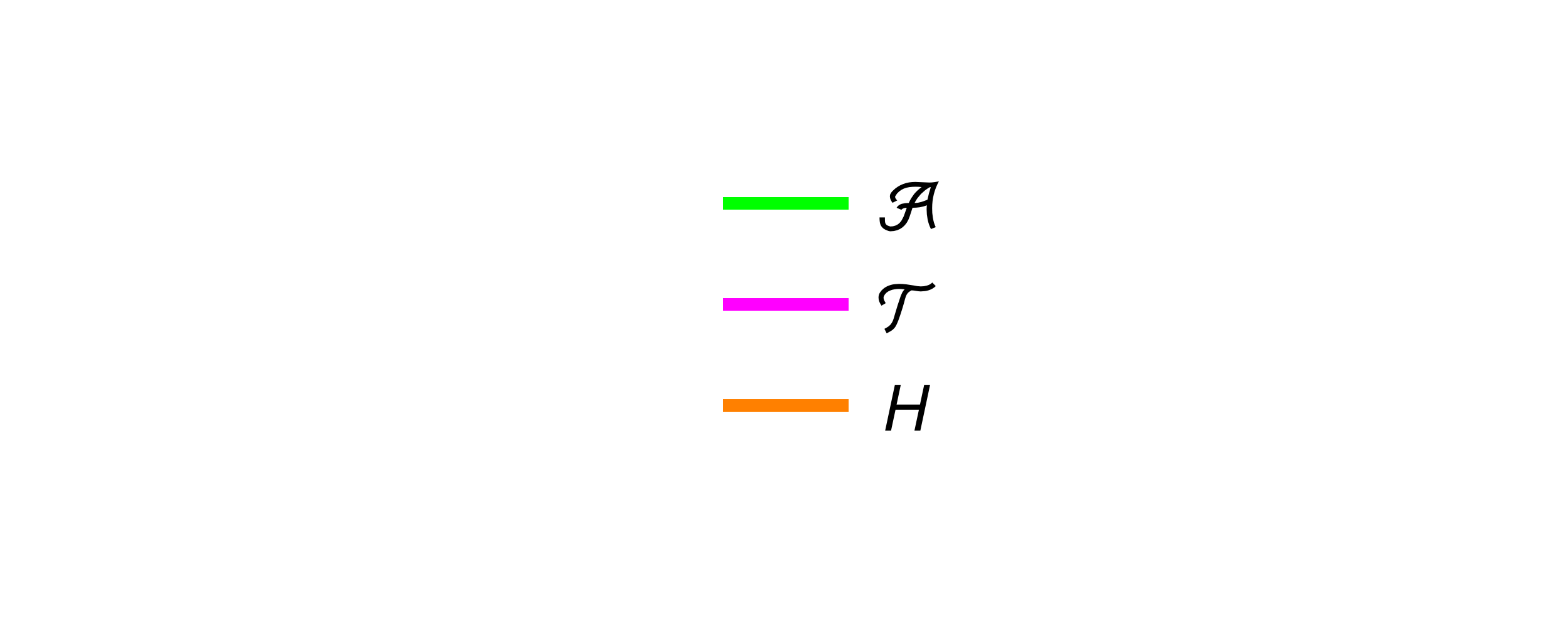}
\includegraphics[width=0.49\textwidth]{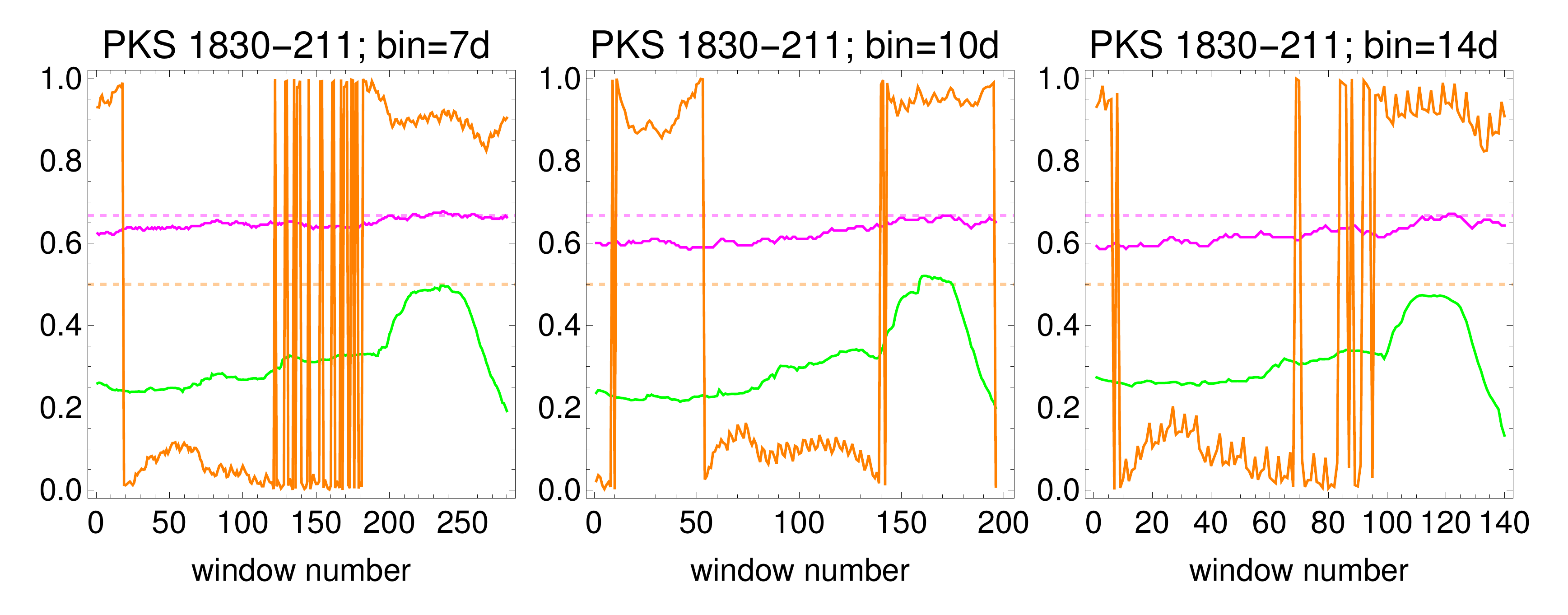}
\caption{Time evolution of the Hurst exponents $H$ estimated with the DFA, Abbe values $\mathcal{A}$, and fractions of turning points $\mathcal{T}$. The legend applies to all panels. The horizontal dashed orange line denotes $H=0.5$, and the magenta one is $\mathcal{T}=2/3$. {\it Left three columns:} BL Lacs; {\it right three columns:} FSRQs.}
\label{fig_HurstAT_results_time_evolution}
\end{figure*}

We undertook the following approach in order to try to differentiate between $H>0.5$ and $H<0.5$. Given a time series of length $N$, it was divided into windows of size $\floor*{n/2}$. The first window starts at $x_1$, the second advances one points and starts at $x_2$, etc. We end with $\ceil*{n/2}$ such chunks, to each we apply the DFA algorithm, and eventually obtain the time evolution of $H$. The $H$ values for each LC and binning will vary over time, but we aim at seeking a general answer whether $H>0.5$ or $H<0.5$, or it oscillates discontinuously between $H\approx 0$ and $H\approx 1$. The results are gatherdd in Fig.~\ref{fig_HurstAT_results_time_evolution}, where also the time evolution of $\mathcal{A}$ and $\mathcal{T}$ is displayed (see Sect.~\ref{results::AT}).

We indeed confirm that for blazars characterized by $\beta\approx 1$, the $H$ values behave incoherently in time. For the most part, Mrk~501 and TXS~0506+056 actually exhibit $H>0.5$, implying long term memory. PKS 2155$-$304 is the best behaving object, with the time evolution of $H$ consistent in all three binnings. We observe jumps between $H\approx 0$ and $H\approx 1$ for B2 1520+31 and B2 1633+38, suggesting that the $\beta$ estimates obtained in Sect.~\ref{results::fourier} and \ref{results::LSP} do not capture the true stochastic dynamics of the system. We can safely conclude that long range dependence is rather confidently detected in the LC of PKS 2155$-$304, and likely also for Mrk~501 and TXS~0506+056. On the other hand, 3C~454.3 is the only object for which $H<0.5$ is consistently returned, in agreement with $\beta\lesssim 2$, suggesting its variability is fBm-like with short-term memory.

\subsection{The $\mathcal{A-T}$ plane}
\label{results::AT}

\begin{figure*}
\centering
\includegraphics[width=0.49\textwidth]{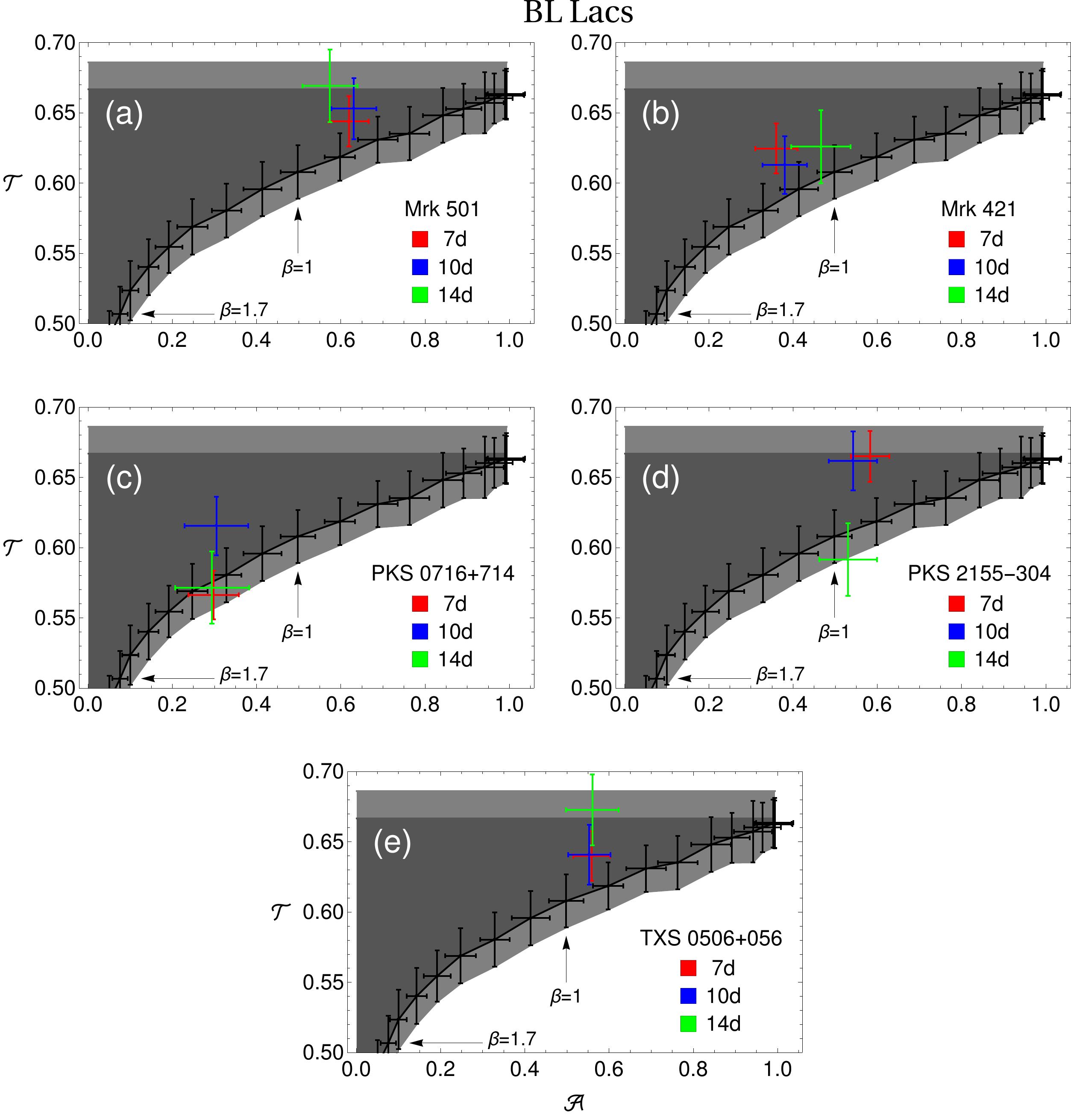}
\includegraphics[width=0.49\textwidth]{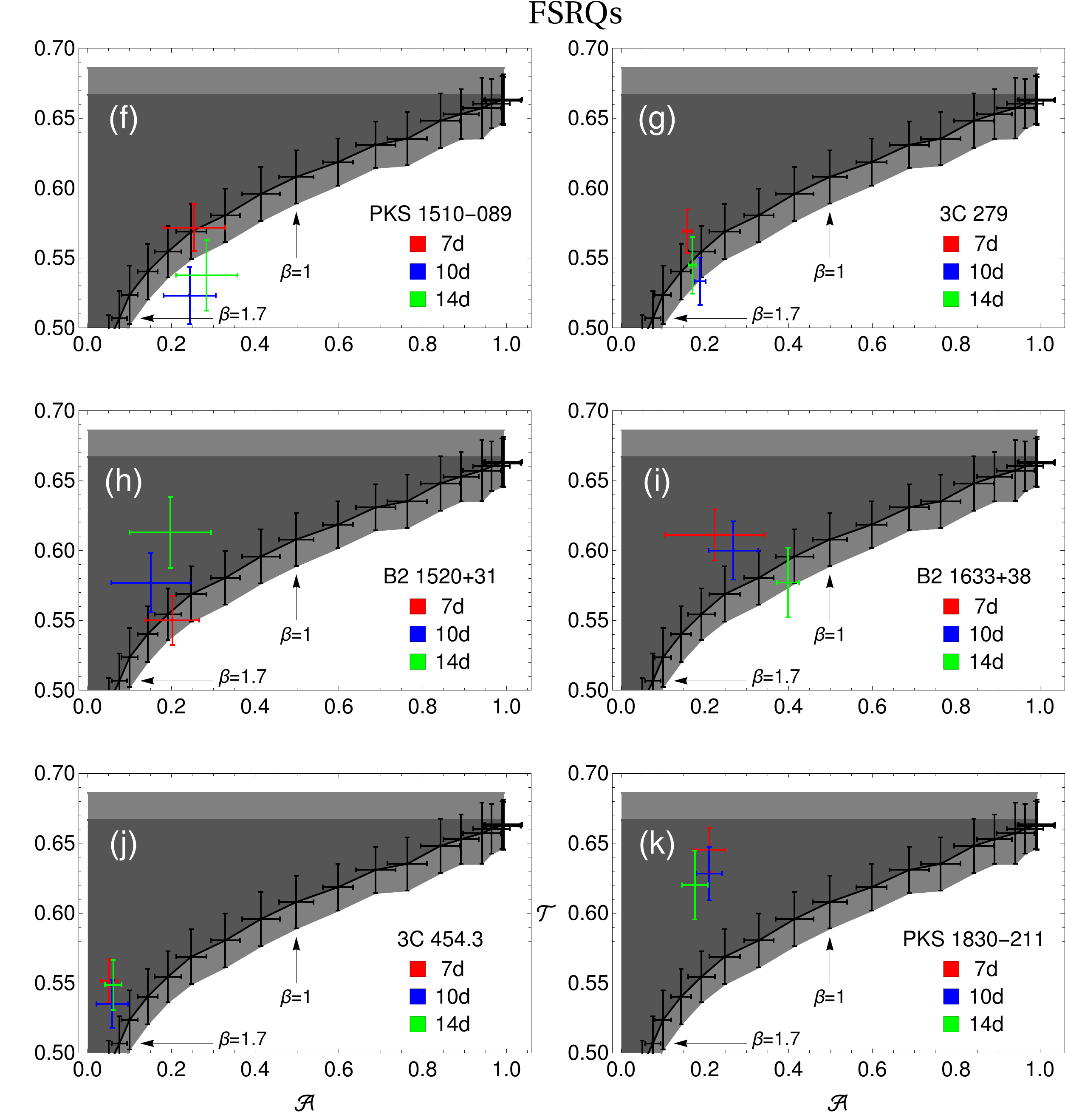}\\
\includegraphics[width=0.8\textwidth]{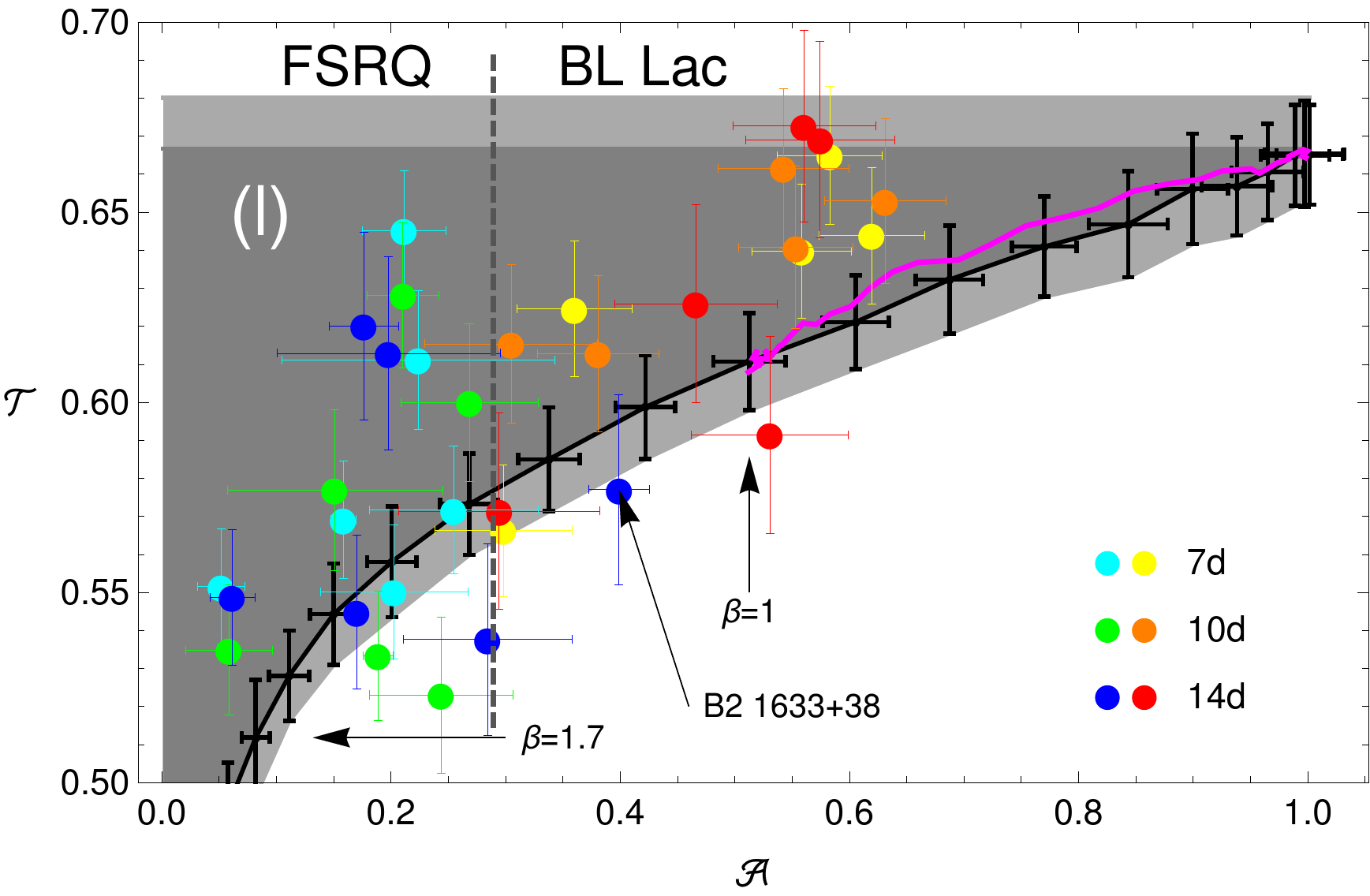}
\caption{Locations in the $\mathcal{A}-\mathcal{T}$ plane of (a)--(e) BL Lacs and (f)--(k) FSRQs. The dark gray area is the region between the pure PL line (traced by the red points from Fig.~\ref{fig_Zunino}, repeated here---in black---for convenience) and $\mathcal{T}=2/3$, and the light gray regions represent the error bars of the simulations. The red, blue and green points correspond to different binnings, indicated in the legends. Locations of $\beta=1$ and $\beta=1.7$ noise are also displayed for reference. (l) Separation of FSRQs and BL Lacs in the $\mathcal{A}-\mathcal{T}$ plane. The FSRQs are denoted with cold colors, while BL Lacs---with warm ones. Note a different color-coding than in panels (a)--(k). The vertical dashed line marks the rough separation. The outlier, B2~1633+38 in a 14 d binning, is indicated with an arrow. The magenta line is the PLC with $\beta=1$ and increasing Poisson level, starting from $C=0$. }
\label{fig_AT_separation}
\end{figure*}

The locations in the $\mathcal{A}-\mathcal{T}$ plane were calculated according to Sect.~\ref{sect::ATplane}. To constrain the errors, from each LC we generated 1000 MC realizations by drawing the observations randomly from $\mathcal{N}\left(\mu,\sigma^2\right)$, where $\mu$ is the observed value, and $\sigma$ is its standard error, and for each realization the $(\mathcal{A},\mathcal{T})$ location was computed. The dispersion is assessed by the standard deviation of the 1000 MC samples. The results are displayed in Fig.~\ref{fig_AT_separation}, where the region available to generic PLC models is highlighted. As per Sect.~\ref{results::fourier} and \ref{results::LSP}, the majority of the examined blazars are well described by a pure PL: Mrk 421, PKS 0716+714, PKS 2155$-$304, PKS 1510$-$089, and 3C 279, using both Fourier spectra and LSPs. Four of them are close to a pink noise PL with $\beta\approx 1$, except for PKS 2155$-$304, which yields $\beta\approx 0.7-0.8$. The $(\mathcal{A},\mathcal{T})$ locations are roughly consistent with $\beta\in (1,1.5)$, but generally shifted towards slightly steeper PL PSDs. This is in agreement with the underestimation of the $\beta$ indices observed in the benchmark tests in Sect.~\ref{sect::FourLS}. Note however that in case of Mrk 421 we observed a slight, and in case of PKS 2155$-$304 a strong shift upwards the $\mathcal{A}-\mathcal{T}$ plane.

Mrk~501 is a source that is most shifted to the right and upwards, which is due to a high level of Poisson noise dominating the PL component in a substantial part of its PSD. The result of increasing the Poisson noise (Fig.~\ref{fig_Zunino}) is that the path for different $\beta$ is raised and shortened. Hence the $(\mathcal{A},\mathcal{T})$ locations are shifted closer to the point $(1,2/3)$ corresponding to white noise. Therefore, overall the description of Mrk 501 in the $\mathcal{A}-\mathcal{T}$ plane is reasonable.

Finally, B2 1520+31 and B2 1633+38 yield $\beta\gtrsim 1$, but for some binnings the PSDs were better described by a PLC rather than a pure PL. They behave differently in the $\mathcal{A}-\mathcal{T}$ plane: for higher binnings, B2 1520+31 moves vertically upwards, while 1633+38 moves diagonally to the right and downwards. The latter appears to be more consistent with a pure PL location; for the 10 binning there is strong support for the pure PL model in the LSP approach ($\Delta_i \approx 3.5$). For the 7 d binning of B2 1520+31, though, the pure PL is out of question, with $\Delta_i>10$ in the Fourier approach, and $\Delta_i\approx 4.5$ in case of LSP, what means inconsiderable support for the pure PL model.

Recall that the LCs, despite spanning a long period in time, contain a moderate number of points, especially for the 10 d and 14 d binnings. On the other hand, the 7 d binning allows for a denser sampling, but increases the uncertainty of the measurements.

\section{Discussion}
\label{discussion}

\subsection{Summary of results}
\label{summary_of_results}

We analyzed with a number of techniques the \textit{Fermi}-LAT $\gamma$-ray LCs of 11 blazars. The global PSDs were obtained via Fourier spectrum and the LSP, and subsequently used to compute the PL indices $\beta$. The values resulting from these methods are consistent with each other (see Tables~\ref{table1} and \ref{table3}). There is a small discrepancy only in case of 3C~454.3, for which the Fourier PSD is described by a pure PL, and the LSP by PLC---the latter being slightly steeper than the former. Overall, the resultant shapes indicate a colored noise with $1\lesssim\beta\lesssim 2$.

We employed wavelet scalograms to detect transient phenomena. We confirmed the well-known QPO in PKS~2155$-$304, with a period of $612\pm 42$~days. We did not find any other similarly unambiguous QPOs at the $3\sigma$ confidence level. B2~1633+38 exhibits a peculiar, significant feature that evolves from $P\sim 500$~days to $P>1000$~days over the span of the LC, that might be associated with the previously reported presumed QPOs with $P\sim 600$~days \citep{Oter20}. Additionally, there is a peak in the global wavelet PSD in the 7~d binned LC of PKS~1830$-$211. This, however, appears to be a dubious detection, as the statistically significant features in the scalogram are located exactly at the border of the COI, at two different frequencies, and can be identified with at most two cycles in the LC. Moreover, there is no significant detection in 10 and 14~d binned LCs of this source. We therefore conclude a lack of significant QPOs in the remaining sample of blazars. In particular, we cannot confirm the presence of QPOs in Mrk~501 \citep{Bhat19}, Mrk~421 and PKS~0716+714 \citep{bhat20}, PKS~1510$-$089 and 3C~279 \citep{Sandrinelli16}, and B2~1520+31 \citep{gupta19} (compare with Table~\ref{BlazarSummary}). Finally, we observe hints of a significant, persistent structure in the scalograms of PKS~0716+714 at $P\gtrsim 1000$~days. This might be a candidate QPO, but since the ratio of the period to the length of the LCs allows for 2--3 cycles only, it does not constitute a detection yet. Similar conclusions were drawn by \citet{bhat20}, while \citet{penil20} reported a low-significance QPO detection at a period of $\sim $1000~days. Further monitoring of this source is therefore required.

We modeled the LCs as discrete stochastic processes, ARMA and FARIMA, fitting them with the method of moments and by maximum likelihood estimation. The latter yielded simple, low order $p\leqslant 2$ models. The method of moments, however, in some instances (Mrk~501, PKS~0716+714, PKS~2155$-$304, B2~1520+31, B2~1633+38) returns higher orders. For most of them (Mrk~501, PKS~2155$-$304, B2~1520+31, B2~1633+38) the increased model complexity is caused by overfitting at the Poisson noise level. On the other hand, PKS~0716+714 shows nontrivial features significantly above the Poisson noise level, and hence hints at a more complicated process underlying the observed variability.

The LCs were modeled as CARMA processes as well (Table~\ref{table7}). We searched for the best CARMA$(p,q)$ model with $1\leqslant p\leqslant 7$ and $0\leqslant q\leqslant 6$. In almost every instance we obtained the orders $(1,0)$ or $(2,1)$. In only a few cases the $BIC$ pointed at more complex models, but their PSDs are consistent with these simple instances (see Fig.~\ref{fig_CARMA_compare} and \ref{fig_CARMA_examples}), as the additional parameters introduced by higher orders were utilized for overfitting below the Poisson noise level. We are therefore in agreement with the results of \citet{ryan19} when it comes to the eight blazars common to both works. Additionally, the CARMA fits to Mrk~501, TXS~0506+056, and PKS~1830$-$211 are consistent with a PL description, i.e. CARMA$(1,0)$ with a high $T_{\rm break,L}$.

Nevertheless, motivated by the fact that the proper model to describe the data may not necessarily be a simple PL, we estimated the Hurst exponents with several of the most popular and established methods. For most BL Lacs we obtain $H>0.5$ significantly, signifying long-term memory. The FSRQs, however, kept oscillating between $H\lesssim 1$ and $H\gtrsim 0$, except 3C~454.3 for which $H<0.5$ convincingly. We emphasize, though, that some of the objects (Mrk~421, PKS~0716+714, and most FSRQs) do not allow to formulate an unambiguous claim about their persistence: there is no convincing majority of consistent $H$ estimates to infer whether $H\lessgtr 0.5$. This is in accordance with the fact that their PSDs are highly consistent with a PL description with $\beta\sim 1$: due to the discontinuity of $H$ (see Fig.~\ref{fig_Hurst_disc}) it is difficult to distinguish between stationary cases with $H\lesssim 1$ and nonstationary with $H\gtrsim 0$. We note that it might be proper to model the LCs as a CARFIMA process \citep{tsai05,tsai09,feigelson18}, i.e. a fractionally integrated CARMA process (or, in other words, a continuous-time FARIMA) that allows for long-term memory, as well as CARMA driven by a L\'evy process \citep{brockwell3,brockwell4}. Likewise, the fractional Fokker-Planck equation admits long-term memory solutions \citep{weron08,deng09}, and might be used to model the electron distribution \citep{stawarz08,chen11}, thus constitutes a generalization of the kinetic continuity equation \citep{finke14,finke15}.

In regards to the memory of the process, consider again the orders of the ARMA models. If the observed variability is caused by some (possibly recurring) disturbance in the jet or disk, it will need some time to dissolve. One can then expect that increasing the binning of an LC twice will reduce the ARMA order twice as well, which is indeed the case for some blazars (Sect.~\ref{results::arma}). The relation between the rest and observer frames is $\tau_{\rm rest} = (1+z)^{-1}\tau_{\rm obs}$. In particular, for B2~1633+38 (7~d binning; as per Table~\ref{table5}), $\tau_{\rm obs}=28\,{\rm days}$ transforms to 10 days in the rest frame. A decrease of the order $p$ with the increase of the bin size is also evident, although less clear, for PKS~0716+714, PKS~2155$-$304, B2~1520+31, to which we relate $\tau_{\rm rest} = 21$, 19, and 14 days, assuming $\tau_{\rm obs} = 28,\,21,\,35\,{\rm days}$, according to the orders $p$ for a 7~d binning. Such a relation suggests there might exists a characteristic time scale for the perturbations to die out, i.e. any fluctuation in the jet affects the emission for time $\tau_{\rm rest}$. Physical differences in the blazars, e.g. BH mass, accretion rate etc. could lead to different $\tau_{\rm rest}$.

Finally, we highlight a novel finding made with the recently introduced $\mathcal{A}-\mathcal{T}$ plane: the FSRQs and BL Lacs are clearly separated, as shown in Fig.~\ref{fig_AT_separation}. The FSRQs are characterized by lower values of $\mathcal{A}$ than BL Lacs. The same phenomenon was also discovered by \citet{zywucka20} in the optical (I band) light curves of blazar candidates behind the Magellanic Clouds \citep{zywucka18}, observed by the Optical Gravitational Lensing Experiment (OGLE). This demonstrates that the interrelations between flux values are different for the two blazar populations, and allows to distinguish them based solely on data in the time domain, without referring to PSDs or spectroscopic properties. Moreover, the separation is insensitive to sampling: the blazar candidates from \citet{zywucka20} were observed irregularly, with a time step approximately a multiple of 1 d, while the $\gamma$-ray LCs examined herein were binned into 7 d, 10 d, and 14 d intervals.

The separation, at first glance, might appear to be a result of FSRQs having on average steeper PSDs than BL Lacs. This is not the case: this would be true if all objects were located at the pure PL line in the $\mathcal{A}-\mathcal{T}$ plane, and their locations would be consistent with the $\beta$ estimates. However, the examined sources attain various values of $\mathcal{T}$, too, sometimes placing them far from the PL line. In effect of adding Poisson noise, the PL line is elevated and shortened, i.e. pulled closer to the point $(1,2/3)$ corresponding to white noise. Such a path is depicted in magenta in Fig.~\ref{fig_AT_separation}, and it does not separate the two classes at all. Most of the blazars' PSDs herein are consistent with a pure PL, though, hence the location in the $\mathcal{A}-\mathcal{T}$ plane highlights some hidden properties in the time series' temporal structure, not revealed by Fourier spectra or LSPs. Moreover, the region of availability for PSDs of PLC is also available to ARMA processes (see Fig.~\ref{fig_ARMA_AT}, and \citealt{tarnopolski19b} for an analytic treatment).

\subsection{Comparison with previous studies}
\label{comparison}

\paragraph{PL indices}
\citet{Naka13} analyzed almost 4-years-long LCs of 15 blazars from the \textit{Monitored Source List} (MSL), provided by the \textit{Fermi}-LAT team, using the Fourier transform. Their sample contains five blazars the same as in our work, i.e. Mrk~421, 3C~279, PKS~1510$-$089, PKS~2155$-$304, and 3C~454.3. The PL indices $\beta$ of the considered blazars are compatible with our results within errors, except for Mrk~421, for which we obtained a steeper PSD ($0.93\pm0.12$ vs. $0.38\pm0.21$). 13 blazars, with particular emphasis on blazar types, i.e. eight FSRQs and five BL Lacs, were analyzed by \citet{sobolewska14} with a superposition of OU processes. Their sample also includes a bunch of blazars studied in this work: B2~1633+38, B2~1520+31, 3C~454.3, 3C~279, PKS~1510$-$089, PKS~0716+714, PKS~2155$-$304, and Mrk~421, for which they find PL indices mostly to be $\beta\lesssim 1$. There is also no difference in $\beta$ values among the two blazar types. \citet{Abdo15} studied a 3-years-long LC of PKS~1830$-$211, generated with the unbinned maximum likelihood technique. Our Fourier PSD is slightly flatter than theirs ($1.05\pm0.15$ vs. $1.25\pm0.12$). The 7-years-long LCs (calculated with unbinned maximum likelihood method) of three blazars were analyzed by \citet{Kushwaha17}. Common objects in this case are Mrk~421, B2~1520+31, and PKS~1510$-$089. The PL index of Fourier PSDs is in an agreement only for PKS~1510$-$089, while for Mrk~421 and B2~1520+31 we obtained flatter and steeper PSDs, respectively. \citet{prokhorov17} for PKS~0716+714 obtained $\beta=0.57$, and $\beta=0.67$ for PKS~2155$-$304, while we obtain $\beta\sim 1$ and $\beta\sim 0.75$. They get $\beta$ within $\sim 0.5-0.8$, though, for all their sources. \citet{Alga18} analyzed a 3-years-long LC of B2~1633+382, generated with the adaptive binning method, to study its variability in radio, optical, and HE $\gamma$-rays. Here again the fitted PL is flatter in our case ($1.22\pm0.15$ vs. $1.70\pm0.20$). \citet{Bhat19} investigated a 10-years-long LC of Mrk~501, generated with unbinned maximum likelihood method, finding a PL index of $0.99\pm0.01$, which is much flatter than the value found in this work ($1.83\pm0.20$). \citet{covino19} examined long-term ($\sim$decade), 30~day-binned aperture photometry LCs generated by the {\it Fermi} team of 10 blazars, including PKS~0716+714 and PKS~2155$-$304. Our PL index is comparable in case of the former blazar, but gives a flatter PSD for the latter one. Subsequently, \citet{meyer19} investigated 9.5-years-long, weekly-binned LCs of six FSRQs, including PKS~1510$-$089, 3C~279, and 3C~454.3. Except for PKS~1510$-$089, they obtained clearly flatter PL fits than we do, i.e. $\sim 0.9,\,0.65,\,1.1$ vs. our $1.03,\,1.12,\,1.68$ (Fourier PSDs) or $0.96,\,1.17,\,1.95$ (LSPs), respectively. Our results are consistent with \citet{bhat20}, except for 3C~454.3, for which the authors reported $\beta=1.3\pm 0.17$, while we obtained $\beta\sim 2$. Finally, regarding Mrk~421 and PKS~2155$-$304, we obtained values broadly consistent with those of \citet{goyal20}.

All discrepancies highlighted here are most likely caused by different methods of data analysis, e.g. the methods of generating and fitting the PSDs, as well as differences in data itself, i.e. lengths of the analyzed LCs or method of their generation.

\paragraph{QPOs}
Several authors reported QPOs in different blazars in recent years. \citet{Sandrinelli14} analyzed long-term IR, optical, and HE $\gamma$-ray (MSL \textit{Fermi}-LAT) LCs of PKS~2155$-$304, with LSP and the Date Compensated Discrete Fourier Transform methods, finding a QPO with a period of 650 days in the $0.1-300$~GeV energy range. Afterwards, \citet{Sandrinelli16} investigated similar data of six blazars, including 3C~279, PKS~1510$-$089, and PKS~2155$-$304, using LSPs. All objects showed QPOs in HE $\gamma$-ray LCs, at periods of 24 and 39 days with 3$\sigma$ significance in case of 3C~279, 115 days with 2.58$\sigma$ for PKS~1510$-$089, as well as 642 days (3$\sigma$), 52 and 61 days (2.58$\sigma$) for PKS~2155$-$304. \citet{Sandrinelli17} studied optical and MSL \textit{Fermi}-LAT data of three blazars, i.e. BL~Lac, Mrk~421, and PKS~0716+714, using LSP and claiming no QPOs with significance $>$3$\sigma$. \citet{zhang17a} generated 8-years-long \textit{Fermi}-LAT LCs of PKS~2155$-$304, using two methods: the maximum likelihood optimization and the exposure-weighted aperture photometry. The authors searched for QPOs with LSP and WWZ methods, finding a QPO with the period of $\sim$635 days at $\sim$4.9$\sigma$ significance level, thus confirming the QPO period found by \citet{Sandrinelli14, Sandrinelli16}. We also confirm this QPO at 3$\sigma$ significance level, finding a period of $612\pm 42$~days, consistent with a linear decrease of its value over the years.

The 7-years-long LCs of Mrk 421, PKS 1510-089, and B2 1520+31 were generated with unbinned likelihood analysis method by \citet{gupta19}, and then QPOs were searched for applying LSP and WWZ techniques. A quasi-periodic signal was found only for B2 1520+31, with a period of $\sim$71 days with $>3\sigma$ significance. We did not obtain any peculiar structures in the scalograms for this source. \citet{covino19} studied long-term ($\sim$decade) MSL \textit{Fermi}-LAT LCs, including PKS~0716+714 and PKS~2155$-$304, finding no QPOs at a 3$\sigma$ significance level as well.

\citet{Bhat19} generated a 10-years-long \textit{Fermi}-LAT LC, using unbinned likelihood analysis. The LC was analyzed with the LSP and WWZ, showing a QPO with a period of $\sim$330 days, reaching a significance of 2.58$\sigma$. We did not observe any signs of this QPO at a 3$\sigma$ level in the scalograms, though. Likewise, we did not confirm the QPOs in Mrk~421 and PKS~0716+714, reported by \citet{bhat20} on a <4$\sigma$ level.

\citet{Oter20} analyzed the 11-years-long \textit{Fermi}-LAT LC of B2 1633+38, generated with unbinned likelihood analysis, using three different techniques: the Discrete Correlation Function (DCF), LSP, and WWZ. They found a hint of QPOs with periods of 639 and 646 days at 2$\sigma$ level, with WWZ and LSP methods, respectively, as well as a more significant QPO at 3--4$\sigma$ having a period of 581 days with DCF. We obtained a peculiar structure in the scalograms of this blazar, evolving from $P\sim 500$~days to $P>1000$~days over the course of the LC, with 3$\sigma$ significance.

\paragraph{Discrepancies with previous analyses}
\citet{covino19} analyzed blazars' LCs generated by the \textit{Fermi}-LAT team with the aperture photometry method, available on the LAT 3FGL Catalog Aperture Photometry Lightcurves webpage\footnote{\url{https://fermi.gsfc.nasa.gov/ssc/data/access/lat/4yr\_catalog/ap\_lcs.php}}. These LCs constitute a quick-look facility to easily scan for interesting features in the data. At the same time it is strongly discouraged to use this data directly for the detailed scientific analysis, since the LCs can contain the emission from nearby sources. Also the fluxes may be affected by the background emission (point-like and/or diffuse). Similarly, \citet{Naka13,Sandrinelli14,Sandrinelli16,Sandrinelli17,castignani17,ryan19} studied the LCs accessible on the Monitored Source List Light Curves webpage\footnote{\url{https://fermi.gsfc.nasa.gov/ssc/data/access/lat/msl_lc/}} provided by the \textit{Fermi} Science Support Center. Here again, the LCs are generated in an automatic way, using preliminary calibration and instrument response functions, to help searching for flares in the data and plan the follow-up analysis in other bands. It is not recommended to use this data in a scientific analysis as well. 

Various authors adopted different significance levels for asserting the significance of QPO detections, ranging from $\lesssim 2\sigma$, through arbitrary values like 95\% or 99\%, rarely exceeding $3\sigma$ (99.73\%). E.g., \citet{Bhat19} was the first to try to look for possible QPOs in the \textit{Fermi}-LAT data of Mrk~501, finding a QPO with a period of $\sim$330 days. This is a very interesting result which, however, needs to be further verified, since this possible QPO was reported only on a $2.58\sigma$ significance level. Moreover, \citet{gupta19} claimed a QPO detection in B2~1520+31 merely on a $1.96\sigma$ level. The correct methodology is to set a goal, e.g. for finding a signal on a given significance level (though one that warrants a detection, i.e. at least $>3\sigma$), and not report just any detected feature (especially with low significance). Moreover, sometimes a false alarm probability is reported \citep{Sandrinelli14} or used to set the significance \citep{gupta19}. This, however, tests a peak in the periodogram against a white noise model \citep{vanderplas18}, while blazars exhibit PL-like PSDs, hence peaks exceeding the $3\sigma$ level against white noise will usually not be significant when tested against colored noise. Another issue is an abuse of the wavelet scalogram: while this is a powerful tool sensitive to transient phenomena in a time series, and therefore ideal to search for QPOs in blazar LCs, it is often used---with asserting the significance---as a global periodogram, like an LSP with which it is often highly consistent \citep{zhang17a,zhang17b}. A global periodogram is an average of the power at a given frequency among the whole LC. As was demonstrated in Sect.~\ref{results::wavepal}, a scalogram can reveal that an apparently statistically significant peak in a global periodogram can signal a spurious quasiperiodicity, which is not supported when looking how the power among different frequencies is distributed in time. Unfortunately, in many works the scalogram is merely displayed, without performing any significance estimation \citep{zhang17a,zhang17b,Bhat19,gupta19,Oter20} or even displaying the COI. We performed here such an estimation that allowed to conclude that any qusiperiodicity in our sample that might be claimed based on the global periodogram is spurious. While in some instances the scalogram is very convincing, e.g. in case of PG~1553+113 \citep{Acke15}, still a significance analysis of its local features, not only global, is required. Finally, we stress again that a QPO, when present in the data, is not necessarily a signature of any periodic phenomena occuring in the system. Hence attributing a significant, constant period obtained from a global periodogram, with a true periodicity, is in our opinion unjustified. Indeed, the peaks observed in an LC can be spaced very nonuniformly, therefore a visual inspection of the LC for the presence of equidistant peaks can be misleading.

\paragraph{ARMA}
The only attempt, to the best of our knowledge, to analyze $\gamma$-ray {\it Fermi}-LAT data with an ARMA$(p,q)$ model is the case of a BL Lac type object PKS~0301$-$243, located at $z=0.266$, using a 30~d binned LC \citep{zhang17b}. Fitting returned $(3,2)$ as the order, meaning that three previous observations affect any given flux value in the LC. This implies a characteristic time scale of $\tau_{\rm rest} = 71$ days, which is substantially greater compared to our results. This is intriguing, since it suggests that such a time scale might differ among different blazars. Adding this source to B2~1633+38, PKS~0716+714, PKS~2155$-$304, and B2~1520+31, produces a lack of correlation between $\log M_{\rm BH}$ and $\tau_{\rm rest}$ ($M_{\rm BH} = 8\times 10^8\,M_{\odot}$ of PKS~0301$-$243 taken from \citealt{ghisellini10}). When $\tau_{\rm rest}$ is calculated for all 11 of our blazars, with the best-fit orders for the 7~d binning, no correlation with $M_{\rm BH}$ can be inferred either. It appears that the BH mass is not a crucial parameter in influencing the values of $\tau_{\rm rest}$. Further research on a bigger sample is therefore necessary for assessing the meaningfulness of such time scale regarding its physical origin.

It is worth to mention that the 22 and 37 GHz LCs of 3C~279 and 3C~454.3 were fitted with ARIMA$(p,d,q)$ models \citep{bewketu19}, resulting in $d=1$ (i.e., once differenced to ensure stationarity), with $(p,q)=(1,1)$ for 3C~279, and $(5,1)$, $(1,2)$ for 3C~454.3 at 22 and 37 GHz, respectively. The fits are slightly inadequate, hence suggest that long-term memory processes (FARIMA/CARFIMA) might be a better model. The correlation between radio and $\gamma$-ray bands \citep[][and references therein]{angelakis19} implies a common emission site, hence a multiwavelength approach to $\tau_{\rm rest}$ might provide additional constraints on the radiation mechanism.

\paragraph{CARMA}

\citet{ryan19} analyzed $\gamma$-ray LCs of eight blazars common to our sample in 1 and 7 day binning, with the CARMA models. For two sources they obtained the same $(1,0)$ orders as we did (with the 7 d binning), and constrained the break time scales $T_{\rm break}$ of their Lorentzian PSDs. Those are Mrk~421 with $T_{\rm break} = 389^{+502}_{-107}\,{\rm days}$, and PKS~1510$-$089 with $T_{\rm break} = 186^{+89}_{-31}\,{\rm days}$. We obtain $(389\pm107)\,{\rm days}$ and $(209\pm 24)\,{\rm days}$, respectively, which is in agreement within the errors.

For B2~1520+31, in turn, both \citet{ryan19} and we arrived at CARMA$(2,1)$ as the best fit. The break time scales, corresponding to the widths of the zero-centered Lorentzians that the PSD is composed of, are $>1230\,{\rm days}$ and $68^{+18}_{-27}\,{\rm days}$ \citep{ryan19}, which are in agreement with our results, $(3075\pm 961)\,{\rm days}$ and $(102\pm 33)\,{\rm days}$.

We also calculated $T_{\rm break}$ of CARMA$(1,0)$ models of all objects in our sample, and compared them with values given by \citet{ryan19}. Both sets are consistent within errors. 

\paragraph{Hurst exponents}
The Hurst exponents of blazar LCs have been computed for optical data of blazar candidates behind Magellanic Clouds \citep{zywucka20}, with a much more irregular sampling than the $\gamma$-ray observations utilized herein. It was found that in most instances $H<0.5$, contrary to the $\gamma$-ray case. While this might suggest that optical and $\gamma$-ray emission are governed by different processes, or with different parameters, the computation of Hurst exponents is a delicate matter, and especially with short-term data needs to be done with caution \citep{katsev}. The most straightforward approach is to analyze time series with a greater number of points, which can be achieved in case of {\it Fermi}-LAT data by using a smaller binning for the LCs.

\paragraph{The $\mathcal{A}-\mathcal{T}$ plane}
Likewise, the separation of FSRQ and BL Lac $\gamma$-ray LCs in the $\mathcal{A}-\mathcal{T}$ plane is a novel finding, however it is supported by a separation observed using optical LCs of blazar candidates behind Magellanic Clouds \citep{zywucka20}. Therefore, both optical and $\gamma$-ray observations, with very different sampling, suggest that there are intrinsically different variability patterns between different types of blazars.

\subsection{Interpretation \& context}
\label{interpretation}

There has not been a unified procedure in the literature for detecting QPOs in astronomical data: various methods have been used to produce the PSDs, and no consistent significance level has been adopted (see Table~\ref{BlazarSummary} and Sect.~\ref{comparison}). We performed a homogeneous investigation (similar in spirit to \citealt{covino19}) of wavelet scalograms, confirmed the QPO in PKS~2155$-$304, with a period of $\sim$610~days in the 7~d binned LC, but detected only a dubious quasiperiodicity just above the $3\sigma$ level in the global wavelet periodogram of PKS~1830$-$211 (only in the 7~d binned LC, though). We stress the importance of the subtle meaning of the term {\it quasi} while interpreting any such detection: a QPO need not be inherently linked to any periodic behavior of the system. In fact, it is a well-known fact that red-noise type processes often resemble QPOs for a few cycles \citep{vaughan16,ait20}, and that detections reported so far were only on QPOs lasting 4--6 cycles. X-ray data, despite a much finer sampling of LCs, usually span relatively short periods (compared to optical or $\gamma$-ray LCs), and the number of cycles reported is also not high \citep[e.g. ][]{lachowicz09}. Hence, a QPO, usually understood as a transient phenomenon with a Lorentzian feature in the PSD, is very likely to be a manifestation of correlations governing the variability, inherent e.g. in ARMA or CARMA models. Such correlations may be the result of stochastically occuring disruptions in the relativistic jet itself or the accretion disk coupled to the jet. QPO signatures might come from jet wobbling, its precession and nutation, peculiar plasma injections into the jet, or interaction of the jet with ambient medium \citep{caproni04,bosch12,britzen18,lico20}. One cannot also exclude the possibility that nonlinear physical processes \citep[e.g. intermittency; ][]{garcia17a,garcia17b} lead naturally to coupling that results in such features, or that a combination of all above effects occurs, with various relative strength in different AGNs. Finally, a binary SMBH system is often invoked as a driving mechanism for the QPOs, but such systems must be too rare to be a universal explanation \citep{holgado18}.

The generating mechanism for QPOs has thus not been unambiguously identified. It is therefore not clear whether QPOs are an inherent property of $\gamma$-ray emission of blazars, and that it is just the difficulty of its detection or insufficient data that make them so rare and elusive. On the other hand, also the general shape of the PSD has not been convincingly derived so far. \citet{finke14,finke15} considered synchrotron-self Compton (SSC) and external Compton (EC) as models for the emission, including Klein-Nishina effects. They modeled the evolution of nonthermal plasma with an assumption of PL injection rate to the jet that lead to a PL form of the PSD as well. In particular, a broken PL with low- and high-frequency indices $\beta$ and $\beta+2$ was obtained, and the break frequency might be associated with the electron escape time scale. Recall that we do not observe a break in neither Fourier spectra nor LSPs for any object. On the other hand, we obtained some clear breaks by fitting ARMA and CARMA models, reaching consistent values of a few hundred days with both methods. It is curious to note that similar time scales ($100-400\,{\rm days}$) were obtained with the LSP for a sample of optical LCs of FSRQ candidates behind the Magellanic Clouds \citep{zywucka20}. A rotating helical structure can be formed in the jet due to it being tied to the rotaing BH or accretion disk \citep{raiteri13}. The evolving viewing angle may alter the observed flux, even though the total emitted flux remains constant. Different parts of the jet can attain different orientations in time, thus leading to a complex flux variability. These effects might be the origin of the PL-like shape of blazar PSDs.

\citet{kaulakys06} derived a stochastic differential equation whose solutions yield a PL PSD. Unfortunately, this cannot constitute a model for blazar variability, as these solutions also exhibit a PL distribution of the signal intensity, while the PDFs of observed fluxes are lognormal (Sect.~\ref{sect::rms}). CARMA models, on the other hand, being Gaussian processes, are good candidates for modeling the logarithmic flux variations, and their PSDs are flexible enough to incorporate nearly pure PL, broken PL, and QPOs. The defining Eq.~(\ref{eq36}), however, needs to be directly connected to physical properties of the system. Up to now, only the simplest cases were given a heuristic interpretation in terms of e.g. some characteristic time scales in the jet and disk, or related to BH mass \citep{moreno19,ryan19}. CARMA processes with low orders, $(1,0)$ and $(2,1)$, are in turn easily associated with simple physical models, i.e. the OU process and damped harmonic oscillator \citep{kasliwal17}. This is, however, an inverse-engineering approach, as it is desired to derive the equations from first principles, not to associate observed features with some parameters of phenomenological models. Identification of physical processes occuring in the emission region \citep[SSC, EC, etc.; ][]{finke14,finke15,chen16} with a justified injection rate is one possible direction to pursue. In particular, the particle-in-cells (PIC) simulations \citep{sironi13} suggest that the injection rate's dependence on the Lorentz factor is not necessarily a PL. The PIC simulations are based on microturbulence occuring in the jet shocks. In turn, \citet{narayan12a} considered a jets-in-a-jet model, in which turbulence in the jet fluid gives rise to relativistic random motion in different directions in different subregions of the jet. The sum of flux contributions from each subregion produces the observed variability. The distribution of the Doppler beaming factors adds to the observed PSD. This model works well for PKS~2155$-$304 under sufficiently strong magnetization of the medium, although it again shifts the question: what causes the turbulence, and what are its properties? 

As the presence of a QPO in CARMA (and also ARMA) processes of a given order $(p,q)$ depends on the AR and MA coefficients, it may be possible to identify the broader class of physical processes that generate the variations in the LCs, i.e. to describe the variability by means of some (stochastic) differential equation, in which the coefficients depend on physical parameters of the system. If so, one might recognize which of the physical parameters (e.g., magnetic fields, density of the ambient medium, viscosity of the disk, etc.) govern the steepness of the PSD, whether it contains a break or a QPO, and their corresponding time scales. In particular, the fact that in some objects we observe lower orders $p$ of fitted ARMA$(p,q)$ models for LCs with a higher binning, hints at a characteristic time scale above which the observations become less correlated. Indeed, our ARMA fits suggest that the state of the $\gamma$-emitting region affects future measurements for a few weeks. \citet{zhang17b} considered a 9-years-long LC (30~d binning) of the BL Lac object PKS~0301$-$243. They fitted an ARMA$(3,2)$ process, a relatively high order, implying a characteristic rest-frame time scale of $\sim 71$ days. On the other hand, the relation between order $p$ and binning is not present when fitting CARMA models---in fact, for B2~1520+31 we even obtained an opposite dependence. Overall, all of the CARMA fits are low-order, and higher orders appear as a result of overfitting to account for variations below the Poisson noise level. Therefore, whether such timescales, inferred directly from temporal data, are of any meaning is yet to be determined.

In case of Fourier spectra and LSPs, the PL and PLC fits were performed to binned PSDs, hence there are relatively few points for the fitting algorithm. In PL and PLC, there is by definition no break. As mentioned in Sect.~\ref{results::fourier}, an attempt to model the binned PSDs with a smoothly broken PL was ruled out in all cases, likely owing to the small number of points. On the other hand, ARMA and CARMA models are fitted to the points in the LC, so there are much more points available; second, the PSDs of these processes by definition have a break since they contain a Lorentzian around $f\sim 0$. The ARMA and CARMA fits are therefore more sensitive to detecting a break---i.e., the break will always be present, but (i) it may be located at very low frequencies, related to the length of the time series, so not physical (and practically not detectable), or (ii) be not well constrained in the sense of having a large uncertainty, including within the errors frequencies corresponding to the length of the time series. This is most prominent in case of B2~1520+31 (Table~\ref{table5}) or Mrk~501 (Table~\ref{table7}).

Overall, we note that the utilized methods to model the PSDs (PL fits to Fourier spectra and LSP; ARMA and CARMA modeling) returned outcomes quite similar in shape among the examined blazars. One can see in Tables~\ref{table1} and \ref{table3} that sources with a significant contamination of Poisson noise (Mrk~501 and TXS~0506+056, which are the dimmest among the examined herein) yield steeper PSDs, although $\beta$ falls roughly in the range $\sim 1-2$ for our sample. CARMA models and Hurst exponents are all consistent with each other among our objects. This leads to an observation that the population of blazars is an ensemble, i.e. suggests to treat each blazar as a different realization of {\it the same} underlying process, taking into account the unequivocal differentiation into two classes, FSRQs and BL Lacs, visible clearly also in the $\mathcal{A}-\mathcal{T}$ plane. Hence, a population synthesis study of blazars will likely shed light on the nature of the variability from a global perspective.

Moreover, $H>0.5$, obtained in most instances, suggests to account for the presence of persistence in modeling the variability. In discrete models, FARIMA is a generalization of ARMA that, through a fractional differencing parameter $d$, allows for $H\neq 0.5$. Likewise, a CARFIMA process \citep{tsai05,tsai09,feigelson18} is a candidate description for the blazar variability, whose advantages are both its inherent continuity and long-term memory. Here, too, the reservations about interpretation of CARMA models hold, i.e. it is desired to connect the AR and MA coefficients, as well as $d$, to some physically meaningful properties of the blazar. We generally see in Fig.~\ref{fig_Hurst_results} that $H\gtrsim 0.8$ (sometimes spilling to $H\lesssim 0.2$ due to the discontinuity at the fBm-fGn border; see Fig.~\ref{fig_Hurst_disc}), except for 3C~454.3, which prefers $H\lesssim 0.5$, consistent with PL indices from Tables~\ref{table1} and \ref{table3}. This, on one hand, suggests that the processes governing blazar emission are inherently described by high values of $H$, but on the other hand it requires an explanation for the unusual, compared to other objects in our sample, behavior of 3C~454.3

Finally, the sharp separation of FSRQs and BL Lacs in the $\mathcal{A}-\mathcal{T}$ plane suggests that either the processes governing the variability are different between the two classes, or are qualitatively different realizations of the same physical scenario. We stress that this separation is not trivially connected with FSRQs having steeper PSDs, as the $\beta\lessgtr 1$ division in the $\mathcal{A}-\mathcal{T}$ plane does not divide the classes at all. That the separation is meaningful is supported by a recent finding that FSRQ and BL Lac blazar candidates observed behind Magellanic Clouds also occupy distinct regions of the $\mathcal{A}-\mathcal{T}$ plane \citep{zywucka20}. This finding can therefore serve as an additional tool in classifying blazars, in particular when spectroscopy is not feasible.

\subsection{Possible applications}
\label{applications}

The methodology employed herein to study blazar LCs consists of various time series analysis techniques, and hence is applicable to any temporally (or spatially) ordered data. Below we list some of the potentially interesting applications.
\begin{enumerate}

\item Multiwavelength analysis of blazar LCs. We examined herein the $\gamma$-ray data from {\it Fermi}-LAT, but some of the methods (LSPs, Hurst exponents, the $\mathcal{A}-\mathcal{T}$ plane) were already applied to optical LCs of blazar candidates behind the Magellanic Clouds \citep{zywucka20}, observed by OGLE. Also the densely sampled optical LCs \citep[e.g., ][]{smith18} or X-day data \citep[e.g., ][]{gaur18} are a promising target for such investigations.  

\item Likewise other AGN types, e.g. Seyfert galaxies \citep{gallo18}, radio galaxies \citep{Mars09}, or quasars, including the search for QPOs \citep{li10} in various energy bands. Moreover, the characteristic features as well as differences between jetted and non-jetted AGNs can be studied \citep[e.g., ][]{Pado17,Pado17a}.

\item X-ray binaries exhibit broad-band PSDs \citep{bayless11}, whose shape can be constrained with several techniques, complementing each other, including modeling the LCs with Gaussian stochastic processes. Moreover, chaotic modulation is also a considered possibility \citep{sukova16,mannattil16,phillipson18} that further validates the use of nonstandard, in particular nonlinear, tools of time series analysis.

\item Pulsar spin-down rates can be erratic, or even chaotic \citep{seymour}. One might expect that persistence analysis via Hurst exponents will lead to insights on the properties of the processes governing the spin-down of the periodic pulsation. 

\item Gamma-ray bursts (GRBs). The Hurst exponents were already shown to differ for short and long GRBs \citep{maclachlan13,tarnopolski15c}. The $\mathcal{A}-\mathcal{T}$ plane, operating exclusively in the time domain, might be an insightful addition to other methods, e.g. the minimum variability time scale \citep{golkhou14}, in classifying the prompt LCs. Therefore, it will possibly allow a meaningful inference of the number of classes, i.e. rejecting or supporting the existence of the elusive GRB group with intermediate durations (see \citealt{tarnopolski19a} for an overview of this issue). The PSDs of individual GRBs were also already analyzed with PL models, and it was found that they exhibit high PL indices, $1\lesssim \beta\lesssim 6$, which are anticorrelated with peak energy \citep{dichiara16,guidorzi16}. QPOs in prompt LCs remain undetected \citep{dichiara13}.

\item Variable stars, e.g. of the T~Tauri type which are young, pre-main sequence stars that exhibit significant variability in all wavelength bands, but also spectral and polarization changes, and are surrounded by protoplanetary accretion disks \citep{appenzeller89}. Their long-term, densely sampled LCs \citep{rigon17} could be modeled with CARMA processes. Their different sub-types may be classified in the $\mathcal{A}-\mathcal{T}$ plane as well as with the Hurst exponents. Typical quasiperiodic variations on the order of days make them interesting targets for the time series analysis techniques employed herein, especially for searching QPOs with wavelet scalograms. Likewise in case of, e.g., sdB stars exhibiting stochastic pulsations \citep{ostensen14}.

\item The $\mathcal{A}-\mathcal{T}$ plane of densely sampled optical spectroscopic observations of binary and multiple stellar systems  \cite[e.g.][]{Dimi18}, covering the entire cycle phases, could help to distinguish these systems among single stars.

\item The X-ray flickering of cataclysmic variables yields $H>0.5$ \citep{anzolin10}. However, the employed $R/S$ method was demonstrated herein to be strongly biased, therefore a reanalysis of the XMM-Newton LCs with a variety of $H$ estimation techniques, in order to establish consistency, is desired.

\item Extrasolar planets manifest their existence in optical data by, e.g., small changes in LCs of the stellar-planetary systems that can be detected by transient and/or gravitational microlensing methods \citep[e.g.][]{Seag03,Bond04}. A planet search algorithm utilizing ARIMA models has been already proven to be effective \citep{caceres19}. Moreover, the locations in the $\mathcal{A-T}$ plane, estimated based on properly sampled datasets, could assist in detections of new extrasolar planets.

\item Sunspots have been investigated with ARMA models, in particular to formulate predictions on the solar cycle \citep{brajsa09}, although recently a neural network approach is utilized more often \citep{liu19sun}. Forecasts based on estimated Hurst exponents were also performed \citep{pesnell12}. In regard to solar physics, solar wind proton density fluctuations are characterized by $H \sim 0.8$, placing constraints on the models of kinetic turbulence \citep{carbone18}. Moreover, the solar radio emission\footnote{\url{http://www.oa.uj.edu.pl/slonce/index.html}} variability is perfect for such analyses, and might provide useful insights into the solar dynamics.

\item The background emission, e.g. in X-rays in case of XMM-Newton, consists of several components, i.e. electronic readout noise, high energy particles, particle induced X-rays generated inside the camera, and thermal CCD noise \citep{nevalainen05}. Estimation of the statistical properties of the blank-sky--background is crucial in properly extracting the signal of interest.

\item While magnetically arrested disks \citep{narayan03,tchekhovskoy15} and those with standard and normal evolution accretion flow \citep{narayan12b} have similar spectral density distributions \citep{xie19}, it is curious whether these two states can be distinguished in the time domain, e.g. by analyzing numerically computed accretion rates or the magnetic flux with the presented time series analysis tools.

\item The recently released gravitational wave data \citep{ligo}, in the form of densely sampled strain time series, is a tempting target for various analyses, like the ones performed herein.

\item Time-dependent magnetohydrodynamic numerical simulations of stellar interior \citep{merkin16}, resulting in time series of, e.g., magnetic field or plasma density, are tempting applications for a variety of methods of time series analysis. Spatial evolution of relativistic jet properties are also subject to such analyses, as time series need not be ordered in time only. The subtle variations along the jet might shed light on GRB or AGN physics as well \citep{huarte11,sironi13}. 

\item Observations in the extremely low frequency (ELF) regime of electromagnetic waves. Long-term observations of natural electromagnetic fields (EFs) in the ELF range ($0.03-300$~Hz) are conducted by several ELF stations all over the world\footnote{In particular, by the World ELF Radiolocation Array (WERA) project (\url{http://oa.uj.edu.pl/elf}), which consists of three ELF stations located in Poland, Argentina and USA \citep{Kulak14,Mlynarczyk17} that measure continuously with a sampling frequency of $\sim 1$~kHz.}. The PSD analysis of ELF data reveals several resonances in the Earth-ionosphere cavity: the Schumann resonances, which are the global electromagnetic resonances in Earth-ionosphere waveguide, generated and excited by lightning discharges over the world \citep{Kulak06,Nieckarz09},  and ionospheric Alfven resonances, caused by magnetic waves trapped between different layers of the ionosphere \citep{Odzimek06}. The measured ELF variations of the magnetic field contain colored stochastic noise, which can be modeled by ARMA/CARMA processes as well. 

\item It is well known that many problems in celestial mechanics, like the three-body problem or the stability of the Solar System, are chaotic in its nature. Centaurs---a transient population of small bodies in the outskirts of the Solar System---have chaotic orbits whose evolution resembles different types of stochastic processes. These types can be distinguished using the Hurst exponents \citep{brenae09}. Whether techniques stemming from stochastic analysis can provide even further insight into the dynamics of the Solar and extraterrestrial systems is an interesting open problem.

\end{enumerate}

\section{Conclusions}
\label{conclusions}

\begin{enumerate}
\item From extensive MC testing we found that parametric modeling gives unreliable results. In particular, retrieval of the PL index of the PSD, $\beta$, yields a wide range of outcomes for different realizations of the same underlying stochastic process, and similarly the dispersion of the parameters of autoregressive models is high. Likewise, the Hurst exponent values returned by different algorithms can in principle be inconsistent. We therefore recommend to employ simultaneously various algorithms and examine the consistency of the obtained estimates, as well as to perform MC simulations, e.g. bootstrapping the LC within measurements' uncertainties or resorting to Bayesian methods.

\item Due to very consistent PSD properties of blazars in our sample, we suggest that each object can be treated as one realization of a common stochastic process underlying the observed variability. Therefore, the population of blazars (although possibly differentiated into FSRQs and BL Lacs) should be treated as an ensemble, allowing for a more global insight into the governing mechanisms driving the variablity.

\item Orders of the ARMA models, fitted for different LC bins, hint at some characteristic dissipation time scale, after which the disturbances in the emitting region stop affecting the emission significantly.

\item ARMA and CARMA models imply the existence of a break in the PSDs at time scales $\sim$ hundreds of days.

\item In a majority of cases, the Hurst exponents are greater than 0.5, indicating the presence of long-term memory in the systems, except for 3C~454.3 which yields $H<0.5$. Stochastic processes that allow persistence are therefore promising candidates for modeling blazar LCs.

\item We confirm the presence of a QPO in PKS~2155$-$304, with a period of $612\pm 42$~days. We detect a peculiar structure in the scalogram of B2~1633+38, at periods $\sim 600-1000$~days, and a candidate QPO in PKS~0716+714 at a period $>1000$~days. However, we do not detect any QPOs in the remaining sources; in particular we do not confirm the QPOs in Mrk~501, Mrk~421, PKS~1510$-$089, 3C~279, B2~1520+31.

\item We confirmed a proposition, initially put forward based on optical data of blazar candidates \citep{zywucka20}, that FSRQs and BL Lacs are separated in the $\mathcal{A}-\mathcal{T}$ plane, i.e. can be distinguished based on the LCs only.
\end{enumerate}

\acknowledgments
M.T. acknowledges support by the Polish National Science Center (NSC) through the OPUS grant No. 2017/25/B/ST9/01208. The work of N.\.{Z}. is supported by the South African Research Chairs Initiative (grant no. 64789) of the Department of Science and Innovation and the National Research Foundation\footnote{Any opinion, finding and conclusion or recommendation expressed in this material is that of the authors and the NRF does not accept any liability in this regard.} of South Africa. V.M. is supported by the NSC grant No. 2016/22/E/ST9/00061. J.P.-G. acknowledges financial support from the State Agency for Research of the Spanish MCIU through the ''Center of Excellence Severo Ochoa'' award to the Instituto de Astrof\'isica de Andaluc\'ia (SEV-2017-0709) and from Spanish public funds for research under project ESP2017-87676-C5-5-R.

\software{\textsc{Mathematica} \citep{Mathematica}, \textsc{R} \citep[][\url{http://www.R-project.org}]{RTeam}, \textsc{carma\_pack} \citep[][\url{https://github.com/bckelly80/carma_pack}]{kelly14}, \textsc{wavepal} \citep[][\url{https://github.com/guillaumelenoir/WAVEPAL}]{lenoir18a,lenoir18b}, \textsc{Matlab} \citep{Matlab}}.

\appendix
\section{Maximum likelihood estimation of ARMA and FARIMA}
\label{appendixA}

\begin{figure*}
\centering
\includegraphics[width=0.49\textwidth]{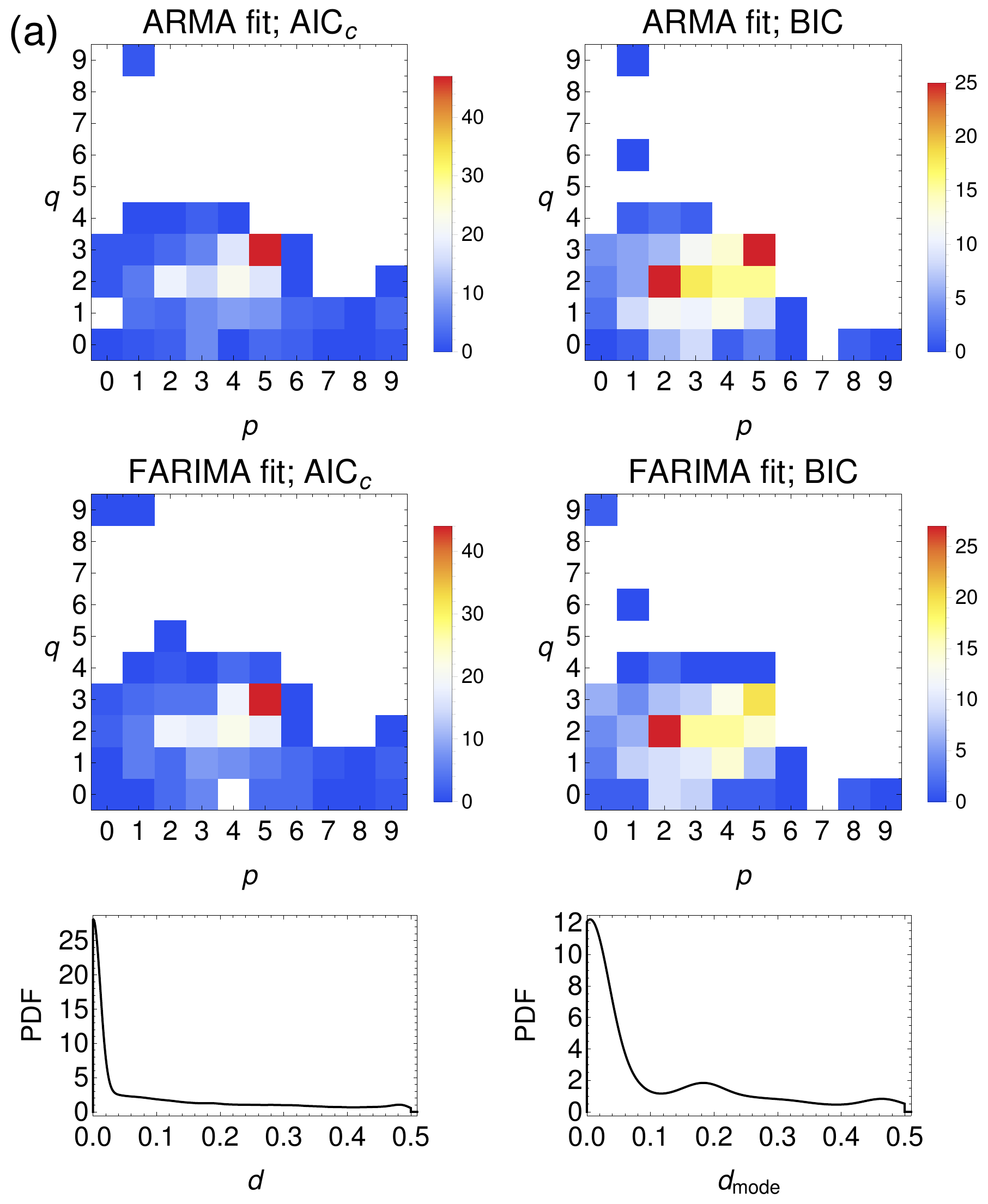}
\includegraphics[width=0.49\textwidth]{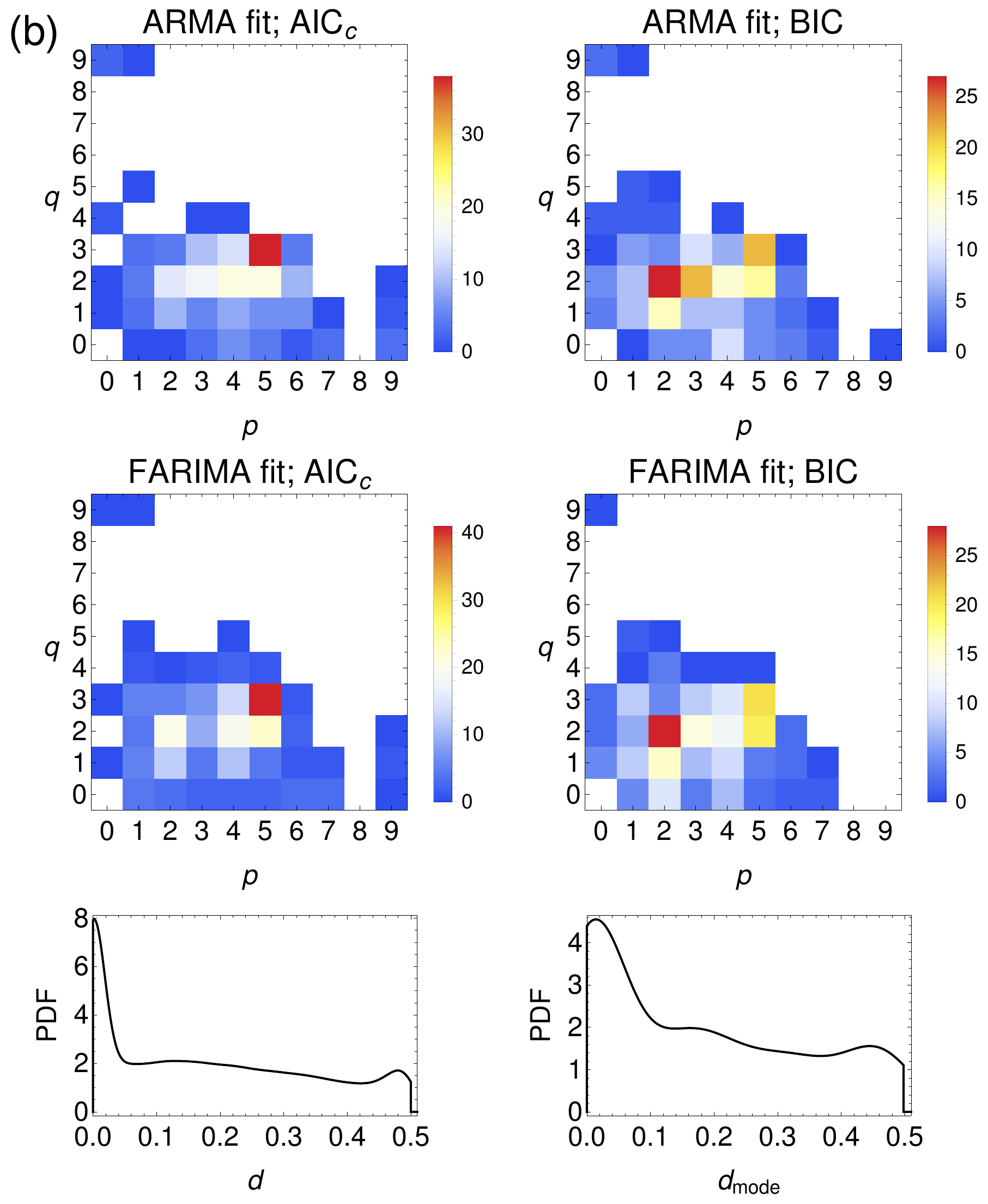}\\
\includegraphics[width=0.49\textwidth]{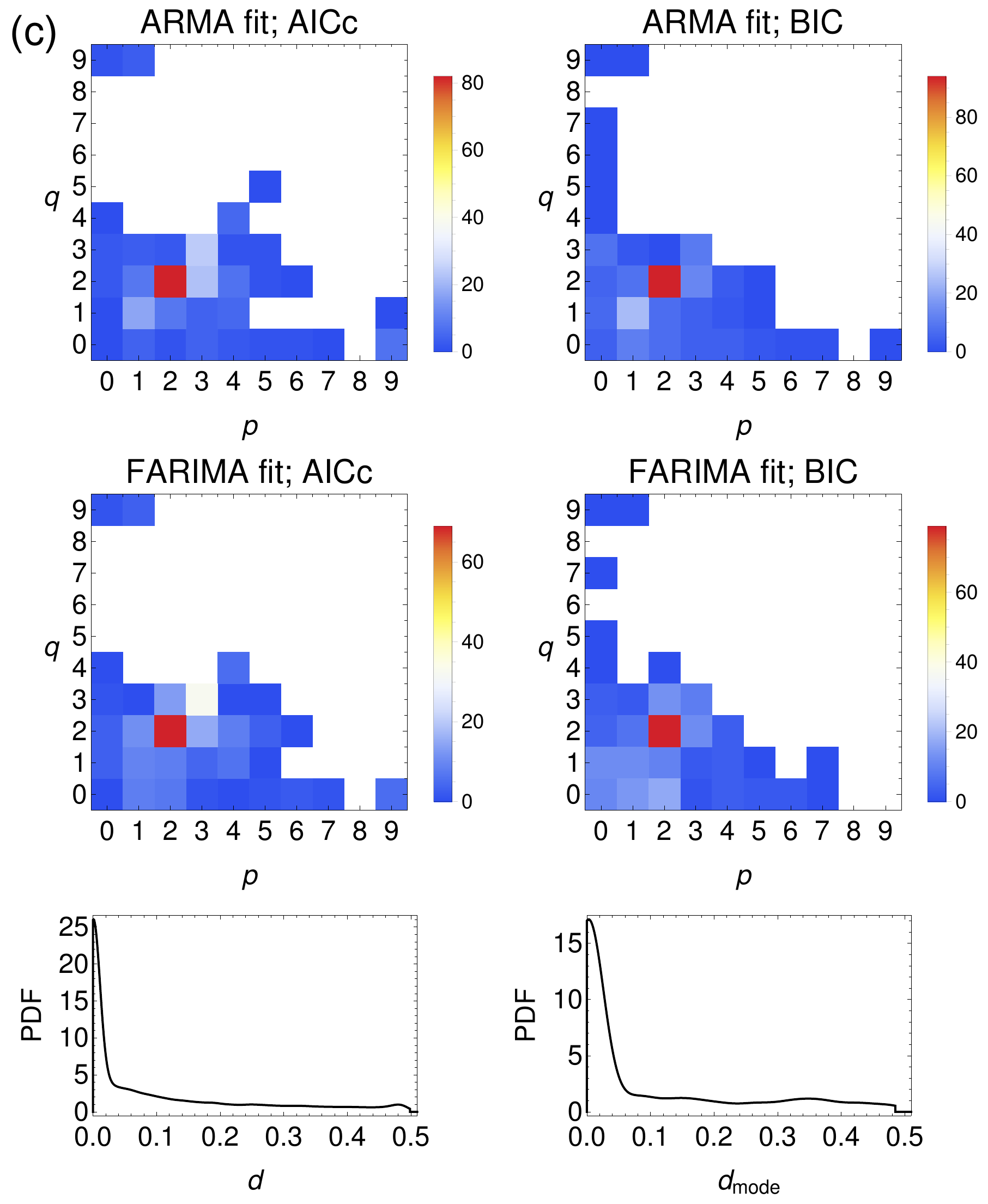}
\includegraphics[width=0.49\textwidth]{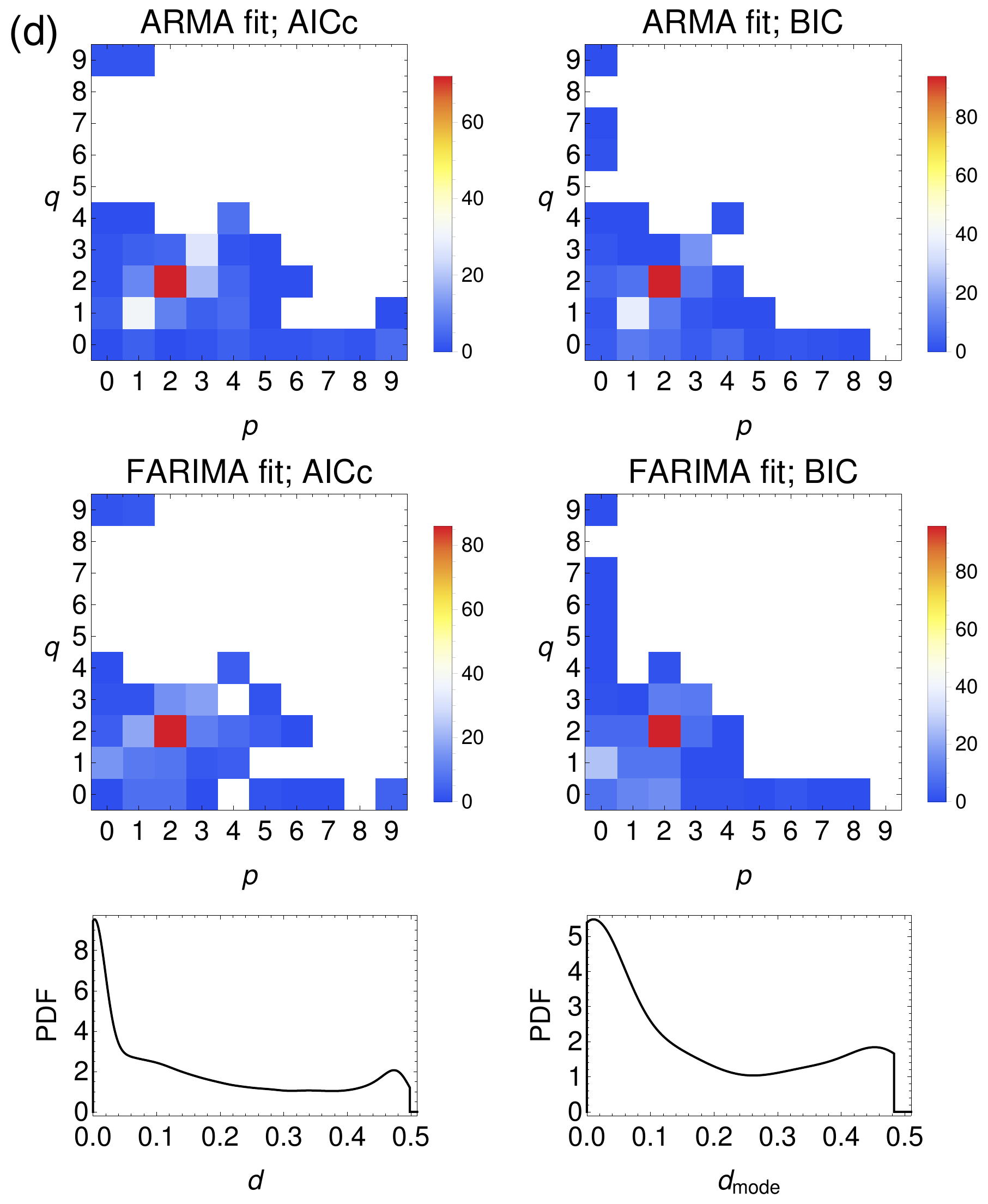}
\caption{Same as Fig.~\ref{clean_noise}, but fitted with \textsc{R}.}
\label{clean_noise_app}
\end{figure*}

The benchmark testing for fitting ARMA and FARIMA processes from Sect.~\ref{sect::arma_bench} was repeated in \textsc{R} in order to utilize a different computer system, implementation, algorithm, and method. Time series were simulated with \texttt{arima.sim} and \texttt{fracdiff.sim}. The commands \texttt{auto.arima} and \texttt{arfima}, with the maximum likelihood method (\texttt{method = "ML"}), were used for the fitting. The results are displayed in Fig.~\ref{clean_noise_app}.

\section{Implementations of the codes}
\label{AppendixB}

\begin{itemize}
\item Fourier spectra and LSPs: they were computed in \textsc{Mathematica} by directly implementing the methodology from Sect.~\ref{methods::fourier} (employing the command \texttt{Fourier}\footnote{https://reference.wolfram.com/language/ref/Fourier.html} with default \texttt{FourierParameters}) and \ref{methods::ls} [Eq.~(\ref{eq12}) and (\ref{eq13})]. To bin the raw PSDs, the function \texttt{BinListsBy}\footnote{\url{https://community.wolfram.com/groups/-/m/t/1081009}} was utilized. Fits to the binned PSDs (in log-log scale) were done with \texttt{NonlinearModelFit}\footnote{\url{https://reference.wolfram.com/language/ref/NonlinearModelFit.html}}.

\item \textsc{wavepal}: the scalograms were computed using the function 
{\tt Wavepal.plot\_scalogram}, with the 99.73\% percentile for the $3\sigma$ confidence level. Before calculating the scalograms, the LCs were detrended by the polynomial trend of degree $n=7$ using function {\tt Wavepal.choose\_trend\_degree}.

\item ARMA: the best-fit was obtained using \textsc{Mathematica} with\footnote{\url{https://reference.wolfram.com/language/ref/TimeSeriesModelFit.html}} \texttt{TimeSeriesModelFit[data, \{"ARMA", \{p, q\}\}]} by iterating over \texttt{p} and \texttt{q}, and choosing according to the $BIC$. The confidence intervals of the resulting PSD were inferred by bootstrapping with the best-fit parameters' standard errors. So were the uncertainties of $f_0$ and the breaks obtained.

\item CARMA: the maximum likelihood estimates for $AIC_c$ were computed using the function {\tt CarmaModel.choose\_order}, and then the values of $AIC_c$ were recalculated to $BIC$ for further estimation of $(p,q)$ order. The MCMC sampler was run using {\tt CarmaModel.run\_mcmc} for $150\,000$ iterations, with the first $50\,000$ of those discarded as burn-in. To check the chain convergence we used the multiple-chain convergence diagnostics \citep{Gelamn92} implemented in the \textsc{PyMC} python module\footnote{\url{https://pymc-devs.github.io/pymc/modelchecking.html}} by the function {\tt pymc.gelman\_rubin}.

\item Hurst exponents and $\mathcal{A}-\mathcal{T}$ plane: the \textsc{Mathematica} implementations are available at \url{https://github.com/mariusz-tarnopolski/Hurst-exponent-and-A-T-plane}.
\end{itemize}

\bibliography{bibliography}{}

\end{document}